\documentclass[aps, prb, reprint, superscriptaddress, amsmath, amssymb,
  longbibliography, floatfix]{revtex4-2}

\usepackage{graphicx}
\usepackage{dcolumn}
\usepackage{bm}
\usepackage{mathtools}
\usepackage{xcolor}
\usepackage{placeins}
\usepackage[colorlinks=true,linkcolor=blue,citecolor=blue,urlcolor=blue]{hyperref}
\allowdisplaybreaks[1]
\begin{document}

\title{Second-order dc conductivity in the velocity-gauge Keldysh formalism: gauge-invariant decomposition into nonlinear Drude, Berry-curvature-dipole, and quantum-metric responses}
\author{Junya Shibata}
\email{j\_shibata@toyo.jp}
\affiliation{Department of Electrical, Electronic and Communications Engineering,
Toyo University, Kawagoe, Saitama, 350-8585, Japan}

\date{\today}
\begin{abstract}
We derive a gauge-invariant clean-limit decomposition of the
second-order dc conductivity in multiband tight-binding systems
using the velocity-gauge Keldysh Green's function formalism.
Within the constant-relaxation-time approximation, the response
contains the leading nonlinear Drude term proportional to
$\tau^{2}$, the Berry-curvature-dipole term proportional to
$\tau$, and intraband and interband quantum-metric-dipole
contributions at order $\tau^{0}$.
The nonlinear Drude sector also contains a subleading,
purely dispersive kinetic contribution at order $\tau^{0}$,
$\sigma^{\mathrm{ND},\tau^{0}}_{ijk}
=-(e^{3}/24\hbar)\sum_{\bm{k},n}f_n^{\prime\prime}
\partial_i\partial_j\partial_k\varepsilon_n$. 
This term, already identified in other finite-broadening
treatments, completes the present lifetime hierarchy through
$\mathcal{O}(\tau^{0})$ within the stated model.
The intraband geometric term is a Fermi-surface dipole of
the ordinary band quantum metric; the interband term is
expressed using a metric weighted by inverse band-energy
differences.
Before integration by parts, the latter is written
as a Fermi-sea-type response in the present representation.
Working directly in the band basis without projectors, we
explicitly track the Peierls contact vertices and
demonstrate the term-by-term cancellation of connection-dependent
commutator structures between the covariant-quantum-connection
and three-Berry-connection sectors.
This cancellation was also demonstrated in the projector-based
formulation of Ulrich \textit{et al.}
[Phys.\ Rev.\ B \textbf{113}, L201107 (2026)].
After accounting for index and convention differences,
the Fermi-surface coefficients agree with that work.
We verify the single-band limit using an exact finite-broadening
expression for the second-order conductivity, extracted from
a constant-$\Gamma$ Keldysh steady state solved nonperturbatively
in the electric field.
In the scalar single-orbital model used for this check,
all Berry-curvature and quantum-metric contributions vanish,
so the kinetic term is the entire $\tau^{0}$ response. 
The clean-limit expansion of this exact expression reproduces
both the leading nonlinear Drude term and the subleading
kinetic contribution, providing a consistency check within
the same dissipative model. 
A real two-band model further illustrates a finite intraband
quantum-metric dipole despite identically vanishing Berry
curvature and Berry-curvature-dipole response.
The dc limit is taken at finite broadening before the
large-$\tau$ expansion.
The $\tau^{0}$ contributions are independent of the lifetime
within this specified model; their independence from the
relaxation mechanism is not established.
\end{abstract}

\maketitle

\section{Introduction}
\label{sec:introduction}

Nonlinear dc transport provides a direct probe of the quantum
geometry of Bloch bands.
The geometry of a band is encoded in the quantum geometric
tensor~\cite{Provost1980}, whose antisymmetric and symmetric
parts are the Berry curvature and the quantum metric,
respectively.
The Berry curvature controls anomalous transport phenomena such
as the anomalous Hall effect~\cite{Nagaosa2010} and the
nonlinear Hall effect driven by the Berry-curvature dipole
(BCD) in time-reversal-invariant noncentrosymmetric
systems~\cite{SodemannFu2015}.
Early material proposals for the BCD mechanism include black
phosphorus with broken inversion symmetry~\cite{Low2015TopologicalCurrents},
and the effect has been observed in bilayer WTe$_{2}$ and related
materials~\cite{MaSong2019,Kang2019,Tokura2018}.
By contrast, the quantum metric has more recently emerged as an
independent source of observable responses, including superfluid
weight in flat bands~\cite{Peotta2015Superfluid,Torma2022Review},
resonant optical responses~\cite{Ahn2022Riemannian}, and
geometric contributions to nonlinear
transport~\cite{GaoYangNiu2014,Das2023,KaplanHolderYan2024,JiangHolderYan2025Review}, with quantum-metric-induced nonlinear
Hall effects observed experimentally in topological
antiferromagnets~\cite{Wang2023Nature,Gao2023Science}.

At second order in a static electric field, the dc current
contains several contributions with distinct dependences on the
quasiparticle lifetime $\tau$.
The nonlinear Drude response scales as $\tau^{2}$ and is
governed by the band dispersion~\cite{Deyo2009,Moore2010}.
The BCD response scales as $\tau$ and gives the standard
semiclassical nonlinear Hall effect in inversion-broken
metals~\cite{SodemannFu2015,Low2015TopologicalCurrents,DuLuXie2021Review}.
In addition, geometric $\mathcal{O}(\tau^{0})$ contributions
controlled by the quantum metric arise from both intraband and
interband channels~\cite{GaoYangNiu2014,Wang2021CuMnAs,Liu2021AFM,Das2023,KaplanHolderYan2024,Watanabe2020,Kusunose2022}.
These $\mathcal{O}(\tau^{0})$ terms are important because they
survive even when the Berry curvature is forced to vanish by
symmetry, for example in $\mathcal{PT}$-symmetric or certain
antiferromagnetic systems~\cite{Watanabe2020,Wang2021CuMnAs,Liu2021AFM,Wang2023Nature,Gao2023Science},
providing a route to nonlinear dc responses that probe the
metric structure of Bloch states rather than the BCD mechanism.

A closely related recent development is the gauge-invariant
projector calculus for quantum state
geometry~\cite{Mitscherling2025ProjectorCalculus,Avdoshkin2025Multistate}, in which quantum-geometric objects are
organized in terms of band projectors and their derivatives,
naturally avoiding gauge-dependent Bloch-state phases.
This framework has been applied to optical polarization,
injection currents, and shift
currents~\cite{Mitscherling2025ProjectorCalculus,Avdoshkin2025Multistate}, and makes clear that intermediate
Berry-connection expressions should not be assigned physical
meaning before all diagrams and contact terms have been
combined.
Jiang, Holder, and Yan~\cite{JiangHolderYan2025Review} review
the nonlinear Drude, BCD, and QMD transport formulas and
discuss the Green's function route to the dc limit.
Ulrich \textit{et al.}~\cite{Ulrich2026IntraQMD} derive a
projector-based Green's function expression that decomposes the
dc response into ND, BCD, intraband-QMD, and interband-QMD
sectors and establishes the cancellation of
connection-dependent terms within that framework.

The present work derives the dc response from the Peierls current
vertices within the velocity-gauge Keldysh--Kubo framework.
Their band-basis expansion contains covariant derivatives of
interband Berry connections and three-connection products,
whose combination must preserve gauge
covariance~\cite{Aversa1995,Sipe2000,Ventura2017,Parker2019}.
The contact vertices $V_{ij}$ and $V_{ijk}$ are essential
to the cancellation of connection-dependent terms.
Earlier velocity-gauge diagrammatic and Keldysh treatments
are given in
Refs.~\cite{Parker2019,JoaoLopes2020,MichishitaPeters2021,Michishita2022,Kusunose2022,Nakazawa2025Interband};
complementary Boltzmann and quantum-kinetic approaches address
disorder-mediated nonlinear Hall
transport~\cite{NandySodemann2019,Du2019DisorderNHE,XiaoDuNiu2019,Du2021QuantumTheoryNHE,DuLuXie2021Review}.
The Peierls interaction and current vertices are already
organized diagrammatically by Jo\~{a}o and Viana Parente
Lopes~\cite{JoaoLopes2020}, and the connection cancellation
is explicit in the projector formulation of
Ulrich \textit{et al.}~\cite{Ulrich2026IntraQMD}; related
geometric structures are discussed in Ref.~\cite{Qiang2025Clarification}.
Here we trace their relation term by term without projectors
for a general $N$-band tight-binding model.
The derivation also retains the known kinetic
contribution~\cite{Guo2026Dissipation,Anan2026Nonreciprocal,Kitamura2026QuasiequilibriumShift,Ulrich2026IntraQMD},
completing the lifetime hierarchy through $\mathcal{O}(\tau^{0})$
within the stated constant-self-energy model.
An exact finite-broadening expression for the single-band
second-order conductivity reproduces the leading
$\tau^{2}$ Drude and subleading $\tau^{0}$ kinetic terms,
checking the fully diagonal sector within the same dissipative model.

The resulting dc response is organized into four sectors:
\begin{align}
  \sigma^{\mathrm{DC}}_{ijk}
  &= \sigma^{\mathrm{ND}}_{ijk}
   + \sigma^{\mathrm{BCD}}_{ijk}
   + \sigma^{\mathrm{intra\text{-}QMD}}_{ijk}
   + \sigma^{\mathrm{inter\text{-}QMD}}_{ijk},
  \label{eq:intro-decomp}
\end{align}
valid in the constant-scattering-time approximation under the
non-degenerate clean condition
$|\varepsilon_{nm}|\tau/\hbar\gg 1$ for $n\neq m$, where
$\varepsilon_{nm}\equiv\varepsilon_{n}-\varepsilon_{m}$.
The nonlinear Drude sector itself contains two orders of the
lifetime expansion,
\begin{align}
  \sigma^{\mathrm{ND}}_{ijk}
  = \sigma^{\mathrm{ND},\tau^{2}}_{ijk}
  + \sigma^{\mathrm{ND},\tau^{0}}_{ijk},
  \label{eq:intro-ND-split}
\end{align}
so that the four sectors scale as $\tau^{2}$, $\tau$,
$\tau^{0}$, and $\tau^{0}$ at leading order, while the
subleading Drude piece
$\sigma^{\mathrm{ND},\tau^{0}}_{ijk}$ contributes at the same
$\mathcal{O}(\tau^{0})$ order as the two
quantum-metric-dipole terms.
The order-$\tau^{0}$ response is therefore not purely
geometric: it contains one purely dispersive term in addition
to the two metric terms.
The geometric contribution follows from cancellation between
the covariant-connection sector $\sigma^{\mathcal{C}}_{ijk}$
and the three-connection sector $\sigma^{\mathcal{T}}_{ijk}$.
The intraband term is a Fermi-surface quantum-metric dipole;
the interband term is written, before integration by parts,
as a Fermi-sea-type response involving a band-normalized metric.
The comparison with the projector-based
result~\cite{Ulrich2026IntraQMD} is given in
Sec.~\ref{subsec:comparison-dc-limit}.

We employ the Keldysh Green's function
formalism~\cite{Keldysh1965,Rammer1986,Haug2008} 
with a constant phenomenological relaxation time. 
Intra-atomic dipole matrix elements, vertex corrections,
skew scattering, side-jump processes, and disorder-induced
corrections to the distribution function lie outside the
present scope; these effects have been analyzed in complementary
nonlinear-Hall transport theories~\cite{NandySodemann2019,Du2019DisorderNHE,XiaoDuNiu2019,Du2021QuantumTheoryNHE}.

The remainder of the paper is organized as follows.
Section~\ref{sec:formalism} introduces the Peierls
velocity-gauge formulation and the band-geometric notation.
Section~\ref{sec:second_order_current} derives the
finite-frequency second-order conductivity and takes the dc
limit.
Section~\ref{sec:dc-conductivity-band} presents the band-basis
decomposition and the clean-limit lifetime expansion.
Section~\ref{sec:results} gives the explicit formulas for all
four sectors and relates them to previous work; it also
demonstrates the cancellation of the connection-dependent
sectors (Sec.~\ref{sec:cancellation}) and delimits the scope of
the constant-$\Gamma$ closure (Sec.~\ref{sec:scope}).
Section~\ref{sec:models} illustrates the decomposition with a
tilted massive Dirac lattice model, a real two-band model
in which the Berry curvature vanishes identically, and an
exact one-band check of the kinetic contribution.
Section~\ref{sec:discussion} summarizes the scope and
implications of the result.

\section{Formalism}
\label{sec:formalism}
\subsection{Peierls coupling and current vertices in a tight-binding representation}

We consider a noninteracting multiband tight-binding Hamiltonian
\begin{align}
  \hat{H}_{0}
  = \sum_{\bm{k}}\sum_{\mu\nu}
    \hat{\Psi}^{\dagger}_{\bm{k}\mu}\,
    H_{\mu\nu}(\bm{k})\,
    \hat{\Psi}_{\bm{k}\nu}
  \equiv
  \sum_{\bm{k}}
    \hat{\Psi}^{\dagger}_{\bm{k}}\,
    H(\bm{k})\,
    \hat{\Psi}_{\bm{k}},
  \label{eq:tight-binding}
\end{align}
where $H_{\mu\nu}(\bm{k})$ is the $N\times N$ Bloch Hamiltonian
matrix in the orbital basis ($\mu,\nu = 1,\ldots,N$),
$\hat{\Psi}^{\dagger}_{\bm{k}\mu}$ ($\hat{\Psi}_{\bm{k}\mu}$)
is the creation (annihilation) operator for a Bloch electron
in orbital $\mu$ at crystal momentum $\bm{k}$,
and $\hat{\Psi}^{\dagger}_{\bm{k}}$ denotes the $N$-component
row vector
$(\hat{\Psi}^{\dagger}_{\bm{k}1},\ldots,
  \hat{\Psi}^{\dagger}_{\bm{k}N})$.
Here we use $e>0$ for the magnitude of the electron charge,
so the electron charge is $-e$.

A spatially uniform electromagnetic field is introduced via
the Peierls velocity-gauge substitution,
\begin{align}
  H_{\bm{A}}(\bm{k},t)
  = H\!\left(\bm{k}+\frac{e\bm{A}(t)}{\hbar}\right),
  \label{eq:peierls-main}
\end{align}
with $\bm{E}(t) = -\partial_{t}\bm{A}(t)$.
An equivalent length-gauge formulation
exists~\cite{Aversa1995,Sipe2000}, but the velocity gauge
is adopted here as it generates the perturbation series
systematically through the Peierls expansion.

Expanding Eq.~\eqref{eq:peierls-main} in powers of $\bm{A}$
and substituting into Eq.~\eqref{eq:tight-binding},
the total Hamiltonian separates as
$\hat{H}_{\bm{A}}(t) = \hat{H}_{0} + \hat{H}_{\mathrm{ext}}(t)$,
where the light--matter interaction $\hat{H}_{\mathrm{ext}}(t)$
reads
\begin{align}
  \hat{H}_{\mathrm{ext}}(t)
  &= e\sum_{i}\sum_{\bm{k}}
     \hat{\Psi}^{\dagger}_{\bm{k}}\,
     V_{i}(\bm{k})\,
     \hat{\Psi}_{\bm{k}}\,A_{i}(t)
  \nonumber\\
  &\quad
   + \frac{e^{2}}{2}\sum_{ij}\sum_{\bm{k}}
     \hat{\Psi}^{\dagger}_{\bm{k}}\,
     V_{ij}(\bm{k})\,
     \hat{\Psi}_{\bm{k}}\,A_{i}(t)A_{j}(t)
  \nonumber\\
  &\quad
   + \frac{e^{3}}{3!}\sum_{ijk}\sum_{\bm{k}}
     \hat{\Psi}^{\dagger}_{\bm{k}}\,
     V_{ijk}(\bm{k})\,
     \hat{\Psi}_{\bm{k}}
  \nonumber\\
  &\qquad\quad
     \times A_{i}(t)A_{j}(t)A_{k}(t)
  + \cdots,
  \label{eq:Hext}
\end{align}
where
\begin{align}
  V_{i}
  = \frac{1}{\hbar}\partial_{i}H,
  \quad
  V_{ij}
  = \frac{1}{\hbar^{2}}\partial_{i}\partial_{j}H,
  \quad
  V_{ijk}
  = \frac{1}{\hbar^{3}}\partial_{i}\partial_{j}\partial_{k}H,
  \label{eq:vertices-main}
\end{align}
are the first-, second-, and third-rank velocity vertices,
respectively, and $\partial_{i}\equiv\partial/\partial k_{i}$.
The first term on the right-hand side of Eq.~\eqref{eq:Hext}
is the paramagnetic coupling to the vector potential,
while the higher-order terms are the two-photon and three-photon
contact interactions that arise naturally from the Peierls
substitution~\eqref{eq:peierls-main}.
These coupling constants are determined entirely by the
band structure through the velocity vertices
[Eq.~\eqref{eq:vertices-main}] and involve no free parameters.

The physical current operator is obtained from
$\hat{\mathcal{J}}_{i}
 = -\delta\hat{H}_{\bm{A}}(t)/\delta A_{i}(t)$ as
\begin{align}
  \hat{\mathcal{J}}_{i}(t)
  &= -e\sum_{\bm{k}}
     \hat{\Psi}^{\dagger}_{\bm{k}}\,V_{i}(\bm{k})\,
     \hat{\Psi}_{\bm{k}}
  \nonumber\\
  &\quad
   - e^{2}\sum_{j}\sum_{\bm{k}}
     \hat{\Psi}^{\dagger}_{\bm{k}}\,V_{ij}(\bm{k})\,
     \hat{\Psi}_{\bm{k}}\,A_{j}(t)
  \nonumber\\
  &\quad
   - \frac{e^{3}}{2}\sum_{jk}\sum_{\bm{k}}
     \hat{\Psi}^{\dagger}_{\bm{k}}\,V_{ijk}(\bm{k})\,
     \hat{\Psi}_{\bm{k}}\,A_{j}(t)A_{k}(t)
   + \cdots,
  \label{eq:current-vertices}
\end{align}
and thus contains the one-photon current vertex $V_{i}$
and the two-photon and three-photon contact-current vertices
$V_{ij}$ and $V_{ijk}$, respectively.
These contact vertices play an essential role in the velocity
gauge: they cancel the apparent low-frequency singularities
that arise when converting the vector potential $A_{j}(\omega)$
to the electric field $E_{j}(\omega) = i\omega A_{j}(\omega)$
in Fourier space, thereby ensuring that the conductivity
remains finite in the dc limit $\omega\to 0$.

\subsection{Contour-ordered Green's function and current expectation value}

To evaluate the nonlinear response to the external field
$\bm{A}(t)$, we compute the expectation value of the current
operator $\hat{\mathcal{J}}_{i}$ in the nonequilibrium
steady state driven by $\bm{A}(t)$.
We adopt the Keldysh formalism~\cite{Keldysh1965,Rammer1986,Haug2008},
in which all nonequilibrium information is encoded in the
contour-ordered Green's function defined on the Keldysh
contour $\mathcal{C}$.
Closely related Keldysh and velocity-gauge formulations of
nonlinear conductivity and its dc limit are given in
Refs.~\cite{JoaoLopes2020,MichishitaPeters2021,Kusunose2022,Nakazawa2025Interband,Ulrich2026IntraQMD}.
A key advantage of this formalism is that it allows a
systematic perturbative expansion of the Green's function to
arbitrary order in the external field $\bm{A}(t)$.

The expectation value of the current operator is
\begin{align}
  \mathcal{J}_{i}(t)
  &\equiv
  \langle\hat{\mathcal{J}}_{i}(t)\rangle_{\hat{H}_{\bm{A}}(t)}
  \nonumber\\
  &= ie\hbar\sum_{\bm{k}}
     \mathrm{Tr}\!\bigl[
       V_{i}(\bm{k})\,G^{<}_{\bm{k},\bm{k}}(t,t)
     \bigr]
  \nonumber\\
  &\quad
  + ie^{2}\hbar\sum_{j}\sum_{\bm{k}}
     \mathrm{Tr}\!\bigl[
       V_{ij}(\bm{k})\,G^{<}_{\bm{k},\bm{k}}(t,t)
     \bigr]A_{j}(t)
  \nonumber\\
  &\quad
  + \frac{ie^{3}\hbar}{2}\sum_{j,k}\sum_{\bm{k}}
     \mathrm{Tr}\!\bigl[
       V_{ijk}(\bm{k})\,G^{<}_{\bm{k},\bm{k}}(t,t)
     \bigr]A_{j}(t)A_{k}(t),
  \label{eq:J-lesser-current}
\end{align}
where $G^{<}_{\bm{k},\bm{k}}(t,t)$ is the lesser Green's
function at equal times.
Expanding $G^{<}_{\bm{k},\bm{k}}(t,t)$ perturbatively in
$\bm{A}$ and retaining terms up to second order, the
second-order current $\mathcal{J}^{(2)}_{i}(t)$ is
\begingroup\small
\begin{align}
  \mathcal{J}^{(2)}_{i}(t)
  &= ie\hbar\sum_{\bm{k}}
     \mathrm{Tr}\!\bigl[
       V_{i}(\bm{k})\,G^{(2)<}_{\bm{k}}(t,t)
     \bigr]
  \nonumber\\
  &\quad
   + ie^{2}\hbar\sum_{j}\sum_{\bm{k}}
     \mathrm{Tr}\!\bigl[
       V_{ij}(\bm{k})\,G^{(1)<}_{\bm{k}}(t,t)
     \bigr]A_{j}(t)
  \nonumber\\
  &\quad
   + \frac{ie^{3}\hbar}{2}\sum_{jk}\sum_{\bm{k}}
     \mathrm{Tr}\!\bigl[
       V_{ijk}(\bm{k})\,G^{(0)<}_{\bm{k}}(t,t)
     \bigr]A_{j}(t)A_{k}(t),
  \label{eq:J2-lesser}
\end{align}
\endgroup
where $G^{(p)<}_{\bm{k}}(t,t)$ denotes the $p$th-order
correction to the lesser Green's function in $\bm{A}$.
The three lines in Eq.~\eqref{eq:J2-lesser} correspond to
the three classes of diagrams entering the second-order
conductivity, illustrated in
Fig.~\ref{fig:second_order_current_diagrams}:
the two one-photon-vertex contributions
[Figs.~\ref{fig:second_order_current_diagrams}(a) and (b)],
in which the current vertex $V_{i}$ couples to the
second-order lesser Green's function $G^{(2)<}$,
the diamagnetic contribution
[Fig.~\ref{fig:second_order_current_diagrams}(c)],
in which the two-photon contact vertex $V_{ij}$ couples to
$G^{(1)<}$, and the tadpole contribution
[Fig.~\ref{fig:second_order_current_diagrams}(d)],
in which the three-photon contact vertex $V_{ijk}$ couples
to the equilibrium Green's function $G^{(0)<}$.
To evaluate $G^{(p)<}_{\bm{k}}(t,t)$, we expand the
contour-ordered Green's function perturbatively in $\bm{A}$
and extract the lesser component of each order by applying
the Langreth rules~\cite{Langreth1976,Haug2008};
the explicit perturbative expansion is given in
Appendix~\ref{app:green-current}, and the detailed
calculations are collected in the Supplemental
Material~\cite{SM}.

\section{Second-order current and nonlinear conductivity}
\label{sec:second_order_current}
\subsection{Second-order nonlinear conductivity tensor}

The Fourier component of the second-order current
$\mathcal{J}^{(2)}_{i}(\omega)$ decomposes as
\begin{align}
  \mathcal{J}^{(2)}_{i}(\omega)
  = \mathcal{J}^{\mathrm{p}(2a)}_{i}(\omega)
  + \mathcal{J}^{\mathrm{p}(2b)}_{i}(\omega)
  + \mathcal{J}^{\mathrm{d}(2)}_{i}(\omega)
  + \mathcal{J}^{\mathrm{t}(2)}_{i}(\omega),
  \label{eq:J2-decomp}
\end{align}
where the four contributions correspond, respectively, to the
paramagnetic current generated by two one-photon vertices,
the paramagnetic current generated by the two-photon contact
vertex $V_{jk}$, the diamagnetic current generated by
$V_{ij}$ or $V_{ik}$ acting on the first-order lesser
Green's function, and the tadpole current generated by
$V_{ijk}$ acting on the equilibrium lesser Green's function.
The Feynman diagrams representing these four contributions
are shown in Fig.~\ref{fig:second_order_current_diagrams},
and the explicit expressions are given in
Appendix~\ref{app:green-current}.

\begin{figure*}[t]
  \centering
  \includegraphics[width=0.96\textwidth]{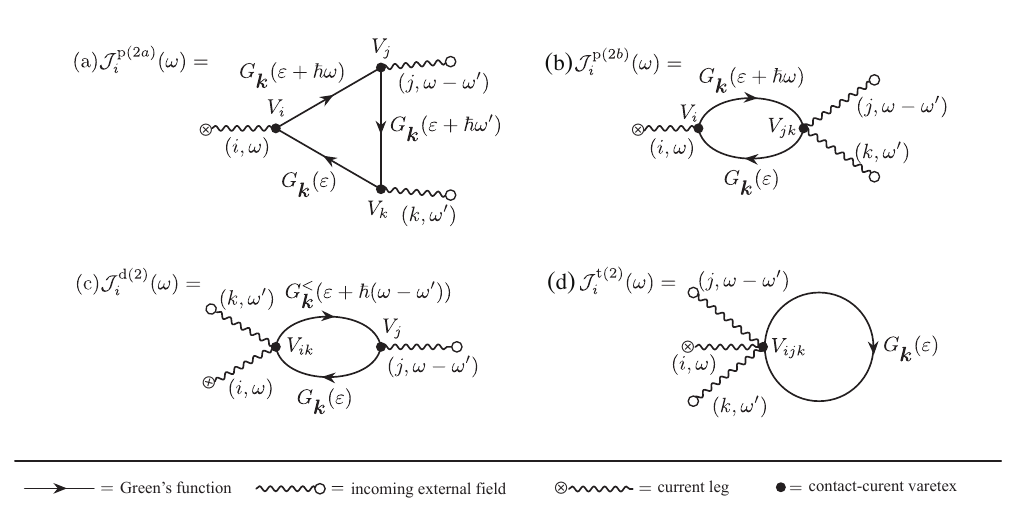}
  \caption{Diagrammatic representation of the four
    second-order current contributions in the Peierls
    velocity gauge:
    (a)~$\mathcal{J}_{i}^{\mathrm{p}(2a)}(\omega)$,
    (b)~$\mathcal{J}_{i}^{\mathrm{p}(2b)}(\omega)$,
    (c)~$\mathcal{J}_{i}^{\mathrm{d}(2)}(\omega)$, and
    (d)~$\mathcal{J}_{i}^{\mathrm{t}(2)}(\omega)$.
    Solid lines with arrows denote equilibrium Green's
    functions, wavy lines ending with open circles denote
    incoming external fields, and wavy lines ending with
    $\otimes$ denote the outgoing current leg $(i,\omega)$.
    Black dots denote current or contact vertices.}
  \label{fig:second_order_current_diagrams}
\end{figure*}

Collecting the four contributions and expressing the result
in terms of the electric field via
$A_{j}(\omega) = E_{j}(\omega)/(i\omega)$,
the second-order current takes the standard convolution form
\begin{align}
  \mathcal{J}_{i}^{(2)}(\omega)
  &= \sum_{jk}
     \int_{-\infty}^{\infty}\frac{d\omega_{1}}{2\pi}
     \int_{-\infty}^{\infty}\frac{d\omega_{2}}{2\pi}\,
  \nonumber\\
  &\quad\times
     2\pi\delta(\omega-\omega_{12})\,
     \sigma_{ijk}(\omega_{1},\omega_{2})\,
     E_{j}(\omega_{1})\,E_{k}(\omega_{2}),
  \label{eq:J2-sigma}
\end{align}
where $\sigma_{ijk}(\omega_{1},\omega_{2})$ is the
second-order nonlinear conductivity tensor and
$\omega_{12}\equiv\omega_{1}+\omega_{2}$.
Collecting all four contributions
Eqs.~\eqref{eq:Jp2a}--\eqref{eq:Jt2},
$\sigma_{ijk}(\omega_{1},\omega_{2})$ is given by
\begin{align}
  &\sigma_{ijk}(\omega_{1},\omega_{2})
  = -\frac{ie^{3}}{2}
     \sum_{\bm{k}}
     \int_{-\infty}^{\infty}\frac{d\varepsilon}{2\pi}\,
     \frac{1}{\omega_{1}\omega_{2}}
  \bigg\{
  \nonumber\\
  &\quad
     \mathrm{Tr}\!\Big[
       V_{i}\,G_{\bm{k}}(\varepsilon+\hbar\omega_{12})\,
       V_{j}\,G_{\bm{k}}(\varepsilon+\hbar\omega_{2})\,
       V_{k}\,G_{\bm{k}}(\varepsilon)
     \Big]^{<}
  \nonumber\\
  &\quad
  +  \mathrm{Tr}\!\Big[
       V_{i}\,G_{\bm{k}}(\varepsilon+\hbar\omega_{12})\,
       V_{k}\,G_{\bm{k}}(\varepsilon+\hbar\omega_{1})\,
       V_{j}\,G_{\bm{k}}(\varepsilon)
     \Big]^{<}
  \nonumber\\
  &\quad
  +  \mathrm{Tr}\!\Big[
       V_{i}\,G_{\bm{k}}(\varepsilon+\hbar\omega_{12})\,
       V_{jk}\,G_{\bm{k}}(\varepsilon)
     \Big]^{<}
  \nonumber\\
  &\quad
  +  \mathrm{Tr}\!\left[
       V_{ij}\,G_{\bm{k}}(\varepsilon+\hbar\omega_{2})\,
       V_{k}\,G_{\bm{k}}(\varepsilon)
     \right]^{<}
  \nonumber\\
  &\quad
  +  \mathrm{Tr}\!\left[
       V_{ik}\,G_{\bm{k}}(\varepsilon+\hbar\omega_{1})\,
       V_{j}\,G_{\bm{k}}(\varepsilon)
     \right]^{<}
  \nonumber\\
  &\quad
  +  \mathrm{Tr}\!\left[
       V_{ijk}\,G^{<}_{\bm{k}}(\varepsilon)
     \right]
  \bigg\}.
  \label{eq:sigma2-full}
\end{align}
The superscript $<$ denotes that the lesser component is
taken after forming the product of contour-ordered Green's
functions and velocity vertices.
The overall factor of $1/2$ ensures that
$\sigma_{ijk}(\omega_{1},\omega_{2})$ is symmetric under
$(j,\omega_{1})\leftrightarrow(k,\omega_{2})$, as required
by the permutation symmetry of the second-order response
tensor~\cite{Boyd2008}.
The first and second traces in Eq.~\eqref{eq:sigma2-full}
map onto each other under this interchange, as do the fourth
and fifth, while the third and sixth are individually
symmetric.

Before taking the dc limit, we specify the single-particle
Green's functions.
The contour-ordered propagator satisfies
$[G_{\bm{k}}(\varepsilon)]^{-1}
 = (\varepsilon+\mu)I - H(\bm{k})$,
and its retarded and advanced components are
\begin{align}
  G_{\bm{k}}^{\mathrm{R}}(\varepsilon)
  &= \bigl[(\varepsilon+\mu+i\gamma)I-H(\bm{k})\bigr]^{-1},
  \label{eq:G-retarded-def}
  \\
  G_{\bm{k}}^{\mathrm{A}}(\varepsilon)
  &= \bigl[(\varepsilon+\mu-i\gamma)I-H(\bm{k})\bigr]^{-1}
   = \bigl[G_{\bm{k}}^{\mathrm{R}}(\varepsilon)\bigr]^{\dagger},
  \label{eq:G-advanced-def}
\end{align}
where $I$ is the identity matrix in orbital space and
$\gamma = \hbar/(2\tau)$ is the phenomenological
quasiparticle broadening.
The equilibrium lesser Green's function is
\begin{align}
  G_{\bm{k}}^{<}(\varepsilon)
  = f(\varepsilon)\,\Lambda_{\bm{k}}(\varepsilon),
  \label{eq:G-lesser-def}
\end{align}
where $f(\varepsilon)
 = [1+\exp(\varepsilon/k_{\mathrm{B}}T)]^{-1}$
is the Fermi-Dirac distribution with the energy variable
measured from the chemical potential, and
\begin{align}
  \Lambda_{\bm{k}}(\varepsilon)
  &= G_{\bm{k}}^{\mathrm{A}}(\varepsilon)
  - G_{\bm{k}}^{\mathrm{R}}(\varepsilon)
  = 2i\gamma\,G_{\bm{k}}^{\mathrm{R}}(\varepsilon)G_{\bm{k}}^{\mathrm{A}}(\varepsilon)
  \label{eq:spectral-matrix-def}
\end{align}
is the spectral matrix.
Applying the Langreth rules~\cite{Langreth1976,Haug2008} to
decompose the lesser component of the products of
contour-ordered Green's functions in
Eq.~\eqref{eq:sigma2-full} and shifting the internal
energy variable $\varepsilon$ so that all Fermi functions
carry the same argument $f(\varepsilon)$, one obtains the
finite-frequency second-order conductivity
\begin{widetext}
\begin{align}
  \sigma_{ijk}(\omega_{1},\omega_{2})
  ={}&-\frac{ie^{3}}{2}
       \sum_{\bm{k}}
       \int_{-\infty}^{\infty}\frac{d\varepsilon}{2\pi}\,
       f(\varepsilon)\,\frac{1}{\omega_{1}\omega_{2}}
  \bigg\{
  \nonumber\\
  &\qquad
     \mathrm{Tr}\!\big[
       V_{i}G^{\mathrm{R}}(\varepsilon+\hbar\omega_{12})\,
       V_{j}G^{\mathrm{R}}(\varepsilon+\hbar\omega_{2})\,
       V_{k}\Lambda(\varepsilon)
     \big]\nonumber\\
 &\quad
  +  \mathrm{Tr}\!\big[
       V_{i}G^{\mathrm{R}}(\varepsilon+\hbar\omega_{12})\,
       V_{k}G^{\mathrm{R}}(\varepsilon+\hbar\omega_{1})\,
       V_{j}\Lambda(\varepsilon)
     \big]
  \nonumber\\
  &\quad
  +  \mathrm{Tr}\!\big[
       V_{i}G^{\mathrm{R}}(\varepsilon+\hbar\omega_{1})\,
       V_{j}\Lambda(\varepsilon)\,
       V_{k}G^{\mathrm{A}}(\varepsilon-\hbar\omega_{2})
     \big]\nonumber\\&\quad
  +  \mathrm{Tr}\!\big[
       V_{i}G^{\mathrm{R}}(\varepsilon+\hbar\omega_{2})\,
       V_{k}\Lambda(\varepsilon)\,
       V_{j}G^{\mathrm{A}}(\varepsilon-\hbar\omega_{1})
     \big]
  \nonumber\\
  &\quad
  +  \mathrm{Tr}\!\big[
       V_{i}\Lambda(\varepsilon)\,
       V_{j}G^{\mathrm{A}}(\varepsilon-\hbar\omega_{1})\,
       V_{k}G^{\mathrm{A}}(\varepsilon-\hbar\omega_{12})
     \big]\nonumber\\&\quad
  +  \mathrm{Tr}\!\big[
       V_{i}\Lambda(\varepsilon)\,
       V_{k}G^{\mathrm{A}}(\varepsilon-\hbar\omega_{2})\,
       V_{j}G^{\mathrm{A}}(\varepsilon-\hbar\omega_{12})
     \big]
  \nonumber\\
  &\quad
  +  \mathrm{Tr}\!\big[
       V_{i}G^{\mathrm{R}}(\varepsilon+\hbar\omega_{12})\,
       V_{jk}\Lambda(\varepsilon)
     \big]
  +  \mathrm{Tr}\!\big[
       V_{i}\Lambda(\varepsilon)\,
       V_{jk}G^{\mathrm{A}}(\varepsilon-\hbar\omega_{12})
     \big]
  \nonumber\\
  &\quad
  +  \mathrm{Tr}\!\big[
       V_{ij}G^{\mathrm{R}}(\varepsilon+\hbar\omega_{2})\,
       V_{k}\Lambda(\varepsilon)
     \big]
  +  \mathrm{Tr}\!\big[
       V_{ik}G^{\mathrm{R}}(\varepsilon+\hbar\omega_{1})\,
       V_{j}\Lambda(\varepsilon)
     \big]
  \nonumber\\
  &\quad
  +  \mathrm{Tr}\!\big[
       V_{ij}\Lambda(\varepsilon)\,
       V_{k}G^{\mathrm{A}}(\varepsilon-\hbar\omega_{2})
     \big]
  +  \mathrm{Tr}\!\big[
       V_{ik}\Lambda(\varepsilon)\,
       V_{j}G^{\mathrm{A}}(\varepsilon-\hbar\omega_{1})
     \big]
  \nonumber\\
  &\quad
  +  \mathrm{Tr}\!\big[
       V_{ijk}\,\Lambda(\varepsilon)
     \big]
  \bigg\},
  \label{eq:sigma2-finite-frequency}
\end{align}
\end{widetext}
where the crystal momentum argument $\bm{k}$ is suppressed
for brevity and all matrices are evaluated at the same
$\bm{k}$.
Equation~\eqref{eq:sigma2-finite-frequency} is the
velocity-gauge second-order conductivity tensor at finite
frequency, valid prior to taking the dc limit.
The prefactor $1/(\omega_{1}\omega_{2})$ originates from
the conversion $A_{j}(\omega) = E_{j}(\omega)/(i\omega)$.
The apparent singularities at $\omega_{1}=0$ or
$\omega_{2}=0$ are unphysical and cancel exactly between
the one-photon-vertex and contact-vertex contributions
once all terms are combined, as demonstrated below.
Equation~\eqref{eq:sigma2-finite-frequency} is equivalent to the
finite-frequency velocity-gauge formulas derived in
Refs.~\cite{JoaoLopes2020,MichishitaPeters2021,Nakazawa2025Interband}.
To compare signs and numerical factors, one has to translate the
vertex and spectral-function conventions. In the present paper we use
$e>0$ for the magnitude of the electron charge, the Peierls
substitution $H(\bm{k}+e\bm{A}/\hbar)$, and the current operator
$\hat{\mathcal{J}}_{i}=-\delta\hat{H}_{\bm{A}}/\delta A_{i}$.
The charge-current vertices used in Refs.~\cite{MichishitaPeters2021,Nakazawa2025Interband}
are related to our velocity vertices by
\begin{align}
  J_{i}=-e\hbar V_{i},
  \qquad
  J_{ij}=e^{2}\hbar^{2}V_{ij},
  \qquad
  J_{ijk}=-e^{3}\hbar^{3}V_{ijk}.
  \label{eq:vertex-convention-comparison}
\end{align}
Moreover, those works often write the spectral factor as
$G^{\mathrm{R}}-G^{\mathrm{A}}$, whereas our convention is
\begin{align}
  \Lambda(\varepsilon)=G^{\mathrm{A}}(\varepsilon)-G^{\mathrm{R}}(\varepsilon)
  =-\left[G^{\mathrm{R}}(\varepsilon)-G^{\mathrm{A}}(\varepsilon)\right].
  \label{eq:spectral-convention-comparison}
\end{align}
After applying Eqs.~\eqref{eq:vertex-convention-comparison} and
\eqref{eq:spectral-convention-comparison}, Eq.~\eqref{eq:sigma2-finite-frequency}
has the same relative signs and numerical coefficients as the
second-order AC conductivity of Jo\~{a}o and Lopes~\cite{JoaoLopes2020},
Michishita and Peters~\cite{MichishitaPeters2021}, and Nakazawa
\textit{et al.}~\cite{Nakazawa2025Interband}. The overall factor
$1/2$ in Eq.~\eqref{eq:sigma2-finite-frequency} is the same
symmetrization over the two incoming electric-field legs that is
alternatively implemented in the literature by writing an explicit
$(j,\omega_{1})\leftrightarrow(k,\omega_{2})$ term.

\subsection{Second-order dc conductivity}
\label{subsec:dc-limit}

The expression for $\sigma_{ijk}(\omega_{1},\omega_{2})$
contains an explicit prefactor $1/(\omega_{1}\omega_{2})$
originating from $A_{j}(\omega)=E_{j}(\omega)/(i\omega)$.
Although this appears to generate unphysical divergences as
$\omega_{1},\omega_{2}\to 0$, these singularities cancel
exactly when all contributions are combined, a consequence
of the Ward--Takahashi identity associated with $U(1)$
gauge invariance~\cite{Ward1950,Takahashi1957},
\begin{align}
  \partial_{i}G^{\mathrm{R/A}}_{\bm{k}}(\varepsilon)
  = G^{\mathrm{R/A}}_{\bm{k}}(\varepsilon)\,
    \hbar V_{i}(\bm{k})\,
    G^{\mathrm{R/A}}_{\bm{k}}(\varepsilon).
  \label{eq:WT-identity}
\end{align}
This identity ensures that each spurious coefficient in
the $\omega$ expansion is a total crystal-momentum
derivative, which vanishes upon Brillouin-zone
integration~\cite{Aversa1995,Sipe2000}.
A crucial role is played by the three-photon
contact-current term
$\mathrm{Tr}[V_{ijk}G^{<}_{\bm{k}}(\varepsilon)]
E_{j}E_{k}$, which is unique to the velocity gauge and
arises naturally from the Peierls substitution.
Without this term the cancellation would be incomplete
and the dc conductivity would be
gauge-dependent~\cite{Aversa1995,Sipe2000,Ventura2017,Parker2019}.
The explicit verification that all spurious poles cancel
order by order in $\omega_{1}$ and $\omega_{2}$ is given
in the Supplemental Material~\cite{SM}.

Taking the dc limit $\omega_{1},\omega_{2}\to 0$
by expanding
\begin{align}
  G^{\mathrm{R/A}}(\varepsilon \pm \hbar\omega)
  &\simeq
  G^{\mathrm{R/A}}(\varepsilon)
  \pm \hbar\omega\,\partial_{\varepsilon}G^{\mathrm{R/A}}(\varepsilon)
  \nonumber\\
  &\quad
  + \frac{(\hbar\omega)^{2}}{2}
    \partial_{\varepsilon}^{2}G^{\mathrm{R/A}}(\varepsilon)
  + \mathcal{O}(\omega^{3})
  \label{eq:GRA-expansion}
\end{align}
and retaining terms through $\mathcal{O}(\omega^{2})$
to cancel the $1/(\omega_{1}\omega_{2})$ prefactor,
the finite remainder is
\begin{widetext}
\begin{align}
  \sigma^{\mathrm{DC}}_{ijk}
  &= e^{3}\hbar^{2}
     \int_{-\infty}^{\infty}\frac{d\varepsilon}{2\pi}\,
     f(\varepsilon)
     \sum_{\bm{k}}
     \mathrm{Im}\,\mathrm{Tr}\bigg[
  \nonumber\\
  &\quad
     \partial_{\varepsilon}\!\left(
       V_{i}\partial_{\varepsilon}G^{\mathrm{R}}
       V_{j}G^{\mathrm{R}}V_{k}G^{\mathrm{A}}
     \right)
  -  V_{i}\partial_{\varepsilon}\!\left(
       \partial_{\varepsilon}G^{\mathrm{R}}V_{j}G^{\mathrm{R}}
     \right)V_{k}G^{\mathrm{R}}
  \nonumber\\
  &\quad
   + \partial_{\varepsilon}\!\left(
       V_{i}\partial_{\varepsilon}G^{\mathrm{R}}
       V_{k}G^{\mathrm{R}}V_{j}G^{\mathrm{A}}
     \right)
  -  V_{i}\partial_{\varepsilon}\!\left(
       \partial_{\varepsilon}G^{\mathrm{R}}V_{k}G^{\mathrm{R}}
     \right)V_{j}G^{\mathrm{R}}
  \nonumber\\
  &\quad
  +  \partial_{\varepsilon}\!\left(
       V_{i}\partial_{\varepsilon}G^{\mathrm{R}}
       V_{jk}G^{\mathrm{A}}
     \right)
  -  V_{i}\partial_{\varepsilon}^{2}G^{\mathrm{R}}
       V_{jk}G^{\mathrm{R}}
  \bigg],
  \label{eq:sigma-dc-trace}
\end{align}
\end{widetext}
where $G^{\mathrm{R}}\equiv G^{\mathrm{R}}_{\bm{k}}(\varepsilon)$
and $G^{\mathrm{A}}\equiv G^{\mathrm{A}}_{\bm{k}}(\varepsilon)$.
The dc second-order response satisfies
\begin{align}
  \mathcal{J}_{i}^{(2)}
  = \sum_{j,k}\sigma_{ijk}^{\mathrm{DC}}E_{j}E_{k},
  \qquad
  \sigma_{ijk}^{\mathrm{DC}} = \sigma_{ikj}^{\mathrm{DC}},
  \label{eq:current-def}
\end{align}
where the symmetrization over the two electric-field
indices is already incorporated.
For comparison with the dc-limit formulas of
Refs.~\cite{MichishitaPeters2021,Nakazawa2025Interband,Ulrich2026IntraQMD},
Eq.~\eqref{eq:sigma-dc-trace} can be integrated by parts in
$\varepsilon$ and organized into retarded--retarded--advanced,
retarded--advanced, retarded--retarded--retarded, and retarded--retarded
parts. This is the same $RRA+RA+RRR+RR$ organization used in
microscopic Keldysh treatments of the dc nonlinear conductivity. The
signs of the $-\partial_{\varepsilon}f$ terms and the $f$ terms agree
after the vertex convention in Eq.~\eqref{eq:vertex-convention-comparison}
is used. The coefficient of the $V_{jk}$ contact-current term also
agrees: in some presentations it appears with a factor $1/2$ together
with an explicit $(j\leftrightarrow k)$ addition, whereas in
Eq.~\eqref{eq:sigma-dc-trace} the symmetry $V_{jk}=V_{kj}$ has already
been used. Therefore Eq.~\eqref{eq:sigma-dc-trace} has no extra sign or
factor discrepancy relative to the previously derived dc-limit
conductivities. In the notation of Ulrich \textit{et al.}~\cite{Ulrich2026IntraQMD},
identifying $\Gamma=\hbar/(2\tau)$ and translating their gap convention
to our $\varepsilon_{nm}=\varepsilon_{n}-\varepsilon_{m}$ gives the same
NLD, BCD, interQMD, and intraQMD coefficients after the Fermi-sea-type
form of the interband term is converted to their Fermi-surface
representation by integration by parts.
Equation~\eqref{eq:sigma-dc-trace} is the central result
of the present formalism and serves as the starting point
for the band-geometric decomposition in
Sec.~\ref{sec:dc-conductivity-band}.

\section{Band-basis representation of the dc conductivity}
\label{sec:dc-conductivity-band}

\subsection{Bloch band basis}

We evaluate the trace sums in Eq.~\eqref{eq:sigma-dc-trace}
by transforming to the band basis, in which the Green's
functions are diagonal.
Let $U(\bm{k})$ be the unitary matrix that diagonalizes
the Bloch Hamiltonian,
\begin{align}
  U^{\dagger}(\bm{k})\,H(\bm{k})\,U(\bm{k})
  = \mathcal{E}(\bm{k})
  \equiv \mathrm{diag}\bigl(
    \varepsilon_{1}(\bm{k}),\ldots,\varepsilon_{N}(\bm{k})
  \bigr),
  \label{eq:band-diagonalization-main}
\end{align}
whose columns $|u_{n}(\bm{k})\rangle$ are the Bloch
eigenstates satisfying
\begin{align}
  H(\bm{k})\,|u_{n}(\bm{k})\rangle
  = \varepsilon_{n}(\bm{k})\,|u_{n}(\bm{k})\rangle,
  \qquad n = 1,\ldots,N.
  \label{eq:bloch-eigen}
\end{align}
In the band basis, the retarded and advanced Green's
functions are diagonal,
\begin{align}
  g^{\mathrm{R}}_{n}(\varepsilon)
  &\equiv \langle u_{n}|G^{\mathrm{R}}(\varepsilon)|u_{n}\rangle
   = \frac{1}{\varepsilon - \varepsilon_{n} + i\gamma},
  \label{eq:gR-band}
  \\
  g^{\mathrm{A}}_{n}(\varepsilon)
  &\equiv \langle u_{n}|G^{\mathrm{A}}(\varepsilon)|u_{n}\rangle
   = \frac{1}{\varepsilon - \varepsilon_{n} - i\gamma}
   = \bigl(g^{\mathrm{R}}_{n}(\varepsilon)\bigr)^{*},
  \label{eq:green-functions-band}
\end{align}
where the chemical potential has been absorbed into the
energy variable $\varepsilon\to\varepsilon+\mu$ and
$\gamma = \hbar/(2\tau) > 0$ is the quasiparticle
broadening.
The corresponding spectral function is
\begin{align}
  \lambda_{n}(\varepsilon)
  \equiv g^{\mathrm{A}}_{n}(\varepsilon)
       - g^{\mathrm{R}}_{n}(\varepsilon)
  = \frac{2i\gamma}
         {(\varepsilon - \varepsilon_{n})^{2} + \gamma^{2}},
  \label{eq:lambda-band}
\end{align}
which reduces to
$\lambda_{n}(\varepsilon)\to 2\pi i\,\delta(\varepsilon
 - \varepsilon_{n})$ in the clean limit $\gamma\to 0^{+}$.

We work in the clean dc limit by retaining contributions
through $\mathcal{O}(\tau^{0})$ under the non-degenerate
band condition
\begin{align}
  \frac{|\varepsilon_{nm}|\,\tau}{\hbar} \gg 1,
  \qquad n \neq m,
  \label{eq:clean-limit-condition}
\end{align}
where $\varepsilon_{nm} \equiv \varepsilon_{n}-\varepsilon_{m}$
is the interband energy difference.
This condition ensures that interband coherences are
suppressed on the timescale $\tau$, allowing each band
to be treated independently.
Degenerate or nearly degenerate subspaces require a
non-Abelian generalization~\cite{Culcer2005} and are
left for future work.

\subsection{Nonlinear dc conductivity in the band basis}

Inserting complete sets of eigenstates
$\sum_{n}|u_{n}\rangle\langle u_{n}| = I$
between each pair of matrices in
Eq.~\eqref{eq:sigma-dc-trace},
the nonlinear dc conductivity takes the form
\begin{align}
  \sigma^{\mathrm{DC}}_{ijk}
  &= \frac{e^{3}}{\hbar}
     \int_{-\infty}^{\infty}\frac{d\varepsilon}{2\pi}\,
     f(\varepsilon)
     \sum_{\bm{k}}
     \mathrm{Im}\!\left[\mathcal{B}_{ijk}\right],
  \label{eq:sigma-dc-band}
\end{align}
where $\mathcal{B}_{ijk}$ is a combination of
the band-basis velocity matrix elements
$v^{nm}_{i} = \langle u_{n}|\hbar V_{i}|u_{m}\rangle$ and
$v^{nm}_{ij} = \langle u_{n}|\hbar^{2}V_{ij}|u_{m}\rangle$
with the Green's function kernels
$\mathcal{K}^{(1)}_{nml}$ and $\mathcal{K}^{(2)}_{nm}$:
\begin{align}
  \mathcal{B}_{ijk}
  &= \sum_{n,m,l}
     \bigl(
       v^{nm}_{i}v^{ml}_{j}v^{ln}_{k}
     + v^{nm}_{i}v^{ml}_{k}v^{ln}_{j}
     \bigr)
     \mathcal{K}^{(1)}_{nml}
  \nonumber\\
  &\quad
  + \sum_{n,m}
    v^{nm}_{i}\,v^{mn}_{jk}\,
    \mathcal{K}^{(2)}_{nm}.
  \label{eq:Bijk-band}
\end{align}
The three-index kernel $\mathcal{K}^{(1)}_{nml}$ and
the two-index kernel $\mathcal{K}^{(2)}_{nm}$ are defined by
\begin{align}
  \mathcal{K}^{(1)}_{nml}
  &\equiv
  \partial_{\varepsilon}
  \!\left[
    (\partial_{\varepsilon}g^{\mathrm{R}}_{m})\,
    g^{\mathrm{R}}_{l}\,g^{\mathrm{A}}_{n}
  \right]
  - \partial_{\varepsilon}
  \!\left[
    (\partial_{\varepsilon}g^{\mathrm{R}}_{m})\,
    g^{\mathrm{R}}_{l}
  \right]g^{\mathrm{R}}_{n},
  \label{eq:K1-def-band}
  \\
  \mathcal{K}^{(2)}_{nm}
  &\equiv
  \partial_{\varepsilon}
  \!\left[
    (\partial_{\varepsilon}g^{\mathrm{R}}_{m})\,
    g^{\mathrm{A}}_{n}
  \right]
  - (\partial_{\varepsilon}^{2}g^{\mathrm{R}}_{m})\,
    g^{\mathrm{R}}_{n},
  \label{eq:K2-def-band}
\end{align}
and encode the energy-resolved Green's function structure
after the band-basis projection.
Their explicit product forms and clean-limit expansion
are given in Appendix~\ref{app:kernel-expansion}.

\subsection{Band geometry in the velocity matrix}

The velocity vertex $V_{i} = \partial_{i}H/\hbar$ is
represented in the band basis by the matrix elements
$v^{nm}_{i} = \langle u_{n}|\hbar V_{i}|u_{m}\rangle$,
which decompose as
\begin{align}
  v^{nm}_{i}
  = u^{n}_{i}\,\delta_{nm}
  + i\varepsilon_{nm}\,\mathcal{A}^{nm}_{i},
  \label{eq:velocity-decomp}
\end{align}
where $u^{n}_{i} \equiv \partial_{i}\varepsilon_{n}$
is the band-energy gradient (the group velocity multiplied by
$\hbar$), and
$\varepsilon_{nm}\equiv\varepsilon_{n}-\varepsilon_{m}$
is the interband energy difference.
The off-diagonal part is governed by the Berry connection
\begin{align}
  \mathcal{A}^{nm}_{i}(\bm{k})
  = i\langle u_{n}(\bm{k})|\partial_{i}u_{m}(\bm{k})\rangle,
  \label{eq:berry-connection-main}
\end{align}
where $\mathcal{A}^{nn}_{i}$ is the diagonal Berry
connection and the off-diagonal elements
$\mathcal{A}^{nm}_{i}$ ($n\neq m$) enter the interband
optical matrix elements.

The second-rank velocity vertex
$V_{ij} = \partial_{i}\partial_{j}H/\hbar^2$ is
represented in the band basis by the matrix elements
$v^{nm}_{ij} = \langle u_{n}|\hbar^{2}V_{ij}|u_{m}\rangle$,
whose diagonal part is
\begin{align}
  v^{nn}_{ij}
  = u^{n}_{ij}
  - 2\sum_{p\neq n}
    \varepsilon_{np}\,\mathcal{G}^{np}_{ij},
  \label{eq:vnn-ij}
\end{align}
where
\begin{align}
  u^{n}_{ij} \equiv \partial_{i}\partial_{j}\varepsilon_{n}
  \label{eq:velocity-2-d-decomp}
\end{align}
is the second derivative of the band dispersion 
and $\mathcal{G}^{np}_{ij}$ is the
band-pair-resolved \textit{quantum metric} defined below. 
The off-diagonal part is 
\begin{align}
  v^{nm}_{ij}
  &= i u^{nm}_{i}\mathcal{A}^{nm}_{j}
   + i u^{nm}_{j}\mathcal{A}^{nm}_{i}
   + i\varepsilon_{nm}\,\mathcal{D}_{j}\mathcal{A}^{nm}_{i}
  \nonumber\\
  &\quad
  - \sum_{p\neq n,m}
    \bigl(
      \varepsilon_{np}\,\mathcal{A}^{np}_{i}\mathcal{A}^{pm}_{j}
    + \varepsilon_{mp}\,\mathcal{A}^{np}_{j}\mathcal{A}^{pm}_{i}
    \bigr),
  \label{eq:vnm-ij}
\end{align}
where
\begin{align}
  u^{nm}_{i}
  \equiv \partial_{i}\varepsilon_{nm}
  = \partial_{i}\varepsilon_{n} - \partial_{i}\varepsilon_{m},
  \label{eq:velocity-2-od-decomp}
\end{align}
and $\mathcal{D}_{j}\mathcal{A}^{nm}_{i}$ is the 
\textit{covariant
derivative} defined below.
Although Eq.~\eqref{eq:vnm-ij} is not manifestly symmetric
under $i\leftrightarrow j$, both orderings are equivalent
since they originate from
$U^{\dagger}\partial_{i}\partial_{j}H\,U$, which is
symmetric by commutativity of partial derivatives.

Here $\mathcal{G}^{nm}_{ij}$ in Eq.~\eqref{eq:vnn-ij} 
denotes the band-pair-resolved
quantum metric, defined as the real part of the
band-pair-resolved \textit{quantum geometric tensor}
\begin{align}
  \mathcal{Q}_{ij}^{nm}
  \equiv \mathcal{A}_{i}^{nm}\mathcal{A}_{j}^{mn}
  = \mathcal{G}_{ij}^{nm} - \frac{i}{2}\Omega_{ij}^{nm},
  \qquad n\neq m.
  \label{eq:pair-qgt}
\end{align}
where $\Omega_{ij}^{nm}$ is the band-pair-resolved 
\textit{Berry curvature}. 
Explicitly,
\begin{align}
  \mathcal{G}_{ij}^{nm}
  &= \mathrm{Re}\,\mathcal{Q}_{ij}^{nm}
   = \frac{1}{2}
     \bigl(
       \mathcal{A}_{i}^{nm}\mathcal{A}_{j}^{mn}
     + \mathcal{A}_{j}^{nm}\mathcal{A}_{i}^{mn}
     \bigr),
  \label{eq:quantum-metric}
  \\
  \Omega_{ij}^{nm}
  &= -2\,\mathrm{Im}\,\mathcal{Q}_{ij}^{nm}
   = i\bigl(
       \mathcal{A}_{i}^{nm}\mathcal{A}_{j}^{mn}
     - \mathcal{A}_{j}^{nm}\mathcal{A}_{i}^{mn}
     \bigr).
  \label{eq:berry-curvature-pair}
\end{align}
The band-resolved quantum metric and Berry curvature are
obtained by summing over partner bands,
\begin{align}
  \mathcal{G}_{ij}^{n}
  = \sum_{m\neq n}\mathcal{G}_{ij}^{nm},
  \qquad
  \Omega_{ij}^{n}
  = \sum_{m\neq n}\Omega_{ij}^{nm}.
  \label{eq:metric-curvature-band}
\end{align}

The covariant derivative appearing in Eq.~\eqref{eq:vnm-ij}
is defined as 
\begin{align}
  \mathcal{D}_{j}\mathcal{A}^{nm}_{i}
  \equiv \partial_{j}\mathcal{A}^{nm}_{i}
  + i\bigl(
      \mathcal{A}^{mm}_{j} - \mathcal{A}^{nn}_{j}
    \bigr)\mathcal{A}^{nm}_{i}.
  \label{eq:covariant-derivative}
\end{align}
Under the band-dependent gauge transformation
$|u_{n}\rangle\to e^{i\chi_{n}}|u_{n}\rangle$,
the off-diagonal Berry connection transforms as
\begin{align}
  \mathcal{A}^{nm}_{i}
  \to e^{i(\chi_{m}-\chi_{n})}\mathcal{A}^{nm}_{i},
  \qquad n\neq m,
  \label{eq:gauge-transformation-offdiag-A}
\end{align}
while the diagonal connections shift as
$\mathcal{A}^{nn}_{j}\to\mathcal{A}^{nn}_{j}
 - \partial_{j}\chi_{n}$.
The ordinary derivative $\partial_{j}\mathcal{A}^{nm}_{i}$
acquires an extra term $\propto\partial_{j}(\chi_{m}-\chi_{n})$
under this transformation, which is cancelled by the
connection difference in Eq.~\eqref{eq:covariant-derivative},
so that $\mathcal{D}_{j}\mathcal{A}^{nm}_{i}$ transforms
covariantly,
\begin{align}
  \mathcal{D}_{j}\mathcal{A}^{nm}_{i}
  \to e^{i(\chi_{m}-\chi_{n})}
      \mathcal{D}_{j}\mathcal{A}^{nm}_{i}.
  \label{eq:gauge-transformation-covariant-derivative}
\end{align}
Consequently, the band-pair-resolved \textit{quantum-connection} 
\begin{align}
  \mathcal{C}^{nm}_{ij|k}
  \equiv
  \mathcal{A}^{nm}_{i}\,\mathcal{D}_{k}\mathcal{A}^{mn}_{j}
  = \Gamma^{nm}_{ij|k} - i\widetilde{\Gamma}^{nm}_{ij|k}
  \label{eq:quantum-connection}
\end{align}
is gauge-invariant.
Although $\mathcal{D}_{k}\mathcal{A}^{mn}_{j}$ itself is
only gauge-covariant, carrying the phase
$e^{i(\chi_{n}-\chi_{m})}$, its product with
$\mathcal{A}^{nm}_{i}$, which carries the opposite phase
$e^{i(\chi_{m}-\chi_{n})}$, yields a band-closed object
whose phase factors cancel identically:
\begin{align}
  \mathcal{A}^{nm}_{i}\,\mathcal{D}_{k}\mathcal{A}^{mn}_{j}
  &\to
  e^{i(\chi_{m}-\chi_{n})}
  e^{i(\chi_{n}-\chi_{m})}
  \mathcal{A}^{nm}_{i}\,\mathcal{D}_{k}\mathcal{A}^{mn}_{j}
  \nonumber\\
  &= \mathcal{A}^{nm}_{i}\,\mathcal{D}_{k}\mathcal{A}^{mn}_{j}.
  \label{eq:quantum-connection-gauge-inv}
\end{align}
This gauge-invariant object provides a natural geometric
building block for nonlinear response.
Since $\sigma^{\mathrm{DC}}_{ijk}$ is a physical observable,
it must be gauge-invariant; the quantum connection
$\mathcal{C}^{nm}_{ij|k}$ is the appropriate
band-pair-resolved object that encodes the covariant
derivative of the Berry connection while remaining
gauge-invariant.
The real and imaginary parts are
\begin{align}
  \Gamma^{nm}_{ij|k}
  = \mathrm{Re}\,\mathcal{C}^{nm}_{ij|k},
  \qquad
  \widetilde{\Gamma}^{nm}_{ij|k}
  = -\mathrm{Im}\,\mathcal{C}^{nm}_{ij|k}.
  \label{eq:quantum-connection-real-imag}
\end{align}
The real and imaginary parts $\Gamma^{nm}_{ij|k}$ and
$\widetilde{\Gamma}^{nm}_{ij|k}$ are related to
momentum-space derivatives of the quantum metric and
Berry curvature; the relevant identities are collected in
Appendix~\ref{app:geometry}.

\subsection{Gauge-invariant decomposition of the dc conductivity}
\label{subsec:band-geom-decomp}

Substituting the band-basis velocity decompositions
Eqs.~\eqref{eq:velocity-decomp}, \eqref{eq:vnn-ij},
and \eqref{eq:vnm-ij} into
$\mathcal{B}_{ijk}$ [Eq.~\eqref{eq:Bijk-band}],
every contribution can be classified by the number of
distinct band indices involved.
We consider the closed three-band product and the
closed two-band product in turn.

The closed three-band product
$v^{nm}_{i}v^{ml}_{j}v^{ln}_{k}$ represents a
three-band interband process in which three one-photon
current vertices $V_{i}$, $V_{j}$, and $V_{k}$ connect
three distinct bands $n$, $m$, and $l$ in a closed cycle;
it corresponds to the triangular electron loop of
Fig.~\ref{fig:second_order_current_diagrams}(a).
The decomposition reads
\begin{align}
  &v_{i}^{nm}v_{j}^{ml}v_{k}^{ln}
  + v_{i}^{nm}v_{k}^{ml}v_{j}^{ln}
  \nonumber\\
  &= 2u_{i}^{n}u_{j}^{n}u_{k}^{n}
     \,\delta_{nm}\delta_{ml}\delta_{ln}
  \nonumber\\
  &\quad
  + \varepsilon_{nl}^{2}u_{i}^{n}
    \bigl(\mathcal{Q}^{nl}_{jk}+\mathcal{Q}^{nl}_{kj}\bigr)\delta_{nm}
  \nonumber\\
  &\quad
  + \varepsilon_{nm}^{2}
    \bigl(u_{j}^{m}\mathcal{Q}^{nm}_{ik}+u_{k}^{m}\mathcal{Q}^{nm}_{ij}\bigr)
    \delta_{ml}
  \nonumber\\
  &\quad
  + \varepsilon_{nm}^{2}
    \bigl(u_{j}^{n}\mathcal{Q}^{nm}_{ik}+u_{k}^{n}\mathcal{Q}^{nm}_{ij}\bigr)
    \delta_{ln}
  \nonumber\\
  &\quad
  - i\varepsilon_{nm}\varepsilon_{ml}\varepsilon_{ln}
    \bigl(\mathcal{T}^{nml}_{ijk}+\mathcal{T}^{nml}_{ikj}\bigr),
  \label{eq:vvv}
\end{align}
where
\begin{align}
  \mathcal{T}^{nml}_{ijk}
  \equiv
  \mathcal{A}_{i}^{nm}\mathcal{A}_{j}^{ml}\mathcal{A}_{k}^{ln}
  \label{eq:three-Berry-connection}
\end{align}
is the \textit{three-Berry-connection} factor, in which the band
indices form a closed cycle $n\to m\to l\to n$.
It is gauge-invariant because the three relative phases
$e^{i(\chi_{m}-\chi_{n})}$,
$e^{i(\chi_{l}-\chi_{m})}$, and
$e^{i(\chi_{n}-\chi_{l})}$ cancel in the product
[cf.~Eq.~\eqref{eq:gauge-transformation-offdiag-A}].
Physically, $\mathcal{T}^{nml}_{ijk}$ represents a
three-band interband process corresponding to the
triangular electron loop of
Fig.~\ref{fig:second_order_current_diagrams}(a).

Each line of Eq.~\eqref{eq:vvv} contributes to a distinct
sector of the dc conductivity.
The first line ($n=m=l$, purely intraband) contains the
product $2u^{n}_{i}u^{n}_{j}u^{n}_{k}$ of three group
velocities of band $n$ and generates the nonlinear Drude
sector $\sigma^{\mathrm{ND}}_{ijk}$.
The second through fourth lines each involve one intraband
and one interband pair, coupling a group velocity to the
quantum geometric tensor $\mathcal{Q}^{nm}_{ij}$; together
they contribute to the
quantum-geometric sector $\sigma^{\mathrm{QG}}_{ijk}$.
The fifth line ($n,m,l$ all distinct) is proportional to
the three-Berry-connection factor $\mathcal{T}^{nml}_{ijk}$
and defines the three-connection sector
$\sigma^{\mathcal{T}}_{ijk}$.

The closed two-band product $v^{nm}_{i}v^{mn}_{jk}$
represents a two-band process in which a one-photon
current vertex $V_{i}$ and a two-photon contact vertex
$V_{jk}$ connect the same band pair $(n,m)$; it
corresponds to the diagrams of
Figs.~\ref{fig:second_order_current_diagrams}(b) and (c).
The decomposition is
\begin{align}
  &v_{i}^{nm}v_{jk}^{mn}
  \nonumber\\
  &= \Bigl(
       u_{i}^{n}u_{jk}^{n}
     - 2u_{i}^{n}
       \sum_{p\neq n}\varepsilon_{np}\mathcal{G}^{np}_{jk}
     \Bigr)\delta_{nm}
  \nonumber\\
  &\quad
  + \varepsilon_{nm}(\partial_{j}\varepsilon_{nm})\,\mathcal{Q}^{nm}_{ik}
  + \varepsilon_{nm}(\partial_{k}\varepsilon_{nm})\,\mathcal{Q}^{nm}_{ij}
  \nonumber\\
  &\quad
  + \varepsilon_{nm}^{2}\,\mathcal{C}^{nm}_{ij|k}
  \nonumber\\
  &\quad
  - i\sum_{p\neq n,m}
    \Bigl(
      \varepsilon_{nm}\varepsilon_{mp}\,\mathcal{T}^{nmp}_{ijk}
    + \varepsilon_{nm}\varepsilon_{np}\,\mathcal{T}^{nmp}_{ikj}
    \Bigr).
  \label{eq:v-v2}
\end{align}
The first line ($\delta_{nm}$, intraband) collects the
single-band Drude term $u^{n}_{i}u^{n}_{jk}$, which
contributes to $\sigma^{\mathrm{ND}}_{ijk}$ together with
the first line of Eq.~\eqref{eq:vvv}, and the
quantum-metric correction
$-2u^{n}_{i}\sum_{p\neq n}\varepsilon_{np}\mathcal{G}^{np}_{jk}$
from the off-diagonal part of the diagonal second-rank
velocity [Eq.~\eqref{eq:vnn-ij}], which contributes to
$\sigma^{\mathrm{QG}}_{ijk}$.
The second line couples the gradient of the interband
energy difference $\partial_{j,k}\varepsilon_{nm}$ to the
quantum geometric tensor $\mathcal{Q}^{nm}_{ij}$ and also
contributes to $\sigma^{\mathrm{QG}}_{ijk}$.
The third line,
$\varepsilon_{nm}^{2}\mathcal{C}^{nm}_{ij|k}
 = \varepsilon_{nm}^{2}
   \mathcal{A}^{nm}_{i}\mathcal{D}_{k}\mathcal{A}^{mn}_{j}$,
is the quantum-connection term that defines the
covariant-connection sector $\sigma^{\mathcal{C}}_{ijk}$.
The fourth line generates an additional three-connection
contribution $\sigma^{\mathcal{T}}_{ijk}$ involving an
intermediate band $p$ ($p\neq n,m$).

Collecting all contributions, the dc conductivity
decomposes first as
\begin{align}
  \sigma^{\mathrm{DC}}_{ijk}
  = \sigma^{\mathrm{ND}}_{ijk}
  + \sigma^{\mathrm{QG}}_{ijk}
  + \sigma^{\mathcal{C}}_{ijk}
  + \sigma^{\mathcal{T}}_{ijk}.
  \label{eq:sigma-dc-intermediate}
\end{align}
The nonlinear Drude sector $\sigma^{\mathrm{ND}}_{ijk}$
originates from the single-band contributions and is
given by
\begin{align}
  \sigma^{\mathrm{ND}}_{ijk}
  &= \frac{e^{3}}{\hbar}
     \int_{-\infty}^{\infty}\frac{d\varepsilon}{2\pi}\,
     f(\varepsilon)
     \sum_{\bm{k}}
     \mathrm{Im}\!\left[\mathcal{B}^{\mathrm{ND}}_{ijk}\right],
  \label{eq:sigma-ND}
  \\
  \mathcal{B}^{\mathrm{ND}}_{ijk}
  &= \sum_{n}
     \biggl[
       2(\partial_{i}\varepsilon_{n})
       (\partial_{j}\varepsilon_{n})
       (\partial_{k}\varepsilon_{n})\,
       \mathcal{K}^{(1)}_{nnn}
  \nonumber\\
  &\qquad
     + (\partial_{i}\varepsilon_{n})
       (\partial_{j}\partial_{k}\varepsilon_{n})\,
       \mathcal{K}^{(2)}_{nn}
     \biggr].
  \label{eq:B-ND}
\end{align}
The quantum-geometric sector $\sigma^{\mathrm{QG}}_{ijk}$
originates from products involving the quantum geometric
tensor $\mathcal{Q}^{nm}_{ij}$ and is given by
\begin{align}
  \sigma^{\mathrm{QG}}_{ijk}
  &= \frac{e^{3}}{\hbar}
     \int_{-\infty}^{\infty}\frac{d\varepsilon}{2\pi}\,
     f(\varepsilon)
     \sum_{\bm{k}}
     \mathrm{Im}\!\left[\mathcal{B}^{\mathrm{QG}}_{ijk}\right],
  \label{eq:sigma-QG}
\end{align}
with
\begin{align}
  \mathcal{B}^{\mathrm{QG}}_{ijk}
  &= {\sum_{n,m}}^{\prime}
     \varepsilon_{nm}(\partial_{i}\varepsilon_{n})
     \bigl(\mathcal{Q}^{nm}_{jk} + \mathcal{Q}^{nm}_{kj}\bigr)
     \bigl[\varepsilon_{nm}\mathcal{K}^{(1)}_{nnm}
           - \mathcal{K}^{(2)}_{nn}\bigr]
  \nonumber\\
  &\quad
  + {\sum_{n,m}}^{\prime}
    \varepsilon_{nm}\mathcal{Q}^{nm}_{ik}
    \Bigl[
      (\partial_{j}\varepsilon_{m})
      \bigl(\varepsilon_{nm}\mathcal{K}^{(1)}_{nmm}
            - \mathcal{K}^{(2)}_{nm}\bigr)
  \nonumber\\
  &\qquad\quad
    + (\partial_{j}\varepsilon_{n})
      \bigl(\varepsilon_{nm}\mathcal{K}^{(1)}_{nmn}
            + \mathcal{K}^{(2)}_{nm}\bigr)
    \Bigr]
  \nonumber\\
  &\quad
  + {\sum_{n,m}}^{\prime}
    \varepsilon_{nm}\mathcal{Q}^{nm}_{ij}
    \Bigl[
      (\partial_{k}\varepsilon_{m})
      \bigl(\varepsilon_{nm}\mathcal{K}^{(1)}_{nmm}
            - \mathcal{K}^{(2)}_{nm}\bigr)
  \nonumber\\
  &\qquad\quad
    + (\partial_{k}\varepsilon_{n})
      \bigl(\varepsilon_{nm}\mathcal{K}^{(1)}_{nmn}
            + \mathcal{K}^{(2)}_{nm}\bigr)
    \Bigr].
  \label{eq:B-QG}
\end{align}
The imaginary part of $\mathcal{B}^{\mathrm{QG}}_{ijk}$
decomposes according to the real and imaginary parts of
the quantum geometric tensor
$\mathcal{Q}^{nm}_{ij}
 = \mathcal{G}^{nm}_{ij} - (i/2)\Omega^{nm}_{ij}$,
giving rise to the BCD and quantum-metric contributions
after the clean-limit kernel expansion.
The covariant-quantum-connection sector
$\sigma^{\mathcal{C}}_{ijk}$ arises from the
$\mathcal{C}=\mathcal{A}^{nm}_{i}\mathcal{D}_{k}\mathcal{A}^{mn}_{j}$ term in
Eq.~\eqref{eq:vnm-ij} and is given by
\begin{align}
  \sigma^{\mathcal{C}}_{ijk}
  &= \frac{e^{3}}{\hbar}
     \int_{-\infty}^{\infty}\frac{d\varepsilon}{2\pi}\,
     f(\varepsilon)
     \sum_{\bm{k}}
     \mathrm{Im}\!\left[\mathcal{B}^{\mathcal{C}}_{ijk}\right],
  \label{eq:sigma-C}
  \\
  \mathcal{B}^{\mathcal{C}}_{ijk}
  &= {\sum_{n,m}}^{\prime}
     \varepsilon_{nm}^{2}\,
     \mathcal{C}^{nm}_{ij|k}\,
     \mathcal{K}^{(2)}_{nm}.
  \label{eq:B-C}
\end{align}
Finally, the three-connection sector
$\sigma^{\mathcal{T}}_{ijk}$ originates from products of
three off-diagonal Berry connections $\mathcal{T}^{nml}_{ijk}=\mathcal{A}^{nm}_{i}\mathcal{A}_{j}^{ml}\mathcal{A}_{k}^{ln}$ and is given by
\begin{align}
  \sigma^{\mathcal{T}}_{ijk}
  &= \frac{e^{3}}{\hbar}
     \int_{-\infty}^{\infty}\frac{d\varepsilon}{2\pi}\,
     f(\varepsilon)
     \sum_{\bm{k}}
     \mathrm{Im}\!\left[\mathcal{B}^{\mathcal{T}}_{ijk}\right],
  \label{eq:sigma-T}
  \\
  \mathcal{B}^{\mathcal{T}}_{ijk}
  &= -i{\sum_{n,m,l}}^{\prime}
     \biggl\{
       \varepsilon_{nm}\varepsilon_{ml}\varepsilon_{ln}
       \mathcal{K}^{(1)}_{nml}
       \bigl(
         \mathcal{T}^{nml}_{ijk}
       + \mathcal{T}^{nml}_{ikj}
       \bigr)
  \nonumber\\
  &\quad
     + \varepsilon_{nm}\mathcal{K}^{(2)}_{nm}
       \bigl(
         \varepsilon_{ml}\mathcal{T}^{nml}_{ijk}
       - \varepsilon_{nl}\mathcal{T}^{nml}_{ikj}
       \bigr)
     \biggr\}.
  \label{eq:B-T}
\end{align}

\subsection{Clean-limit expansion of the kernels}
\label{subsec:clean-limit}

To evaluate $\sigma^{\mathrm{DC}}_{ijk}$ from
Eq.~\eqref{eq:sigma-dc-intermediate}, we expand the
Green's function kernels $\mathcal{K}^{(1)}_{nml}$ and
$\mathcal{K}^{(2)}_{nm}$ in the clean limit, retaining all
terms through $\mathcal{O}(\tau^{0})$ under the
non-degenerate band condition
$|\varepsilon_{nm}|\tau/\hbar\gg 1$ for $n\neq m$.
The expansion relies on the three algebraic identities
\begin{align}
  g^{\mathrm{R}}_{n}g^{\mathrm{A}}_{n}
  &= \frac{\tau}{i\hbar}\bigl(g^{\mathrm{A}}_{n}-g^{\mathrm{R}}_{n}\bigr),
  \label{eq:kernel-identity-RA}
  \\
  g^{\mathrm{R}}_{m}g^{\mathrm{R}}_{n}
  &= \frac{g^{\mathrm{R}}_{n}-g^{\mathrm{R}}_{m}}
          {\varepsilon_{nm}},
  \label{eq:kernel-identity-RR}
  \\
  g^{\mathrm{R}}_{m}g^{\mathrm{A}}_{n}
  &= \frac{g^{\mathrm{A}}_{n}-g^{\mathrm{R}}_{m}}
          {\varepsilon_{nm}+2i\gamma}
   = \frac{g^{\mathrm{A}}_{n}-g^{\mathrm{R}}_{m}}
          {\varepsilon_{nm}}
   + \mathcal{O}(\tau^{-1}),
  \label{eq:kernel-identity-RA-offdiag}
\end{align}
where Eq.~\eqref{eq:kernel-identity-RA} follows from
$g^{\mathrm{A}}_{n}-g^{\mathrm{R}}_{n}
 = 2i\gamma\,g^{\mathrm{R}}_{n}g^{\mathrm{A}}_{n}$
and $r = 1/(2i\gamma)$,
Eq.~\eqref{eq:kernel-identity-RR} is the partial-fraction
decomposition of two retarded propagators, and the last
equality in Eq.~\eqref{eq:kernel-identity-RA-offdiag}
holds because $2i\gamma/\varepsilon_{nm}
= \mathcal{O}(\tau^{-1})$ in the clean limit.

The lifetime hierarchy is determined by applying these
identities and expanding the kernels as distributions under
the energy integral.  The explicit factor $r$ is of order
$\tau$, whereas $g^{\mathrm{R/A}}_{n}$ and
$g^{\mathrm{A}}_{n}-g^{\mathrm{R}}_{n}$ have finite
distributional limits when integrated against a smooth test
function.  The $\mathcal{O}(\tau)$ peak height of the
spectral difference must not be identified with its
distributional order.  Collecting the explicit lifetime
factors together with the required subleading spectral terms
gives the clean-limit expansions of
$\mathcal{K}^{(1)}_{nml}$ and $\mathcal{K}^{(2)}_{nm}$.
The quantum-geometric sector $\sigma^{\mathrm{QG}}_{ijk}$
further separates as
\begin{align}
  \sigma^{\mathrm{QG}}_{ijk}
  = \sigma^{\mathrm{BCD}}_{ijk}
  + \sigma^{\mathcal{G}}_{ijk},
  \label{eq:sigma-QG-split}
\end{align}
where $\sigma^{\mathrm{BCD}}_{ijk} = \mathcal{O}(\tau)$
arises from the imaginary part $\Omega^{nm}_{ij}$ of
$\mathcal{Q}^{nm}_{ij}$, and
$\sigma^{\mathcal{G}}_{ijk} = \mathcal{O}(\tau^{0})$
arises from its real part $\mathcal{G}^{nm}_{ij}$.
As demonstrated in Sec.~\ref{sec:cancellation}, the
covariant-connection sector $\sigma^{\mathcal{C}}_{ijk}$
and the three-connection sector $\sigma^{\mathcal{T}}_{ijk}$
do not vanish individually but cancel their
connection-dependent parts, yielding a gauge-invariant
remainder.

For the fully diagonal kernels $\mathcal{K}^{(1)}_{nnn}$ and
$\mathcal{K}^{(2)}_{nn}$ the leading behavior is
$\mathcal{O}(\tau^{2})$, but the expansion also produces an
explicit $\mathcal{O}(\tau^{0})$ remainder.
Because the present expansion is claimed to be complete
through $\mathcal{O}(\tau^{0})$, this remainder lies within
the stated accuracy and must be evaluated on the same footing
as the interband $\mathcal{O}(\tau^{0})$ terms.
As shown in Appendix~\ref{app:kernel-expansion}, retaining
only the leading Dirac-delta limits of the spectral factors
is insufficient to extract this remainder.  Subleading terms
in these factors can combine with explicit powers of $\tau$
and contribute at $\mathcal{O}(\tau^{0})$ after integration
against a smooth Fermi function.  A systematic distributional
expansion must retain all such terms.  Here we evaluate the
diagonal kernels at finite broadening and use their convergent
moments to expand the energy integrals, making the
$\mathcal{O}(\tau^{0})$ term unambiguous.
The two $\mathcal{O}(\tau^{0})$ QMD contributions emerge
from the combination
\begin{align}
  \sigma^{\mathcal{G}}_{ijk}
  + \sigma^{\mathcal{C}}_{ijk}
  + \sigma^{\mathcal{T}}_{ijk}
  = \sigma^{\mathrm{intra\text{-}QMD}}_{ijk}
  + \sigma^{\mathrm{inter\text{-}QMD}}_{ijk}.
  \label{eq:QMD-from-cancellation}
\end{align}
The final gauge-invariant decomposition is therefore
\begin{align}
  \sigma^{\mathrm{DC}}_{ijk}
  = \sigma^{\mathrm{ND}}_{ijk}
  + \sigma^{\mathrm{BCD}}_{ijk}
  + \sigma^{\mathrm{intra\text{-}QMD}}_{ijk}
  + \sigma^{\mathrm{inter\text{-}QMD}}_{ijk},
  \label{eq:main-decomposition}
\end{align}
with the lifetime hierarchy
\begin{subequations}
\begin{widetext}
\begin{align}
  &\sigma^{\mathrm{ND},\tau^{2}}_{ijk}
  = \mathcal{O}(\tau^{2}), \\
  &\sigma^{\mathrm{BCD}}_{ijk}
  = \mathcal{O}(\tau), \\
  &\sigma^{\mathrm{ND},\tau^{0}}_{ijk}
  = \mathcal{O}(\tau^{0}),
  \quad
  \sigma^{\mathrm{intra\text{-}QMD}}_{ijk}
  = \mathcal{O}(\tau^{0}),
  \quad
  \sigma^{\mathrm{inter\text{-}QMD}}_{ijk}
  = \mathcal{O}(\tau^{0}). 
  \end{align}
\end{widetext}
  \label{eq:lifetime-hierarchy}
\end{subequations}
Three distinct contributions therefore appear at order
$\tau^{0}$.
Two of them, $\sigma^{\mathrm{intra\text{-}QMD}}_{ijk}$ and
$\sigma^{\mathrm{inter\text{-}QMD}}_{ijk}$, are governed
by derivatives of the quantum metric $\mathcal{G}^{nm}_{ij}$
and by its coupling to the interband energy difference
$\varepsilon_{nm}$; they remain finite in systems with
identically vanishing Berry curvature, for which the BCD
mechanism is absent.
The third, $\sigma^{\mathrm{ND},\tau^{0}}_{ijk}$, is the
subleading part of the fully diagonal sector and depends only
on the band dispersion; it carries no geometric content.
All three are independent of $\tau$, but this is independence
of the lifetime within a fixed dissipative closure and not
independence of the relaxation mechanism itself; the scope of
this statement is discussed in Sec.~\ref{sec:scope}.
The explicit expressions for each sector are given in
Sec.~\ref{sec:results}, and the cancellation proof is
presented in Sec.~\ref{sec:cancellation}.
The detailed derivations are collected in
Appendices~\ref{app:geometry}~and~\ref{app:kernel-expansion}
and in the Supplemental Material~\cite{SM}.

\section{Results}
\label{sec:results}

From this section onward we display the chemical potential
explicitly and write the Fermi--Dirac distribution as a
function of the absolute band energy,
$f(\varepsilon)
 =\{\exp[(\varepsilon-\mu)/k_{\mathrm{B}}T]+1\}^{-1}$;
in Sec.~\ref{sec:formalism} the same function was written with
its energy variable measured from the chemical potential, and
the two conventions differ only by the origin of the energy
axis.  We use the shorthand notation
$f_{n}\equiv f(\varepsilon_{n})$ and
$f_{n}^{(p)}\equiv\partial_{\varepsilon}^{p}f(\varepsilon)
 |_{\varepsilon=\varepsilon_{n}}$ for the distribution and its
energy derivatives evaluated at band $n$, with
$f_{n}^{\prime}\equiv f_{n}^{(1)}$,
$f_{n}^{\prime\prime}\equiv f_{n}^{(2)}$ and
$f_{n}^{\prime\prime\prime}\equiv f_{n}^{(3)}$, and define
the Brillouin-zone sum as
\begin{align}
  \sum_{\bm{k}}
  \equiv \int_{\mathrm{BZ}}\frac{d^{d}\bm{k}}{(2\pi)^{d}}.
  \label{eq:BZ-sum}
\end{align}
The terms in the clean-limit decomposition
Eq.~\eqref{eq:main-decomposition} carry the lifetime
hierarchy summarized in Eq.~\eqref{eq:lifetime-hierarchy},
in which the nonlinear Drude sector contributes at both
$\mathcal{O}(\tau^{2})$ and $\mathcal{O}(\tau^{0})$.
This dc transport decomposition is the velocity-gauge
Keldysh counterpart of the general quantum-geometric
organization emphasized in projector-calculus
approaches~\cite{Mitscherling2025ProjectorCalculus,Avdoshkin2025Multistate}: the final observable is expressed
in terms of gauge-invariant band energies and band-geometric
quantities.  The explicit three-connection structures cancel
after the appropriate sectors are combined.
A closely related lifetime hierarchy and the separation into
ND, BCD, intraband-QMD, and interband-QMD sectors were
obtained recently in a projector-based Green's function
calculation~\cite{Ulrich2026IntraQMD}, whose published
version also retains the subleading dispersive
$\mathcal{O}(\tau^{0})$ contribution discussed below;
the comparison of signs and coefficients is given in
Sec.~\ref{subsec:comparison-dc-limit}.

\subsection{\texorpdfstring{Nonlinear Drude contribution: $\mathcal{O}(\tau^{2})$ and $\mathcal{O}(\tau^{0})$}{Nonlinear Drude contribution: O(tau2) and O(tau0)}}
\label{subsec:ND-contrib}

The fully diagonal sector contributes at two orders of the
lifetime expansion.
Using the diagonal-kernel integrals derived in
Appendix~\ref{app:kernel-expansion},
\begin{align}
  \int\!\frac{d\varepsilon}{2\pi}\,f(\varepsilon)\,
  \mathrm{Im}\,\mathcal{K}^{(2)}_{nn}
  &= -\frac{\tau^{2}}{\hbar^{2}}\,f_{n}^{\prime}
   + \frac{1}{24}\,f_{n}^{\prime\prime\prime}
   + \mathcal{O}(\tau^{-1}),
  \label{eq:K2-eint}
  \\
  \int\!\frac{d\varepsilon}{2\pi}\,f(\varepsilon)\,
  \mathrm{Im}\,\mathcal{K}^{(1)}_{nnn}
  &= -\frac{\tau^{2}}{2\hbar^{2}}\,f_{n}^{\prime\prime}
   + 0 + \mathcal{O}(\tau^{-1}),
  \label{eq:K1-eint}
\end{align}
the nonlinear Drude sector splits as
\begin{align}
  \sigma^{\mathrm{ND}}_{ijk}
  = \sigma^{\mathrm{ND},\tau^{2}}_{ijk}
  + \sigma^{\mathrm{ND},\tau^{0}}_{ijk}.
  \label{eq:ND-split}
\end{align}
The most divergent clean-limit contribution is the
nonlinear Drude term,
\begin{align}
  \sigma_{ijk}^{\mathrm{ND},\tau^{2}}
  = -\frac{e^{3}\tau^{2}}{\hbar^{3}}
    \sum_{\bm{k}}\sum_{n}
    f_{n}\,\partial_{i}\partial_{j}\partial_{k}\varepsilon_{n}.
  \label{eq:ND}
\end{align}
This is a purely intraband, single-band contribution.
After integration by parts over the Brillouin zone,
Eq.~\eqref{eq:ND} is equivalent to the
constant-relaxation-time Boltzmann
result~\cite{Deyo2009,Moore2010}.
This term scales as $\tau^{2}$ and gives the leading
clean-limit contribution whenever it is allowed by
symmetry; since it depends only on the band dispersion,
it is not a quantum-geometric contribution.

The subleading part of the same sector is the kinetic
contribution
\begin{align}
  \sigma_{ijk}^{\mathrm{ND},\tau^{0}}
  &= -\frac{e^{3}}{24\hbar}
    \sum_{\bm{k}}\sum_{n}
    f_{n}^{\prime\prime}\,
    \partial_{i}\partial_{j}\partial_{k}\varepsilon_{n}
  \nonumber\\
  &= -\frac{e^{3}}{48\hbar}
    \sum_{\bm{k}}\sum_{n}
    u_{i}^{n}u_{j}^{n}u_{k}^{n}\,f_{n}^{(4)},
  \label{eq:ND-tau0}
\end{align}
where $u_{i}^{n}\equiv\partial_{i}\varepsilon_{n}$ and the
second equality follows from two integrations by parts.
Equation~\eqref{eq:ND-tau0} is independent of $\tau$ and
therefore contributes at the same order as the two
quantum-metric-dipole terms of
Sec.~\ref{subsec:QMD-contrib}, but it is controlled entirely
by the band dispersion and carries no geometric content.
It is a Fermi-surface quantity in the sense that
$f_{n}^{\prime\prime}$ and $f_{n}^{(4)}$ are peaked within a
few $k_{\mathrm{B}}T$ of the chemical potential; the
large-$\tau$ expansion that produces it requires
$\gamma\ll k_{\mathrm{B}}T$, that is,
$\hbar/\tau\ll k_{\mathrm{B}}T$, in addition to the
non-degeneracy condition Eq.~\eqref{eq:clean-limit-condition}.

This kinetic term also appears in
Refs.~\cite{Ulrich2026IntraQMD,Guo2026Dissipation,Anan2026Nonreciprocal,Kitamura2026QuasiequilibriumShift}.
Section~\ref{subsec:comparison-dc-limit} compares these formulations;
Sec.~\ref{sec:oneband} verifies the present result against
an exact solution of the same dissipative model.

It is natural to ask why no analogous kinetic term appears in
the linear dc conductivity.  The relevant comparison is between
the combined linear-response kernel
$\mathrm{Re}\,\mathcal{K}_{nn}$ and the second-order kernel
$\mathrm{Im}\,\mathcal{K}^{(2)}_{nn}$.
Both are odd in the shifted energy variable, but their
broadening prefactors differ by one power of $\gamma^{-1}$.
The convergent moment controlling $\mathcal{O}(\tau^{0})$
is therefore the $p=2$ moment in linear response, which
vanishes by parity, and the $p=3$ moment at second order,
which equals $\pi/2$ for the scaled kernel
$\gamma^{4}\mathrm{Im}\,\mathcal{K}^{(2)}_{nn}$.
The latter produces the $f_n^{\prime\prime\prime}/24$ term
in Eq.~\eqref{eq:K2-eint}.
The other second-order diagonal kernel,
$\mathrm{Im}\,\mathcal{K}^{(1)}_{nnn}$, is even; its
$p=4$ moment vanishes by explicit evaluation.
The exact kernels, their moments, and the cancellation of the
Hadamard finite-part contributions in linear response are given
at the end of Appendix~\ref{app:diagonal-kernels} and in the
Supplemental Material~\cite{SM}.

\subsection{\texorpdfstring{Berry-curvature-dipole contribution: $\mathcal{O}(\tau)$}
{Berry-curvature-dipole contribution: O(tau)}}

The next term in the lifetime expansion is the
Berry-curvature-dipole contribution,
\begin{align}
  \sigma_{ijk}^{\mathrm{BCD}}
  = \frac{e^{3}\tau}{2\hbar^{2}}
    \sum_{\bm{k}}\sum_{n,l}
    f_{n}
    \bigl[
      \epsilon_{ijl}\,D^{n}_{kl}
    + \epsilon_{ikl}\,D^{n}_{jl}
    \bigr],
  \label{eq:BCD}
\end{align}
where
\begin{align}
  D^{n}_{al} \equiv \partial_{a}\Omega^{n}_{l},
  \qquad
  \Omega^{n}_{ab} = \sum_{l}\epsilon_{abl}\Omega^{n}_{l}
  \label{eq:BCD-defs}
\end{align}
is the Berry-curvature dipole density and Berry curvature
of band $n$, respectively.
Equation~\eqref{eq:BCD} is the standard Berry-curvature-dipole
nonlinear Hall response of Sodemann and
Fu~\cite{SodemannFu2015}, with early material and transport
extensions discussed in
Refs.~\cite{Low2015TopologicalCurrents,NandySodemann2019,DuLuXie2021Review}.
It originates from the imaginary part of the quantum
geometric tensor, namely the Berry curvature, combined with
the retarded--advanced pinch singularity in the clean-limit
kernel, and scales linearly with $\tau$.
Representative experimental realizations include bilayer
WTe$_{2}$~\cite{MaSong2019,Kang2019}.
In the language of quantum state geometry,
Eq.~\eqref{eq:BCD} is the dc transport realization of a
Berry-curvature dipole, which also appears as a
gauge-invariant projector derivative in the
projector-calculus
classification~\cite{Mitscherling2025ProjectorCalculus}.
Time-reversal symmetry may be present, whereas inversion
symmetry alone forces the Brillouin-zone integral of the
BCD to vanish.

\subsection{\texorpdfstring{Quantum-metric-dipole contributions: $\mathcal{O}(\tau^{0})$}{Quantum-metric-dipole contributions: O(tau0)}}
\label{subsec:QMD-contrib}

The geometric part of the $\mathcal{O}(\tau^{0})$ response
is controlled by the quantum metric.
Separating this part into an intraband Fermi-surface term and an
interband term, written as a Fermi-sea term before integration by parts, clarifies the relation to previous
work on quantum-metric-induced nonlinear
transport~\cite{GaoYangNiu2014,Wang2021CuMnAs,Liu2021AFM,Das2023,KaplanHolderYan2024,Michishita2022,JiangHolderYan2025Review,Qiang2025Clarification,Ulrich2026IntraQMD} and to the projector-calculus
organization of quantum metric dipoles~\cite{Mitscherling2025ProjectorCalculus,Avdoshkin2025Multistate}.

The intraband contribution is the Fermi-surface
quantum-metric dipole
\begin{align}
  \sigma_{ijk}^{\mathrm{intra\text{-}QMD}}
  = \frac{e^{3}}{2\hbar}
    \sum_{\bm{k}}\sum_{n}
    f_{n}^{\prime}\,\partial_{i}\mathcal{G}_{jk}^{n}.
  \label{eq:intraQMD-band}
\end{align}
At zero temperature, $f_{n}^{\prime} = -\delta(\varepsilon_{n}-\mu)$,
so Eq.~\eqref{eq:intraQMD-band} is explicitly a Fermi-surface
dipole of the ordinary band quantum metric.
The same intraband quantum-metric-dipole structure has been
emphasized in Ref.~\cite{Ulrich2026IntraQMD}.
The contribution of the present work is to obtain
Eq.~\eqref{eq:intraQMD-band} directly from the
velocity-gauge Keldysh--Kubo formula as part of the same
clean-limit expansion that produces the nonlinear Drude
and BCD terms, and to show that $\mathcal{G}_{jk}^{nm}$
is the real, metric part of the pair-resolved quantum
geometric tensor $\mathcal{Q}^{nm}_{jk}$.

The interband contribution is written, before integration by parts, as a Fermi-sea term
in terms of the band-normalized metric
\begin{align}
  \widetilde{\mathcal{G}}_{ij}^{n}
  \equiv 2\sum_{m\neq n}
         \frac{\mathcal{G}_{ij}^{nm}}{\varepsilon_{nm}},
  \label{eq:normalized-metric}
\end{align}
as
\begin{align}
  \sigma_{ijk}^{\mathrm{inter\text{-}QMD}}
  &= -\frac{e^{3}}{\hbar}
     \sum_{\bm{k}}\sum_{n}
     f_{n}
     \bigg[
       \partial_{i}\widetilde{\mathcal{G}}_{jk}^{n}
     - \frac{1}{2}
       \bigl(
         \partial_{j}\widetilde{\mathcal{G}}_{ki}^{n}
       + \partial_{k}\widetilde{\mathcal{G}}_{ij}^{n}
       \bigr)
     \bigg].
  \label{eq:interQMD-band}
\end{align}
This interband quantum-metric contribution is related to
the gauge-invariant organization of Fermi-sea-type metric
responses and to the clarification of connection-dependent
terms discussed in
Refs.~\cite{Qiang2025Clarification,Mitscherling2025ProjectorCalculus}.
In the present derivation, Eq.~\eqref{eq:interQMD-band}
is obtained together with Eq.~\eqref{eq:intraQMD-band}
from the cancellation mechanism demonstrated in
Sec.~\ref{sec:cancellation}.

Equivalently, Eq.~\eqref{eq:interQMD-band} can be
written in the pair-resolved form
\begin{align}
  \sigma_{ijk}^{\mathrm{inter\text{-}QMD}}
  &= \frac{e^{3}}{\hbar}
     \sum_{\bm{k}}{\sum_{n,m}}^{\prime}
     f_{n}
     \bigg[
       -2\partial_{i}
       \!\left(\frac{\mathcal{G}_{jk}^{nm}}{\varepsilon_{nm}}\right)
  \nonumber\\
  &\quad
     + \partial_{j}
       \!\left(\frac{\mathcal{G}_{ki}^{nm}}{\varepsilon_{nm}}\right)
     + \partial_{k}
       \!\left(\frac{\mathcal{G}_{ij}^{nm}}{\varepsilon_{nm}}\right)
     \bigg].
  \label{eq:interQMD-pair}
\end{align}
No Berry-connection commutator survives in
Eq.~\eqref{eq:interQMD-pair}; the only geometric input
is the gauge-invariant metric $\mathcal{G}_{ij}^{nm}$.
The energy denominator $\varepsilon_{nm}^{-1}$ and the
Fermi factor $f_{n}$ encode the Fermi-sea character of
this representation and distinguish it from the Fermi-surface
intraband QMD in Eq.~\eqref{eq:intraQMD-band}.

\subsection{Comparison with recent dc-limit formulas}
\label{subsec:comparison-dc-limit}

It is useful to rewrite
Eqs.~\eqref{eq:ND}, \eqref{eq:BCD},
\eqref{eq:intraQMD-band}, and \eqref{eq:interQMD-pair}
in the zero-temperature Fermi-surface form used in the
projector-based Green's function derivation of
Ref.~\cite{Ulrich2026IntraQMD}.
Let $\xi_{n}\equiv\varepsilon_{n}-\mu$,
$u_{i}^{n}\equiv\partial_{i}\varepsilon_{n}$, and
$u_{jk}^{n}\equiv\partial_{j}\partial_{k}\varepsilon_{n}$.
Using $f_{n} = \theta(-\xi_{n})$,
$f_{n}^{\prime} = -\delta(\xi_{n})$, and integration by
parts, we reproduce the Fermi-surface expression of
Eq.~(S109) of Ref.~\cite{Ulrich2026IntraQMD} in our notation.
To keep it distinct from the complete dc response
$\sigma^{\mathrm{DC}}_{ijk}$ of
Eq.~\eqref{eq:main-decomposition}, which also contains the
kinetic term, we denote the quoted expression by
$\widetilde{\sigma}^{\mathrm{DC}}_{ijk}$; the kinetic
contribution displayed separately in that reference is
discussed immediately below:
\begingroup\small
\begin{align}
  \widetilde{\sigma}_{ijk}^{\mathrm{DC}}
  &= \frac{e^{3}}{\hbar}
     \sum_{\bm{k}}\sum_{n}
     \delta(\xi_{n})
  \bigg\{
    -\frac{\tau^{2}}{\hbar^{2}}
     u_{i}^{n}u_{jk}^{n}\nonumber\\
     &\quad
  + \frac{\tau}{2\hbar}
    \bigl(u_{k}^{n}\Omega_{ij}^{n}+u_{j}^{n}\Omega_{ik}^{n}\bigr)
  \nonumber\\
  &\quad
  - {\sum_{m}}^{\prime}
    \biggl(
      2u_{i}^{n}\frac{\mathcal{G}_{jk}^{nm}}{\varepsilon_{nm}}
    - u_{k}^{n}\frac{\mathcal{G}_{ij}^{nm}}{\varepsilon_{nm}}
    - u_{j}^{n}\frac{\mathcal{G}_{ik}^{nm}}{\varepsilon_{nm}}
    \biggr)
  - \frac{1}{2}\partial_{i}\mathcal{G}_{jk}^{n}
  \bigg\}.
  \label{eq:dc-limit-ulrich-form}
\end{align}
\endgroup
Equation~\eqref{eq:dc-limit-ulrich-form} is the direct
counterpart of Eq.~(S109) of Ref.~\cite{Ulrich2026IntraQMD}.
The equality is exact after the following convention
translation: Ref.~\cite{Ulrich2026IntraQMD} uses the
index order $\sigma^{a;bc}$, corresponding to our
$\sigma_{abc}$; its band gap is
$E_{m}-E_{n} = -\varepsilon_{nm}$; and its off-diagonal
two-state metric satisfies
$g_{mn}^{bc} = \mathrm{Re}\,Q_{mn}^{bc} = -\mathcal{G}_{bc}^{nm}$,
differing by a sign from our positive pair metric.
With these identifications, the coefficients
$-\tau^{2}/\hbar^{2}$ (ND),
$\tau/(2\hbar)$ (BCD),
$(-2,+1,+1)$ (interband-QMD), and
$-1/2$ (intraband-QMD) in
Eq.~\eqref{eq:dc-limit-ulrich-form} all agree with
Ref.~\cite{Ulrich2026IntraQMD}.
The remaining difference between the two presentations is
not a sign or numerical discrepancy but a difference in the
organization between Fermi-sea-type and Fermi-surface terms
and between Berry-connection and projector conventions.

Equation~\eqref{eq:dc-limit-ulrich-form} displays only the
leading $\mathcal{O}(\tau^{2})$ part of the diagonal sector;
the kinetic term Eq.~\eqref{eq:ND-tau0} must be added to it
for the response to be complete through
$\mathcal{O}(\tau^{0})$.
The published version of Ref.~\cite{Ulrich2026IntraQMD}
includes this contribution in its Supplemental Material.
Through order $\Gamma^{0}$, the expansion of the
trivial-geometry coefficient multiplying
$E^{abc}_{n}\,\mathrm{tr}[\hat{P}_{n}]$ gives
$\frac{1-2\theta(-E_{n})}{8\Gamma^{2}}
 +\frac{1}{24}\delta^{\prime}(E_{n})$,
where $E^{abc}_{n}=\partial_{a}\partial_{b}\partial_{c}E_{n}$.
The expansion example in Sec.~S2.3 of that reference displays
both terms.  Its Eq.~(S74) retains only the leading dispersive
contribution, while Eqs.~(S75)--(S80) derive the kinetic term
separately.  The first term gives the leading nonlinear Drude
response, and the second gives the subleading $\tau^{0}$
kinetic contribution.  The latter is treated separately because
it does not mix with the quantum geometric invariants, and its
presence is also stated in the main text of the Letter.
At zero temperature,
$f_n^{\prime\prime}=-\delta^{\prime}(\varepsilon_n-\mu)$,
so Eq.~\eqref{eq:ND-tau0} becomes
$\sigma^{\mathrm{ND},\tau^{0}}_{ijk}
 =(e^{3}/24\hbar)\sum_{\bm{k}}\sum_n
 \delta^{\prime}(\varepsilon_n-\mu)
 \partial_i\partial_j\partial_k\varepsilon_n$.
After matching the conventions, this agrees with the published
expression in both coefficient and sign for nondegenerate bands,
for which $\mathrm{tr}[\hat{P}_n]=1$%
\footnote{The following three Fermi-surface forms are equivalent
through integration by parts over the periodic Brillouin zone:
\begin{align*}
\sum_{\bm{k}}\delta^{\prime}(\xi_n)\,
 \partial_a\partial_b\partial_c\varepsilon_n
 &=-\sum_{\bm{k}}\delta^{\prime\prime}(\xi_n)\,
 u_a^n\partial_b\partial_c\varepsilon_n\\
 &=\frac12\sum_{\bm{k}}\delta^{\prime\prime\prime}(\xi_n)\,
 u_a^n u_b^n u_c^n,
\end{align*}
where $\xi_n=\varepsilon_n-\mu$ and $u_a^n=\partial_a\varepsilon_n$.
The last expression is the zero-temperature velocity-product
form of the kinetic contribution in
Ref.~\cite{Guo2026Dissipation}.}. 
The four displayed coefficients in
Eq.~\eqref{eq:dc-limit-ulrich-form} are unchanged by the
addition of the kinetic contribution.  Including this term on
both sides gives agreement of the complete responses through
order $\tau^{0}$ after the convention translation above.
The complete $\mathcal{O}(\tau^{0})$ response contains the two
quantum-metric-dipole terms and the kinetic term.

Each of the following recent works provides a related formulation.
Reference~\cite{Guo2026Dissipation} derives the second-order
conductivity for a system coupled to a wide-band fermionic bath
and identifies a purely kinetic $\Gamma^{0}$ contribution
proportional to $v_i v_j v_k f^{(4)}$.
This corresponds to the velocity-product form of
Eq.~\eqref{eq:ND-tau0}.
Reference~\cite{Anan2026Nonreciprocal} obtains the same kinetic
structure in a constant-self-energy Keldysh calculation and
explains how coupling to a fermionic bath with a constant
density of states gives rise to the assumed self-energy.
These are the same self-energy assumptions specified explicitly
in Sec.~\ref{sec:scope}.
Reference~\cite{Kitamura2026QuasiequilibriumShift} establishes
the equivalence of the kinetic term across these formulations
and emphasizes in its concluding discussion that independence
from the relaxation time $\tau$ does not imply independence
from the relaxation mechanism.

The corresponding single-band check is presented in
Sec.~\ref{sec:oneband}.

The review by Jiang, Holder, and
Yan~\cite{JiangHolderYan2025Review} collects the
nonlinear-transport formulas in the convention
$j^{c} = \sigma^{ab;c}E^{a}E^{b}$.
Under the relabeling $(a,b;c)\to(j,k;i)$, the ND and
BCD formulas of that reference have the same signs and
coefficients as Eqs.~\eqref{eq:ND} and \eqref{eq:BCD}.
For the metric sector, the same review writes a compact
band-normalized QMD formula
\begin{align}
  \sigma_{\mathrm{QMD}}^{ab;c}
  &= -\frac{e^{3}}{\hbar}\sum_{\bm{k}}\sum_{n}
     f_{n}
     \biggl[
       2\partial_{c}G_{n}^{ab}
     - \frac{1}{2}\bigl(
         \partial_{a}G_{n}^{bc}
       + \partial_{b}G_{n}^{ac}
       \bigr)
     \biggr],
  \label{eq:JHY-QMD-compact}
\end{align}
where $G_{n}^{ab}$ is the band-normalized quantum
metric, corresponding to our $\widetilde{\mathcal{G}}_{ab}^{n}$
in Eq.~\eqref{eq:normalized-metric}.
Equation~\eqref{eq:JHY-QMD-compact} should therefore not
be identified term by term with only our interband formula
Eq.~\eqref{eq:interQMD-band}: our result separates the
$\mathcal{O}(\tau^{0})$ sector into the interband term with
coefficients $(-1, +1/2, +1/2)$ in the band-normalized
notation and the additional Fermi-surface intraband-QMD
term Eq.~\eqref{eq:intraQMD-band}.
The coefficient difference between Eq.~\eqref{eq:JHY-QMD-compact}
and Eq.~\eqref{eq:interQMD-band} reflects the sensitivity
of the QMD sector that has been discussed in recent
comparisons~\cite{JiangHolderYan2025Review,Qiang2025Clarification,Ulrich2026IntraQMD}; the projector-based result of
Ref.~\cite{Ulrich2026IntraQMD} fixes the convention
adopted in the present paper.

Two earlier non-projector treatments deserve a separate comparison, since
they address the same response by Green's function methods in which no
projector decomposition is introduced.
Michishita and Nagaosa~\cite{Michishita2022} analyze the nonlinear
response at finite dissipation using Green's functions and explicitly
identify Berry-curvature-dipole and quantum-metric-dipole contributions
(Appendix~D, in particular Eqs.~(D11)--(D14) of that reference).
Their treatment includes smeared and dissipative geometric quantities
associated with finite broadening.  The present formulation makes the
Peierls contact vertices explicit and displays the term-by-term
cancellation of the connection-dependent structures between
$\sigma^{\mathcal{C}}_{ijk}$ and $\sigma^{\mathcal{T}}_{ijk}$
(Sec.~\ref{sec:cancellation}), organizing the resulting dc response
by powers of $\tau$ through $\mathcal{O}(\tau^{0})$.
Jo\~{a}o and Viana Parente Lopes~\cite{JoaoLopes2020} develop
a basis-independent spectral framework for nonlinear optical
response in arbitrary noninteracting tight-binding models,
using the velocity gauge and Keldysh Green's functions.
Their framework already includes the higher-order interaction
and current vertices generated by the Peierls substitution
and organizes their contributions diagrammatically.
It does not, however, develop the band-basis geometric
decomposition and large-$\tau$ dc hierarchy considered here.
Our finite-frequency expression,
Eq.~\eqref{eq:sigma2-finite-frequency}, agrees with theirs
after translating the vertex and spectral conventions in
Eqs.~\eqref{eq:vertex-convention-comparison} and
\eqref{eq:spectral-convention-comparison}.
This agreement is a consistency check of the finite-frequency formula.

The connection cancellation is also explicit in the published
Supplemental Material of Ref.~\cite{Ulrich2026IntraQMD}:
Eq.~(S97) and the discussion immediately following it identify
the cancellation with the connection terms in Eq.~(S83)
of that reference.
The present derivation shows how the corresponding
cancellation arises from the velocity-gauge vertices in the
band basis without using projectors.
These comparisons distinguish the established geometric and
kinetic contributions from the explicit relation between
the Peierls vertices, the band-basis connection cancellation,
and the lifetime hierarchy obtained in the present formulation.

Microscopic Keldysh treatments of nonlinear dc conductivity
provide a complementary
perspective~\cite{MichishitaPeters2021,Nakazawa2025Interband}.
In particular, Nakazawa \textit{et al.}~\cite{Nakazawa2025Interband}
derive a dc second-order Keldysh formula in which the
potentially divergent $\omega_{1}^{-1}\omega_{2}^{-1}$
terms cancel before the limit $\omega_{1},\omega_{2}\to 0$
is taken.
In the clean limit, their intraband term reduces to the
Fermi-surface form of our nonlinear Drude
Eq.~\eqref{eq:dc-limit-ulrich-form}, and their BCD term
has the same sign as ours after symmetrization.
The quantum-metric contribution in the dirty limit of
Ref.~\cite{Nakazawa2025Interband} scales as $\tau^{2}$
under $|\varepsilon_{nm}|\tau\ll 1$ and should not be
identified with the clean-limit $\mathcal{O}(\tau^{0})$
QMD terms; the metric sector is sensitive to the
scattering model and to the order in which the dc and
clean limits are taken~\cite{JiangHolderYan2025Review,Ulrich2026IntraQMD,Nakazawa2025Interband}.

\subsection{Relation to projector calculus and optical nonlinear responses}

The structures in Eqs.~\eqref{eq:BCD},
\eqref{eq:intraQMD-band}, and \eqref{eq:interQMD-band}
are naturally interpreted in the language of quantum state
geometry.
Gauge-invariant projector calculus organizes the quantum
metric, Berry curvature, and their dipoles in terms of
band projectors and their momentum
derivatives~\cite{Mitscherling2025ProjectorCalculus},
and has been applied to polarization and optical
injection/shift
currents~\cite{Avdoshkin2025Multistate,Mitscherling2025ProjectorCalculus},
where resonant interband denominators select two-state
geometric tensors and quantum geometric connections.
The present result addresses a different limit: the
metallic second-order dc conductivity, in which the
clean-limit Green's function kernels separate the response
by powers of $\tau$.
The present formulas should therefore be viewed as a
dc Keldysh--transport realization of the same
gauge-invariant geometric principle, with the additional
information of lifetime scaling and Fermi-surface versus
Fermi-sea-type weighting before integration by parts.

The explicit formulas in
Eqs.~\eqref{eq:ND}--\eqref{eq:interQMD-band} separate the
dispersive and geometric contributions.
In particular, Eq.~\eqref{eq:intraQMD-band} permits a finite
intraband-QMD contribution when the Berry curvature vanishes
but the Fermi-surface quantum-metric dipole is nonzero.

\subsection{Cancellation of connection-dependent terms}
\label{sec:cancellation}

The compact QMD expressions in
Eqs.~\eqref{eq:intraQMD-band} and \eqref{eq:interQMD-band}
are not obtained term by term.
In the band-basis expansion of the velocity-gauge Keldysh
kernel, the geometric $\mathcal{O}(\tau^{0})$ sector first
contains explicit connection-dependent structures.
Their reduction to metric derivatives is the dc-transport
counterpart of the organization of geometric invariants in
projector calculus~\cite{Mitscherling2025ProjectorCalculus,Avdoshkin2025Multistate}.
Ulrich \textit{et al.}\ explicitly organize the same
cancellation in projector language~\cite{Ulrich2026IntraQMD};
here we carry it out term by term in the velocity-gauge
band basis.

The first connection-dependent structure is the
covariant-quantum-connection tensor
\begin{align}
  \mathcal{C}_{ij|k}^{nm}
  \equiv \mathcal{A}_{i}^{nm}\mathcal{D}_{k}\mathcal{A}_{j}^{mn},
  \label{eq:CalC-def}
\end{align}
and the second is the three-Berry-connection tensor
\begin{align}
  \mathcal{T}_{ijk}^{nml}
  \equiv \mathcal{A}_{i}^{nm}\mathcal{A}_{j}^{ml}\mathcal{A}_{k}^{ln}.
  \label{eq:Tcal-def}
\end{align}
Both closed objects are individually invariant under
bandwise $U(1)$ phase transformations, as shown in
Eqs.~\eqref{eq:quantum-connection-gauge-inv} and
\eqref{eq:three-Berry-connection}.  The cancellation below
removes the explicit three-connection structures; it does not
establish band-phase invariance for the first time.
This algebraic reduction is distinct from the electromagnetic
gauge consistency enforced by the Peierls contact vertices.

The covariant-quantum-connection sector contains a metric-derivative
part and a connection-dependent commutator part.
Defining
\begin{align}
  X_{ijk}^{nm}
  &\equiv
  \mathrm{Im}\!\left[
    \mathcal{A}_{i}^{nm}[\mathcal{A}_{j},\mathcal{A}_{k}]'_{mn}
  \right]
  + \mathrm{Im}\!\left[
    \mathcal{A}_{j}^{nm}[\mathcal{A}_{k},\mathcal{A}_{i}]'_{mn}
  \right]
  \nonumber\\
  &\quad
  - \mathrm{Im}\!\left[
    \mathcal{A}_{k}^{nm}[\mathcal{A}_{i},\mathcal{A}_{j}]'_{mn}
  \right],
  \label{eq:X-comm-main}
\end{align}
where the prime excludes the external bands from the
intermediate-band sum, the covariant-quantum-connection contribution
is 
\begingroup\small
\begin{align}
  \sigma_{ijk}^{\mathcal{C}}
  &= \frac{e^{3}}{\hbar}
     \sum_{\bm{k}}{\sum_{n,m}}^{\prime}
     \frac{f_{n}}{\varepsilon_{nm}}
     \bigl(
       \partial_{k}\mathcal{G}_{ij}^{nm}
     + \partial_{j}\mathcal{G}_{ki}^{nm}
     - \partial_{i}\mathcal{G}_{jk}^{nm}
     + X_{ijk}^{nm}
     \bigr),
  \label{eq:CalC-main}
\end{align}
\endgroup
and the three-Berry-connection sector gives the opposite
commutator,
\begin{align}
  \sigma_{ijk}^{\mathcal{T}}
  = -\frac{e^{3}}{\hbar}
    \sum_{\bm{k}}{\sum_{n,m}}^{\prime}
    \frac{f_{n}}{\varepsilon_{nm}}
    X_{ijk}^{nm}.
  \label{eq:T-main}
\end{align}
Adding these two sectors, the commutator terms cancel
exactly:
\begingroup\small
\begin{align}
  \sigma_{ijk}^{\mathcal{C}}
  + \sigma_{ijk}^{\mathcal{T}}
  = \frac{e^{3}}{\hbar}
    \sum_{\bm{k}}{\sum_{n,m}}^{\prime}
    \frac{f_{n}}{\varepsilon_{nm}}
    \bigl(
      \partial_{k}\mathcal{G}_{ij}^{nm}
    + \partial_{j}\mathcal{G}_{ki}^{nm}
    - \partial_{i}\mathcal{G}_{jk}^{nm}
    \bigr).
  \label{eq:CalC-T-cancel-main}
\end{align}
\endgroup
This removes all explicit Berry-connection commutators
from the final dc response.
The remaining terms combine with the quantum-metric
sector $\sigma^{\mathcal{G}}_{ijk}$
[Eq.~\eqref{eq:sigma-QG-split}] to give
Eqs.~\eqref{eq:intraQMD-band} and \eqref{eq:interQMD-pair}.
The full algebra, including the clean-limit kernels for
the three-distinct-band sector, is given in
Appendices~\ref{app:kernel-expansion} and
\ref{app:dc-calculations} and the Supplemental Material~\cite{SM}.

The quantum-geometric sector separates into BCD and
metric pieces,
\begin{align}
  \sigma^{\mathrm{QG}}_{ijk}
  = \sigma^{\mathrm{BCD}}_{ijk}
  + \sigma^{\mathcal{G}}_{ijk},
  \label{eq:QG-BCD-QM-main}
\end{align}
where
\begin{align}
  \sigma^{\mathcal{G}}_{ijk}
  &= \sigma_{ijk}^{\mathrm{intra\text{-}QMD}}
  \nonumber\\
  &\quad
  + \frac{e^{3}}{\hbar}
    \sum_{\bm{k}}{\sum_{n,m}}^{\prime}
    f_{n}
    \left(
      -\frac{\partial_{i}\mathcal{G}_{jk}^{nm}}{\varepsilon_{nm}}
      + 2\frac{\partial_{i}\varepsilon_{nm}}{\varepsilon_{nm}^{2}}
        \mathcal{G}_{jk}^{nm}
    \right)
  \nonumber\\
  &\quad
  - \frac{e^{3}}{\hbar}
    \sum_{\bm{k}}{\sum_{n,m}}^{\prime}
    f_{n}
    \left(
      \frac{\partial_{k}\varepsilon_{nm}}{\varepsilon_{nm}^{2}}
      \mathcal{G}_{ij}^{nm}
    + \frac{\partial_{j}\varepsilon_{nm}}{\varepsilon_{nm}^{2}}
      \mathcal{G}_{ki}^{nm}
    \right).
  \label{eq:QM-before-cancellation-main}
\end{align}
Combining Eqs.~\eqref{eq:QM-before-cancellation-main}
and \eqref{eq:CalC-T-cancel-main},
\begin{align}
  \sigma^{\mathcal{G}}_{ijk}
  + \sigma_{ijk}^{\mathcal{C}}
  + \sigma_{ijk}^{\mathcal{T}}
  = \sigma_{ijk}^{\mathrm{intra\text{-}QMD}}
  + \sigma_{ijk}^{\mathrm{inter\text{-}QMD}}.
  \label{eq:QM-C-T-result}
\end{align}
Equation~\eqref{eq:QM-C-T-result} is the explicit
velocity-gauge Keldysh--Kubo cancellation underlying the
geometric part of the final $\mathcal{O}(\tau^{0})$ response.

\subsection{\texorpdfstring{Scope of the constant-$\Gamma$ steady state}
{Scope of the constant-Gamma steady state}}
\label{sec:scope}

The decomposition derived above uses a constant self-energy
within the bare-bubble approximation.
The practical role of the Keldysh formulation here is to retain
both the spectral broadening and the relaxation prescription
that determines the steady-state occupation in the same
Green's function calculation.
The dc limit is taken with these inputs explicit, before the
large-$\tau$ expansion.
The self-energy assumptions underlying
Eqs.~\eqref{eq:G-retarded-def}--\eqref{eq:spectral-matrix-def}
are
\begin{align}
  \Sigma^{\mathrm{R}}(\varepsilon)=-i\gamma I,
  \quad
  \Sigma^{\mathrm{A}}(\varepsilon)=+i\gamma I,
  \quad
  \Sigma^{<}(\varepsilon)=2i\gamma f(\varepsilon)I,
  \label{eq:self-energies}
\end{align}
where $\gamma=\hbar/(2\tau)$, $I$ is the identity matrix in
orbital space, and $f(\varepsilon)$ is the equilibrium
Fermi function.  The lesser component specifies relaxation
toward this equilibrium distribution and makes
Eq.~\eqref{eq:G-lesser-def} consistent with
Eqs.~\eqref{eq:G-retarded-def} and \eqref{eq:G-advanced-def}.
These self-energies are assumed to be independent of momentum,
proportional to the identity in orbital space, and independent
of the applied field in the velocity gauge used here.
These assumptions specify the model used in applying the
Ward identity, Eq.~\eqref{eq:WT-identity}, and in performing
the response expansion.
A momentum-dependent self-energy introduces additional terms
involving $\partial_i\Sigma^{\mathrm{R/A}}$ into the momentum
derivative of the Green function.
A field-dependent self-energy generates additional response
vertices through its field derivatives.
The cancellation of the apparent $1/(\omega_1\omega_2)$
singularities would therefore have to be re-examined with
these additional terms included.

Two consequences follow.
First, the dc limit is taken at finite $\gamma>0$, and the
resulting dc conductivity is then expanded for large $\tau$.
The two quantum-metric contributions and the kinetic term
appear as finite contributions at order $\tau^{0}$ in this
expansion.  Their independence from $\tau$ holds within the
specified dissipative model.
It does not establish that these contributions persist in an
isolated system with the relaxation mechanism removed.
A different treatment of relaxation can modify the
self-energies, including the lesser component, and thereby
alter the steady-state distribution and the $\tau^{0}$ response.
This distinction between independence from the lifetime and
independence from the relaxation mechanism has been emphasized
in Refs.~\cite{Guo2026Dissipation,Kitamura2026QuasiequilibriumShift}.

Second, comparisons with other dc-response treatments must
account for both the relaxation model and the limiting procedure.
Earlier treatments reviewed in
Ref.~\cite{WatanabeYanase2024Review} include interband-QMD-type
terms without an analogous intraband-QMD term.
In the present formulation, the intraband contribution
Eq.~\eqref{eq:intraQMD-band} is a Fermi-surface term weighted
by $f_n^{\prime}$.
Within the specified dissipative model, it is obtained by
taking the dc limit at finite $\gamma>0$ and then expanding
the resulting expression for large $\tau$.
For a comparison between the velocity and length gauges,
the Hamiltonian, current operator, and self-energies must be
transformed consistently.  In particular, the lesser
self-energy must describe the same physical relaxation
prescription in both gauges.
The comparison must also use the same order of limits and
consistent approximations.
Under these conditions, the electromagnetic gauge choice
alone does not change the physical response.

The kinetic term Eq.~\eqref{eq:ND-tau0} illustrates a separate
issue concerning the accuracy of the clean-limit expansion.
It is present in the finite-broadening expression but is lost
if only the leading spectral contributions of the diagonal
kernels are retained before energy integration.
This omission is a truncation error within a fixed dissipative
model and should be distinguished from dependence on the choice
of relaxation mechanism.
Retaining the self-energies and the required subleading kernel
moments explicitly makes both the model assumptions and the
accuracy of the expansion accessible to verification.

\section{Model examples}
\label{sec:models}
We illustrate the physical content of the decomposition
Eq.~\eqref{eq:main-decomposition} with three minimal models,
each chosen so that a different sector of the decomposition
is isolated.
The first is a tilted massive Dirac lattice model, which
serves as a minimal benchmark for the Berry-curvature-dipole
mechanism~\cite{SodemannFu2015,MaSong2019,Kang2019,Tokura2018}.
The second is a real two-band model in which the Berry
curvature vanishes identically, providing a diagnostic for
quantum-metric nonlinear transport in symmetry settings where
the BCD channel is absent~\cite{GaoYangNiu2014,Wang2021CuMnAs,Liu2021AFM,Das2023,KaplanHolderYan2024,Wang2023Nature,Gao2023Science,Qiang2025Clarification,Ulrich2026IntraQMD}.
The third is a one-band model, in which every
Berry-curvature and quantum-metric sector vanishes by
construction, so that the entire $\mathcal{O}(\tau^{0})$
response reduces to the kinetic term
Eq.~\eqref{eq:ND-tau0}.
For a single band the constant-$\Gamma$ Keldysh steady state
can be solved in closed form and nonperturbatively in the
applied field, and this model therefore provides a check of
the $\mathcal{O}(\tau^{2})$ and $\mathcal{O}(\tau^{0})$
coefficients of Sec.~\ref{subsec:ND-contrib} that is
independent of the kernel expansion of
Appendix~\ref{app:kernel-expansion}.

\subsection{Tilted massive Dirac lattice model}
\label{subsec:tilted-dirac}

As a benchmark for the BCD response, consider the lattice
Hamiltonian
\begin{align}
  H_{\mathrm{tD}}(\bm{k})
  &= t\sin k_{x}\,\sigma_{0}
   + v_{x}\sin k_{x}\,\sigma_{x}
  \nonumber\\
  &\quad
   + v_{y}\sin k_{y}\,\sigma_{y}
   + M(\bm{k})\,\sigma_{z}
  \nonumber\\
  &\equiv t\sin k_{x}\,\sigma_{0} + \bm{d}(\bm{k})\cdot\bm{\sigma},
  \label{eq:tilted-dirac-model}
\end{align}
where the $\bm{d}$ vector is
\begin{align}
  \bm{d}(\bm{k})
  = \bigl(v_{x}\sin k_{x},\; v_{y}\sin k_{y},\; M(\bm{k})\bigr),
  \label{eq:tilted-dirac-dvec}
\end{align}
and
\begin{align}
  M(\bm{k})
  = m + b_{x}(1-\cos k_{x}) + b_{y}(1-\cos k_{y}).
  \label{eq:tilted-dirac-mass}
\end{align}
This is a lattice-regularized massive Dirac model of the
type commonly used to illustrate Berry-curvature physics in
two-band systems~\cite{Nagaosa2010}.
The scalar tilt $t\sin k_{x}\,\sigma_{0}$ shifts the two
eigenvalues by the same momentum-dependent amount and does not
affect the Bloch eigenvectors.
Writing $\bm{d}(\bm{k}) = d(\bm{k})\,\hat{\bm{d}}(\bm{k})$ with
$d(\bm{k}) = |\bm{d}(\bm{k})|$ and
$\hat{\bm{d}} = (\sin\theta\cos\phi,\,\sin\theta\sin\phi,\,\cos\theta)$,
the normalized eigenvectors of $H_{\mathrm{tD}}(\bm{k})$ are
\begin{align}
&  |u_{+}(\bm{k})\rangle
  = \begin{pmatrix}
      \cos(\theta/2) \\
      \sin(\theta/2)\,e^{i\phi}
    \end{pmatrix},\\  
&  |u_{-}(\bm{k})\rangle
  = \begin{pmatrix}
      -\sin(\theta/2)\,e^{-i\phi} \\
      \cos(\theta/2)
    \end{pmatrix},
  \label{eq:tilted-dirac-eigenvectors}
\end{align}
with eigenvalues
\begin{align}
  \varepsilon_{\pm}(\bm{k}) = t\sin k_{x} \pm d(\bm{k}),
  \label{eq:tilted-dirac-eigenvalues}
\end{align}
where
\begin{align}
  \cos\theta = \frac{M(\bm{k})}{d(\bm{k})},
  \qquad
  \tan\phi = \frac{v_{y}\sin k_{y}}{v_{x}\sin k_{x}}.
  \label{eq:tilted-dirac-angles}
\end{align}
The interband Berry connection $\mathcal{A}^{+-}_{i}(\bm{k})$
[Eq.~\eqref{eq:berry-connection-main}] is given by
\begingroup\small
\begin{align}
  \mathcal{A}^{+-}_{i}(\bm{k})
  = -\frac{i}{2}
    \left[
      (\partial_{i}\theta) - i\sin\theta\,(\partial_{i}\phi)
    \right]e^{-i\phi},
  \quad
  \mathcal{A}^{-+}_{i} = \bigl(\mathcal{A}^{+-}_{i}\bigr)^{*},
  \label{eq:tilted-dirac-berry-connection}
\end{align}
\endgroup
and determines the band-pair-resolved quantum geometric tensor
$\mathcal{Q}^{+-}_{ij} = \mathcal{A}^{+-}_{i}\mathcal{A}^{-+}_{j}$.
Substituting Eq.~\eqref{eq:tilted-dirac-berry-connection} and expanding,
\begin{align}
  \mathcal{Q}^{+-}_{ij}
  &= \frac{1}{4}
     \left[
       (\partial_{i}\theta)(\partial_{j}\theta)
       + \sin^{2}\theta\,(\partial_{i}\phi)(\partial_{j}\phi)
     \right]
  \nonumber\\
  &\quad
     + \frac{i}{4}\sin\theta
       \left[
         (\partial_{i}\theta)(\partial_{j}\phi)
         - (\partial_{i}\phi)(\partial_{j}\theta)
       \right].
  \label{eq:tilted-dirac-Q}
\end{align}
Taking the real part gives the quantum metric,
\begin{align}
  \mathcal{G}^{+-}_{ij}
  = \mathrm{Re}\,\mathcal{Q}^{+-}_{ij}
  = \frac{1}{4}
    \left[
      (\partial_{i}\theta)(\partial_{j}\theta)
      + \sin^{2}\theta\,(\partial_{i}\phi)(\partial_{j}\phi)
    \right],
  \label{eq:tilted-dirac-metric}
\end{align}
which is the pullback of the round metric on the Bloch sphere.
Taking $-2$ times the imaginary part gives the Berry curvature,
\begin{align}
  \Omega^{+-}_{ij}
  = -2\,\mathrm{Im}\,\mathcal{Q}^{+-}_{ij}
  = -\frac{1}{2}\sin\theta
    \left[
      (\partial_{i}\theta)(\partial_{j}\phi)
      - (\partial_{i}\phi)(\partial_{j}\theta)
    \right],
  \label{eq:tilted-dirac-berry-curvature}
\end{align}
or explicitly for the $xy$ component,
\begin{align}
  \Omega^{+-}_{xy}
  = -\frac{1}{2}\sin\theta
    \left[
      (\partial_{x}\theta)(\partial_{y}\phi)
      - (\partial_{y}\theta)(\partial_{x}\phi)
    \right].
  \label{eq:tilted-dirac-berry-curvature-xy}
\end{align}
Since $\mathcal{G}^{+-}_{ij}=\mathcal{G}^{-+}_{ij}$ and
$\Omega^{+-}_{ij}=-\Omega^{-+}_{ij}$, the total Berry curvature of
band $+$ in this two-band model coincides with the pair-resolved
value, $\Omega^{+}_{ij}=\Omega^{+-}_{ij}$.  Equation~\eqref{eq:tilted-dirac-berry-curvature}
is equivalent to the familiar $\bm{d}$-vector expression
\begin{align}
  \Omega^{\pm}_{xy}
  = \mp\frac{1}{2}\,
    \hat{\bm{d}}\cdot
    \bigl(\partial_{x}\hat{\bm{d}}\times\partial_{y}\hat{\bm{d}}\bigr),
  \label{eq:tilted-dirac-berry-curvature-dvec}
\end{align}
with $\hat{\bm{d}}=\bm{d}/d$.

For the numerical diagnostics in Fig.~\ref{fig:model-summary-2x4},
we evaluate the BCD and intraband-QMD quantities
\begin{align}
  D_{xz}&=\sum_{\bm{k},n}f_{n}\partial_{x}\Omega_{z}^{n},
  \qquad
  S_{xyy}=\frac{1}{2}\sum_{\bm{k},n}f_{n}^{\prime}
    \partial_{x}\mathcal{G}_{yy}^{n},
  \label{eq:tilted-dirac-observables}
\end{align}
as functions of the chemical potential, using the parameters
\begin{align}
  v_{x}=v_{y}=1,\quad b_{x}=b_{y}=1,\quad m=0.4,\quad t=0.3.
  \label{eq:tilted-dirac-parameters}
\end{align}
With the conventions of Eqs.~\eqref{eq:BCD} and
\eqref{eq:intraQMD-band}, these diagnostics correspond to
\begin{align*}
  \frac{\sigma_{yxx}^{\mathrm{BCD}}}{\tau}
  &=-\frac{e^{3}}{\hbar^{2}}D_{xz},
  \qquad
  \sigma_{xyy}^{\mathrm{intra\text{-}QMD}}
  =\frac{e^{3}}{\hbar}S_{xyy}.
\end{align*}
The indices of the QMD diagnostic are $xyy$, not $yxx$:
for this model the latter intraband component vanishes by parity in
$k_y$.  These panels do not display the nonlinear-Drude or
interband-QMD channels.  The quantum metric and Berry curvature can
be evaluated from the band-basis velocity matrix elements summarized
in Appendix~\ref{app:model-details}, or equivalently from the
$\bm d$-vector expressions given above.

The upper row of Fig.~\ref{fig:model-summary-2x4}
summarizes the tilted massive Dirac model.  Panel (a) shows the
three-dimensional dispersion $\varepsilon_{\pm}(\bm{k})$, in
which the scalar term $t\sin k_{x}$ tilts both bands along the
$k_{x}$ direction, while panel (b) shows the corresponding
high-symmetry-path dispersion.  Panels (c) and (d) show the BCD
diagnostic $D_{xz}$ and the intraband-QMD diagnostic $S_{xyy}$,
respectively, as functions of chemical potential.  Placing the
two diagnostics side by side makes explicit that the tilted model
supports both a finite BCD response and a finite quantum-metric
contribution. 

\subsection{Real two-band model: vanishing BCD with finite intraband QMD}
\label{subsec:real-qmd-model}

To isolate the quantum-metric contribution, we introduce the
real two-band Hamiltonian
\begin{align}
  H_{\mathrm{rQMD}}(\bm{k})
  = d_{0}(\bm{k})\,\sigma_{0}
  + d_{x}(\bm{k})\,\sigma_{x}
  + d_{z}(\bm{k})\,\sigma_{z},
  \label{eq:real-qmd-model}
\end{align}
with
\begin{align}
  d_{0}(\bm{k}) &= t_{0}\sin k_{x},
  \nonumber\\
  d_{x}(\bm{k}) &= m + t_{x}\cos k_{x} + t_{y}\cos k_{y},
  \nonumber\\
  d_{z}(\bm{k}) &= \lambda\sin k_{y}.
  \label{eq:real-qmd-dvec}
\end{align}
The purpose of this model is to provide a controlled minimal
setting in which the Berry curvature is identically absent
while the ordinary band quantum metric remains finite,
thereby isolating quantum-metric nonlinear transport from
the BCD mechanism~\cite{GaoYangNiu2014,Wang2021CuMnAs,Liu2021AFM,Das2023,KaplanHolderYan2024,Qiang2025Clarification,Ulrich2026IntraQMD,Wang2023Nature,Gao2023Science}.

The eigenvalues are
\begin{align}
  \varepsilon_{\pm}(\bm{k})
  = d_{0}(\bm{k}) \pm d(\bm{k}),
  \qquad
  d(\bm{k}) = \sqrt{d_{x}^{2}(\bm{k}) + d_{z}^{2}(\bm{k})}.
  \label{eq:real-qmd-eigenvalues}
\end{align}
For $d(\bm{k})>0$, write
\begin{align}
  \hat{\bm d}(\bm{k})
  = \frac{(d_x,0,d_z)}{d}
  = (\sin\theta,0,\cos\theta),
\end{align}
where $\theta$ is the corresponding two-argument angle, defined by
\begin{align}
  \sin\theta = \frac{d_x(\bm{k})}{d(\bm{k})},
  \qquad
  \cos\theta = \frac{d_z(\bm{k})}{d(\bm{k})}.
  \label{eq:real-qmd-angles}
\end{align}
A convenient local real gauge for the normalized eigenvectors is
\begin{align}
  |u_{+}(\bm{k})\rangle
  &= \begin{pmatrix}
       \cos(\theta/2) \\
       \sin(\theta/2)
     \end{pmatrix},
  \nonumber\\
  |u_{-}(\bm{k})\rangle
  &= \begin{pmatrix}
       -\sin(\theta/2) \\
       \cos(\theta/2)
     \end{pmatrix}.
  \label{eq:real-qmd-eigenvectors}
\end{align}
The corresponding interband Berry connections are purely imaginary,
\begin{align}
  \mathcal{A}^{+-}_{i}(\bm{k})
  &= i\langle u_{+}|\partial_{i}u_{-}\rangle
   = -\frac{i}{2}\,\partial_{i}\theta,
  \nonumber\\
  \mathcal{A}^{-+}_{i}(\bm{k})
  &= \bigl(\mathcal{A}^{+-}_{i}\bigr)^{*}
   = \frac{i}{2}\,\partial_{i}\theta
   = -\mathcal{A}^{+-}_{i}.
  \label{eq:real-qmd-berry-connection}
\end{align}
Consequently, the band-pair-resolved quantum geometric tensor is
\begin{align}
  \mathcal{Q}^{+-}_{ij}
  = \mathcal{A}^{+-}_{i}\mathcal{A}^{-+}_{j}
  = \frac{1}{4}(\partial_{i}\theta)(\partial_{j}\theta),
  \label{eq:real-qmd-Q}
\end{align}
which is purely real.  Thus the ordinary band quantum metric is
generally finite,
\begin{align}
  \mathcal{G}^{+}_{ij}
  = \mathcal{G}^{-}_{ij}
  = \frac{1}{4}\,\partial_{i}\hat{\bm d}\cdot\partial_{j}\hat{\bm d}
  = \frac{1}{4}(\partial_{i}\theta)(\partial_{j}\theta),
  \label{eq:real-qmd-metric}
\end{align}
whereas the Berry curvature vanishes identically.  Equivalently,
\begin{align}
  \Omega_{xy}^{\pm}
  = \mp\frac{1}{2}\,
    \hat{\bm d}\cdot
    \bigl(\partial_{x}\hat{\bm d}\times\partial_{y}\hat{\bm d}\bigr)
  = 0,
  \label{eq:real-qmd-berry-zero}
\end{align}
away from band degeneracies.  Therefore
\begin{align}
  \sigma_{ijk}^{\mathrm{BCD}} = 0
  \label{eq:real-qmd-bcd-zero}
\end{align}
for this model.

The scalar term $d_{0}(\bm{k})\sigma_{0}$ shifts the two
eigenvalues by the same momentum-dependent amount and does not
affect the Bloch eigenvectors.  It therefore leaves
$\Omega_{ij}^{\pm}$ and $\mathcal{G}_{ij}^{\pm}$ unchanged,
but the odd-in-$k_x$ term $t_{0}\sin k_x$ deforms the Fermi
surface and breaks the cancellation of the Fermi-surface
integral in Eq.~\eqref{eq:intraQMD-band}.  Hence, for suitable
tensor components and chemical potentials,
\begin{align}
  \sigma_{ijk}^{\mathrm{BCD}} = 0,
  \qquad
  \sigma_{ijk}^{\mathrm{intra\text{-}QMD}} \neq 0.
  \label{eq:bcd-zero-qmd-nonzero}
\end{align}

\begin{figure*}[t]
  \centering
  \includegraphics[width=\textwidth]{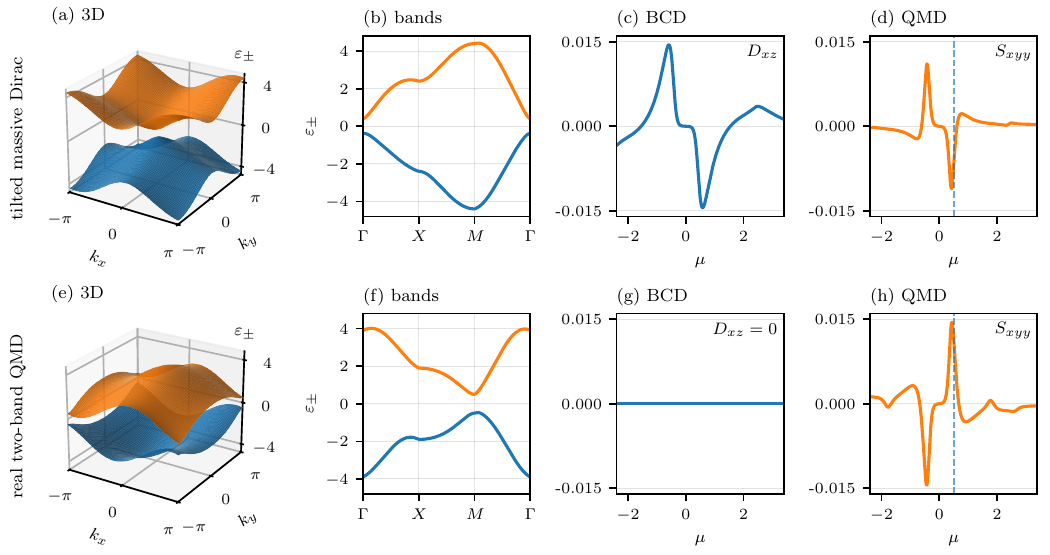}
  \caption{Two-model summary of the band structures and nonlinear
    geometric diagnostics.  The first row shows the tilted massive
    Dirac lattice model Eq.~\eqref{eq:tilted-dirac-model}, and the
    second row shows the real two-band QMD model
    Eq.~\eqref{eq:real-qmd-model}.  Panels (a,e) show the
    three-dimensional dispersions $\varepsilon_{\pm}(\bm{k})$;
    panels (b,f) show the dispersions along
    $\Gamma$--$X$--$M$--$\Gamma$; panels (c,g) show the
    dimensionless BCD diagnostic $D_{xz}$; and panels (d,h) show
    the intraband-QMD diagnostic
    $S_{xyy}\equiv \frac{1}{2}\sum_{\bm{k},n}
    f_{n}^{\prime}\partial_{x}\mathcal{G}_{yy}^{n}$.
    For the real model, $d_{y}=0$ makes the Berry curvature and
    hence $D_{xz}$ identically zero, while $S_{xyy}$ remains
    finite.  The response curves use a finite-temperature
    smearing $T=0.05$ in lattice units.  The dashed lines in
    panels (d) and (h) mark $\mu=0.5$, used in
    Fig.~\ref{fig:qmd-integrand-comparison}.}
  \label{fig:model-summary-2x4}
\end{figure*}

A concrete parameter set used for the numerical diagnostics in
Fig.~\ref{fig:model-summary-2x4} and
Appendix~\ref{app:model-details} is
\begin{align}
  (m,\,t_{x},\,t_{y},\,\lambda,\,t_{0})
  = (2.2,\,1.0,\,0.7,\,0.9,\,0.5).
  \label{eq:real-qmd-parameters}
\end{align}
For this choice,
$d(\bm{k})\ge m-t_x-t_y=0.5$, so the direct band separation
satisfies $2d(\bm{k})\ge1.0$ throughout the Brillouin zone.
Using the Brillouin-zone normalization employed throughout,
a Fermi function with $T=0.05$, and a $501\times501$ mesh, the
representative chemical potential $\mu=0.5$ gives
\begin{align}
  \sum_{\bm{k},n}
  f_{n}^{\prime}\,\partial_{x}\mathcal{G}_{yy}^{n}
  &\simeq 2.32\times10^{-2},
  \nonumber\\
  S_{xyy}
  \equiv \frac{1}{2}\sum_{\bm{k},n}
  f_{n}^{\prime}\,\partial_{x}\mathcal{G}_{yy}^{n}
  &\simeq 1.16\times10^{-2},
  \nonumber\\
  \sigma_{xyy}^{\mathrm{intra\text{-}QMD}}
  &\simeq 1.16\times10^{-2}\,\frac{e^{3}}{\hbar},
  \label{eq:real-qmd-numerical-check}
\end{align}
whereas $\sigma_{xyy}^{\mathrm{BCD}}=0$ identically by
Eq.~\eqref{eq:real-qmd-bcd-zero}.  This explicitly demonstrates
that the intraband QMD is an independent Fermi-surface
quantum-metric effect rather than a reformulation of the BCD
response.

\begin{figure*}[t]
  \centering
  \includegraphics[width=\textwidth]{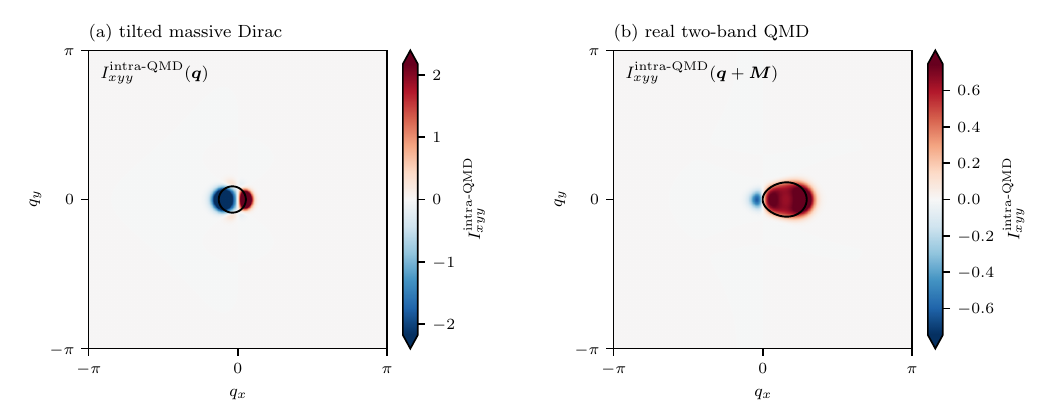}
  \caption{Momentum-resolved intraband-QMD integrand
    $I_{xyy}^{\mathrm{intra\text{-}QMD}}$ defined in
    Eq.~\eqref{eq:qmd-kresolved-integrand}, evaluated at
    $\mu=0.5$ and $T=0.05$.  (a)~Tilted massive Dirac lattice
    model Eq.~\eqref{eq:tilted-dirac-model}, displayed in
    coordinates $\bm{q}=\bm{k}$ centered at $\Gamma$.
    (b)~Real two-band QMD model Eq.~\eqref{eq:real-qmd-model},
    displayed in shifted coordinates
    $\bm{q}=\bm{k}-\bm{M}$ with $\bm{M}=(\pi,\pi)$.
    Solid and dashed black contours denote upper- and lower-band
    Fermi contours, respectively; only contours present at the
    selected chemical potential are shown.  Each panel uses an
    independent color range centered at zero.  To reveal the
    localized sign structure, the extreme $0.5\%$ of
    $|I_{xyy}^{\mathrm{intra\text{-}QMD}}|$ values are saturated,
    as indicated by the colorbar extensions.  The Brillouin-zone
    averages correspond to the $S_{xyy}$ curves in panels (d)
    and (h) of Fig.~\ref{fig:model-summary-2x4}.}
  \label{fig:qmd-integrand-comparison}
\end{figure*}

The lower row of Fig.~\ref{fig:model-summary-2x4} gives the
corresponding compact diagnostics.  Panel (e) shows the full
three-dimensional dispersion: the scalar tilt
$d_{0}(\bm{k})=t_{0}\sin k_x$ breaks Fermi-surface inversion
symmetry while the Hamiltonian remains real.  Consequently,
the Berry curvature and BCD vanish identically, as shown by the
flat $D_{xz}=0$ curve in panel (g), whereas the quantum metric
remains finite and produces the nonzero intraband-QMD curve in
panel (h).  The dashed lines in panels (d) and (h) mark $\mu=0.5$,
the chemical potential used for the momentum-resolved QMD maps in
Fig.~\ref{fig:qmd-integrand-comparison}.

Both models also allow a useful momentum-resolved visualization
of the intraband-QMD contribution before the Brillouin-zone
integration, in the same spirit as the $k$-resolved
quantum-geometric diagnostics discussed in
Ref.~\cite{Ulrich2026IntraQMD}.  For either model, we plot the
gauge-invariant integrand
\begin{align}
  I_{xyy}^{\mathrm{intra\text{-}QMD}}(\bm{k})
  = \frac{1}{2}\sum_{n=\pm}
    f_{n}^{\prime}(\bm{k})\,
    \partial_{k_{x}}\mathcal{G}_{yy}^{n}(\bm{k}),
  \label{eq:qmd-kresolved-integrand}
\end{align}
whose Brillouin-zone average gives the diagnostic $S_{xyy}$
shown in panels (d) and (h) of
Fig.~\ref{fig:model-summary-2x4}.  This map is not an additional
response formula---only its Brillouin-zone integral enters the
conductivity---but it identifies the momentum-space regions
that generate the finite intra-QMD signal in the final
metric representation.

Figure~\ref{fig:qmd-integrand-comparison} compares the two models
at the same $\mu=0.5$ and $T=0.05$.  In the tilted massive Dirac
model, the response is concentrated around the
$\Gamma$-centered upper-band Fermi contour and coexists with a
finite BCD response.  In the real two-band model, the response
is localized near the tilted Fermi contour around the $M$ point,
whereas the Berry curvature and hence the BCD response vanish
identically.  The side-by-side maps therefore distinguish the
momentum-space localization of the QMD integrand from the
presence or absence of Berry curvature.

\subsection{\texorpdfstring{Exact one-band check of the $\tau^{0}$ hierarchy}
{Exact one-band check of the tau0 hierarchy}}
\label{sec:oneband}
In this subsection we consider the single-orbital, spinless
one-band Hamiltonian
\begin{align}
  H_{\mathrm{1b}}(k)=\varepsilon_{k}
  &=-2h\cos k+2h^{\prime}\sin 2k,
  \qquad -\pi\le k\le\pi.
  \label{eq:oneband-eps}
\end{align}
The lattice constant is set to unity.
Here, $h$ is a real nearest-neighbor hopping amplitude,
and $ih^{\prime}$ is a purely imaginary next-nearest-neighbor
hopping amplitude.  Momentum arguments are written as
subscripts throughout this subsection, so that $\varepsilon$
without a subscript denotes the energy variable.
Because the Bloch Hamiltonian is a $1\times1$ matrix, its
eigenvector can be chosen independent of $k$.  The Berry
connection then vanishes globally, together with every
Berry-curvature and quantum-metric sector of
Eq.~\eqref{eq:main-decomposition}.
The kinetic term in Eq.~\eqref{eq:ND-tau0} therefore gives the
entire response at order $\tau^{0}$ within the
constant-$\Gamma$ model used here.

\begin{figure*}[t]
  \centering
  \includegraphics[width=\textwidth]{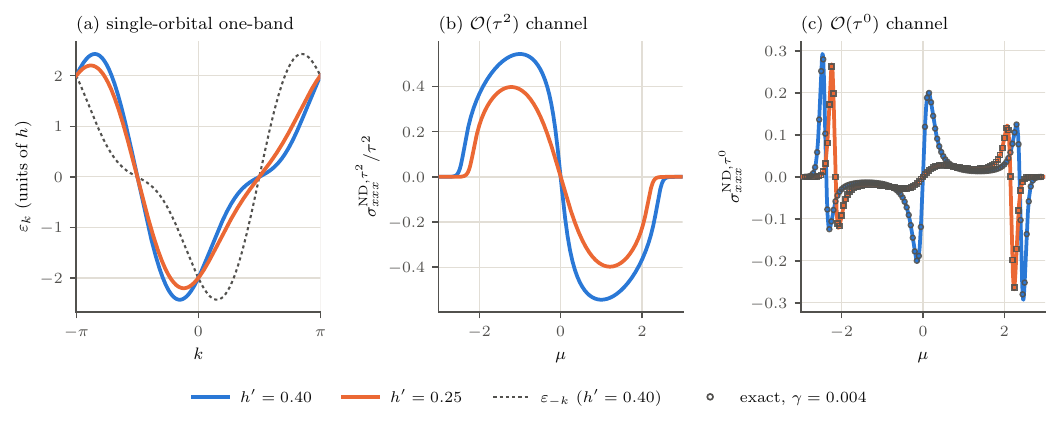}
  \caption{The one-band model of Eq.~\eqref{eq:oneband-eps}
    with $h=1$, in the same format as
    Fig.~\ref{fig:model-summary-2x4}.
    (a) Band $\varepsilon_{k}$ for $h^{\prime}=0.40$ and $0.25$.
    The dotted curve is $\varepsilon_{-k}$ for $h^{\prime}=0.40$
    and displays the asymmetry about $k=0$; the absence of any
    shifted inversion center follows from
    Eq.~\eqref{eq:oneband-inversion-test}.
    (b) Nonlinear Drude channel
    $\sigma^{\mathrm{ND},\tau^{2}}_{xxx}/\tau^{2}$
    [Eq.~\eqref{eq:oneband-tau2}] versus chemical potential.
    (c) Kinetic channel $\sigma^{\mathrm{ND},\tau^{0}}_{xxx}$
    [Eq.~\eqref{eq:oneband-tau0}] (lines), compared with the
    finite-$\gamma$ residual
    $\sigma_{\mathrm{exact}}-\sigma^{\mathrm{ND},\tau^{2}}_{xxx}$
    at $\gamma=0.004$ (open symbols at selected chemical
    potentials).  The residual is given by
    Eq.~\eqref{eq:oneband-residual-exact} and contains
    $\mathcal{O}(\gamma^{2})$ corrections to the kinetic term.
    The response panels use $T=0.05$, as in
    Fig.~\ref{fig:model-summary-2x4}.  Every Berry-curvature and
    quantum-metric sector vanishes identically in this model;
    the entire second-order response vanishes as
    $h^{\prime}\to0$.  Energies are in units of $h$, with
    $e=\hbar=k_{\mathrm{B}}=1$, the Brillouin-zone average
    $\int_{-\pi}^{\pi}d\kappa/(2\pi)$, and no spin-degeneracy
    factor.}
  \label{fig:oneband-model}
\end{figure*}

The dispersion is shown in Fig.~\ref{fig:oneband-model}(a).
For $hh^{\prime}\neq0$, it has no inversion center, including
a center shifted away from $k=0$.  Indeed,
\begin{align}
  \varepsilon_{k_{0}+k}-\varepsilon_{k_{0}-k}
  &=4h\sin k_{0}\sin k
   +4h^{\prime}\cos(2k_{0})\sin 2k.
  \label{eq:oneband-inversion-test}
\end{align}
The identity
$\varepsilon_{k_{0}+k}\equiv\varepsilon_{k_{0}-k}$ would
require both $\sin k_{0}=0$ and $\cos(2k_{0})=0$, which
cannot hold simultaneously.  In real space the hopping
convention is
\begin{align}
  \hat{H}_{\mathrm{1b}}
  &=\sum_{j}\bigl[
      -h\hat{c}^{\dagger}_{j+1}\hat{c}_{j}
      +i h^{\prime}\hat{c}^{\dagger}_{j+2}\hat{c}_{j}
      +\mathrm{H.c.}\bigr].
  \label{eq:oneband-real-space}
\end{align}
A transformation $\hat{c}_{j}\to e^{i\theta j}\hat{c}_{j}$
only shifts the momentum origin and cannot remove both
hopping phases.  For the oriented loop
$j\to j+1\to j+2\to j$, the product of hopping amplitudes is
$(-h)(-h)(-i h^{\prime})=-i h^{2}h^{\prime}$, whose phase is
$-\pi/2$ for $h,h^{\prime}>0$.  This is a gauge-invariant loop
phase on the hopping graph, not a magnetic flux through a
finite-area plaquette of the straight one-dimensional chain.
Complex second-neighbor hoppings provide a familiar mechanism
for time-reversal breaking~\cite{Haldane1988}.

For a single spinless band, time-reversal symmetry would
require $\varepsilon_{-k}=\varepsilon_{k}$, in which case the
nonlinear Drude response of Eq.~\eqref{eq:ND} vanishes by
parity.  The asymmetric dispersion is the microscopic source
of nonreciprocal transport~\cite{Rikken2001,Tokura2018}; at
fixed $h$, $h^{\prime}$ controls this asymmetry, and the entire
second-order response vanishes as $h^{\prime}\to0$.

A general expression for the current in the nonequilibrium
steady state driven by $\bm{A}(t)$ is given in
Eq.~\eqref{eq:J-lesser-current}.  Its three terms are the
expansion of a single current vertex through second order in
$\bm{A}$.  Keeping the Peierls-shifted vertex unexpanded gives
\begin{align}
  \mathcal{J}(t)
  &=ie\sum_{k}
    \left.\bigl(\partial_{k}\varepsilon_{k}\bigr)
    \right|_{k\to k+a(t)}G^{<}_{k}(t,t),
  \label{eq:oneband-J-lesser}
\end{align}
where $a(t)=eA(t)/\hbar$ and
$G^{<}_{k}(t,t)\equiv G^{<}_{k,k}(t,t)$.
Thus the contact vertices $V_{i},V_{ij},V_{ijk}$ generated by
the Peierls substitution are already included in the momentum
shift; no separate contact-vertex expansion is required in
this representation.  This statement is made within the
self-energy closure specified below and does not include
additional interaction or disorder vertex corrections beyond
that closure.

To evaluate the lesser Green's function, we first solve the
retarded equation of motion.  Under the Peierls substitution
of Eq.~\eqref{eq:peierls-main},
\begin{align}
  \biggl[i\hbar\frac{\partial}{\partial t}
    -\varepsilon_{k+a(t)}+i\gamma\biggr]
  G^{\mathrm{R}}_{k}(t,t^{\prime})
  &=\delta(t-t^{\prime}).
  \label{eq:oneband-eom}
\end{align}
Since $\varepsilon_{k+a(t)}$ is a scalar, it commutes with
itself at different times.  The time-ordered exponential
therefore reduces to an ordinary exponential,
\begingroup\small
\begin{align}
  G^{\mathrm{R}}_{k}(t,t^{\prime})
  &=-\frac{i}{\hbar}\theta(t-t^{\prime})
  \exp\biggl[
      -\frac{i}{\hbar}\int_{t^{\prime}}^{t}ds\,
        \varepsilon_{k+a(s)}
      -\frac{\gamma}{\hbar}(t-t^{\prime})\biggr],
  \label{eq:oneband-GR}
\end{align}
\endgroup
with $G^{\mathrm{A}}_{k}(t^{\prime},t)
=[G^{\mathrm{R}}_{k}(t,t^{\prime})]^{*}$.
At this stage no expansion in the applied field has been made.

Once the memory of the initial state has decayed, the lesser
Green's function is~\cite{Langreth1976,Haug2008,Stefanucci2013}
\begingroup\small
\begin{align}
  G^{<}_{k}(t,t^{\prime})
  &=\int_{-\infty}^{t}dt_{1}
    \int_{-\infty}^{t^{\prime}}dt_{2}\,
    G^{\mathrm{R}}_{k}(t,t_{1})\Sigma^{<}(t_{1},t_{2})
    G^{\mathrm{A}}_{k}(t_{2},t^{\prime}).
  \label{eq:G-lesser-t}
\end{align}
\endgroup
The integration ranges follow from the retarded and advanced
step functions.  We adopt the same dissipative closure as in
Eq.~\eqref{eq:self-energies},
$\Sigma^{<}(\varepsilon)=2i\gamma f(\varepsilon)$,
with $\Sigma^{\mathrm{R/A}}=\mp i\gamma$.  These self-energies
are momentum independent, scalar in orbital space, and
independent of the applied field.  In the time domain,
\begin{align}
  \Sigma^{<}(t_{1},t_{2})
  &=\int_{-\infty}^{\infty}\frac{d\varepsilon}{2\pi\hbar}\,
    e^{-i\varepsilon(t_{1}-t_{2})/\hbar}
    \Sigma^{<}(\varepsilon)
  \nonumber\\
  &=2i\gamma\int_{-\infty}^{\infty}
    \frac{d\varepsilon}{2\pi\hbar}\,
    e^{-i\varepsilon(t_{1}-t_{2})/\hbar}f(\varepsilon).
  \label{eq:Sigma-lesser-eps}
\end{align}
At fixed $\varepsilon$, the dependence on $t_{1}$ and $t_{2}$
factorizes, so that the equal-time result becomes
\begingroup\small
\begin{align}
  &G^{<}_{k}(t,t)
  =2i\gamma\int_{-\infty}^{\infty}
    \frac{d\varepsilon}{2\pi\hbar}\,f(\varepsilon)
  \left|
      \int_{-\infty}^{t}dt_{1}\,
      G^{\mathrm{R}}_{k}(t,t_{1})
      e^{-i\varepsilon t_{1}/\hbar}\right|^{2}.
  \label{eq:oneband-factorized}
\end{align}
\endgroup

For a static uniform field, $a(t)=qt$ with
$q=\dot{a}=-eE/\hbar$.  The shifted crystal momentum
$\kappa=k+a(t)$ absorbs the explicit time dependence through
$k+a(s)=\kappa-q(t-s)$.  Setting $u=t-t_{1}$, the factor
$e^{-i\varepsilon t/\hbar}$ cancels against its conjugate,
and we obtain
\begin{align}
  &G^{<}_{k}(t,t)
  =\frac{2i\gamma}{\hbar^{2}}
    \int_{-\infty}^{\infty}\frac{d\varepsilon}{2\pi\hbar}\,
    f(\varepsilon)|I(\varepsilon,\kappa;q)|^{2},
  \label{eq:oneband-Gless}\\
  &I(\varepsilon,\kappa;q)
  =\int_{0}^{\infty}du\,
    \exp\biggl[\frac{i\varepsilon-\gamma}{\hbar}u
      -\frac{i}{\hbar}\int_{0}^{u}dx\,
        \varepsilon_{\kappa-qx}\biggr].
  \label{eq:oneband-I}
\end{align}
These expressions are exact and nonperturbative in the field
within the chosen closure.  Their dependence on $k$ and $t$
is entirely through $\kappa$, so we write the equal-time
lesser Green's function as $G^{<}(\kappa;q)$.  Relabeling the
Brillouin-zone sum as $\sum_{k}\to\sum_{\kappa}$ then makes
the current time independent.  Convergence of the $u$ integral
requires $\gamma>0$.

With $n(\kappa;q)=-i\hbar G^{<}(\kappa;q)$, the current and
occupation are
\begin{align}
  \mathcal{J}
  &=ie\sum_{\kappa}
    (\partial_{\kappa}\varepsilon_{\kappa})G^{<}(\kappa;q)
  \nonumber\\
  &=-\frac{e}{\hbar}\sum_{\kappa}
    (\partial_{\kappa}\varepsilon_{\kappa})n(\kappa;q),
  \label{eq:J-exact}\\
  n(\kappa;q)
  &=\frac{2\gamma}{\hbar^{2}}
    \int_{-\infty}^{\infty}\frac{d\varepsilon}{2\pi}\,
    f(\varepsilon)|I(\varepsilon,\kappa;q)|^{2}.
  \label{eq:oneband-n}
\end{align}
Although Eq.~\eqref{eq:J-exact} has the form of a Boltzmann
current, the occupation is obtained from the exact steady
state of the constant-broadening Keldysh model, rather than
from a relaxation-time kinetic equation.  At $q=0$,
$I=i\hbar G^{\mathrm{R}}(\varepsilon)$, and
\begin{align}
  n(\kappa;0)
  &=\int_{-\infty}^{\infty}\frac{d\varepsilon}{\pi}\,
    \frac{\gamma f(\varepsilon)}
         {(\varepsilon-\varepsilon_{\kappa})^{2}+\gamma^{2}}
   \xrightarrow{\gamma\to0^{+}}f_{\kappa}.
  \label{eq:oneband-equilibrium}
\end{align}
Here $f_{\kappa}=f(\varepsilon_{\kappa})$.  This fixes the
normalization.  The equilibrium current vanishes because
$n(\kappa;0)$ depends on $\kappa$ only through the periodic
band energy.

To extract $\sigma^{(2)}\equiv\sigma_{xxx}$, we now expand
the exact occupation in $q$ at fixed $\gamma>0$,
\begin{align}
  n(\kappa;q)
  &=n(\kappa;0)+q\nu^{(1)}(\kappa)
   +q^{2}\nu^{(2)}(\kappa)+\mathcal{O}(q^{3}).
  \label{eq:n-expansion}
\end{align}
The zeroth-order term carries no current and the first-order
term gives the linear conductivity.  Consistently with
Eq.~\eqref{eq:J2-sigma} evaluated at
$\omega_{1}=\omega_{2}=0$, we write
$\mathcal{J}^{(2)}=\sigma^{(2)}E^{2}$ without an additional
$1/2!$ and include no spin-degeneracy factor.  Since
$q^{2}=e^{2}E^{2}/\hbar^{2}$,
\begin{align}
  \sigma^{(2)}
  &=-\frac{e^{3}}{\hbar^{3}}\sum_{\kappa}
    (\partial_{\kappa}\varepsilon_{\kappa})\nu^{(2)}(\kappa).
  \label{eq:oneband-sigma}
\end{align}

The accumulated phase has the expansion
\begin{align}
  \int_{0}^{u}dx\,\varepsilon_{\kappa-qx}
  &=\varepsilon_{\kappa}u
   -\frac{u^{2}}{2}(\partial_{\kappa}\varepsilon_{\kappa})q
  +\frac{u^{3}}{6}
    (\partial_{\kappa}^{2}\varepsilon_{\kappa})q^{2}
   +\mathcal{O}(q^{3}).
  \label{eq:oneband-phase-exp}
\end{align}
Defining
\begin{align}
  z&=\gamma-i(\varepsilon-\varepsilon_{\kappa}),
  \quad
  \bar{z}=\gamma+i(\varepsilon-\varepsilon_{\kappa}),
  \label{eq:oneband-z}
\end{align}
and using
$\int_{0}^{\infty}du\,u^{n}e^{-zu/\hbar}
 =n!(\hbar/z)^{n+1}$, we find
\begingroup\small
\begin{align}
  I
  &=\frac{\hbar}{z}
   +i\hbar^{2}\frac{\partial_{\kappa}\varepsilon_{\kappa}}{z^{3}}q
  -i\hbar^{3}
    \frac{\partial_{\kappa}^{2}\varepsilon_{\kappa}}{z^{4}}q^{2}
   -3\hbar^{3}
    \frac{(\partial_{\kappa}\varepsilon_{\kappa})^{2}}{z^{5}}q^{2}
   +\mathcal{O}(q^{3}).
  \label{eq:oneband-Iexp}
\end{align}
\endgroup
Collecting the coefficient of $q^{2}$ in $|I|^{2}$ gives
\begin{align}
  \nu^{(2)}(\kappa)
  &=(\partial_{\kappa}^{2}\varepsilon_{\kappa})\mathcal{G}_{1}
   +(\partial_{\kappa}\varepsilon_{\kappa})^{2}\mathcal{G}_{2},
  \label{eq:oneband-nu2}
\end{align}
where the argument $\varepsilon_{\kappa}$ of
$\mathcal{G}_{1}$ and $\mathcal{G}_{2}$ is understood:
\begin{align}
  &\mathcal{G}_{1}
  =2\gamma\hbar^{2}
    \int_{-\infty}^{\infty}\frac{d\varepsilon}{2\pi}\,
    f(\varepsilon)i\biggl[
      \frac{1}{z\bar{z}^{4}}-\frac{1}{z^{4}\bar{z}}\biggr],
  \nonumber\\
  &\mathcal{G}_{2}
  =2\gamma\hbar^{2}
    \int_{-\infty}^{\infty}\frac{d\varepsilon}{2\pi}\,
    f(\varepsilon)
  \biggl[
      \frac{1}{z^{3}\bar{z}^{3}}
      -3\biggl(\frac{1}{z\bar{z}^{5}}
              +\frac{1}{z^{5}\bar{z}}\biggr)\biggr].
  \label{eq:oneband-G12}
\end{align}
Equations~\eqref{eq:oneband-sigma}--\eqref{eq:oneband-G12}
determine the exact second-order coefficient at finite
$\gamma$, denoted below by
$\sigma_{\mathrm{exact}}\equiv\sigma^{(2)}$.
The steady-state solution was obtained before expanding in
the field; the second-order coefficient itself was extracted
by the explicit expansion in $q$.

We next take $\gamma\to0^{+}$ at fixed $T>0$, after the
weak-field expansion.  Here
$f(\varepsilon)=[e^{(\varepsilon-\mu)/T}+1]^{-1}$ with
$k_{\mathrm{B}}=1$.  The zero-temperature limit, if required,
must be treated separately; the remainder estimates below
are not claimed to be uniform in a simultaneous
$T\to0$ limit.  Setting
$\varepsilon=\varepsilon_{\kappa}+\gamma s$ gives
\begin{align}
  \mathcal{G}_{1}
  &=\frac{\hbar^{2}}{\pi\gamma^{3}}
    \int_{-\infty}^{\infty}ds\,
    f(\varepsilon_{\kappa}+\gamma s)g_{1}(s),
  \nonumber\\
  \mathcal{G}_{2}
  &=\frac{\hbar^{2}}{\pi\gamma^{4}}
    \int_{-\infty}^{\infty}ds\,
    f(\varepsilon_{\kappa}+\gamma s)g_{2}(s),
  \label{eq:oneband-G12-scaled}
\end{align}
with
\begin{align}
  g_{1}(s)
  &=-2\,\mathrm{Im}\,
    \frac{1}{(1-is)(1+is)^{4}}
   =\frac{2s(3-s^{2})}{(1+s^{2})^{4}},
  \label{eq:oneband-g1}\\
  g_{2}(s)
  &=\frac{1}{(1+s^{2})^{3}}
   -6\,\mathrm{Re}\,
    \frac{1}{(1-is)(1+is)^{5}}
  \nonumber\\
  &=\frac{-5s^{4}+38s^{2}-5}{(1+s^{2})^{5}}.
  \label{eq:oneband-g2}
\end{align}
The first kernel is odd and the second is even.  With
$f^{(p)}_{\kappa}
 =\left.\partial_{\varepsilon}^{p}f(\varepsilon)
  \right|_{\varepsilon=\varepsilon_{\kappa}}$, the local terms
through order $\gamma^{0}$ follow from the convergent moments
\begin{widetext}
\begin{align}
  &\int_{-\infty}^{\infty}ds\,s g_{1}(s)
  =\frac{\pi}{4},
  \quad
  \int_{-\infty}^{\infty}ds\,s^{3}g_{1}(s)
  =-\frac{\pi}{4},
  \label{eq:oneband-moments-g1}\\
  &\int_{-\infty}^{\infty}ds\,g_{2}(s)
  =0,
  \quad
  \int_{-\infty}^{\infty}ds\,s^{2}g_{2}(s)
  =\frac{\pi}{2},
  \quad
  \int_{-\infty}^{\infty}ds\,s^{4}g_{2}(s)
  =0.
  \label{eq:oneband-moments-g2}
\end{align}
\end{widetext}
The zeroth moment of $g_{2}$ removes a would-be
$\mathcal{O}(\tau^{4})$ contribution, while its fourth moment
removes the $\mathcal{O}(\tau^{0})$ contribution of
$\mathcal{G}_{2}$.  However, parity alone does not imply that
the remainder in either individual kernel starts at
$\gamma^{2}$.  Their large-$|s|$ behavior is
\begin{align}
  g_{1}(s)&=-\frac{2}{s^{5}}+\mathcal{O}(s^{-7}),
  \quad
  g_{2}(s)=-\frac{5}{s^{6}}+\mathcal{O}(s^{-8}),
  \label{eq:oneband-kernel-tails}
\end{align}
so that $\int ds\,s^{5}g_{1}(s)$ and
$\int ds\,s^{6}g_{2}(s)$ diverge.  An unrestricted Taylor
expansion of $f(\varepsilon_{\kappa}+\gamma s)$ cannot be
integrated term by term over the whole real axis.
With $\tau=\hbar/(2\gamma)$, the individual asymptotic
expansions at fixed $T>0$ are instead
\begin{align}
  \mathcal{G}_{1}
  &=\tau^{2}f^{(1)}_{\kappa}
   -\frac{\hbar^{2}}{24}f^{(3)}_{\kappa}
   +\mathcal{O}(\gamma),
  \label{eq:oneband-Gexpand-G1}\\
  \mathcal{G}_{2}
  &=\tau^{2}f^{(2)}_{\kappa}+\mathcal{O}(\gamma).
  \label{eq:oneband-Gexpand-G2}
\end{align}
The $\mathcal{O}(\gamma)$ terms need not vanish in either
kernel separately.  Their cancellation in the current will be
established below using an exact finite-$\gamma$ identity,
rather than inferred from a divergent higher moment.

For a smooth periodic band, integration by parts over the
Brillouin zone gives
\begin{align}
  \sum_{\kappa}f^{(p)}_{\kappa}
    \partial_{\kappa}^{3}\varepsilon_{\kappa}
  &=-\sum_{\kappa}f^{(p+1)}_{\kappa}
    (\partial_{\kappa}\varepsilon_{\kappa})
    (\partial_{\kappa}^{2}\varepsilon_{\kappa})
  \nonumber\\
  &=\frac{1}{2}\sum_{\kappa}f^{(p+2)}_{\kappa}
    (\partial_{\kappa}\varepsilon_{\kappa})^{3},
  \label{eq:oneband-bz-identity}
\end{align}
where $p$ is a nonnegative integer and all boundary terms
vanish.  Substituting the individual kernel expansions into
the current, the leading terms combine as
\begin{align}
  \sigma^{\mathrm{ND},\tau^{2}}_{xxx}
  &=-\frac{e^{3}\tau^{2}}{\hbar^{3}}\sum_{\kappa}
    \bigl[f^{(1)}_{\kappa}
      (\partial_{\kappa}\varepsilon_{\kappa})
      (\partial_{\kappa}^{2}\varepsilon_{\kappa})
  +f^{(2)}_{\kappa}
      (\partial_{\kappa}\varepsilon_{\kappa})^{3}\bigr]
  \nonumber\\
  &=-\frac{e^{3}\tau^{2}}{\hbar^{3}}\sum_{\kappa}
    f_{\kappa}\partial_{\kappa}^{3}\varepsilon_{\kappa},
  \label{eq:oneband-tau2}
\end{align}
while the constant term is
\begin{align}
  \sigma^{\mathrm{ND},\tau^{0}}_{xxx}
  &=\frac{e^{3}}{24\hbar}\sum_{\kappa}
    f^{(3)}_{\kappa}
    (\partial_{\kappa}\varepsilon_{\kappa})
    (\partial_{\kappa}^{2}\varepsilon_{\kappa})
  \nonumber\\
  &=-\frac{e^{3}}{24\hbar}\sum_{\kappa}
    f^{(2)}_{\kappa}\partial_{\kappa}^{3}\varepsilon_{\kappa}
  \nonumber\\
  &=-\frac{e^{3}}{48\hbar}\sum_{\kappa}
    f^{(4)}_{\kappa}(\partial_{\kappa}\varepsilon_{\kappa})^{3}.
  \label{eq:oneband-tau0}
\end{align}
These are the one-band nonlinear Drude and kinetic terms
associated with Eqs.~\eqref{eq:ND} and \eqref{eq:ND-tau0},
respectively, in the single-spinless-band normalization and
current convention stated above.  The leading term agrees
with the Boltzmann result when $\tau=\hbar/(2\gamma)$.
With the response, spin-counting, and Brillouin-zone
normalization conventions fixed, the coefficient $1/24$
follows directly without any additional rescaling.
The coefficients $1/24$ and $1/48$ are related by integration
by parts and are therefore not independent.
The quoted sign and prefactor follow the response and charge
conventions stated above.  In the individual-kernel
representation, the constant term comes entirely from the
third moment of $g_{1}$, whereas $\mathcal{G}_{2}$ has no
constant contribution.  This is the one-band counterpart of
the parity selection rule discussed in
Sec.~\ref{subsec:ND-contrib}.

To determine the corrections to the second-order conductivity,
we first combine the two finite-$\gamma$ contributions before
taking the clean limit.  Since $\mathcal{G}_{1}$ depends on $\kappa$
only through $\varepsilon_{\kappa}$, periodicity gives
\begin{align}
  &\sum_{\kappa}
    (\partial_{\kappa}\varepsilon_{\kappa})
    (\partial_{\kappa}^{2}\varepsilon_{\kappa})\mathcal{G}_{1}
  =-\frac{1}{2}\sum_{\kappa}
    (\partial_{\kappa}\varepsilon_{\kappa})^{3}
    \partial_{\varepsilon_{\kappa}}\mathcal{G}_{1}.
  \label{eq:oneband-current-ibp}
\end{align}
Consequently,
\begin{align}
  \sigma_{\mathrm{exact}}
  &=-\frac{e^{3}}{\hbar^{3}}\sum_{\kappa}
    (\partial_{\kappa}\varepsilon_{\kappa})^{3}
    \biggl[\mathcal{G}_{2}
      -\frac{1}{2}\partial_{\varepsilon_{\kappa}}\mathcal{G}_{1}
    \biggr].
  \label{eq:oneband-sigma-combined}
\end{align}
In Eq.~\eqref{eq:oneband-G12-scaled}, differentiating with
respect to $\varepsilon_{\kappa}$ and integrating once by
parts in $s$ produces the combined kernel
\begin{align}
  g_{2}(s)+\frac{1}{2}\frac{dg_{1}(s)}{ds}
  &=\frac{2(7s^{2}-1)}{(1+s^{2})^{5}}
  \nonumber\\
  &=\frac{1}{3}\frac{d^{2}}{ds^{2}}
    \frac{1}{(1+s^{2})^{3}}.
  \label{eq:oneband-combined-kernel}
\end{align}
Two further integrations by parts in $s$ therefore give
\begin{align}
  \mathcal{G}_{2}
   -\frac{1}{2}\partial_{\varepsilon_{\kappa}}\mathcal{G}_{1}
  &=\frac{\hbar^{2}}{\pi\gamma^{4}}
    \int_{-\infty}^{\infty}ds\,
    f(\varepsilon_{\kappa}+\gamma s)
    \frac{2(7s^{2}-1)}{(1+s^{2})^{5}}
  \nonumber\\
  &=\frac{\hbar^{2}}{3\pi\gamma^{2}}
    \int_{-\infty}^{\infty}ds\,
    \frac{f^{(2)}(\varepsilon_{\kappa}+\gamma s)}{(1+s^{2})^{3}}.
  \label{eq:oneband-combined-integral}
\end{align}
Here $f^{(2)}$ is the second derivative with respect to its
energy argument.  Applying the Brillouin-zone identity
\eqref{eq:oneband-bz-identity} at each fixed $s$ yields
\begin{align}
  \sigma_{\mathrm{exact}}
  &=-\frac{e^{3}}{3\pi\hbar\gamma^{2}}\sum_{\kappa}
    (\partial_{\kappa}\varepsilon_{\kappa})^{3}
    \int_{-\infty}^{\infty}ds\,
    \frac{f^{(2)}(\varepsilon_{\kappa}+\gamma s)}{(1+s^{2})^{3}}
  \nonumber\\
  &=-\frac{2e^{3}}{3\pi\hbar\gamma^{2}}\sum_{\kappa}
    \partial_{\kappa}^{3}\varepsilon_{\kappa}
    \int_{-\infty}^{\infty}ds\,
    \frac{f(\varepsilon_{\kappa}+\gamma s)}{(1+s^{2})^{3}}.
  \label{eq:oneband-exact-gamma}
\end{align}
Equation~\eqref{eq:oneband-exact-gamma} is exact for every
$\gamma>0$ within the specified closure, and it holds for any
smooth periodic scalar band, not only for
Eq.~\eqref{eq:oneband-eps}.  It is an exact formula for the
second-order coefficient, not a truncation of that coefficient
in $\gamma$ and not the full current at arbitrary field
strength.

The two finite-$\gamma$ contributions have thus been combined
into a single integral with the weight $(1+s^2)^{-3}$,
which is positive for every real $s$.
To obtain the clean-limit expansion at fixed $T>0$, we expand
$f(\varepsilon_\kappa+\gamma s)$ locally in $\gamma s$ and
integrate the terms with convergent moments.
The required moments of this weight are
\begin{widetext}
\begin{align}
  \int_{-\infty}^{\infty}\frac{ds}{(1+s^{2})^{3}}
  =\frac{3\pi}{8},
  \quad
  \int_{-\infty}^{\infty}\frac{s^{2}ds}{(1+s^{2})^{3}}
  =\frac{\pi}{8},
  \quad
  \int_{-\infty}^{\infty}\frac{s^{4}ds}{(1+s^{2})^{3}}
  =\frac{3\pi}{8}.
  \label{eq:oneband-combined-moments}
\end{align}
\end{widetext}
They give the controlled expansion, at fixed $T>0$,
\begin{align}
  \sigma_{\mathrm{exact}}
  &=-\frac{e^{3}\tau^{2}}{\hbar^{3}}\sum_{\kappa}
    f_{\kappa}\partial_{\kappa}^{3}\varepsilon_{\kappa}
  \nonumber\\
  &\quad-\frac{e^{3}}{24\hbar}\sum_{\kappa}
    f^{(2)}_{\kappa}\partial_{\kappa}^{3}\varepsilon_{\kappa}
  \nonumber\\
  &\quad-\frac{e^{3}\gamma^{2}}{96\hbar}\sum_{\kappa}
    f^{(4)}_{\kappa}\partial_{\kappa}^{3}\varepsilon_{\kappa}
   +\mathcal{O}(\gamma^{3}).
  \label{eq:oneband-current-expansion}
\end{align}
In particular, the $\mathcal{O}(\gamma)$ contributions of the
individual kernels cancel after combination in the current
and integration over the periodic Brillouin zone.  The first
correction beyond the kinetic $\tau^{0}$ term is therefore
$\mathcal{O}(\gamma^{2})=\mathcal{O}(\tau^{-2})$, even
though Eqs.~\eqref{eq:oneband-Gexpand-G1} and
\eqref{eq:oneband-Gexpand-G2} separately have
$\mathcal{O}(\gamma)$ remainders.  The sixth moment of
$(1+s^{2})^{-3}$ diverges, so this argument does not justify
an all-orders expansion containing only even powers of
$\gamma$.  The remainder after the displayed $\gamma^{2}$
term is generally $\mathcal{O}(\gamma^{3})$, rather than
$\mathcal{O}(\gamma^{4})$.

This one-band check does not use the diagrammatic expansion
of the main derivation.
The Peierls-shifted current vertex is kept intact while the
steady state is solved nonperturbatively in the field.
The effects of the contact vertices are included in the
momentum shift, so no separate contact-vertex expansion is
required.  In this direct time-domain calculation, no explicit
$1/(\omega_1\omega_2)$ singularities arise, and thus the
Ward identity need not be invoked to cancel such singularities.
The calculation requires no band-basis decomposition and does
not use the diagrammatic kernels $\mathcal{K}^{(1)}$ and
$\mathcal{K}^{(2)}$ of Appendix~\ref{app:kernel-expansion}.
After obtaining the steady state, we extract the second-order
conductivity by expanding in $q=-eE/\hbar$ at fixed $\gamma>0$.
We then obtain the clean-limit behavior by an asymptotic
expansion in $\gamma$ at fixed $T>0$, using kernel moments
and the relation $\tau=\hbar/(2\gamma)$.
The agreement therefore checks the fully diagonal kernel
algebra in Appendix~\ref{app:kernel-expansion} and its
clean-limit evaluation.
Both calculations use the self-energies in
Eq.~\eqref{eq:self-energies}, so their agreement establishes
consistency within this dissipative model.
It does not establish independence from the relaxation
mechanism or determine the higher-order corrections in the
other multiband geometric channels.

\begin{figure*}[t]
  \centering
  \includegraphics[width=\textwidth]{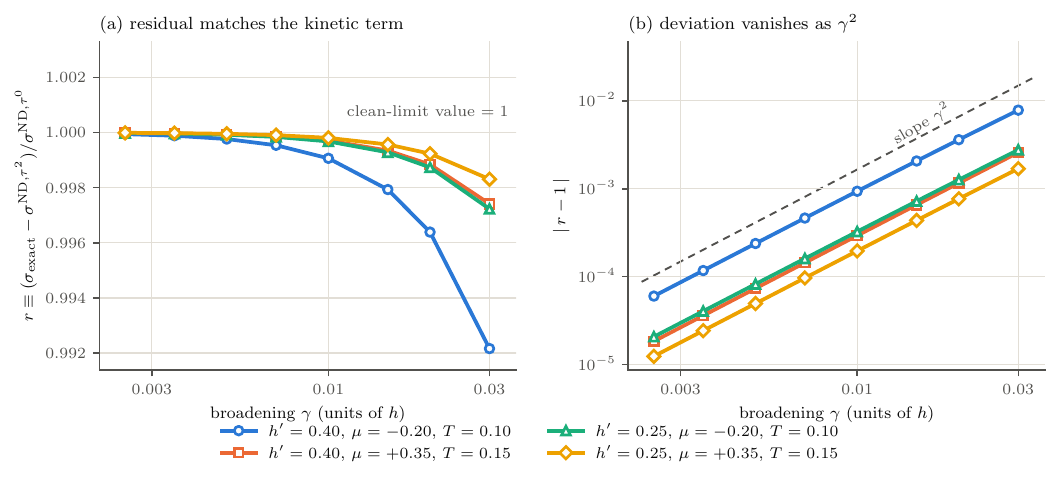}
  \caption{Approach to the clean limit for the model of
    Fig.~\ref{fig:oneband-model}.
    (a) Ratio $r$ [Eq.~\eqref{eq:oneband-ratio}] of the
    finite-$\gamma$ residual
    $\sigma_{\mathrm{exact}}-\sigma^{\mathrm{ND},\tau^{2}}_{xxx}$
    to the kinetic term of Eq.~\eqref{eq:oneband-tau0}, for
    $h^{\prime}=0.40$ and $0.25$ at two choices of $(\mu,T)$.
    All four curves approach the clean-limit value $r=1$ as
    $\gamma\to0^{+}$; equality is not required at finite
    $\gamma$.
    (b) Deviation $|r-1|$ on logarithmic axes; the dashed line
    indicates a $\gamma^{2}$ slope, consistent with
    Eq.~\eqref{eq:oneband-current-expansion}.
    Energies are in units of $h$, with
    $e=\hbar=k_{\mathrm{B}}=1$ and the same Brillouin-zone
    normalization as Fig.~\ref{fig:oneband-model}.}
  \label{fig:oneband}
\end{figure*}

Figures~\ref{fig:oneband-model} and \ref{fig:oneband} show the 
numerical verification for $h=1$ and $h^{\prime}=0.40,0.25$. 
In the numerical calculations we set $e=\hbar=k_{\mathrm{B}}=1$ 
and use the Brillouin-zone average
$\sum_{\kappa}=\int_{-\pi}^{\pi}d\kappa/(2\pi)$, approximated by a uniform grid of 4096 points, 
without a spin-degeneracy factor.
Figure~\ref{fig:oneband-model} follows the presentation of
the two-band models in Secs.~\ref{subsec:tilted-dirac} and
\ref{subsec:real-qmd-model}: 
panel~(a) shows the band as a function of $k$, 
while panels~(b) and (c) show
$\sigma^{\mathrm{ND},\tau^{2}}_{xxx}/\tau^{2}$ and
$\sigma^{\mathrm{ND},\tau^{0}}_{xxx}$ as functions of $\mu$
at $T=0.05$ over $-3\le\mu\le3$.

Although the leading channel in Eq.~\eqref{eq:oneband-tau2}
is written with the Fermi-sea weight $f_{\kappa}$, it also
admits the Fermi-surface representations in
Eq.~\eqref{eq:oneband-bz-identity}.  Thus the use of
$f_{\kappa}$ for one channel and $f^{(2)}_{\kappa}$ for the
other should not be interpreted as a representation-independent
separation into Fermi-sea and Fermi-surface mechanisms. 
A direct relation between the two curves, 
at fixed band and fixed $T$, is
\begin{align}
  \sigma^{\mathrm{ND},\tau^{0}}_{xxx}
  &=\frac{\hbar^{2}}{24}\frac{\partial^{2}}{\partial\mu^{2}}
    \biggl(\frac{\sigma^{\mathrm{ND},\tau^{2}}_{xxx}}{\tau^{2}}\biggr).
  \label{eq:oneband-mu-curvature}
\end{align}
This explains the sharper chemical-potential dependence of the kinetic channel.  
An additional check follows from
\begin{align}
  \varepsilon_{\pi-k}&=-\varepsilon_{k},
  \qquad
  \sigma_{\mathrm{exact}}(-\mu)
  =-\sigma_{\mathrm{exact}}(\mu),
  \label{eq:oneband-mu-symmetry}
\end{align}
where $T$ and $\gamma$ are held fixed in the second equality.
The same oddness in $\mu$ holds separately for the
$\tau^{2}$ and $\tau^{0}$ coefficients.  In particular,
these second-order quantities vanish at $\mu=0$, as well as
in the inversion-symmetric limit $h^{\prime}=0$.

The open symbols in Fig.~\ref{fig:oneband-model}(c) show the
finite-$\gamma$ residual
$\sigma_{\mathrm{exact}}-\sigma^{\mathrm{ND},\tau^{2}}_{xxx}$
at $\gamma=0.004$, compared with the kinetic term (solid curves).
The residual approaches the kinetic term as $\gamma\to0$,
with corrections of order $\gamma^{2}$ given in
Eq.~\eqref{eq:oneband-current-expansion}.
For each of the two values of $h^{\prime}$, the maximum absolute
difference over the 601-point grid
$\mu=-3,-2.99,\ldots,3$ is approximately $0.13\%$ of the
maximum absolute amplitude of the kinetic contribution.
The comparison includes points near the sharp features and
zero crossings of the kinetic contribution.
This normalization keeps the error measure well defined even
where the kinetic term vanishes, whereas a pointwise relative
error would be undefined at those zeros.

Figure~\ref{fig:oneband} examines the clean-limit convergence
at fixed $\mu$ and $T$.  For each value of $h^{\prime}$, we use
$(\mu,T)=(-0.20,0.10)$ and $(0.35,0.15)$.
For these four parameter sets,
$\sigma^{\mathrm{ND},\tau^{0}}_{xxx}\neq0$, and we define
\begin{align}
  r&=\frac{\sigma_{\mathrm{exact}}
           -\sigma^{\mathrm{ND},\tau^{2}}_{xxx}}
          {\sigma^{\mathrm{ND},\tau^{0}}_{xxx}}.
  \label{eq:oneband-ratio}
\end{align}
For all four parameter sets, panel~(a) shows $r\to1$ as
$\gamma\to0^{+}$.  The horizontal reference line marks this
limiting value.  At $\gamma=0.0025$, the maximum deviation
across the four sets satisfies $\max|r-1|<6.1\times10^{-5}$.
Panel~(b) shows $|r-1|$ on logarithmic axes.
Over $0.0025\le\gamma\le0.03$, power-law fits give exponents
$1.96$--$1.99$, consistent with the first correction beyond
the kinetic term being $\mathcal{O}(\gamma^{2})$ as derived in
Eq.~\eqref{eq:oneband-current-expansion}.
The small deviations of the fitted exponents from $2$ are
compatible with higher-order corrections over the finite
fitting range.
Omitting the kinetic term leaves only the leading nonlinear
Drude contribution in this model, since all geometric
contributions vanish.  Such an expression predicts a zero
residual in Fig.~\ref{fig:oneband-model}(c) and $r=0$ in
Fig.~\ref{fig:oneband}(a), whereas the exact calculation gives
a nonzero limiting residual and $r\to1$.

For a stable numerical evaluation, the leading
$\mathcal{O}(\tau^{2})$ term is removed inside the energy
integral rather than by subtracting two separately computed
quantities of order $\gamma^{-2}$.  An exact form particularly
suited to this subtraction follows directly from
Eq.~\eqref{eq:oneband-exact-gamma}:
\begin{align}
  \sigma_{\mathrm{exact}}
   -\sigma^{\mathrm{ND},\tau^{2}}_{xxx}
  &=-\frac{2e^{3}}{3\pi\hbar\gamma^{2}}\sum_{\kappa}
    \partial_{\kappa}^{3}\varepsilon_{\kappa}
    \int_{0}^{\infty}\frac{ds}{(1+s^{2})^{3}}\nonumber\\
    &\quad \times
  \bigl[
      f(\varepsilon_{\kappa}+\gamma s)
      +f(\varepsilon_{\kappa}-\gamma s)-2f_{\kappa}\bigr].
  \label{eq:oneband-residual-exact}
\end{align}
The symmetric Fermi-function difference must itself be
evaluated with sufficient precision, since subtracting
nearly equal function values can otherwise obscure the
small $\gamma^{2}$ deviation; in the calculations reported
here it is evaluated from a hyperbolic-function identity
rather than by direct cancellation.
Equation~\eqref{eq:oneband-residual-exact}
also avoids taking the difference of the two individual
kernels after their integration and provides a direct
cross-check of the original kernel representation.

\FloatBarrier
\section{Discussion and outlook}
\label{sec:discussion}

We have derived the clean-limit second-order dc conductivity
in the velocity-gauge Peierls approximation, tracing the
relation between contact vertices and connection cancellation
in the band basis without projectors.
Alongside the leading nonlinear Drude, BCD, and two QMD
contributions, the diagonal sector yields the known kinetic
term~\cite{Guo2026Dissipation,Anan2026Nonreciprocal,Kitamura2026QuasiequilibriumShift,Ulrich2026IntraQMD}
of Eq.~\eqref{eq:ND-tau0}, completing the response through
$\mathcal{O}(\tau^{0})$ within the specified model.
The leading Drude response depends only on the band dispersion,
whereas the BCD term reflects Berry curvature and a
retarded--advanced pinch singularity.
The two QMD contributions are governed by band geometry.
The shared order $\tau^{0}$ therefore groups physically distinct
kinetic and geometric mechanisms.
The exact single-band expression
Eq.~\eqref{eq:oneband-exact-gamma} reproduces the $\tau^{2}$
and $\tau^{0}$ coefficients and checks the fully diagonal
sector within the same dissipative model.
The dc limit is taken at finite broadening before the
large-$\tau$ expansion;
lifetime independence does not establish independence from
the relaxation mechanism (Sec.~\ref{sec:scope}).

The intermediate combinations $\mathcal{C}^{nm}_{ij|k}$ and
$\mathcal{T}^{nml}_{ijk}$ are individually invariant under
bandwise $U(1)$ phase transformations.
Combining their contributions cancels the explicit
three-connection structures and expresses the geometric part
through derivatives of the ordinary and band-normalized
quantum metrics (Sec.~\ref{sec:cancellation},
Appendix~\ref{app:dc-calculations}, and the Supplemental
Material~\cite{SM}).
This organization is consistent with the projector-calculus
description of gauge-invariant
observables~\cite{Mitscherling2025ProjectorCalculus,Avdoshkin2025Multistate}.
As detailed in Sec.~\ref{subsec:comparison-dc-limit}, the
Fermi-surface coefficients agree with
Ref.~\cite{Ulrich2026IntraQMD} after translating conventions;
the compact QMD expression of
Ref.~\cite{JiangHolderYan2025Review} uses a different
organization of the metric sector, whose relation to the
present intraband--interband separation is clarified by
Eqs.~\eqref{eq:intraQMD-band} and \eqref{eq:interQMD-band}.
The Peierls vertices and connection cancellation have
precedents in Refs.~\cite{JoaoLopes2020,Ulrich2026IntraQMD};
the present derivation makes their relation explicit in the
velocity-gauge band basis while retaining the kinetic term.

The real two-band model in Sec.~\ref{subsec:real-qmd-model}
illustrates the diagnostic value of the decomposition.
Because the Hamiltonian is chosen real, the Berry curvature
is identically zero and the BCD response vanishes
[Eq.~\eqref{eq:real-qmd-bcd-zero}].
Nevertheless, the ordinary quantum metric remains finite,
and the Fermi-surface asymmetry introduced by the scalar
tilt produces a nonzero intraband QMD
[Eq.~\eqref{eq:bcd-zero-qmd-nonzero}].
The vanishing Berry curvature excludes the BCD mechanism.
A finite $\mathcal{O}(\tau^{0})$ response, however, need not
be entirely geometric because the kinetic contribution may
remain, as the one-band example explicitly demonstrates.
Identifying the intraband quantum-metric dipole therefore
requires separating both the kinetic and interband-QMD
contributions.  The displayed $S_{xyy}$ evaluates the
intraband-QMD channel itself rather than the full
$\mathcal{O}(\tau^{0})$ response.

The present formulas should be interpreted within their
stated scope.
We have employed Peierls velocity-gauge coupling in a
localized tight-binding representation, neglected
intra-atomic dipole matrix elements beyond the Peierls
approximation, and used a constant phenomenological
relaxation time in the bare-bubble approximation.
Microscopic impurity scattering, vertex corrections,
side-jump processes, skew scattering, and disorder-induced
changes to the distribution function can generate additional
nonlinear transport contributions and lie outside the
present calculation.  In particular, the BCD term retained here
is the clean constant-relaxation-time contribution, whereas the
additional disorder-mediated nonlinear Hall mechanisms derived in
Refs.~\cite{NandySodemann2019,Du2019DisorderNHE,XiaoDuNiu2019,Du2021QuantumTheoryNHE}
are not included.

Several extensions are natural.
First, the velocity-gauge Keldysh result could be reformulated
directly in projector language, which would make the
connection to the gauge-invariant projector
calculus~\cite{Mitscherling2025ProjectorCalculus} even more
transparent and potentially extend the present results to
degenerate bands.
Second, including a microscopic impurity model would
determine which parts of the quantum-metric response are
renormalized by vertex corrections and side-jump
contributions.
Third, the real two-band diagnostic suggests a route for
material screening: systems with suppressed Berry curvature
but sizable Fermi-surface quantum-metric dipoles should
display a quantum-metric contribution to the nonlinear dc
response distinct from the
BCD mechanism, as has been observed in topological
antiferromagnets~\cite{Wang2023Nature,Gao2023Science}.
Finally, the formulation provides a route to higher-order
dc responses.  The present analysis shows why subleading
dispersive terms must be retained alongside geometric terms.
At third order, explicit evaluation of the diagonal kernels
gives a subleading nonlinear Drude contribution
$\sigma^{\mathrm{ND},\tau^{1}}_{ijkl}$ at the same order as
quantum-metric-quadrupole structures; power counting alone
does not establish its nonzero coefficient.
Related third-order responses have been studied through
Berry-curvature multipoles, quantum-metric quadrupoles, and
quantum-kinetic
approaches~\cite{Zhang2023BerryMultipoles,Mandal2024ThirdOrder,Liu2025QuantumMetricQuadrupole,JiangHolderYan2025Review}.
We have also verified cancellation of the purely static
vector-potential contribution at the current-vertex level.
A complete low-frequency reduction, including cancellation
of all apparent frequency singularities and the clean-limit
geometric decomposition, remains for future work.

\appendix
\section{Perturbative expansion and second-order current}
\label{app:green-current}

This appendix presents the perturbative expansion of the
contour-ordered Green's function to second order in the vector
potential and the resulting second-order current contributions.
The detailed calculations are collected in the Supplemental
Material~\cite{SM}.

\subsection{Perturbative Green's functions}

The Fourier components of the zeroth-, first-, and
second-order contour-ordered Green's functions, obtained by
standard perturbation theory~\cite{Rammer1986,Haug2008}
and the Langreth rules~\cite{Langreth1976,Haug2008}, are
\begin{align}
  G^{(0)}_{\bm{k}}(\omega)
  &= 2\pi\delta(\omega)
     \int_{-\infty}^{\infty}
     \frac{d\varepsilon}{2\pi\hbar}\,
     G_{\bm{k}}(\varepsilon),
  \label{eq:G0-omega}
  \\
  G^{(1)}_{\bm{k}}(\omega)
  &= e\sum_{j}
     \int_{-\infty}^{\infty}\frac{d\varepsilon}{2\pi\hbar}\,
     G_{\bm{k}}(\varepsilon+\hbar\omega)\,
     V_{j}(\bm{k})\,G_{\bm{k}}(\varepsilon)\,
     A_{j}(\omega),
  \label{eq:G1-freq}
\end{align}
and, at second order, the two-vertex contribution
\begin{align}
  G^{(2a)}_{\bm{k}}(\omega)
  &= e^{2}\sum_{j,k}
     \int_{-\infty}^{\infty}\frac{d\varepsilon}{2\pi\hbar}
     \int_{-\infty}^{\infty}\frac{d\omega'}{2\pi}\,
     A_{j}(\omega-\omega')\,A_{k}(\omega')
  \nonumber\\
  &\quad\times
     G_{\bm{k}}(\varepsilon+\hbar\omega)\,
     V_{j}(\bm{k})\,
     G_{\bm{k}}(\varepsilon+\hbar\omega')\,
     V_{k}(\bm{k})\,
     G_{\bm{k}}(\varepsilon),
  \label{eq:G2a-freq}
\end{align}
and the contact contribution from the two-photon vertex
$V_{jk}$,
\begin{align}
  G^{(2b)}_{\bm{k}}(\omega)
  &= \frac{e^{2}}{2}\sum_{j,k}
     \int_{-\infty}^{\infty}\frac{d\varepsilon}{2\pi\hbar}
     \int_{-\infty}^{\infty}\frac{d\omega'}{2\pi}\,
     A_{j}(\omega-\omega')\,A_{k}(\omega')
  \nonumber\\
  &\quad\times
     G_{\bm{k}}(\varepsilon+\hbar\omega)\,
     V_{jk}(\bm{k})\,
     G_{\bm{k}}(\varepsilon).
  \label{eq:G2b-freq}
\end{align}
Here $G_{\bm{k}}(\varepsilon)$ is the equilibrium
contour-ordered propagator; its retarded and advanced
components are defined in
Eqs.~\eqref{eq:G-retarded-def} and \eqref{eq:G-advanced-def}.

\subsection{Second-order current}

The Fourier component of the second-order current
$\mathcal{J}^{(2)}_{i}(\omega)$ decomposes as
\begin{align}
  \mathcal{J}^{(2)}_{i}(\omega)
  &= \mathcal{J}^{\mathrm{p}(2a)}_{i}(\omega)
   + \mathcal{J}^{\mathrm{p}(2b)}_{i}(\omega)
   + \mathcal{J}^{\mathrm{d}(2)}_{i}(\omega)
   + \mathcal{J}^{\mathrm{t}(2)}_{i}(\omega),
  \label{eq:app-J2-decomp}
\end{align}
where the four contributions correspond to the paramagnetic
current from $G^{(2a)<}(\omega)$ [Eq.~\eqref{eq:G2a-freq}],
the paramagnetic current from $G^{(2b)<}(\omega)$
[Eq.~\eqref{eq:G2b-freq}], the diamagnetic current from
$G^{(1)<}(\omega)$ [Eq.~\eqref{eq:G1-freq}], and the tadpole
current from $G^{(0)<}(\omega)$ [Eq.~\eqref{eq:G0-omega}],
respectively.
Using $A_{j}(\omega) = E_{j}(\omega)/(i\omega)$, each
contribution reads
\begin{widetext}
\begin{align}
  \mathcal{J}^{\mathrm{p}(2a)}_{i}(\omega)
  &= ie\hbar\sum_{\bm{k}}
     \mathrm{Tr}\!\left[
       V_{i}(\bm{k})\,G^{(2a)<}_{\bm{k}}(\omega)
     \right]
  \nonumber\\
  &= -ie^{3}\sum_{j,k,\bm{k}}
     \int_{-\infty}^{\infty}\frac{d\varepsilon}{2\pi}
     \int_{-\infty}^{\infty}\frac{d\omega'}{2\pi}\,
     \frac{E_{j}(\omega-\omega')\,E_{k}(\omega')}
          {(\omega-\omega')\omega'}
  \nonumber\\
  &\quad\times
     \mathrm{Tr}\!\Big[
       V_{i}(\bm{k})\,
       G_{\bm{k}}(\varepsilon+\hbar\omega)\,
       V_{j}(\bm{k})\,
       G_{\bm{k}}(\varepsilon+\hbar\omega')\,
       V_{k}(\bm{k})\,
       G_{\bm{k}}(\varepsilon)
     \Big]^{<},
  \label{eq:Jp2a}
  \\[6pt]
  \mathcal{J}^{\mathrm{p}(2b)}_{i}(\omega)
  &= ie\hbar\sum_{\bm{k}}
     \mathrm{Tr}\!\left[
       V_{i}(\bm{k})\,G^{(2b)<}_{\bm{k}}(\omega)
     \right]
  \nonumber\\
  &= -i\frac{e^{3}}{2}\sum_{j,k,\bm{k}}
     \int_{-\infty}^{\infty}\frac{d\varepsilon}{2\pi}
     \int_{-\infty}^{\infty}\frac{d\omega'}{2\pi}\,
     \frac{E_{j}(\omega-\omega')\,E_{k}(\omega')}
          {(\omega-\omega')\omega'}
  \nonumber\\
  &\quad\times
     \mathrm{Tr}\!\Big[
       V_{i}(\bm{k})\,
       G_{\bm{k}}(\varepsilon+\hbar\omega)\,
       V_{jk}(\bm{k})\,
       G_{\bm{k}}(\varepsilon)
     \Big]^{<},
  \label{eq:Jp2b}
  \\[6pt]
  \mathcal{J}^{\mathrm{d}(2)}_{i}(\omega)
  &= ie^{2}\hbar\sum_{j}\sum_{\bm{k}}
     \int_{-\infty}^{\infty}\frac{d\omega'}{2\pi}\,
     \mathrm{Tr}\!\left[
       V_{ij}(\bm{k})\,G^{(1)<}_{\bm{k}}(\omega')
     \right]A_{j}(\omega-\omega')
  \nonumber\\
  &= -ie^{3}\sum_{j,k,\bm{k}}
     \int_{-\infty}^{\infty}\frac{d\varepsilon}{2\pi}
     \int_{-\infty}^{\infty}\frac{d\omega'}{2\pi}\,
     \frac{E_{j}(\omega-\omega')\,E_{k}(\omega')}
          {(\omega-\omega')\omega'}
  \nonumber\\
  &\quad\times
     \mathrm{Tr}\!\left[
       V_{ij}(\bm{k})\,
       G_{\bm{k}}(\varepsilon+\hbar\omega')\,
       V_{k}(\bm{k})\,
       G_{\bm{k}}(\varepsilon)
     \right]^{<},
  \label{eq:Jd2}
  \\[6pt]
  \mathcal{J}^{\mathrm{t}(2)}_{i}(\omega)
  &= \frac{ie^{3}\hbar}{2}\sum_{j,k,\bm{k}}
     \int_{-\infty}^{\infty}\frac{d\omega'}{2\pi}
     \int_{-\infty}^{\infty}\frac{d\omega''}{2\pi}\,
     \mathrm{Tr}\!\left[
       V_{ijk}(\bm{k})\,G^{(0)<}_{\bm{k}}(\omega')
     \right]A_{j}(\omega'')A_{k}(\omega-\omega'-\omega'')
  \nonumber\\
  &= -i\frac{e^{3}}{2}\sum_{j,k,\bm{k}}
     \int_{-\infty}^{\infty}\frac{d\varepsilon}{2\pi}
     \int_{-\infty}^{\infty}\frac{d\omega'}{2\pi}\,
     \frac{E_{j}(\omega-\omega')\,E_{k}(\omega')}
          {(\omega-\omega')\omega'}
     \mathrm{Tr}\!\left[
       V_{ijk}(\bm{k})\,G^{<}_{\bm{k}}(\varepsilon)
     \right].
  \label{eq:Jt2}
\end{align}
\end{widetext}
The diagrammatic representation of these four contributions
is shown in Fig.~\ref{fig:second_order_current_diagrams}.
The paramagnetic diagram
[Fig.~\ref{fig:second_order_current_diagrams}(a)]
involves a triangular electron loop with three one-photon
velocity vertices, while the bubble diagram
[Fig.~\ref{fig:second_order_current_diagrams}(b)]
couples the one-photon vertex $V_{i}$ to the two-photon
contact vertex $V_{jk}$ via a single propagator loop.
The contact diagrams
[Figs.~\ref{fig:second_order_current_diagrams}(c) and (d)]
contain the two-photon vertex $V_{ij}$ or the three-photon
vertex $V_{ijk}$ at a single node.

\section{Gauge-covariant band-geometry identities}
\label{app:geometry}

This appendix summarizes the band-geometric identities used
to convert the band-basis Kubo kernel into the
gauge-invariant form used in the main
text~\cite{Provost1980,Sipe2000,Ventura2017,Parker2019}.

\subsection{Covariant derivative and quantum connection}
For $n\neq m$, the off-diagonal Berry connection transforms
under a band-dependent gauge transformation as
\begin{align}
  \mathcal{A}_{i}^{nm}
  \longrightarrow
  e^{i(\chi_{m}-\chi_{n})}\mathcal{A}_{i}^{nm}.
  \label{eq:app-offdiag-A-gauge}
\end{align}
The covariant derivative is defined by
\begin{align}
  \mathcal{D}_{j}\mathcal{A}_{i}^{nm}
  = \partial_{j}\mathcal{A}_{i}^{nm}
  + i\bigl(
      \mathcal{A}_{j}^{mm} - \mathcal{A}_{j}^{nn}
    \bigr)\mathcal{A}_{i}^{nm}
  \label{eq:app-covariant-derivative}
\end{align}
and transforms with the same relative phase as
$\mathcal{A}_{i}^{nm}$, namely
$\mathcal{D}_{j}\mathcal{A}_{i}^{nm}
  \to e^{i(\chi_{m}-\chi_{n})}\mathcal{D}_{j}\mathcal{A}_{i}^{nm}$.
Consequently, $\mathcal{D}_{k}\mathcal{A}_{j}^{mn}$ carries the
opposite phase $e^{i(\chi_{n}-\chi_{m})}$, which is canceled by
the phase of $\mathcal{A}_{i}^{nm}$ in the product.  Therefore
the band-pair-resolved quantum connection
\begin{align}
  \mathcal{C}^{nm}_{ij|k}
  \equiv \mathcal{A}_{i}^{nm}\mathcal{D}_{k}\mathcal{A}_{j}^{mn}
  \label{eq:app-Ccal-def}
\end{align}
is gauge-invariant.
We decompose it as
\begin{align}
  \mathcal{C}^{nm}_{ij|k}
  = \Gamma^{nm}_{ij|k} - i\widetilde{\Gamma}^{nm}_{ij|k},
  \label{eq:app-Ccal-real-imag}
\end{align}
where
\begin{align}
  \Gamma^{nm}_{ij|k}
  = \mathrm{Re}\,\mathcal{C}^{nm}_{ij|k},
  \qquad
  \widetilde{\Gamma}^{nm}_{ij|k}
  = -\mathrm{Im}\,\mathcal{C}^{nm}_{ij|k}.
  \label{eq:app-Gamma-defs}
\end{align}
\subsection{Quantum geometric tensor and its derivatives}

The pair-resolved quantum geometric tensor is
\begin{align}
  \mathcal{Q}_{ij}^{nm}
  \equiv \mathcal{A}_{i}^{nm}\mathcal{A}_{j}^{mn}
  = \mathcal{G}_{ij}^{nm} - \frac{i}{2}\Omega_{ij}^{nm},
  \label{eq:app-pair-qgt}
\end{align}
where $\mathcal{G}_{ij}^{nm} = \mathrm{Re}\,\mathcal{Q}_{ij}^{nm}$
is the band-pair quantum metric and
$\Omega_{ij}^{nm} = -2\,\mathrm{Im}\,\mathcal{Q}_{ij}^{nm}$
is the band pair-resolved Berry curvature.
A direct product differentiation gives
\begin{align}
  \partial_{k}\mathcal{Q}_{ij}^{nm}
  = \mathcal{C}^{nm}_{ij|k} + \mathcal{C}^{mn}_{ji|k}.
  \label{eq:app-dC-direct}
\end{align}
To verify this, use Eq.~\eqref{eq:app-covariant-derivative}
to expand each term:
\begin{align}
  \mathcal{C}^{nm}_{ij|k}
  &= \mathcal{A}_{i}^{nm}\partial_{k}\mathcal{A}_{j}^{mn}
   + i\bigl(
       \mathcal{A}_{k}^{nn} - \mathcal{A}_{k}^{mm}
     \bigr)
     \mathcal{A}_{i}^{nm}\mathcal{A}_{j}^{mn},
  \label{eq:app-Ccal-expand-1}
  \\
  \mathcal{C}^{mn}_{ji|k}
  &= \mathcal{A}_{j}^{mn}\partial_{k}\mathcal{A}_{i}^{nm}
   + i\bigl(
       \mathcal{A}_{k}^{mm} - \mathcal{A}_{k}^{nn}
     \bigr)
     \mathcal{A}_{j}^{mn}\mathcal{A}_{i}^{nm}.
  \label{eq:app-Ccal-expand-2}
\end{align}
The diagonal-connection terms cancel in the sum and
Eq.~\eqref{eq:app-dC-direct} follows.
Taking real and imaginary parts yields
\begin{align}
  \partial_{k}\mathcal{G}_{ij}^{nm}
  &= \Gamma^{nm}_{ij|k} + \Gamma^{nm}_{ji|k},
  \label{eq:dG-Gamma}
  \\
  \partial_{k}\Omega_{ij}^{nm}
  &= 2\bigl(
       \widetilde{\Gamma}^{nm}_{ij|k}
     - \widetilde{\Gamma}^{nm}_{ji|k}
     \bigr).
  \label{eq:dOmega-Gamma}
\end{align}
Thus $\Gamma^{nm}_{ij|k}$ and $\widetilde{\Gamma}^{nm}_{ij|k}$
are the elementary band-pair building blocks from which
momentum-space gradients of the quantum metric and Berry
curvature are assembled.

\subsection{\texorpdfstring{Relation between $\Gamma^{nm}_{ij|k}$ and metric derivatives}{Relation between Gamma and metric derivatives}}

For the cancellation proof in Sec.~\ref{sec:cancellation},
we also need the identity that expresses
$\Gamma^{nm}_{ij|k}$ in terms of metric derivatives and a
primed commutator.
It follows from the pure-gauge (zero-field curvature) relation
\begin{align}
  \partial_{i}\mathcal{A}_{j}^{nm}
  - \partial_{j}\mathcal{A}_{i}^{nm}
  = i\bigl[\mathcal{A}_{i},\mathcal{A}_{j}\bigr]_{nm},
  \label{eq:app-pure-gauge-identity}
\end{align}
which, after absorbing the external-band contributions into
the covariant derivatives, becomes
\begin{align}
  \mathcal{D}_{i}\mathcal{A}_{j}^{nm}
  - \mathcal{D}_{j}\mathcal{A}_{i}^{nm}
  = i\bigl[\mathcal{A}_{i},\mathcal{A}_{j}\bigr]'_{nm},
  \label{eq:app-covariant-curl}
\end{align}
where the primed commutator
\begin{align}
  \bigl[\mathcal{A}_{i},\mathcal{A}_{j}\bigr]'_{nm}
  = \sum_{p\neq n,m}
    \bigl(
      \mathcal{A}_{i}^{np}\mathcal{A}_{j}^{pm}
    - \mathcal{A}_{j}^{np}\mathcal{A}_{i}^{pm}
    \bigr)
  \label{eq:app-primed-commutator}
\end{align}
excludes the external bands $n$ and $m$ from the
intermediate-band sum.

Using Eqs.~\eqref{eq:dG-Gamma} and
\eqref{eq:app-covariant-curl}, one finds
\begin{align}
  2\Gamma^{nm}_{ij|k}
  &= \partial_{k}\mathcal{G}_{ij}^{nm}
   + \partial_{j}\mathcal{G}_{ki}^{nm}
   - \partial_{i}\mathcal{G}_{jk}^{nm}
   + X_{ijk}^{nm},
  \label{eq:Gamma-metric-X}
\end{align}
where
\begin{align}
  X_{ijk}^{nm}
  &= \mathrm{Im}\!\left[
       \mathcal{A}_{i}^{nm}
       \bigl[\mathcal{A}_{j},\mathcal{A}_{k}\bigr]'_{mn}
     \right]
  \nonumber\\
  &\quad
   + \mathrm{Im}\!\left[
       \mathcal{A}_{j}^{nm}
       \bigl[\mathcal{A}_{k},\mathcal{A}_{i}\bigr]'_{mn}
     \right]
  \nonumber\\
  &\quad
   - \mathrm{Im}\!\left[
       \mathcal{A}_{k}^{nm}
       \bigl[\mathcal{A}_{i},\mathcal{A}_{j}\bigr]'_{mn}
     \right].
  \label{eq:X-comm-app}
\end{align}
To derive Eq.~\eqref{eq:Gamma-metric-X}, apply
Eq.~\eqref{eq:dG-Gamma} to the three metric derivatives:
\begin{align}
  \partial_{k}\mathcal{G}_{ij}^{nm}
  &= \Gamma^{nm}_{ij|k} + \Gamma^{nm}_{ji|k},
  \label{eq:app-dG1}
  \\
  \partial_{j}\mathcal{G}_{ki}^{nm}
  &= \Gamma^{nm}_{ki|j} + \Gamma^{nm}_{ik|j},
  \label{eq:app-dG2}
  \\
  \partial_{i}\mathcal{G}_{jk}^{nm}
  &= \Gamma^{nm}_{jk|i} + \Gamma^{nm}_{kj|i}.
  \label{eq:app-dG3}
\end{align}
Combining these and isolating $2\Gamma^{nm}_{ij|k}$ gives
\begin{align}
  &\partial_{k}\mathcal{G}_{ij}^{nm}
  + \partial_{j}\mathcal{G}_{ki}^{nm}
  - \partial_{i}\mathcal{G}_{jk}^{nm}
  - 2\Gamma^{nm}_{ij|k}
  \nonumber\\
  &= \bigl(\Gamma^{nm}_{ji|k} - \Gamma^{nm}_{jk|i}\bigr)
   + \bigl(\Gamma^{nm}_{ki|j} - \Gamma^{nm}_{kj|i}\bigr)
   + \bigl(\Gamma^{nm}_{ik|j} - \Gamma^{nm}_{ij|k}\bigr).
  \label{eq:app-Gamma-difference}
\end{align}
Each difference is converted via
Eq.~\eqref{eq:app-covariant-curl}:
\begin{align}
  \Gamma^{nm}_{ji|k} - \Gamma^{nm}_{jk|i}
  &= -\mathrm{Im}\!\left[
       \mathcal{A}_{j}^{nm}
       \bigl[\mathcal{A}_{k},\mathcal{A}_{i}\bigr]'_{mn}
     \right],
  \label{eq:app-Gamma-diff-1}
  \\
  \Gamma^{nm}_{ki|j} - \Gamma^{nm}_{kj|i}
  &= \mathrm{Im}\!\left[
       \mathcal{A}_{k}^{nm}
       \bigl[\mathcal{A}_{i},\mathcal{A}_{j}\bigr]'_{mn}
     \right],
  \label{eq:app-Gamma-diff-2}
  \\
  \Gamma^{nm}_{ik|j} - \Gamma^{nm}_{ij|k}
  &= -\mathrm{Im}\!\left[
       \mathcal{A}_{i}^{nm}
       \bigl[\mathcal{A}_{j},\mathcal{A}_{k}\bigr]'_{mn}
     \right].
  \label{eq:app-Gamma-diff-3}
\end{align}
Substituting
Eqs.~\eqref{eq:app-Gamma-diff-1}--\eqref{eq:app-Gamma-diff-3}
into Eq.~\eqref{eq:app-Gamma-difference} and rearranging
gives Eq.~\eqref{eq:Gamma-metric-X}.

\section{Clean-limit expansion of the Green's function kernels}
\label{app:kernel-expansion}
\begin{widetext}

This appendix derives the clean-limit expansion of the
Green's function kernels $\mathcal{K}^{(1)}_{nml}$ and
$\mathcal{K}^{(2)}_{nm}$ defined in
Eqs.~\eqref{eq:K1-def-band} and \eqref{eq:K2-def-band}. 
The detailed calculations are collected in the Supplemental
Material~\cite{SM}. 

Throughout we set $\mu = 0$ for notational brevity,
suppress the crystal-momentum argument $\bm{k}$, and
introduce the shorthand
\begin{align}
  R_{n} \equiv g^{\mathrm{R}}_{n}(\varepsilon)
  &= \frac{1}{\varepsilon - \varepsilon_{n} + i\gamma},
  \label{eq:Rn-def}
  \\
  A_{n} \equiv g^{\mathrm{A}}_{n}(\varepsilon)
  &= \frac{1}{\varepsilon - \varepsilon_{n} - i\gamma},
  \qquad \gamma = \frac{\hbar}{2\tau},
  \label{eq:An-def}
\end{align}
together with
\begin{align}
  r   &\equiv \frac{\tau}{i\hbar} = \frac{1}{2i\gamma},
  \label{eq:r-def}
  \\
  \delta_{n} &\equiv \delta(\varepsilon - \varepsilon_{n}).
  \label{eq:delta-n-def}
\end{align}

\subsection{Algebraic identities for products of Green's functions}

Three identities are used repeatedly to decompose products
of scalar retarded and advanced propagators.

For the same-band retarded-advanced product,
\begin{align}
  R_{n}A_{n} = r(A_{n} - R_{n}),
  \label{eq:RA-same}
\end{align}
which follows from $A_{n} - R_{n} = 2i\gamma R_{n}A_{n}$. 
For two retarded propagators at different band energies,
partial fractions give
\begin{align}
  R_{m}R_{n} = \frac{R_{n} - R_{m}}{\varepsilon_{nm}}.
  \label{eq:RR-diff}
\end{align}
For a mixed retarded-advanced product at different bands,
\begin{align}
  R_{m}A_{n}
  &= \frac{A_{n} - R_{m}}{\varepsilon_{nm} + 2i\gamma}
  \nonumber\\
  &= \frac{A_{n} - R_{m}}{\varepsilon_{nm}}
   - \frac{2i\gamma}{\varepsilon_{nm}^{2}}(A_{n} - R_{m})
   + \mathcal{O}(\tau^{-2}).
  \label{eq:RA-diff}
\end{align}
The second line holds in the non-degenerate clean limit
$|\varepsilon_{nm}|\tau/\hbar \gg 1$, under which the
$2i\gamma$ correction is $\mathcal{O}(\tau^{-1})$.

\subsection{Clean-limit behavior of powers of Green's functions}

In the clean limit $\gamma \to 0^{+}$, the Sokhotski--Plemelj
formula gives
\begin{align}
  \mathrm{Im}\,R_{n}^{p}
  &= -\frac{\pi(-1)^{p-1}}{(p-1)!}
     \partial_{\varepsilon}^{p-1}\delta_{n},
  \label{eq:Im-Rp-clean}
  \\
  \mathrm{Im}\,A_{n}^{p}
  &= +\frac{\pi(-1)^{p-1}}{(p-1)!}
     \partial_{\varepsilon}^{p-1}\delta_{n}.
  \label{eq:Im-Ap-clean}
\end{align}
In particular,
\begin{align}
  A_{n} - R_{n}
  &\xrightarrow{\gamma\to 0^{+}} 2\pi i\,\delta_{n},
  \label{eq:AminusR-clean}
  \\
  A_{n}^{2} - R_{n}^{2}
  &\xrightarrow{\gamma\to 0^{+}} -2\pi i\,\partial_{\varepsilon}\delta_{n}.
  \label{eq:A2minusR2-clean}
\end{align}
Note that $r^{2} = -\tau^{2}/\hbar^{2}$ since
$r = \tau/(i\hbar)$; this sign is important for the
$\tau$-power counting below.

\subsection{Product forms of the kernels}

Using $\partial_{\varepsilon}R_{m} = -R_{m}^{2}$ and
$\partial_{\varepsilon}^{2}R_{m} = 2R_{m}^{3}$, the
kernels defined in
Eqs.~\eqref{eq:K1-def-band} and \eqref{eq:K2-def-band}
take the product forms
\begin{align}
  \mathcal{K}^{(1)}_{nml}
  &= \bigl(2R_{m}^{3}R_{l} + R_{m}^{2}R_{l}^{2}\bigr)
     (A_{n} - R_{n})
   + R_{m}^{2}R_{l}\,A_{n}^{2},
  \label{eq:K1-product}
  \\
  \mathcal{K}^{(2)}_{nm}
  &= 2R_{m}^{3}(A_{n} - R_{n}) + R_{m}^{2}\,A_{n}^{2}.
  \label{eq:K2-product}
\end{align}
The following subsections expand each kernel in powers of
$\tau$ using the identities
Eqs.~\eqref{eq:RA-same}--\eqref{eq:RA-diff}.

\subsection{\texorpdfstring{Fully diagonal kernels: $\mathcal{K}^{(1)}_{nnn}$ and $\mathcal{K}^{(2)}_{nn}$}{Fully diagonal kernels}}
\label{app:diagonal-kernels}

Setting $m=l=n$ in Eq.~\eqref{eq:K1-product} and $m=n$ in
Eq.~\eqref{eq:K2-product}, we obtain
\begin{align}
  \mathcal{K}^{(1)}_{nnn}
  &= 3R_{n}^{4}(A_{n}-R_{n})+R_{n}^{3}A_{n}^{2},
  \label{eq:K1-nnn-first}
  \\
  \mathcal{K}^{(2)}_{nn}
  &= 2R_{n}^{3}(A_{n}-R_{n})+R_{n}^{2}A_{n}^{2}.
  \label{eq:K2-nn-first}
\end{align}
Applying $R_{n}A_{n}=r(A_{n}-R_{n})$ repeatedly, with
$r=1/(2i\gamma)=\tau/(i\hbar)$ and $\tau=\hbar/(2\gamma)$,
gives
\begin{align}
  \mathcal{K}^{(2)}_{nn}
  &= r^{2}(A_{n}^{2}-R_{n}^{2})
   -2rR_{n}^{3}-2R_{n}^{4},
  \label{eq:K2-nn-expanded}
  \\
  \mathcal{K}^{(1)}_{nnn}
  &= r^{3}(A_{n}^{2}-R_{n}^{2})
   -2r^{2}R_{n}^{3}-3rR_{n}^{4}-3R_{n}^{5}.
  \label{eq:K1-nnn-expanded}
\end{align}
Their leading clean-limit imaginary parts, consistent with
Eqs.~\eqref{eq:A2minusR2-clean} and \eqref{eq:Im-Rp-clean},
are
\begin{align}
  \mathrm{Im}\,\mathcal{K}^{(2)}_{nn}
  &= \frac{2\pi\tau^{2}}{\hbar^{2}}
     \partial_{\varepsilon}\delta_{n}
   +\mathcal{O}(\tau^{0}),
  \label{eq:Im-K2-nn-leading}
  \\
  \mathrm{Im}\,\mathcal{K}^{(1)}_{nnn}
  &= -\frac{\pi\tau^{2}}{\hbar^{2}}
     \partial_{\varepsilon}^{2}\delta_{n}
   +\mathcal{O}(\tau^{0}).
  \label{eq:Im-K1-nnn-leading}
\end{align}
Here $\delta_{n}=\delta(\varepsilon-\varepsilon_{n})$, and
these asymptotic expressions are understood in the
distributional sense, after integration against a smooth
energy-dependent function.  The
$\mathcal{O}(\tau^{2})$ terms generate the leading nonlinear
Drude contribution $\sigma^{\mathrm{ND},\tau^{2}}_{ijk}$.

The $\mathcal{O}(\tau^{0})$ remainders in
Eqs.~\eqref{eq:Im-K2-nn-leading} and
\eqref{eq:Im-K1-nnn-leading} lie within the accuracy claimed
here and must be retained.  They cannot be obtained by
replacing every finite-width spectral factor in
Eqs.~\eqref{eq:K2-nn-expanded} and
\eqref{eq:K1-nnn-expanded} by its leading Dirac-delta limit
and retaining only the terms without explicit powers of
$\tau$.  Subleading contributions from the spectral factors
can combine with explicit powers of $\tau$ and contribute
at $\mathcal{O}(\tau^{0})$ after the energy integral.
Throughout this discussion, $\delta_{n}$ denotes the Dirac
delta distribution.  The finite-$\gamma$ spectral function
has a Lorentzian profile; $\delta_{n}$ itself is not a
finite-width function.

We therefore evaluate the imaginary parts exactly at finite
broadening.  Introducing the dimensionless energy variable
\begin{align}
  s&\equiv\frac{\varepsilon-\varepsilon_{n}}{\gamma},
  \qquad
  R_{n}=\frac{1}{\gamma(s+i)},
  \qquad
  A_{n}=\frac{1}{\gamma(s-i)},
  \label{eq:diag-s-variable}
\end{align}
a direct computation from
Eqs.~\eqref{eq:K2-nn-first} and \eqref{eq:K1-nnn-first}
gives
\begin{align}
  \mathrm{Im}\,\mathcal{K}^{(2)}_{nn}
  &=\frac{1}{\gamma^{4}}
    \frac{4s(s^{2}-3)}{(1+s^{2})^{4}},
  \label{eq:ImK2-exact}
  \\
  \mathrm{Im}\,\mathcal{K}^{(1)}_{nnn}
  &=\frac{1}{\gamma^{5}}
    \frac{5s^{4}-38s^{2}+5}{(1+s^{2})^{5}}.
  \label{eq:ImK1-exact}
\end{align}
These expressions are exact for every $\gamma>0$.
The first kernel is odd in $s$ and the second is even.
Consequently, among the convergent moments, only odd moments
of $\gamma^{4}\mathrm{Im}\,\mathcal{K}^{(2)}_{nn}$ and even
moments of $\gamma^{5}\mathrm{Im}\,\mathcal{K}^{(1)}_{nnn}$
can be nonzero.

The required integrals can be evaluated using Euler's beta
function, defined for real $x>0$ and $y>0$ by
\begin{align}
  B(x,y)\equiv\int_{0}^{1}dt\,t^{x-1}(1-t)^{y-1}.
\end{align}
In particular, the substitution $t=s^{2}/(1+s^{2})$ on the
positive half-axis gives
\begin{align}
  \int_{-\infty}^{\infty}ds\,
  \frac{s^{2a}}{(1+s^{2})^{b}}
  &=B\biggl(a+\frac{1}{2},b-a-\frac{1}{2}\biggr),
\end{align}
where $a$ is a nonnegative integer and $b>a+1/2$.
The latter inequality ensures convergence at large $|s|$.
The moments needed through $\mathcal{O}(\tau^{0})$, with
their integrands written explicitly, are
\begin{subequations}
\begin{align}
  &\int_{-\infty}^{\infty}ds\,
  \frac{4s^{2}(s^{2}-3)}{(1+s^{2})^{4}}
  =-\frac{\pi}{2},
  \\
  &\int_{-\infty}^{\infty}ds\,
  \frac{4s^{4}(s^{2}-3)}{(1+s^{2})^{4}}
  =\frac{\pi}{2},
  \\
  &\int_{-\infty}^{\infty}ds\,
  \frac{5s^{4}-38s^{2}+5}{(1+s^{2})^{5}}
  =0,
  \\
  &\int_{-\infty}^{\infty}ds\,
  \frac{s^{2}(5s^{4}-38s^{2}+5)}{(1+s^{2})^{5}}
  =-\frac{\pi}{2},
  \\
  &\int_{-\infty}^{\infty}ds\,
  \frac{s^{4}(5s^{4}-38s^{2}+5)}{(1+s^{2})^{5}}
  =0.
\end{align}
  \label{eq:diag-moments}
\end{subequations}
The first two lines are the first and third moments of
$\gamma^{4}\mathrm{Im}\,\mathcal{K}^{(2)}_{nn}$.
The last three lines are the zeroth, second, and fourth
moments of
$\gamma^{5}\mathrm{Im}\,\mathcal{K}^{(1)}_{nnn}$,
respectively.

We now perform the energy integrals at fixed chemical
potential $\mu$ and fixed temperature $T>0$, using
\begin{align}
  f(\varepsilon)
  &=\frac{1}{\exp[(\varepsilon-\mu)/(k_{\mathrm B}T)]+1},
  \qquad
  f_{n}^{(p)}
  =\left.\partial_{\varepsilon}^{p}f(\varepsilon)
    \right|_{\varepsilon=\varepsilon_{n}}.
\end{align}
Here $f_{n}=f_{n}^{(0)}$, and primes denote derivatives with
respect to the energy argument, evaluated at
$\varepsilon=\varepsilon_{n}$.
With $\varepsilon=\varepsilon_{n}+\gamma s$, the exact
integrals become
\begin{align}
  \int_{-\infty}^{\infty}\frac{d\varepsilon}{2\pi}\,
  f(\varepsilon)\mathrm{Im}\,\mathcal{K}^{(2)}_{nn}
  &=\frac{1}{2\pi\gamma^{3}}
    \int_{-\infty}^{\infty}ds\,
    f(\varepsilon_{n}+\gamma s)
    \frac{4s(s^{2}-3)}{(1+s^{2})^{4}},
  \\
  \int_{-\infty}^{\infty}\frac{d\varepsilon}{2\pi}\,
  f(\varepsilon)\mathrm{Im}\,\mathcal{K}^{(1)}_{nnn}
  &=\frac{1}{2\pi\gamma^{4}}
    \int_{-\infty}^{\infty}ds\,
    f(\varepsilon_{n}+\gamma s)
    \frac{5s^{4}-38s^{2}+5}{(1+s^{2})^{5}}.
\end{align}
For fixed $s$, the local Taylor expansion is
\begin{align}
  f(\varepsilon_{n}+\gamma s)
  &=\sum_{p=0}^{4}\frac{(\gamma s)^{p}}{p!}f_{n}^{(p)}
    +\mathcal{O}\bigl((\gamma s)^{5}\bigr).
\end{align}
This expansion is used only to extract the local terms
associated with the convergent moments in
Eq.~\eqref{eq:diag-moments}.  It is not an unrestricted
term-by-term expansion of the integrals over the entire
real axis.  The large-$|s|$ remainder is discussed below.
Using these moments and $\gamma=\hbar/(2\tau)$ gives
\begin{align}
  \int_{-\infty}^{\infty}\frac{d\varepsilon}{2\pi}\,
  f(\varepsilon)\mathrm{Im}\,\mathcal{K}^{(2)}_{nn}
  &=-\frac{\tau^{2}}{\hbar^{2}}f_{n}^{\prime}
    +\frac{1}{24}f_{n}^{\prime\prime\prime}
    +\mathcal{O}(\tau^{-1}),
  \label{eq:K2-nn-eint}
  \\
  \int_{-\infty}^{\infty}\frac{d\varepsilon}{2\pi}\,
  f(\varepsilon)\mathrm{Im}\,\mathcal{K}^{(1)}_{nnn}
  &=-\frac{\tau^{2}}{2\hbar^{2}}f_{n}^{\prime\prime}
    +\mathcal{O}(\tau^{-1}).
  \label{eq:K1-nnn-eint}
\end{align}
In Eq.~\eqref{eq:K2-nn-eint}, the first and third moments
produce the $\tau^{2}$ and $\tau^{0}$ terms, respectively.
In Eq.~\eqref{eq:K1-nnn-eint}, the zeroth moment removes a
possible $\tau^{4}$ term, the second moment produces the
$\tau^{2}$ term, and the fourth moment vanishes exactly.
Thus the coefficient of $\tau^{0}$ in
Eq.~\eqref{eq:K1-nnn-eint} is zero: only the
$\mathcal{K}^{(2)}_{nn}$ channel contributes at this order.
Inserting Eqs.~\eqref{eq:K2-nn-eint} and
\eqref{eq:K1-nnn-eint} into the diagonal sector and
integrating by parts over the periodic Brillouin zone gives
Eqs.~\eqref{eq:ND} and \eqref{eq:ND-tau0}; the details are
given in Appendix~\ref{app:dc-calculations}.

Two remarks clarify the remainder and the regime of
validity.  First, the large-$|s|$ behavior of the scaled
kernels is
\begin{align}
  \gamma^{4}\mathrm{Im}\,\mathcal{K}^{(2)}_{nn}
  &=\frac{4}{s^{5}}+\mathcal{O}(s^{-7}),
  \quad
  \gamma^{5}\mathrm{Im}\,\mathcal{K}^{(1)}_{nnn}
  =\frac{5}{s^{6}}+\mathcal{O}(s^{-8}).
\end{align}
The next moments allowed by parity, namely the fifth moment
of $\gamma^{4}\mathrm{Im}\,\mathcal{K}^{(2)}_{nn}$ and the
sixth moment of
$\gamma^{5}\mathrm{Im}\,\mathcal{K}^{(1)}_{nnn}$, therefore
diverge.  Consequently, parity alone does not imply that
the remainder in each energy integral starts at
$\gamma^{2}$.  Each integral generally has a nonlocal
contribution of order $\gamma$, equivalent to
$\mathcal{O}(\tau^{-1})$.  Such a contribution depends on
the Fermi function away from $\varepsilon_{n}$ and cannot be
determined from a finite set of its derivatives at
$\varepsilon_{n}$.
In the fully diagonal sector, these order-$\gamma$ corrections
cancel when the contributions to the second-order conductivity
are combined and integrated over the periodic Brillouin zone.
For the one-band response, Sec.~\ref{sec:oneband} demonstrates
this cancellation using the exact identity in
Eq.~\eqref{eq:oneband-combined-kernel}.
The first correction beyond the kinetic term is
$\mathcal{O}(\gamma^{2})=\mathcal{O}(\tau^{-2})$, as shown in
Eq.~\eqref{eq:oneband-current-expansion} and checked numerically
in Fig.~\ref{fig:oneband}.
This conclusion applies to the fully diagonal sector.
Higher-order corrections in the other multiband geometric
channels require separate analysis.

Second, the finite-temperature local expansion is controlled
when $\gamma\ll k_{\mathrm B}T$, because the characteristic
scale of $f_{n}^{(p)}$ is $(k_{\mathrm B}T)^{-p}$.
For a multiband application, this thermal condition is
additional to the non-degeneracy condition in
Eq.~\eqref{eq:clean-limit-condition}.
The remainders above refer to $\gamma\to0^{+}$ at fixed
$T>0$ and are not asserted to be uniform in a simultaneous
$T\to0$ limit; the zero-temperature limit must be treated
separately.  

\medskip
\noindent\textit{Parity selection and the absence of a kinetic term in linear
response.}\ \ The exact finite-broadening forms obtained above make explicit
why the linear dc conductivity carries no analogue of the
kinetic term in Eq.~\eqref{eq:ND-tau0}.
The Drude contribution to the linear dc conductivity is
obtained from the real part of the fully diagonal kernel
$\mathcal{K}_{nn}$ defined in the Supplemental Material~\cite{SM}.

In linear response, the derivative representation of
$\mathcal{K}_{nn}$ contains two terms with nonzero real parts.
Both are individually odd in $s$.  Each carries a Hadamard
finite-part contribution at $\mathcal{O}(\tau^{0})$, but these
contributions are equal and opposite and cancel upon combination.
The cancellation improves the large-$|s|$ decay from $s^{-3}$
for the individual terms to $s^{-5}$ for their sum.
The combined kernel remains odd and is given by
\begin{equation}
  \mathrm{Re}\,\mathcal{K}_{nn}
  = -\frac{4s}{\gamma^{3}(1+s^{2})^{3}},
  \label{eq:ReKnn-exact}
\end{equation}
so its convergent even-order moments vanish.
With the prefactor $\gamma^{-3}$ and the measure
$d\varepsilon=\gamma\,ds$, the moment of order $p$
contributes at order $\gamma^{p-2}$.
The Drude term comes from the $p=1$ moment, whereas the
$\mathcal{O}(\tau^{0})$ term would come from the $p=2$
moment.  The latter is convergent for the combined kernel
and vanishes by parity.
Parity constrains the local expansion only while the required
moments converge; it does not imply an all-orders expansion
containing only odd powers of $\tau$.

At second order the corresponding kernel is
$\mathrm{Im}\,\mathcal{K}^{(2)}_{nn}$ of Eq.~\eqref{eq:ImK2-exact}, which is
odd as well, but carries one further power of $\gamma^{-1}$, so the term of
order $p$ now contributes at $\gamma^{p-3}$: the nonlinear Drude term is again
the $p=1$ moment, while the $\mathcal{O}(\tau^{0})$ contribution is the $p=3$
moment.  This odd-order moment equals $\pi/2$ by the second line of
Eq.~\eqref{eq:diag-moments} and produces the
$\tfrac{1}{24}f^{\prime\prime\prime}_{n}$ term of
Eq.~\eqref{eq:K2-nn-eint}.  The shift by one order in the lifetime expansion
moves the $\mathcal{O}(\tau^{0})$ contribution from an even to an odd moment
and thereby reverses the parity selection.  This is why
Eq.~\eqref{eq:ND-tau0} has no linear-response counterpart.

\subsection{\texorpdfstring{Kernel $\mathcal{K}^{(1)}_{nmn}$ with $n\neq m$}
{Kernel K1 nmn with n not equal m}}

Setting $l = n$ ($n\neq m$) in Eq.~\eqref{eq:K1-product}
and applying
Eqs.~\eqref{eq:RR-diff} and \eqref{eq:RA-diff},
\begin{align}
  \mathcal{K}^{(1)}_{nmn}
  &= -\frac{r}{\varepsilon_{nm}^{2}}
     \partial_{\varepsilon}(A_{n}-R_{n})
   - \frac{2A_{n}^{2}}{\varepsilon_{nm}^{3}}
   - \frac{R_{n}^{3}}{\varepsilon_{nm}^{2}}
   - \frac{R_{m}^{2}}{\varepsilon_{nm}^{3}}
   + \frac{3(R_{n}-R_{m})}{\varepsilon_{nm}^{4}}
   + \mathcal{O}(\tau^{-1}).
  \label{eq:K1-nmn-expanded}
\end{align}
The leading real and imaginary parts are
\begin{align}
  \mathrm{Re}\,\mathcal{K}^{(1)}_{nmn}\big|_{\mathcal{O}(\tau)}
  &= -\frac{2\pi\tau}{\hbar\,\varepsilon_{nm}^{2}}
     \partial_{\varepsilon}\delta_{n},
  \label{eq:Re-K1-nmn-tau1}
  \\
  \mathrm{Im}\,\mathcal{K}^{(1)}_{nmn}\big|_{\mathcal{O}(\tau^{0})}
  &= \pi\left[
       \frac{\partial_{\varepsilon}^{2}\delta_{n}}
            {2\varepsilon_{nm}^{2}}
     + \frac{2\partial_{\varepsilon}\delta_{n}}
            {\varepsilon_{nm}^{3}}
     - \frac{\partial_{\varepsilon}\delta_{m}}
            {\varepsilon_{nm}^{3}}
     + \frac{3(\delta_{m}-\delta_{n})}
            {\varepsilon_{nm}^{4}}
     \right].
  \label{eq:Im-K1-nmn-tau0}
\end{align}
The $\mathcal{O}(\tau)$ real part contributes to the BCD
response $\sigma^{\mathrm{BCD}}_{ijk}$, while the
$\mathcal{O}(\tau^{0})$ imaginary part contributes to
the QMD sector.

\subsection{\texorpdfstring{Combination $\varepsilon_{nm}\mathcal{K}^{(1)}_{nnm}-\mathcal{K}^{(2)}_{nn}$ with $n\neq m$}{Combination epsilon K1 minus K2}}

Setting $m = n$, $l = m$ in Eq.~\eqref{eq:K1-product}
gives $\mathcal{K}^{(1)}_{nnm}$, and the combination
$\varepsilon_{nm}\mathcal{K}^{(1)}_{nnm}
- \mathcal{K}^{(2)}_{nn}$ that appears in the
band-basis decomposition evaluates to
\begin{align}
  &\varepsilon_{nm}\mathcal{K}^{(1)}_{nnm}
  - \mathcal{K}^{(2)}_{nn}
  = \frac{r}{\varepsilon_{nm}}
     \partial_{\varepsilon}(A_{n}-R_{n})
   + \frac{A_{n}^{2}}{\varepsilon_{nm}^{2}}
   + \frac{R_{n}^{3}}{\varepsilon_{nm}}
   + \frac{R_{m}-R_{n}}{\varepsilon_{nm}^{3}}
   + \mathcal{O}(\tau^{-1}),
  \label{eq:K1-nnm-minus-K2nn}
\end{align}
with imaginary part
\begin{align}
  &\mathrm{Im}\!\left[
    \varepsilon_{nm}\mathcal{K}^{(1)}_{nnm}
  - \mathcal{K}^{(2)}_{nn}
  \right]
  = \pi\left[
       -\frac{\partial_{\varepsilon}\delta_{n}}
             {\varepsilon_{nm}^{2}}
       -\frac{\partial_{\varepsilon}^{2}\delta_{n}}
             {2\varepsilon_{nm}}
       +\frac{\delta_{n}-\delta_{m}}
             {\varepsilon_{nm}^{3}}
     \right]
   + \mathcal{O}(\tau^{-1}).
  \label{eq:Im-K1-nnm-minus-K2nn}
\end{align}

\subsection{\texorpdfstring{Kernel $\mathcal{K}^{(1)}_{nmm}$ with $n\neq m$}{Kernel K1 nmm with n not equal m}}

Setting $m = l$ ($n\neq m$) in Eq.~\eqref{eq:K1-product}
and expanding,
\begin{align}
  \mathcal{K}^{(1)}_{nmm}
  &= \frac{A_{n}^{2}}{\varepsilon_{nm}^{3}}
   + \frac{R_{m}^{3}}{\varepsilon_{nm}^{2}}
   + \frac{2R_{m}^{2}}{\varepsilon_{nm}^{3}}
   + \frac{3(R_{m}-R_{n})}{\varepsilon_{nm}^{4}}
   + \mathcal{O}(\tau^{-1}),
  \label{eq:K1-nmm-expanded}
\end{align}
with imaginary part
\begin{align}
  \mathrm{Im}\,\mathcal{K}^{(1)}_{nmm}
  &= \pi\left[
       -\frac{\partial_{\varepsilon}\delta_{n}}
             {\varepsilon_{nm}^{3}}
       -\frac{\partial_{\varepsilon}^{2}\delta_{m}}
             {2\varepsilon_{nm}^{2}}
       +\frac{2\partial_{\varepsilon}\delta_{m}}
             {\varepsilon_{nm}^{3}}
       +\frac{3(\delta_{n}-\delta_{m})}
             {\varepsilon_{nm}^{4}}
     \right]
   + \mathcal{O}(\tau^{-1}).
  \label{eq:Im-K1-nmm}
\end{align}

\subsection{\texorpdfstring{Off-diagonal two-point kernel $\mathcal{K}^{(2)}_{nm}$ with $n\neq m$}{Off-diagonal two-point kernel}}

For $n\neq m$, applying
Eqs.~\eqref{eq:RR-diff} and \eqref{eq:RA-diff} to
Eq.~\eqref{eq:K2-product},
\begin{align}
  \mathcal{K}^{(2)}_{nm}
  &= \frac{A_{n}^{2} + R_{m}^{2}}{\varepsilon_{nm}^{2}}
   + \frac{2(R_{m}-R_{n})}{\varepsilon_{nm}^{3}}
   + \mathcal{O}(\tau^{-1}),
  \label{eq:K2-nm-expanded}
\end{align}
with imaginary part
\begin{align}
  \mathrm{Im}\,\mathcal{K}^{(2)}_{nm}
  &= \pi\left[
       \frac{\partial_{\varepsilon}\delta_{m}
           - \partial_{\varepsilon}\delta_{n}}
            {\varepsilon_{nm}^{2}}
     + \frac{2(\delta_{n}-\delta_{m})}
            {\varepsilon_{nm}^{3}}
     \right]
   + \mathcal{O}(\tau^{-1}).
  \label{eq:Im-K2-nm}
\end{align}
This $\mathcal{O}(\tau^{0})$ kernel enters the interband
QMD contribution $\sigma^{\mathrm{inter\text{-}QMD}}_{ijk}$.

\subsection{\texorpdfstring{Three-distinct-band kernel $\mathcal{K}^{(1)}_{nml}$ with $n,m,l$ all distinct}{Three-distinct-band kernel}}

For $n$, $m$, $l$ all distinct, the expansion yields
\begin{align}
  \mathcal{K}^{(1)}_{nml}
  &= \left(
       \frac{2}{\varepsilon_{nm}^{3}\varepsilon_{nl}}
     + \frac{1}{\varepsilon_{nm}^{2}\varepsilon_{nl}^{2}}
     \right)(A_{n}-R_{n})
  \nonumber\\
  &\quad
   + \frac{A_{n}^{2}}{\varepsilon_{nm}^{2}\varepsilon_{nl}}
   - \frac{\varepsilon_{nm}+2\varepsilon_{nl}}
          {\varepsilon_{nm}^{3}\varepsilon_{nl}^{2}}A_{n}
  \nonumber\\
  &\quad
   + \frac{R_{m}^{2}}{\varepsilon_{nm}^{2}\varepsilon_{ml}}
   + \frac{2\varepsilon_{nl}-3\varepsilon_{nm}}
          {\varepsilon_{nm}^{3}\varepsilon_{ml}^{2}}R_{m}
   + \frac{R_{l}}{\varepsilon_{nl}^{2}\varepsilon_{ml}^{2}}
   + \mathcal{O}(\tau^{-1}),
  \label{eq:K1-nml-distinct-expanded}
\end{align}
with imaginary part
\begin{align}
  \mathrm{Im}\,\mathcal{K}^{(1)}_{nml}
  &= \pi\bigg[
       -\frac{\partial_{\varepsilon}\delta_{n}}
             {\varepsilon_{nm}^{2}\varepsilon_{nl}}
     + \left(
         \frac{2}{\varepsilon_{nm}^{3}\varepsilon_{nl}}
       + \frac{1}{\varepsilon_{nm}^{2}\varepsilon_{nl}^{2}}
       \right)\delta_{n}
  \nonumber\\
  &\qquad
     + \frac{\partial_{\varepsilon}\delta_{m}}
            {\varepsilon_{nm}^{2}\varepsilon_{ml}}
     + \frac{3\varepsilon_{nm}-2\varepsilon_{nl}}
            {\varepsilon_{nm}^{3}\varepsilon_{ml}^{2}}\delta_{m}
     - \frac{\delta_{l}}
            {\varepsilon_{nl}^{2}\varepsilon_{ml}^{2}}
     \bigg]
   + \mathcal{O}(\tau^{-1}).
  \label{eq:Im-K1-nml-distinct}
\end{align}
This kernel contributes to the covariant-connection sector
$\sigma^{\mathcal{C}}_{ijk}$ and the three-connection sector
$\sigma^{\mathcal{T}}_{ijk}$, whose connection-dependent
parts cancel exactly as demonstrated in
Sec.~\ref{sec:cancellation}.
\end{widetext}

\section{Detailed calculations of the dc conductivity contributions}
\label{app:dc-calculations}

In this appendix we present the detailed calculation of
each sector of the dc conductivity by substituting the
band-basis velocity decompositions
Eqs.~\eqref{eq:velocity-decomp} and \eqref{eq:vnn-ij}
and the clean-limit kernels of
Appendix~\ref{app:kernel-expansion} into
Eq.~\eqref{eq:sigma-dc-band}. 
The detailed calculations are collected in the Supplemental
Material~\cite{SM}. 

Throughout, the band energy measured from the chemical potential is
$\xi_{n}\equiv\varepsilon_{n}-\mu$, and
\begin{align}
  f_{n} &\equiv f(\xi_{n}),
  \nonumber\\
  f_{n}^{\prime} &\equiv \partial_{\xi_{n}}f(\xi_{n}),
  \nonumber\\
  f_{n}^{\prime\prime} &\equiv \partial_{\xi_{n}}^{2}f(\xi_{n}).
  \label{eq:app-fermi-derivative-defs}
\end{align}
The integration variable $\varepsilon$ in the Green's functions is also
measured from the chemical potential, so the Fermi function is written
as $f(\varepsilon)$. The symbol ${\sum_{n,m}}^{\prime}$ excludes $n=m$, while
${\sum_{n,m,l}}^{\prime}$ sums over mutually distinct
indices.

\subsection{Nonlinear Drude contribution}
\label{app:subsec-ND}

The nonlinear Drude sector is defined by
\begin{align}
  \sigma^{\mathrm{ND}}_{ijk}
  &= \frac{e^{3}}{\hbar}
     \int_{-\infty}^{\infty}\frac{d\varepsilon}{2\pi}\,
     f(\varepsilon)
     \sum_{\bm{k}}
     \mathrm{Im}\,\mathcal{B}^{\mathrm{ND}}_{ijk},
  \label{eq:app-BND-def}
  \\
  \mathcal{B}^{\mathrm{ND}}_{ijk}
  &= \sum_{n}
     \biggl[
       2(\partial_{i}\varepsilon_{n})
       (\partial_{j}\varepsilon_{n})
       (\partial_{k}\varepsilon_{n})\,
       \mathcal{K}^{(1)}_{nnn}
  \nonumber\\
  &\qquad
     + (\partial_{i}\varepsilon_{n})
       (\partial_{j}\partial_{k}\varepsilon_{n})\,
       \mathcal{K}^{(2)}_{nn}
     \biggr],
  \label{eq:app-BND}
\end{align}
so that
\begin{align}
  \mathrm{Im}\,\mathcal{B}^{\mathrm{ND}}_{ijk}
  &= \sum_{n}
     \biggl[
       2(\partial_{i}\varepsilon_{n})
       (\partial_{j}\varepsilon_{n})
       (\partial_{k}\varepsilon_{n})\,
       \mathrm{Im}\,\mathcal{K}^{(1)}_{nnn}
  \nonumber\\
  &\qquad
     + (\partial_{i}\varepsilon_{n})
       (\partial_{j}\partial_{k}\varepsilon_{n})\,
       \mathrm{Im}\,\mathcal{K}^{(2)}_{nn}
     \biggr].
  \label{eq:app-ImBND}
\end{align}
Substituting the diagonal-kernel energy integrals
[Eqs.~\eqref{eq:K2-nn-eint} and \eqref{eq:K1-nnn-eint}],
\begin{align}
  \int\!\frac{d\varepsilon}{2\pi}\,f(\varepsilon)\,
    \mathrm{Im}\,\mathcal{K}^{(2)}_{nn}
  &= -\frac{\tau^{2}}{\hbar^{2}}f_{n}^{\prime}
   + \frac{1}{24}f_{n}^{\prime\prime\prime},
  \label{eq:app-ImK2nn}
  \\
  \int\!\frac{d\varepsilon}{2\pi}\,f(\varepsilon)\,
    \mathrm{Im}\,\mathcal{K}^{(1)}_{nnn}
  &= -\frac{\tau^{2}}{2\hbar^{2}}f_{n}^{\prime\prime},
  \label{eq:app-ImK1nnn}
\end{align}
into Eq.~\eqref{eq:app-ImBND} and integrating by parts in
$\bm{k}$ via
\begin{align}
  \partial_{j}\partial_{k}f_{n}
  = (\partial_{j}\partial_{k}\varepsilon_{n})\,f_{n}^{\prime}
  + (\partial_{j}\varepsilon_{n})
    (\partial_{k}\varepsilon_{n})\,f_{n}^{\prime\prime},
  \label{eq:app-djdk-fn}
\end{align}
one obtains the leading term
\begin{align}
  \sigma^{\mathrm{ND},\tau^{2}}_{ijk}
  = -\frac{e^{3}\tau^{2}}{\hbar^{3}}
    \sum_{\bm{k}}\sum_{n}
    f_{n}\,\partial_{i}\partial_{j}\partial_{k}\varepsilon_{n},
  \label{eq:app-ND-result}
\end{align}
which coincides with the constant-relaxation-time Boltzmann
expression~\cite{Deyo2009,Moore2010}.
The $\mathcal{O}(\tau^{0})$ part of
Eq.~\eqref{eq:app-ImK2nn} contributes only through the
$(\partial_{i}\varepsilon_{n})
 (\partial_{j}\partial_{k}\varepsilon_{n})$ vertex of
Eq.~\eqref{eq:app-ImBND}, since the corresponding
$\mathcal{O}(\tau^{0})$ term of
Eq.~\eqref{eq:app-ImK1nnn} vanishes.
Using
$\partial_{i}f_{n}^{\prime\prime}
 = (\partial_{i}\varepsilon_{n})f_{n}^{\prime\prime\prime}$
and integrating by parts once,
\begin{align}
  \sigma^{\mathrm{ND},\tau^{0}}_{ijk}
  &= \frac{e^{3}}{24\hbar}
     \sum_{\bm{k}}\sum_{n}
     (\partial_{i}\varepsilon_{n})
     (\partial_{j}\partial_{k}\varepsilon_{n})\,
     f_{n}^{\prime\prime\prime}
  \nonumber\\
  &= -\frac{e^{3}}{24\hbar}
     \sum_{\bm{k}}\sum_{n}
     f_{n}^{\prime\prime}\,
     \partial_{i}\partial_{j}\partial_{k}\varepsilon_{n},
  \label{eq:app-ND-tau0-result}
\end{align}
and a further pair of integrations by parts gives the
equivalent Fermi-surface form
\begin{align}
  \sigma^{\mathrm{ND},\tau^{0}}_{ijk}
  = -\frac{e^{3}}{48\hbar}
    \sum_{\bm{k}}\sum_{n}
    u_{i}^{n}u_{j}^{n}u_{k}^{n}\,f_{n}^{(4)},
  \label{eq:app-ND-tau0-alt}
\end{align}
with $u_{i}^{n}=\partial_{i}\varepsilon_{n}$.

\subsection{Quantum-geometric contribution}
\label{app:subsec-QG}

The quantum-geometric sector collects all two-band
contributions from the velocity decomposition
Eq.~\eqref{eq:velocity-decomp} and is defined by
\begin{align}
  \sigma^{\mathrm{QG}}_{ijk}
  &= \frac{e^{3}}{\hbar}
     \int_{-\infty}^{\infty}\frac{d\varepsilon}{2\pi}\,
     f(\varepsilon)
     \sum_{\bm{k}}
     \mathrm{Im}\!\left[\mathcal{B}^{\mathrm{QG}}_{ijk}\right],
  \label{eq:app-BQG-def}
\end{align}
with
\begin{align}
  \mathcal{B}^{\mathrm{QG}}_{ijk}
  &= {\sum_{n,m}}^{\prime}
     \varepsilon_{nm}(\partial_{i}\varepsilon_{n})
     \bigl(\mathcal{Q}^{nm}_{jk} + \mathcal{Q}^{nm}_{kj}\bigr)
     \bigl[\varepsilon_{nm}\mathcal{K}^{(1)}_{nnm}
           - \mathcal{K}^{(2)}_{nn}\bigr]
  \nonumber\\
  &\quad
  + {\sum_{n,m}}^{\prime}
    \varepsilon_{nm}\mathcal{Q}^{nm}_{ik}
    \Bigl[
      (\partial_{j}\varepsilon_{m})
      \bigl(\varepsilon_{nm}\mathcal{K}^{(1)}_{nmm}
            - \mathcal{K}^{(2)}_{nm}\bigr)
  \nonumber\\
  &\qquad\quad
    + (\partial_{j}\varepsilon_{n})
      \bigl(\varepsilon_{nm}\mathcal{K}^{(1)}_{nmn}
            + \mathcal{K}^{(2)}_{nm}\bigr)
    \Bigr]
  \nonumber\\
  &\quad
  + {\sum_{n,m}}^{\prime}
    \varepsilon_{nm}\mathcal{Q}^{nm}_{ij}
    \Bigl[
      (\partial_{k}\varepsilon_{m})
      \bigl(\varepsilon_{nm}\mathcal{K}^{(1)}_{nmm}
            - \mathcal{K}^{(2)}_{nm}\bigr)
  \nonumber\\
  &\qquad\quad
    + (\partial_{k}\varepsilon_{n})
      \bigl(\varepsilon_{nm}\mathcal{K}^{(1)}_{nmn}
            + \mathcal{K}^{(2)}_{nm}\bigr)
    \Bigr].
  \label{eq:app-BQG}
\end{align}
The imaginary part of $\mathcal{B}^{\mathrm{QG}}_{ijk}$
decomposes according to the real and imaginary parts of
the quantum geometric tensor
$\mathcal{Q}^{nm}_{ij}
 = \mathcal{G}^{nm}_{ij} - (i/2)\Omega^{nm}_{ij}$.

Substituting the kernel results
Eqs.~\eqref{eq:Im-K1-nnm-minus-K2nn},
\eqref{eq:Im-K1-nmm}, \eqref{eq:Im-K1-nmn-tau0},
\eqref{eq:Im-K2-nm}, and \eqref{eq:Re-K1-nmn-tau1}
and performing the energy integrals, the quantum-geometric
sector decomposes as
\begin{align}
  \sigma^{\mathrm{QG}}_{ijk}
  = \sigma^{\mathrm{BCD}}_{ijk}
  + \sigma^{\mathcal{G}}_{ijk},
  \label{eq:app-QG-decomp}
\end{align}
where the Berry-curvature part is
\begin{align}
  \sigma^{\mathrm{BCD}}_{ijk}
  &= \frac{e^{3}\tau}{2\hbar^{2}}
     \sum_{\bm{k}}\sum_{n,l}
     f_{n}
     \left[
       \varepsilon_{ijl}\,D^{n}_{kl}
     + \varepsilon_{ikl}\,D^{n}_{jl}
     \right],
  \label{eq:app-BCD-from-QG}
\end{align}
with $D^{n}_{jl}\equiv\partial_{j}\Omega^{n}_{l}$
the Berry-curvature dipole density, and the quantum-metric
part is
\begin{align}
  \sigma^{\mathcal{G}}_{ijk}
  &= \frac{e^{3}}{2\hbar}
     \sum_{\bm{k}}{\sum_{n,m}}^{\prime}
     f_{n}^{\prime}\,\partial_{i}\mathcal{G}^{nm}_{jk}
  \nonumber\\
  &\quad
  + \frac{e^{3}}{\hbar}
    \sum_{\bm{k}}{\sum_{n,m}}^{\prime}
    f_{n}
    \left(
      -\frac{\partial_{i}\mathcal{G}^{nm}_{jk}}{\varepsilon_{nm}}
      + 2\frac{\partial_{i}\varepsilon_{nm}}{\varepsilon_{nm}^{2}}
        \mathcal{G}^{nm}_{jk}
    \right)
  \nonumber\\
  &\quad
  - \frac{e^{3}}{\hbar}
    \sum_{\bm{k}}{\sum_{n,m}}^{\prime}
    f_{n}
    \left(
      \frac{\partial_{k}\varepsilon_{nm}}{\varepsilon_{nm}^{2}}
      \mathcal{G}^{nm}_{ij}
    + \frac{\partial_{j}\varepsilon_{nm}}{\varepsilon_{nm}^{2}}
      \mathcal{G}^{nm}_{ik}
    \right).
  \label{eq:app-G-from-QG}
\end{align}
The $\mathcal{O}(\tau)$ term $\sigma^{\mathrm{BCD}}_{ijk}$
gives the BCD response, while $\sigma^{\mathcal{G}}_{ijk}$
is $\mathcal{O}(\tau^{0})$ and combines with the
connection-geometry sectors below.

\subsection{Covariant-connection contribution}
\label{app:subsec-C}

The covariant-connection sector arises from the
$\mathcal{D}_{k}\mathcal{A}^{mn}_{j}$ term in the
off-diagonal velocity matrix element
Eq.~\eqref{eq:vnm-ij} and is defined by
\begin{align}
  \sigma^{\mathcal{C}}_{ijk}
  &= \frac{e^{3}}{\hbar}
     \int_{-\infty}^{\infty}\frac{d\varepsilon}{2\pi}\,
     f(\varepsilon)
     \sum_{\bm{k}}
     \mathrm{Im}\!\left[\mathcal{B}^{\mathcal{C}}_{ijk}\right],
  \label{eq:app-BC-def}
  \\
  \mathcal{B}^{\mathcal{C}}_{ijk}
  &= {\sum_{n,m}}^{\prime}
     \varepsilon_{nm}^{2}\,
     \mathcal{C}^{nm}_{ij|k}\,
     \mathcal{K}^{(2)}_{nm},
  \label{eq:app-BC}
  \\
  \mathcal{C}^{nm}_{ij|k}
  &\equiv \mathcal{A}^{nm}_{i}\,
         \mathcal{D}_{k}\mathcal{A}^{mn}_{j}.
  \label{eq:app-C-def}
\end{align}
Using the decomposition
$\mathcal{C}^{nm}_{ij|k}
 = \Gamma^{nm}_{ij|k} - i\widetilde{\Gamma}^{nm}_{ij|k}$
[Eq.~\eqref{eq:quantum-connection-real-imag}],
the identity
\begin{align}
  2\Gamma^{nm}_{ij|k}
  = \partial_{k}\mathcal{G}^{nm}_{ij}
  + \partial_{j}\mathcal{G}^{nm}_{ki}
  - \partial_{i}\mathcal{G}^{nm}_{jk}
  + X^{nm}_{ijk}
  \label{eq:app-Gamma-metric-X}
\end{align}
[Eq.~\eqref{eq:Gamma-metric-X}], and substituting
Eq.~\eqref{eq:Im-K2-nm}, one obtains
\begin{align}
  &\sigma^{\mathcal{C}}_{ijk}\nonumber\\
  &= \frac{e^{3}}{\hbar}
     \sum_{\bm{k}}{\sum_{n,m}}^{\prime}
     \frac{f_{n}}{\varepsilon_{nm}}
     \bigl(
       \partial_{k}\mathcal{G}^{nm}_{ij}
     + \partial_{j}\mathcal{G}^{nm}_{ki}
     - \partial_{i}\mathcal{G}^{nm}_{jk}
     + X^{nm}_{ijk}
     \bigr),
  \label{eq:app-C-result}
\end{align}
where $X^{nm}_{ijk}$ is the primed Berry-connection
commutator [Eq.~\eqref{eq:X-comm-app}].

\subsection{Three-connection contribution}
\label{app:subsec-T}

The three-connection sector originates from products of
three off-diagonal Berry connections and is defined by
\begin{align}
  \sigma^{\mathcal{T}}_{ijk}
  &= \frac{e^{3}}{\hbar}
     \int_{-\infty}^{\infty}\frac{d\varepsilon}{2\pi}\,
     f(\varepsilon)
     \sum_{\bm{k}}
     \mathrm{Im}\!\left[\mathcal{B}^{\mathcal{T}}_{ijk}\right],
  \label{eq:app-BT-def}
  \\
  \mathcal{B}^{\mathcal{T}}_{ijk}
  &= -i{\sum_{n,m,l}}^{\prime}
     \biggl\{
       \varepsilon_{nm}\varepsilon_{ml}\varepsilon_{ln}
       \mathcal{K}^{(1)}_{nml}
       \bigl(
         \mathcal{T}^{nml}_{ijk}
       + \mathcal{T}^{nml}_{ikj}
       \bigr)
  \nonumber\\
  &\quad
     + \varepsilon_{nm}\mathcal{K}^{(2)}_{nm}
       \bigl(
         \varepsilon_{ml}\mathcal{T}^{nml}_{ijk}
       - \varepsilon_{nl}\mathcal{T}^{nml}_{ikj}
       \bigr)
     \biggr\},
  \label{eq:app-BT}
\end{align}
where $\mathcal{T}^{nml}_{ijk}
\equiv \mathcal{A}^{nm}_{i}\mathcal{A}^{ml}_{j}
\mathcal{A}^{ln}_{k}$
[Eq.~\eqref{eq:three-Berry-connection}].
Substituting Eq.~\eqref{eq:K1-nml-distinct-expanded}
and performing the energy integral, one finds
\begin{align}
  \sigma^{\mathcal{T}}_{ijk}
  = -\frac{e^{3}}{\hbar}
    \sum_{\bm{k}}{\sum_{n,m}}^{\prime}
    \frac{f_{n}}{\varepsilon_{nm}}
    X^{nm}_{ijk}.
  \label{eq:app-T-result}
\end{align}

\subsection{Cancellation and final decomposition}
\label{app:subsec-cancellation}

Comparing Eqs.~\eqref{eq:app-C-result} and
\eqref{eq:app-T-result}, the commutator terms $X^{nm}_{ijk}$
appear with opposite signs and cancel exactly:
\begin{align}
  &\sigma^{\mathcal{C}}_{ijk}
  + \sigma^{\mathcal{T}}_{ijk}\nonumber\\
  &= \frac{e^{3}}{\hbar}
     \sum_{\bm{k}}{\sum_{n,m}}^{\prime}
     \frac{f_{n}}{\varepsilon_{nm}}
     \bigl(
       \partial_{k}\mathcal{G}^{nm}_{ij}
     + \partial_{j}\mathcal{G}^{nm}_{ki}
     - \partial_{i}\mathcal{G}^{nm}_{jk}
     \bigr).
  \label{eq:app-cancellation-result}
\end{align}
No Berry-connection commutators survive; the result
depends only on derivatives of the gauge-invariant
quantum metric $\mathcal{G}^{nm}_{ij}$.

Combining Eq.~\eqref{eq:app-cancellation-result} with
$\sigma^{\mathcal{G}}_{ijk}$
[Eq.~\eqref{eq:app-G-from-QG}], the geometric part of the
$\mathcal{O}(\tau^{0})$ contribution is
\begin{widetext}
\begin{align}
  &\sigma^{\mathcal{G}}_{ijk}
  + \sigma^{\mathcal{C}}_{ijk}
  + \sigma^{\mathcal{T}}_{ijk}
  \nonumber\\
  &= \frac{e^{3}}{2\hbar}
     \sum_{\bm{k}}{\sum_{n,m}}^{\prime}
     f_{n}^{\prime}\,\partial_{i}\mathcal{G}^{nm}_{jk}
  + \frac{e^{3}}{\hbar}
    \sum_{\bm{k}}{\sum_{n,m}}^{\prime}
    f_{n}
    \biggl[
      -2\partial_{i}
      \!\left(\frac{\mathcal{G}^{nm}_{jk}}{\varepsilon_{nm}}\right)
      + \partial_{j}
      \!\left(\frac{\mathcal{G}^{nm}_{ki}}{\varepsilon_{nm}}\right)
      + \partial_{k}
      \!\left(\frac{\mathcal{G}^{nm}_{ij}}{\varepsilon_{nm}}\right)
    \biggr].
  \label{eq:app-tau0-combined}
\end{align}
\end{widetext}
Identifying the intraband and interband QMD sectors,
\begin{align}
  \sigma^{\mathrm{intra\text{-}QMD}}_{ijk}
  &= \frac{e^{3}}{2\hbar}
     \sum_{\bm{k}}{\sum_{n,m}}^{\prime}
     f_{n}^{\prime}\,\partial_{i}\mathcal{G}^{nm}_{jk},
  \label{eq:app-intraQMD-result}
  \\[4pt]
  \sigma^{\mathrm{inter\text{-}QMD}}_{ijk}
  &= \frac{e^{3}}{\hbar}
     \sum_{\bm{k}}{\sum_{n,m}}^{\prime}
     f_{n}
     \biggl[
       -2\partial_{i}
       \!\left(
         \frac{\mathcal{G}^{nm}_{jk}}{\varepsilon_{nm}}
       \right)
  \nonumber\\
  &\quad
     + \partial_{j}
       \!\left(
         \frac{\mathcal{G}^{nm}_{ki}}{\varepsilon_{nm}}
       \right)
     + \partial_{k}
       \!\left(
         \frac{\mathcal{G}^{nm}_{ij}}{\varepsilon_{nm}}
       \right)
     \biggr],
  \label{eq:app-interQMD-result}
\end{align}
the full dc conductivity takes the gauge-invariant form
\begin{align}
  \sigma^{\mathrm{DC}}_{ijk}
  = \sigma^{\mathrm{ND}}_{ijk}
  + \sigma^{\mathrm{BCD}}_{ijk}
  + \sigma^{\mathrm{intra\text{-}QMD}}_{ijk}
  + \sigma^{\mathrm{inter\text{-}QMD}}_{ijk},
  \label{eq:app-final-decomp}
\end{align}
which establishes the manifestly gauge-invariant
decomposition of the second-order dc nonlinear
conductivity.

\section{Numerical evaluation of the model responses}
\label{app:model-details}

For numerical calculations it is convenient to avoid derivatives of the
phases of eigenvectors.  The pair-resolved quantum metric can be evaluated
from velocity matrix elements,
\begin{equation}  \mathcal{G}^{nm}_{ab}
  =\operatorname{Re}\left[
  \frac{v_{a}^{nm}v_{b}^{mn}}{\varepsilon_{nm}^{2}}
  \right],
  \qquad n\ne m,
  \label{eq:metric-from-velocity}
\end{equation}
where $v_{a}^{nm}=\langle u_{n}|\partial_{a} H|u_{m}\rangle$.  The Berry curvature
can be computed as
\begin{equation}  \Omega^{nm}_{ab}
  =-2\operatorname{Im}\left[
  \frac{v_{a}^{nm}v_{b}^{mn}}{\varepsilon_{nm}^{2}}
  \right],
  \qquad
  \Omega^{n}_{ab}=\sum_{m\ne n}\Omega^{nm}_{ab}.
  \label{eq:berry-from-velocity}
\end{equation}
At finite temperature, $f_{n}^{\prime}$ is evaluated as
\begin{equation}
  f_{n}^{\prime}=-\frac{1}{4T}\operatorname{sech}^{2}\left(\frac{\xi_{n}}{2T}\right),
  \label{eq:fermi-derivative-finite-T}
\end{equation}
where $k_{B}=1$.  At zero temperature, a narrow Gaussian or Lorentzian
representation of $-\delta(\varepsilon_{n}-\mu)$ can be used for the Fermi-surface
term.

For the tilted Dirac lattice model we use the parameters in
Eq.~\eqref{eq:tilted-dirac-parameters}.  For the real two-band QMD model,
we use
\begin{equation}  (m,t_{x},t_{y},\lambda,t_{0})=(2.2,1.0,0.7,0.9,0.5),
  \label{eq:app-real-parameters}
\end{equation}
with the chemical potential varied in Fig.~\ref{fig:model-summary-2x4}(h).
The direct gap of the real model is finite over the Brillouin zone for this
parameter set because $d_{x}=m+t_{x}\cos k_{x}+t_{y}\cos k_{y}\ge0.5$ and the only
zeros of $d_{z}=\lambda\sin k_{y}$ occur where $d_{x}$ remains nonzero.  The real
condition $d_{y}=0$ makes the Berry curvature vanish identically, while the
metric diagnostic $S_{xyy}=(1/2)\sum_{\bm{k},n}f_{n}^{\prime}\partial_{x}\mathcal{G}_{yy}^{n}$ remains
finite when the tilt $d_{0}=t_{0}\sin k_{x}$ distorts the Fermi surface.

\nocite{Foreman2002,Kubo1957,Greenwood1958,Berry1984,XiaoChangNiu2010,Resta2011}

\end{document}


\title{\texorpdfstring{Supplemental Material for\\
``Second-order dc conductivity in the velocity-gauge Keldysh formalism:
 gauge-invariant decomposition into nonlinear Drude, Berry-curvature-dipole,
 and quantum-metric responses''}{Supplemental Material for Second-order dc conductivity in the velocity-gauge Keldysh formalism: gauge-invariant decomposition into nonlinear Drude, Berry-curvature-dipole, and quantum-metric responses}}

\author{Junya Shibata}
\affiliation{Department of Electrical, Electronic and Communications Engineering, 
Toyo University, Kawagoe, Saitama, 350-8585, Japan}
\date{\today}

\maketitle

\begingroup
\footnotesize
\makeatletter
\let\l@subsubsection\@gobble@tw@
\makeatother
\tableofcontents
\endgroup
\clearpage

\section{Peierls coupling and current vertices in a tight-binding representation}
We consider a noninteracting multiband tight-binding Hamiltonian
\begin{align}
  \hat{H}_{0}
  = \sum_{\bm{k}}\sum_{\mu,\nu}
    \hat{\Psi}^{\dagger}_{\bm{k}\mu}\,
    H_{\mu\nu}(\bm{k})\,
    \hat{\Psi}_{\bm{k}\nu}
  \equiv
  \sum_{\bm{k}}
    \hat{\Psi}^{\dagger}_{\bm{k}}\,
    H(\bm{k})\,
    \hat{\Psi}_{\bm{k}},
  \label{eq:tight-binding}
\end{align}
where $H_{\mu\nu}(\bm{k})$ is the $N\times N$ Bloch Hamiltonian
matrix in the orbital basis ($\mu,\nu = 1,\ldots,N$),
$\hat{\Psi}^{\dagger}_{\bm{k}\mu}$ ($\hat{\Psi}_{\bm{k}\mu}$)
is the creation (annihilation) operator for a Bloch electron
in orbital $\mu$ at crystal momentum $\bm{k}$,
and $\hat{\Psi}^{\dagger}_{\bm{k}}$ denotes the $N$-component
row vector
$(\hat{\Psi}^{\dagger}_{\bm{k}1},\ldots,
  \hat{\Psi}^{\dagger}_{\bm{k}N})$.
We use $e>0$ for the magnitude of the electron charge, so
that the electron charge is $-e$.

A spatially uniform electromagnetic field is introduced via
the Peierls velocity-gauge substitution
\begin{align}
  H_{\bm{A}}(\bm{k},t)
  =
  H\left(
    \bm{k}
    +
    \frac{e\bm{A}(t)}{\hbar}
  \right),
  \label{eq:peierls-main}
\end{align}
with
\begin{align}
  \bm{E}(t)=-\partial_t\bm{A}(t).
\end{align}
Thus the external field carries zero wave vector from the
outset, and the response derived below is the homogeneous
${\bm q}=0$ response.
In Fourier space we use
\begin{align}
  A_i(t)
  =
  \int\frac{d\omega}{2\pi}
  e^{-i\omega t}A_i(\omega),
  \qquad
  E_i(\omega)=i\omega A_i(\omega).
  \label{eq:fourier-AE}
\end{align}
The dc limit is obtained by taking the external frequencies
to zero after the cancellation of the apparent
velocity-gauge singularities.
An equivalent length-gauge formulation exists
\cite{Aversa1995,Sipe2000}, but the velocity gauge is
adopted here because the Peierls expansion generates the
current vertices and contact vertices in a systematic way.

Expanding Eq.~\eqref{eq:peierls-main} in powers of $\bm A$ gives 
\begin{align}
  H_{\bm A}({\bm k},t)
  &=H({\bm k})\nonumber\\
  &\quad+e\sum_{i}V_i({\bm k})A_i(t)\nonumber\\
  &\quad+\frac{e^2}{2}\sum_{ij}V_{ij}({\bm k})A_i(t)A_j(t)\nonumber\\
  &\quad+\frac{e^3}{3!}\sum_{ijk}V_{ijk}({\bm k})A_i(t)A_j(t)A_k(t)+\cdots, 
  \label{eq:HA-expansion-main}
\end{align}
where
\begin{align}
  &V_i
  =\frac{1}{\hbar}\partial_iH, \\
  &V_{ij}
  =\frac{1}{\hbar^2}\partial_i\partial_jH,\\
  &V_{ijk}
  =\frac{1}{\hbar^3}\partial_i\partial_j\partial_kH,
  \label{eq:vertices-main}
\end{align}
are the first-, second-, and third-order velocity vertices, respectively, 
and $\partial_i\equiv\partial/\partial k_i$.  
Substituting Eq.~\eqref{eq:HA-expansion-main} into
$\hat{H}_{\bm{A}}(t)
 = \sum_{\bm{k}}\hat{\Psi}^{\dagger}_{\bm{k}}\,
   H_{\bm{A}}(\bm{k},t)\,\hat{\Psi}_{\bm{k}}$,
the total Hamiltonian separates as
$\hat{H}_{\bm{A}}(t) = \hat{H}_{0} + \hat{H}_{\mathrm{ext}}(t)$,
where the light--matter interaction $\hat{H}_{\mathrm{ext}}(t)$
reads
\begin{align}
  \hat{H}_{\mathrm{ext}}(t) = 
  \hat{H}^{(1)}_{\mathrm{ext}}(t)
  +
  \hat{H}^{(2)}_{\mathrm{ext}}(t)
  +
  \hat{H}^{(3)}_{\mathrm{ext}}(t),
  \label{eq:Hext-expanded}
\end{align}
where
\begin{align}
  \hat{H}^{(1)}_{\mathrm{ext}}(t)
  &=
  e\sum_{i}\sum_{\bm{k}}
  \hat{\Psi}^{\dagger}_{\bm{k}}\,
  V_i(\bm{k})\,
  \hat{\Psi}_{\bm{k}}\,
  A_i(t),
  \label{eq:Hext1-time}
  \\
  \hat{H}^{(2)}_{\mathrm{ext}}(t)
  &=
  \frac{e^2}{2}
  \sum_{ij}\sum_{\bm{k}}
  \hat{\Psi}^{\dagger}_{\bm{k}}\,
  V_{ij}(\bm{k})\,
  \hat{\Psi}_{\bm{k}}\,
  A_i(t)A_j(t),
  \label{eq:Hext2-time}
  \\
  \hat{H}^{(3)}_{\mathrm{ext}}(t)
  &=
  \frac{e^3}{3!}\sum_{ijk}
  \sum_{\bm{k}}
  \hat{\Psi}^{\dagger}_{\bm{k}}\,
  V_{ijk}(\bm{k})\,
  \hat{\Psi}_{\bm{k}}\,
  A_i(t)A_j(t)A_k(t).
  \label{eq:Hext3}
\end{align}
The first term is the paramagnetic coupling to the vector potential,
while the higher-order terms are the two-photon and three-photon
contact interactions that arise naturally from the Peierls
substitution~\eqref{eq:peierls-main}.
Note that the coupling constants are determined entirely by the
band structure through the velocity vertices
Eq.~\eqref{eq:vertices-main}, with no free parameters.

The physical current operator is obtained from
\begin{align}
  \hat{\mathcal{J}}_i(t)
  =
  -\frac{\delta\hat{H}_{\bm{A}}(t)}
        {\delta A_i(t)}.
\end{align}
Keeping the terms that contribute to the current up to second
order in $\bm{A}$, we obtain
\begin{align}
  \hat{\mathcal{J}}_i(t)
  =
  \hat{\mathcal{J}}^{(0)}_i(t)
  +
  \hat{\mathcal{J}}^{(1)}_i(t)
  +
  \hat{\mathcal{J}}^{(2)}_i(t),
  \label{eq:current-decomp}
\end{align}
with
\begin{align}
  \hat{\mathcal{J}}^{(0)}_i(t)
  &=
  -e
  \sum_{\bm{k}}
  \hat{\Psi}^{\dagger}_{\bm{k}}\,
  V_i(\bm{k})\,
  \hat{\Psi}_{\bm{k}},
  \label{eq:current0}
  \\
  \hat{\mathcal{J}}^{(1)}_i(t)
  &=
  -e^2\sum_{j}
  \sum_{\bm{k}}
  \hat{\Psi}^{\dagger}_{\bm{k}}\,
  V_{ij}(\bm{k})\,
  \hat{\Psi}_{\bm{k}}\,
  A_j(t),
  \label{eq:current1}
  \\
  \hat{\mathcal{J}}^{(2)}_i(t)
  &=
  -\frac{e^3}{2}\sum_{jk}
  \sum_{\bm{k}}
  \hat{\Psi}^{\dagger}_{\bm{k}}\,
  V_{ijk}(\bm{k})\,
  \hat{\Psi}_{\bm{k}}\,
  A_j(t)A_k(t).
  \label{eq:current2}
\end{align}
Thus the current contains the one-photon current vertex
$V_i$, the two-photon contact-current vertex $V_{ij}$, and
the three-photon contact-current vertex $V_{ijk}$.
These contact vertices are essential in the velocity gauge.
When the vector-potential response is rewritten in terms of
electric fields by using Eq.~\eqref{eq:fourier-AE},
individual terms contain apparent low-frequency
singularities.  Gauge invariance requires these singularities
to cancel between the paramagnetic and contact-current
contributions, leaving a finite conductivity in the
homogeneous dc limit
\cite{Ward1950,Takahashi1957,Aversa1995,Sipe2000,Parker2019}.

\paragraph*{Validity of the Peierls coupling.}
The velocity-gauge Peierls substitution used throughout
this work neglects intra-atomic dipole matrix elements
that arise in a multi-orbital Wannier representation.
The resulting expressions are therefore exact for
single-orbital models and capture the dominant Peierls
contribution in multi-orbital systems where the Wannier
functions are sufficiently localized~\cite{Foreman2002,
Ventura2017}. 
\section{Analysis of the Current Expectation Value}
\label{Sec:ACEV}

\subsection{Time contour-ordered Green's function}
We evaluate the second-order response using the Keldysh
Green-function formalism
\cite{Keldysh1965,Rammer1986,Haug2008,Stefanucci2013}.
Because the external vector potential is spatially uniform,
crystal momentum is conserved and the Green's function is
diagonal in $\bm{k}$.
The contour-ordered Green's function is defined on the
Keldysh contour $\mathcal{C}$ as
\begin{align}
  G_{\bm{k},\mu\nu}(z,z')
  =
  -\frac{i}{\hbar}
  \left\langle
  T_{\mathcal{C}}
  \hat{\Psi}_{\bm{k}\mu}(z)
  \hat{\Psi}^{\dagger}_{\bm{k}\nu}(z')
  \right\rangle , 
  \label{eq:contour-G}
\end{align}
where $T_{\mathrm{C}}$ is the contour-ordering operator,
which orders field operators along $\mathcal{C}$ with a
minus sign for each transposition of fermionic operators. 
The Keldysh contour $\mathcal{C}$ runs from $t=-\infty$ forward
to $t=+\infty$ and then back to $t=-\infty$, thereby capturing
the time evolution under the full Hamiltonian
$\hat{H}_{\bm{A}}(t)=\hat{H}_{0}+\hat{H}_{\mathrm{ext}}(t)$
without the need for an adiabatic switch-off. 
From the contour-ordered Green's function one extracts the
physically relevant real-time components via the Langreth rules.
The retarded, advanced, lesser, and greater Green's functions are
defined as
\begin{align}
  G^{\mathrm{R}}_{\bm{k},\bm{k}'}(t,t')
  &= -\frac{i}{\hbar}\theta(t-t')
     \bigl\langle\bigl\{
       \hat{\Psi}_{\bm{k}}(t),\,
       \hat{\Psi}^{\dagger}_{\bm{k}'}(t')
     \bigr\}\bigr\rangle_{\hat{H}_{\bm{A}}},
  \label{eq:GR}\\
  G^{\mathrm{A}}_{\bm{k},\bm{k}'}(t,t')
  &= \bigl[G^{\mathrm{R}}_{\bm{k}',\bm{k}}(t',t)\bigr]^{*},
  \label{eq:GA}\\
  G^{<}_{\bm{k},\bm{k}'}(t,t')
  &= \frac{i}{\hbar}
     \bigl\langle
       \hat{\Psi}^{\dagger}_{\bm{k}'}(t')\,
       \hat{\Psi}_{\bm{k}}(t)
     \bigr\rangle_{\hat{H}_{\bm{A}}},
  \label{eq:Gless}\\
  G^{>}_{\bm{k},\bm{k}'}(t,t')
  &= -\frac{i}{\hbar}
     \bigl\langle
       \hat{\Psi}_{\bm{k}}(t)\,
       \hat{\Psi}^{\dagger}_{\bm{k}'}(t')
     \bigr\rangle_{\hat{H}_{\bm{A}}},
  \label{eq:Ggreater}
\end{align}
where $\theta(t-t')$ is the Heaviside step function and
$\{\cdot,\cdot\}$ denotes the anticommutator.
These components satisfy the identity
$G^{\mathrm{R}}-G^{\mathrm{A}}=G^{>}-G^{<}$.

\subsection{Current expectation value}
The lesser Green's function $G^{<}$ plays a central role in
transport theory: for any single-particle observable with matrix
$M(\bm{k})$ in the orbital basis, its nonequilibrium expectation
value is
\begin{align}
  \langle\hat{O}\rangle
  &= \left\langle
      \sum_{\bm{k}}
      \hat{\Psi}^{\dagger}_{\bm{k}}\,
      O(\bm{k})\,
      \hat{\Psi}_{\bm{k}}
    \right\rangle_{\hat{H}_{\bm{A}}} = -i\hbar\sum_{\bm{k}}
    \mathrm{Tr}\!\left[O(\bm{k})\,G^{<}_{\bm{k}}(t,t)\right],
  \label{eq:bilinear-lesser}
\end{align}
where the trace runs over the orbital indices. 

Using Eq.~\eqref{eq:bilinear-lesser}, the expectation value
of the current is
\begin{align}
  \mathcal{J}_i(t)
  =
  \left\langle
  \hat{\mathcal{J}}_i(t)
  \right\rangle_{\hat{H}_{\bm{A}}}
  =
  \mathcal{J}^{[0]}_i(t)
  +
  \mathcal{J}^{[1]}_i(t)
  +
  \mathcal{J}^{[2]}_i(t),
  \label{eq:J-total-current}
\end{align}
where the superscript in square brackets denotes the order of
the current operator in the vector potential.
The three contributions are
\begin{align}
  \mathcal{J}^{[0]}_i(t)
  &=
  ie\hbar
  \sum_{\bm{k}}
  \mathrm{Tr}
  \left[
  V_i(\bm{k})G^<_{\bm{k}}(t,t)
  \right],
  \label{eq:J-current0-time}
  \\
  \mathcal{J}^{[1]}_i(t)
  &=
  ie^2\hbar\sum_{j}
  \sum_{\bm{k}}
  \mathrm{Tr}
  \left[
  V_{ij}(\bm{k})G^<_{\bm{k}}(t,t)
  \right]
  A_j(t),
  \label{eq:J-current1-time}
  \\
  \mathcal{J}^{[2]}_i(t)
  &=
  \frac{ie^3\hbar}{2}\sum_{jk}
  \sum_{\bm{k}}
  \mathrm{Tr}
  \left[
  V_{ijk}(\bm{k})G^<_{\bm{k}}(t,t)
  \right]
  A_j(t)A_k(t).
  \label{eq:J-current2-time}
\end{align}

The Green's function is expanded perturbatively in powers of
the vector potential as
\begin{align}
  G_{\bm{k}}
  = G^{(0)}_{\bm{k}}
  + G^{(1)}_{\bm{k}}
  + G^{(2)}_{\bm{k}},
  \label{eq:G-expansion}
\end{align}
where $G^{(p)}_{\bm{k}}$ denotes the $p$th-order correction
in $\bm{A}$.
Substituting Eq.~\eqref{eq:G-expansion} into
Eqs.~\eqref{eq:J-current0-time}--\eqref{eq:J-current2-time} and
collecting all terms up to order $A_{j}A_{k}$, the zeroth-,
first-, and second-order currents are
\begin{align}
  \mathcal{J}^{(0)}_{i}(t)
  &= ie\hbar\sum_{\bm{k}}
     \mathrm{Tr}\!\left[
       V_{i}(\bm{k})\,G^{(0)<}_{\bm{k}}(t,t)
     \right],
  \label{eq:J0-lesser}
  \\
\mathcal{J}^{(1)}_{i}(t)
  &= ie\hbar\sum_{\bm{k}}
     \mathrm{Tr}\!\left[
       V_{i}(\bm{k})\,G^{(1)<}_{\bm{k}}(t,t)
     \right]
  \nonumber\\
  &\quad
  + ie^{2}\hbar\sum_{j}\sum_{\bm{k}}
    \mathrm{Tr}\!\left[
      V_{ij}(\bm{k})\,G^{(0)<}_{\bm{k}}(t,t)
    \right]A_{j}(t),
  \label{eq:J1-lesser}
  \\
  \mathcal{J}^{(2)}_{i}(t)
  &= ie\hbar\sum_{\bm{k}}
     \mathrm{Tr}\!\left[
       V_{i}(\bm{k})\,G^{(2)<}_{\bm{k}}(t,t)
     \right]
  \nonumber\\
  &\quad
  + ie^{2}\sum_{j}\hbar\sum_{\bm{k}}
    \mathrm{Tr}\!\left[
      V_{ij}(\bm{k})\,G^{(1)<}_{\bm{k}}(t,t)
    \right]A_{j}(t)
  \nonumber\\
  &\quad
  + \frac{ie^{3}\hbar}{2}\sum_{jk}\sum_{\bm{k}}
    \mathrm{Tr}\!\left[
      V_{ijk}(\bm{k})\,G^{(0)<}_{\bm{k}}(t,t)
    \right]A_{j}(t)A_{k}(t).
  \label{eq:J2-lesser}
  \end{align}
  Equation~\eqref{eq:J0-lesser} is the equilibrium current in
the absence of the external field $\bm{A}(t)$.
Equation~\eqref{eq:J1-lesser} is the linear response to
$\bm{A}(t)$: the first line contains the one-photon current
vertex $V_{i}$ combined with the first-order Green's function
$G^{(1)<}$, while the second line contains the two-photon
contact vertex $V_{ij}$ combined with the equilibrium Green's
function $G^{(0)<}$.
Equation~\eqref{eq:J2-lesser} is the second-order response:
the first line contains the one-photon current vertex $V_{i}$
combined with the second-order Green's function $G^{(2)<}$,
which itself receives contributions from two first-order
perturbation vertices $V_{j}$, $V_{k}$ and from the
two-photon Hamiltonian vertex $V_{jk}$; the second line
contains the contact-current vertex $V_{ij}$ combined with
the first-order Green's function $G^{(1)<}$; and the third
line is the three-photon contact-current term involving
$V_{ijk}$ and the equilibrium lesser Green's function
$G^{(0)<}$.
 
In Fourier space we define
\begin{align}
  \mathcal{J}_{i}(\omega)
  = \int_{-\infty}^{\infty}dt\,e^{i\omega t}\,
    \mathcal{J}_{i}(t).
  \label{eq:J-fourier}
\end{align}
Since the equilibrium current is time-independent, its
Fourier transform is
\begin{align}
  \mathcal{J}^{(0)}_{i}(\omega)
  = \mathcal{J}^{\mathrm{eq}}_{i}\,\delta(\omega).
  \label{eq:J0-fourier}
\end{align}
The first-order current defines the linear optical
conductivity $\sigma_{ij}(\omega)$ through
\begin{align}
  \mathcal{J}^{(1)}_{i}(\omega)
  = \sum_{j}\sigma_{ij}(\omega)\,E_{j}(\omega),
  \label{eq:conductivity-definition-1}
\end{align}
and the second-order current defines the second-order
conductivity tensor $\sigma_{ijk}(\omega_{1},\omega_{2})$
through
\begin{align}
  \mathcal{J}^{(2)}_{i}(\omega)
  &= \sum_{jk}
     \int_{-\infty}^{\infty}\frac{d\omega_{1}}{2\pi}
     \int_{-\infty}^{\infty}\frac{d\omega_{2}}{2\pi}\,
     2\pi\delta(\omega-\omega_{1}-\omega_{2})\,
     \sigma_{ijk}(\omega_{1},\omega_{2})\,
     E_{j}(\omega_{1})\,E_{k}(\omega_{2}).
  \label{eq:conductivity-definition-2}
\end{align}
In this Supplemental Material, we evaluate the dc linear
and second-order conductivities defined in
Eqs.~\eqref{eq:conductivity-definition-1} and
\eqref{eq:conductivity-definition-2} by first taking the
homogeneous limit $\bm{q}=0$ and then taking $\omega\to 0$
or $\omega_{1},\omega_{2}\to 0$.
The perturbative expansion of $G^{(1)<}_{\bm{k}}$ and
$G^{(2)<}_{\bm{k}}$, the application of the Langreth rules,
and the explicit frequency kernels are given in
Secs.~\ref{sec:sigma-1} and \ref{sec:sigma-2}.

\section{Perturbative Expansion of the Contour-Ordered Green's Function}
\label{sec:expand-G}
\subsection{Perturbative Hamiltonian}
To evaluate the perturbative expansion of the contour-ordered
Green's function, we decompose the full Hamiltonian
$\hat{H}_{\bm{A}}(t)$ into the free-particle part and the
light--matter interaction as
\begin{align}
  \hat{H}_{\bm{A}}(t)
  &= \hat{H}_{0}
   + \hat{H}_{\mathrm{ext},1}(t)
   + \hat{H}_{\mathrm{ext},2}(t)
   + \cdots,
  \\
  \hat{H}_{0}
  &= \sum_{\bm{k}}
     \hat{\Psi}^{\dagger}_{\bm{k}}\,H(\bm{k})\,
     \hat{\Psi}_{\bm{k}},
  \\
  \hat{H}_{\mathrm{ext},1}(t)
  &= \sum_{i}\sum_{\bm{k}}
     \hat{\Psi}^{\dagger}_{\bm{k}}\,eV_{i}(\bm{k})\,
     \hat{\Psi}_{\bm{k}}\,A_{i}(t),
  \label{eq:Hext1}
  \\
  \hat{H}_{\mathrm{ext},2}(t)
  &= \frac{1}{2}\sum_{ij}\sum_{\bm{k}}
     \hat{\Psi}^{\dagger}_{\bm{k}}\,e^{2}V_{ij}(\bm{k})\,
     \hat{\Psi}_{\bm{k}}\,A_{i}(t)A_{j}(t),
  \label{eq:Hext2}
\end{align}
where $\hat{H}_{0}$ is the free-particle Hamiltonian, and
$\hat{H}_{\mathrm{ext},1}(t)$ and $\hat{H}_{\mathrm{ext},2}(t)$
are the first- and second-rank perturbation Hamiltonians arising
from the Peierls velocity-gauge substitution
[Eq.~\eqref{eq:HA-expansion-main}].

\subsection{Expansion in powers of
            \texorpdfstring{$\bm{A}$}{A}}

We expand the contour-ordered Green's function
$G_{\bm{k},\bm{k}}(t,t)$ perturbatively in $\bm{A}$ up to
second order, using the perturbation Hamiltonians
Eqs.~\eqref{eq:Hext1} and \eqref{eq:Hext2}:
\begin{align}
  G_{\bm{k},\bm{k}}(t,t)
  &= G^{(0)}_{\bm{k},\bm{k}}(t,t)
   + G^{(1)}_{\bm{k},\bm{k}}(t,t)
   + G^{(2)}_{\bm{k},\bm{k}}(t,t)
   + \cdots.
  \label{eq:G-tt-expansion}
\end{align}
The Fourier component $G_{\bm{k}}(\omega)$ is defined by
\begin{align}
  G_{\bm{k},\bm{k}}(t,t)
  = \int_{-\infty}^{\infty}\frac{d\omega}{2\pi}\,
    e^{-i\omega t}\,G_{\bm{k}}(\omega),
  \label{eq:G-omega-def}
\end{align}
so that
\begin{align}
  G_{\bm{k}}(\omega)
  &= \int_{-\infty}^{\infty}dt\,e^{i\omega t}\,
     G_{\bm{k},\bm{k}}(t,t)
  \nonumber\\
  &= G^{(0)}_{\bm{k}}(\omega)
   + G^{(1)}_{\bm{k}}(\omega)
   + G^{(2)}_{\bm{k}}(\omega)
   + \cdots.
  \label{eq:G-omega-expansion}
\end{align}
Each Fourier component $G^{(p)}_{\bm{k}}(\omega)$ is obtained
by substituting the corresponding perturbative correction
$G^{(p)}_{\bm{k},\bm{k}}(t,t)$ into
Eq.~\eqref{eq:G-omega-def}.
The explicit expressions for $G^{(0)}_{\bm{k}}(\omega)$,
$G^{(1)}_{\bm{k}}(\omega)$, and $G^{(2)}_{\bm{k}}(\omega)$
are derived in the following subsections.  

\subsection{Single-particle Green's function}

The basic building block of the perturbative expansion is the
single-particle Green's function, defined by
\begin{align}
  G_{\bm{k}}(\varepsilon)
  = \frac{1}{\varepsilon + \mu - H(\bm{k})},
  \label{eq:1p-G}
\end{align}
where $\mu$ is the chemical potential and the identity matrix
is implicit.
The retarded and advanced Green's functions are obtained by
displacing the energy argument into the upper and lower complex
half-planes, respectively,
\begin{align}
  G^{\mathrm{R}}_{\bm{k}}(\varepsilon)
  &= G_{\bm{k}}(\varepsilon + i0^{+})
   = \frac{1}{\varepsilon + \mu - H(\bm{k}) + i0^{+}},
  \label{eq:GR-def}
  \\
  G^{\mathrm{A}}_{\bm{k}}(\varepsilon)
  &= G_{\bm{k}}(\varepsilon - i0^{+})
   = \frac{1}{\varepsilon + \mu - H(\bm{k}) - i0^{+}}
   = \bigl[G^{\mathrm{R}}_{\bm{k}}(\varepsilon)\bigr]^{\dagger},
  \label{eq:GA-def}
\end{align}
where $0^{+}$ denotes a positive infinitesimal.
In practice, we adopt the constant-scattering-time
approximation, in which $0^{+}$ is replaced by a finite
broadening $\gamma = \hbar/(2\tau)$, where $\tau$ is the
quasiparticle lifetime.
The retarded and advanced Green's functions then take the
explicit forms
\begin{align}
  G^{\mathrm{R}}_{\bm{k}}(\varepsilon)
  &= \frac{1}{\varepsilon + \mu - H(\bm{k}) + i\gamma},
  \nonumber\\
  G^{\mathrm{A}}_{\bm{k}}(\varepsilon)
  &= \frac{1}{\varepsilon + \mu - H(\bm{k}) - i\gamma}.
  \label{eq:GRA-lifetime}
\end{align}
The difference of the two defines the spectral function
\begin{align}
  \Lambda_{\bm{k}}(\varepsilon)
  \equiv G^{\mathrm{A}}_{\bm{k}}(\varepsilon)
       - G^{\mathrm{R}}_{\bm{k}}(\varepsilon)
  = \frac{2i\gamma}
         {(\varepsilon + \mu - H(\bm{k}))^{2} + \gamma^{2}},
  \label{eq:Lambda-def}
\end{align}
which reduces to $2\pi i\,\delta(\varepsilon + \mu - H(\bm{k}))$
in the clean limit $\gamma \to 0^{+}$.
In thermal equilibrium, the lesser Green's function is related
to the spectral function via the fluctuation--dissipation
theorem,
\begin{align}
  G^{<}_{\bm{k}}(\varepsilon)
  = f(\varepsilon)\,\Lambda_{\bm{k}}(\varepsilon),
  \label{eq:Gless-equil}
\end{align}
where
$f(\varepsilon)
 = \bigl(1 + e^{\varepsilon/k_{\mathrm{B}}T}\bigr)^{-1}$
is the Fermi--Dirac distribution function.

\subsection{Zeroth-order term}
The zeroth-order contour-ordered Green's function is defined with
respect to the unperturbed Hamiltonian $\hat{H}_{0}$ as
\begin{align}
  G^{(0)}_{\bm{k},\bm{k}'}(t,t')
  = -\frac{i}{\hbar}
    \bigl\langle T_{\mathrm{C}}\,
      \hat{\Psi}_{\bm{k}}(t)\,
      \hat{\Psi}^{\dagger}_{\bm{k}'}(t')
    \bigr\rangle_{\hat{H}_{0}}.
  \label{eq:G0-def}
\end{align}
Since $\hat{H}_{0}$ is diagonal in crystal momentum, the
Green's function is diagonal,
$G^{(0)}_{\bm{k},\bm{k}'}(t,t')
 = \delta_{\bm{k},\bm{k}'}\,G^{(0)}_{\bm{k}}(t-t')$,
and depends only on the time difference $t-t'$.
Its Fourier transform with respect to $t-t'$ defines the
single-particle Green's function
\begin{align}
  G_{\bm{k}}(\varepsilon)
  = \int_{-\infty}^{\infty}
    d(t-t')\,e^{i\varepsilon(t-t')/\hbar}\,
    G^{(0)}_{\bm{k}}(t-t'),
  \label{eq:G0-fourier-def}
\end{align}
which is explicitly given by Eq.~\eqref{eq:1p-G}.
The inverse relation,
\begin{align}
  G^{(0)}_{\bm{k}}(t-t')
  = \int_{-\infty}^{\infty}
    \frac{d\varepsilon}{2\pi\hbar}\,
    e^{-i\varepsilon(t-t')/\hbar}\,
    G_{\bm{k}}(\varepsilon),
  \label{eq:G0-inverse}
\end{align}
yields the equal-time expression upon setting $t = t'$,
\begin{align}
  G^{(0)}_{\bm{k},\bm{k}}(t,t)
  = \int_{-\infty}^{\infty}
    \frac{d\varepsilon}{2\pi\hbar}\,
    G_{\bm{k}}(\varepsilon).
  \label{eq:G0-tt}
\end{align}
Taking the Fourier transform of Eq.~\eqref{eq:G0-tt} with
respect to $t$ then gives
\begin{align}
  G^{(0)}_{\bm{k}}(\omega)
  = \int_{-\infty}^{\infty}dt\,e^{i\omega t}\,
    G^{(0)}_{\bm{k},\bm{k}}(t,t)
  = 2\pi\delta(\omega)
    \int_{-\infty}^{\infty}
    \frac{d\varepsilon}{2\pi\hbar}\,
    G_{\bm{k}}(\varepsilon).
  \label{eq:G0-omega}
\end{align}
The factor $2\pi\delta(\omega)$ reflects the fact that, in the
absence of an external field, the equal-time Green's function
is time-independent, and its Fourier transform is therefore
concentrated at $\omega = 0$.

\subsection{First-order term}
\label{sec:sigma-1}

The first-order correction to the contour-ordered Green's function
is obtained by inserting the first-order perturbation term,
\begin{align}
  -\frac{i}{\hbar}
  \int_{-\infty}^{\infty}dt'\,\hat{H}_{\mathrm{ext},1}(t')
  = -\frac{i}{\hbar}
    \int_{-\infty}^{\infty}dt'
    \sum_{j}\sum_{\bm{k}'}
    \hat{\Psi}^{\dagger}_{\bm{k}'}(t')\,
    eV_{j}(\bm{k}')\,
    \hat{\Psi}_{\bm{k}'}(t')\,A_{j}(t'),
\end{align}
into the contour-ordered Green's function $G_{\bm{k},\bm{k}}(t,t)$,
which gives
\begin{align}
  G^{(1)}_{\bm{k},\bm{k}}(t,t)
  &= -\frac{i}{\hbar}
     \left\langle T_{\mathrm{C}}\,\hat{\Psi}_{\bm{k}}(t)
     \left(
       -\frac{i}{\hbar}
       \int_{-\infty}^{\infty}dt'\,\hat{H}_{\mathrm{ext},1}(t')
     \right)
     \hat{\Psi}^{\dagger}_{\bm{k}}(t)
     \right\rangle
  \nonumber\\
  &= -\frac{i}{\hbar}
     \left\langle T_{\mathrm{C}}\,\hat{\Psi}_{\bm{k}}(t)
     \left(
       -\frac{i}{\hbar}
       \int_{-\infty}^{\infty}dt'
       \sum_{j}\sum_{\bm{k}'}
       \hat{\Psi}^{\dagger}_{\bm{k}'}(t')\,
       eV_{j}(\bm{k}')\,
       \hat{\Psi}_{\bm{k}'}(t')\,A_{j}(t')
     \right)
     \hat{\Psi}^{\dagger}_{\bm{k}}(t)
     \right\rangle.
  \label{eq:G1-def}
\end{align}
Applying Wick's theorem to take the connected contraction of
$\hat{\Psi}\hat{\Psi}^{\dagger}$ between times $t$ and $t'$,
and using the translational invariance in momentum
$G^{(0)}_{\bm{k},\bm{k}'} = \delta_{\bm{k},\bm{k}'}
G^{(0)}_{\bm{k}}(t-t')$, one obtains
\begin{align}
  G^{(1)}_{\bm{k},\bm{k}}(t,t)
  &= \left(-\frac{i}{\hbar}\right)^{\!2}
     \sum_{j}\sum_{\bm{k}'}
     \int_{-\infty}^{\infty}dt'\,
     \langle T_{\mathrm{C}}\,
       \hat{\Psi}_{\bm{k}}(t)\hat{\Psi}^{\dagger}_{\bm{k}'}(t')
     \rangle\,
     eV_{j}(\bm{k}')\,
     \langle T_{\mathrm{C}}\,
       \hat{\Psi}_{\bm{k}'}(t')\hat{\Psi}^{\dagger}_{\bm{k}}(t)
     \rangle\,A_{j}(t')
  \nonumber\\
  &= e\sum_{j}
     \int_{-\infty}^{\infty}dt'\,
     G^{(0)}_{\bm{k}}(t-t')\,
     V_{j}(\bm{k})\,
     G^{(0)}_{\bm{k}}(t'-t)\,A_{j}(t').
  \label{eq:G1-wick}
\end{align}
Taking the Fourier transform with respect to time and performing
the $t'$ integration via the identity 
\begin{align}
\int_{-\infty}^{\infty}dt'\,
e^{i(-\hbar\omega+\varepsilon'-\varepsilon)t'/\hbar}
= 2\pi\hbar\,\delta(\varepsilon'-\varepsilon-\hbar\omega),
\end{align}
one finds
\begin{align}
  G^{(1)}_{\bm{k},\bm{k}}(t,t)
  &= e\sum_{j}
     \int_{-\infty}^{\infty}\frac{d\omega}{2\pi}\,e^{-i\omega t}
     \int_{-\infty}^{\infty}\frac{d\varepsilon'}{2\pi\hbar}
     \int_{-\infty}^{\infty}\frac{d\varepsilon}{2\pi\hbar}\,
     (2\pi\hbar)\,\delta(\varepsilon'-\varepsilon-\hbar\omega)\,
     G_{\bm{k}}(\varepsilon')\,V_{j}(\bm{k})\,
     G_{\bm{k}}(\varepsilon)\,A_{j}(\omega)
  \nonumber\\
  &= \int_{-\infty}^{\infty}\frac{d\omega}{2\pi}\,
     e^{-i\omega t}\,G^{(1)}_{\bm{k}}(\omega),
\end{align}
where the $\varepsilon'$ integration is eliminated by the
delta function, yielding the first-order Fourier component
\begin{align}
  G^{(1)}_{\bm{k}}(\omega)
  = e\sum_{j}
    \int_{-\infty}^{\infty}\frac{d\varepsilon}{2\pi\hbar}\,
    G_{\bm{k}}(\varepsilon+\hbar\omega)\,
    V_{j}(\bm{k})\,G_{\bm{k}}(\varepsilon)\,A_{j}(\omega).
  \label{eq:G1-freq}
\end{align}

\subsection{Second-order terms}
\label{sec:sigma-2}

The second-order correction to the contour-ordered Green's function
receives two distinct contributions: the second-order expansion of
$\hat{H}_{\mathrm{ext},1}(t)$, denoted $G^{(2a)}_{\bm{k}}(\omega)$,
and the first-order expansion of $\hat{H}_{\mathrm{ext},2}(t)$,
denoted $G^{(2b)}_{\bm{k}}(\omega)$.

\subsubsection*{\texorpdfstring{Contribution from
  $\hat{H}_{\mathrm{ext},1}^{2}$}{Contribution from Hext1 squared}}

The second-order perturbation term generated by
$\hat{H}_{\mathrm{ext},1}$ is
\begin{align}
  &\frac{1}{2!}
   \left(
     -\frac{i}{\hbar}
     \int_{-\infty}^{\infty}dt\,\hat{H}_{\mathrm{ext},1}(t)
   \right)^{\!2}
  \nonumber\\
  &= \frac{1}{2}\!\left(-\frac{i}{\hbar}\right)^{\!2}
     \sum_{jk}
     \int_{-\infty}^{\infty}dt_{1}
     \sum_{\bm{k}_{1}}
     \hat{\Psi}^{\dagger}_{\bm{k}_{1}}(t_{1})\,
     eV_{j}(\bm{k}_{1})\,
     \hat{\Psi}_{\bm{k}_{1}}(t_{1})\,
     A_{j}(t_{1})
  \nonumber\\
  &\quad\times
     \int_{-\infty}^{\infty}dt_{2}
     \sum_{\bm{k}_{2}}
     \hat{\Psi}^{\dagger}_{\bm{k}_{2}}(t_{2})\,
     eV_{k}(\bm{k}_{2})\,
     \hat{\Psi}_{\bm{k}_{2}}(t_{2})\,
     A_{k}(t_{2}).
\end{align}
Inserting this into the contour-ordered Green's function gives
\begin{align}
  G^{(2a)}_{\bm{k},\bm{k}}(t,t)
  &= \frac{1}{2}\!\left(-\frac{i}{\hbar}\right)^{\!3}
     \sum_{jk}
     \sum_{\bm{k}_{1},\bm{k}_{2}}
     \int_{-\infty}^{\infty}dt_{1}
     \int_{-\infty}^{\infty}dt_{2}\,
     A_{j}(t_{1})A_{k}(t_{2})
  \nonumber\\
  &\quad\times
     \Bigl\langle T_{\mathrm{C}}\,
       \hat{\Psi}_{\bm{k}}(t)\,
       \hat{\Psi}^{\dagger}_{\bm{k}_{1}}(t_{1})\,
       eV_{j}(\bm{k}_{1})\,
       \hat{\Psi}_{\bm{k}_{1}}(t_{1})\,
       \hat{\Psi}^{\dagger}_{\bm{k}_{2}}(t_{2})\,
       eV_{k}(\bm{k}_{2})\,
       \hat{\Psi}_{\bm{k}_{2}}(t_{2})\,
       \hat{\Psi}^{\dagger}_{\bm{k}}(t)
     \Bigr\rangle.
\end{align}
Applying Wick's theorem to take the connected contractions of
$\hat{\Psi}\hat{\Psi}^{\dagger}$, one obtains
\begin{align}
  &\Bigl\langle T_{\mathrm{C}}\,
    \hat{\Psi}_{\bm{k}}(t)\,
    \hat{\Psi}^{\dagger}_{\bm{k}_{1}}(t_{1})\,
    eV_{j}(\bm{k}_{1})\,
    \hat{\Psi}_{\bm{k}_{1}}(t_{1})\,
    \hat{\Psi}^{\dagger}_{\bm{k}_{2}}(t_{2})\,
    eV_{k}(\bm{k}_{2})\,
    \hat{\Psi}_{\bm{k}_{2}}(t_{2})\,
    \hat{\Psi}^{\dagger}_{\bm{k}}(t)
  \Bigr\rangle_{\mathrm{con.}}
  \nonumber\\
  &=
  \langle T_{\mathrm{C}}\,
    \hat{\Psi}_{\bm{k}}(t)\hat{\Psi}^{\dagger}_{\bm{k}_{1}}(t_{1})
  \rangle\,
  eV_{j}(\bm{k}_{1})\,
  \langle T_{\mathrm{C}}\,
    \hat{\Psi}_{\bm{k}_{1}}(t_{1})\hat{\Psi}^{\dagger}_{\bm{k}_{2}}(t_{2})
  \rangle\,
  eV_{k}(\bm{k}_{2})\,
  \langle T_{\mathrm{C}}\,
    \hat{\Psi}_{\bm{k}_{2}}(t_{2})\hat{\Psi}^{\dagger}_{\bm{k}}(t)
  \rangle
  \nonumber\\
  &\quad
  +\langle T_{\mathrm{C}}\,
    \hat{\Psi}_{\bm{k}}(t)\hat{\Psi}^{\dagger}_{\bm{k}_{2}}(t_{2})
  \rangle\,
  eV_{k}(\bm{k}_{2})\,
  \langle T_{\mathrm{C}}\,
    \hat{\Psi}_{\bm{k}_{2}}(t_{2})\hat{\Psi}^{\dagger}_{\bm{k}_{1}}(t_{1})
  \rangle\,
  eV_{j}(\bm{k}_{1})\,
  \langle T_{\mathrm{C}}\,
    \hat{\Psi}_{\bm{k}_{1}}(t_{1})\hat{\Psi}^{\dagger}_{\bm{k}}(t)
  \rangle.
  \nonumber
\end{align}
Since these two terms are related by the simultaneous interchange
$t_{1}\leftrightarrow t_{2}$,
$\bm{k}_{1}\leftrightarrow\bm{k}_{2}$,
$j\leftrightarrow k$, they are identical and the combinatorial
factor $1/2!$ cancels.
Taking the Fourier transform of the resulting time convolution,
using the momentum-diagonal structure
$G^{(0)}_{\bm{k},\bm{k}'}
 = \delta_{\bm{k},\bm{k}'}G^{(0)}_{\bm{k}}(t-t')$,
one finds
\begin{align}
  G^{(2a)}_{\bm{k},\bm{k}}(t,t)
  &= \sum_{jk}
     \sum_{\bm{k}_{1},\bm{k}_{2}}
     \int_{-\infty}^{\infty}dt_{1}
     \int_{-\infty}^{\infty}dt_{2}\,
     A_{j}(t_{1})A_{k}(t_{2})\,
     G^{(0)}_{\bm{k},\bm{k}_{1}}(t-t_{1})\,
     eV_{j}(\bm{k}_{1})\,
     G^{(0)}_{\bm{k}_{1},\bm{k}_{2}}(t_{1}-t_{2})\,
     eV_{k}(\bm{k}_{2})\,
     G^{(0)}_{\bm{k}_{2},\bm{k}}(t_{2}-t)
  \nonumber\\
  &= \sum_{jk}
     \sum_{\bm{k}_{1},\bm{k}_{2}}
     \delta_{\bm{k}\bm{k}_{1}}
     \delta_{\bm{k}_{1}\bm{k}_{2}}
     \delta_{\bm{k}_{2}\bm{k}}
     \int_{-\infty}^{\infty}dt_{1}
     \int_{-\infty}^{\infty}dt_{2}
     \int_{-\infty}^{\infty}\frac{d\omega_{1}}{2\pi}
     \int_{-\infty}^{\infty}\frac{d\omega_{2}}{2\pi}
     \int_{-\infty}^{\infty}\frac{d\varepsilon_{1}}{2\pi\hbar}
     \int_{-\infty}^{\infty}\frac{d\varepsilon_{2}}{2\pi\hbar}
     \int_{-\infty}^{\infty}\frac{d\varepsilon_{3}}{2\pi\hbar}
  \nonumber\\
  &\quad\times
     e^{-i\omega_{1}t_{1}-i\omega_{2}t_{2}
        -i\varepsilon_{1}(t-t_{1})/\hbar
        -i\varepsilon_{2}(t_{1}-t_{2})/\hbar
        -i\varepsilon_{3}(t_{2}-t)/\hbar}
  \nonumber\\
  &\quad\times
     G_{\bm{k}}(\varepsilon_{1})\,
     eV_{j}(\bm{k})\,
     G_{\bm{k}}(\varepsilon_{2})\,
     eV_{k}(\bm{k})\,
     G_{\bm{k}}(\varepsilon_{3})\,
     A_{j}(\omega_{1})A_{k}(\omega_{2})
  \nonumber\\
  &=
     \sum_{jk}
     \int_{-\infty}^{\infty}\frac{d\omega_{1}}{2\pi}
     \int_{-\infty}^{\infty}\frac{d\omega_{2}}{2\pi}
     \int_{-\infty}^{\infty}\frac{d\varepsilon_{1}}{2\pi\hbar}
     \int_{-\infty}^{\infty}\frac{d\varepsilon_{2}}{2\pi\hbar}
     \int_{-\infty}^{\infty}\frac{d\varepsilon_{3}}{2\pi\hbar}\,
     e^{-i(\varepsilon_{1}-\varepsilon_{3})t/\hbar}
  \nonumber\\
  &\quad\times
     (2\pi\hbar)^{2}
     \delta(\varepsilon_{1}-\varepsilon_{2}-\hbar\omega_{1})\,
     \delta(\varepsilon_{2}-\varepsilon_{3}-\hbar\omega_{2})\,
     G_{\bm{k}}(\varepsilon_{1})\,
     eV_{j}(\bm{k})\,
     G_{\bm{k}}(\varepsilon_{2})\,
     eV_{k}(\bm{k})\,
     G_{\bm{k}}(\varepsilon_{3})\,
     A_{j}(\omega_{1})A_{k}(\omega_{2})
  \nonumber\\
  &= \sum_{jk}
     \int_{-\infty}^{\infty}\frac{d\omega_{1}}{2\pi}
     \int_{-\infty}^{\infty}\frac{d\omega_{2}}{2\pi}
     \int_{-\infty}^{\infty}\frac{d\varepsilon_{3}}{2\pi\hbar}\,
     e^{-i(\omega_{1}+\omega_{2})t}
  \nonumber\\
  &\quad\times
     G_{\bm{k}}(\varepsilon_{3}+\hbar(\omega_{1}+\omega_{2}))\,
     eV_{j}(\bm{k})\,
     G_{\bm{k}}(\varepsilon_{3}+\hbar\omega_{2})\,
     eV_{k}(\bm{k})\,
     G_{\bm{k}}(\varepsilon_{3})\,
     A_{j}(\omega_{1})A_{k}(\omega_{2}).
\end{align}
Making the substitutions $\varepsilon_{3}\to\varepsilon$,
$\omega_{1}\to\omega-\omega'$, and $\omega_{2}\to\omega'$,
we identify the Fourier component
\begin{align}
  G^{(2a)}_{\bm{k},\bm{k}}(t,t)
  &= \int_{-\infty}^{\infty}\frac{d\omega}{2\pi}\,
     e^{-i\omega t}\,G^{(2a)}_{\bm{k}}(\omega),
  \\
  G^{(2a)}_{\bm{k}}(\omega)
  &= e^{2}\sum_{jk}
     \int_{-\infty}^{\infty}\frac{d\varepsilon}{2\pi\hbar}
     \int_{-\infty}^{\infty}\frac{d\omega'}{2\pi}\,
     G_{\bm{k}}(\varepsilon+\hbar\omega)\,
     V_{j}(\bm{k})\,
     G_{\bm{k}}(\varepsilon+\hbar\omega')\,
     V_{k}(\bm{k})\,
     G_{\bm{k}}(\varepsilon)\,
     A_{j}(\omega-\omega')\,A_{k}(\omega').
  \label{eq:G2a-freq}
\end{align}

\subsubsection*{\texorpdfstring{Contribution from $\hat{H}_{\mathrm{ext},2}$}{Contribution from Hext2}}

The first-order perturbation term from $\hat{H}_{\mathrm{ext},2}$
is
\begin{align}
  -\frac{i}{\hbar}
  \int_{-\infty}^{\infty}dt_{1}\,\hat{H}_{\mathrm{ext},2}(t_{1})
  &= -\frac{i}{2\hbar}
     \sum_{jk}
     \int_{-\infty}^{\infty}dt_{1}\,
     e^{2}V_{jk}(t_{1})\,A_{j}(t_{1})A_{k}(t_{1})
  \nonumber\\
  &= -\frac{i}{2\hbar}
     \sum_{jk}
     \sum_{\bm{k}_{1}}
     \int_{-\infty}^{\infty}dt_{1}\,
     \hat{\Psi}^{\dagger}_{\bm{k}_{1}}(t_{1})\,
     e^{2}V_{jk}(\bm{k}_{1})\,
     \hat{\Psi}_{\bm{k}_{1}}(t_{1})\,
     A_{j}(t_{1})A_{k}(t_{1}).
\end{align}
Inserting this into the Green's function gives
\begin{align}
  G^{(2b)}_{\bm{k},\bm{k}}(t,t)
  &= \frac{1}{2}\!\left(-\frac{i}{\hbar}\right)^{\!2}
     \sum_{jk}
     \int_{-\infty}^{\infty}dt_{1}\,
     \Bigl\langle T_{\mathrm{C}}\,
       \hat{\Psi}_{\bm{k}}(t)\,
       \hat{\Psi}^{\dagger}_{\bm{k}_{1}}(t_{1})\,
       e^{2}V_{jk}(\bm{k}_{1})\,
       \hat{\Psi}_{\bm{k}_{1}}(t_{1})\,
       \hat{\Psi}^{\dagger}_{\bm{k}}(t)
     \Bigr\rangle
     A_{j}(t_{1})A_{k}(t_{1})
  \nonumber\\
  &= \frac{1}{2}
     \sum_{jk}
     \int_{-\infty}^{\infty}dt_{1}\,
     G_{\bm{k},\bm{k}_{1}}(t-t_{1})\,
     e^{2}V_{jk}(\bm{k}_{1})\,
     G_{\bm{k}_{1},\bm{k}}(t_{1}-t)\,
     A_{j}(t_{1})A_{k}(t_{1})
  \nonumber\\
  &= \frac{1}{2}
     \sum_{jk}
     \int_{-\infty}^{\infty}dt_{1}
     \int_{-\infty}^{\infty}\frac{d\varepsilon_{1}}{2\pi\hbar}
     \int_{-\infty}^{\infty}\frac{d\varepsilon_{2}}{2\pi\hbar}\,
     G_{\bm{k},\bm{k}_{1}}(\varepsilon_{1})\,
     e^{2}V_{jk}(\bm{k}_{1})\,
     G_{\bm{k}_{1},\bm{k}}(\varepsilon_{2})
  \nonumber\\
  &\quad\times
     e^{-i\varepsilon_{1}(t-t_{1})/\hbar
        -i\varepsilon_{2}(t_{1}-t)/\hbar}
     \int_{-\infty}^{\infty}\frac{d\omega_{1}}{2\pi}
     \int_{-\infty}^{\infty}\frac{d\omega_{2}}{2\pi}\,
     e^{-i\omega_{1}t_{1}}e^{-i\omega_{2}t_{1}}
     A_{j}(\omega_{1})A_{k}(\omega_{2})
  \nonumber\\
  &= \frac{1}{2}
     \sum_{jk}
     \int_{-\infty}^{\infty}\frac{d\varepsilon_{1}}{2\pi\hbar}
     \int_{-\infty}^{\infty}\frac{d\varepsilon_{2}}{2\pi\hbar}
     \int_{-\infty}^{\infty}\frac{d\omega_{1}}{2\pi}
     \int_{-\infty}^{\infty}\frac{d\omega_{2}}{2\pi}\,
     e^{-i(\varepsilon_{1}-\varepsilon_{2})t/\hbar}\,
     \delta_{\bm{k},\bm{k}_{1}}
  \nonumber\\
  &\quad\times
     2\pi\hbar\,
     \delta(\varepsilon_{1}-\varepsilon_{2}-\hbar(\omega_{1}+\omega_{2}))\,
     G_{\bm{k}}(\varepsilon_{1})\,
     e^{2}V_{jk}(\bm{k})\,
     G_{\bm{k}}(\varepsilon_{2})\,
     A_{j}(\omega_{1})A_{k}(\omega_{2})
  \nonumber\\
  &= \frac{1}{2}
     \sum_{jk}
     \int_{-\infty}^{\infty}\frac{d\varepsilon_{2}}{2\pi\hbar}
     \int_{-\infty}^{\infty}\frac{d\omega_{1}}{2\pi}
     \int_{-\infty}^{\infty}\frac{d\omega_{2}}{2\pi}\,
     e^{-i(\omega_{1}+\omega_{2})t}
  \nonumber\\
  &\quad\times
     G_{\bm{k}}(\varepsilon_{2}+\hbar(\omega_{1}+\omega_{2}))\,
     e^{2}V_{jk}(\bm{k})\,
     G_{\bm{k}}(\varepsilon_{2})\,
     A_{j}(\omega_{1})A_{k}(\omega_{2}).
\end{align}
Making the substitutions $\omega_{1}+\omega_{2}=\omega$,
$\omega_{2}=\omega'$, and $\varepsilon_{2}=\varepsilon$,
we identify the Fourier component
\begin{align}
  G^{(2b)}_{\bm{k},\bm{k}}(t,t)
  &= \int_{-\infty}^{\infty}\frac{d\omega}{2\pi}\,
     e^{-i\omega t}\,G^{(2b)}_{\bm{k}}(\omega),
  \\
  G^{(2b)}_{\bm{k}}(\omega)
  &= \frac{e^{2}}{2}
     \sum_{jk}
     \int_{-\infty}^{\infty}\frac{d\varepsilon}{2\pi\hbar}
     \int_{-\infty}^{\infty}\frac{d\omega'}{2\pi}\,
     G_{\bm{k}}(\varepsilon+\hbar\omega)\,
     V_{jk}(\bm{k})\,
     G_{\bm{k}}(\varepsilon)\,
     A_{j}(\omega-\omega')\,A_{k}(\omega').
  \label{eq:G2b-freq}
\end{align}
The total second-order correction is therefore
\begin{align}
  G^{(2)}_{\bm{k}}(\omega)
  = G^{(2a)}_{\bm{k}}(\omega) + G^{(2b)}_{\bm{k}}(\omega).
  \label{eq:G2-total}
\end{align}

\section{Expectation value of the current operator and optical conductivity}
\label{sec:esp-sigma}

Using the Green's functions derived in the preceding section,
the expectation value of the total current operator is expanded
in powers of the vector potential in Fourier space as
\begin{align}
  \mathcal{J}_{i}(\omega)
  = \mathcal{J}^{[0]}_{i}(\omega)
  + \mathcal{J}^{[1]}_{i}(\omega)
  + \mathcal{J}^{[2]}_{i}(\omega),
  \label{eq:J-total-current-omega}
\end{align}
where the superscript in square brackets denotes the order
in the vector potential arising from the current operator
itself.
The three contributions are
\begin{align}
  \mathcal{J}^{[0]}_{i}(\omega)
  &= ie\hbar
     \sum_{\bm{k}}
     \mathrm{Tr}\!\left[
       V_{i}(\bm{k})\,G^{<}_{\bm{k}}(\omega)
     \right],
  \label{eq:J-current0}
  \\
  \mathcal{J}^{[1]}_{i}(\omega)
  &= ie^{2}\hbar
     \sum_{\bm{k}}\sum_{j}
     \int_{-\infty}^{\infty}\frac{d\omega'}{2\pi}\,
     \mathrm{Tr}\!\left[
       V_{ij}(\bm{k})\,G^{<}_{\bm{k}}(\omega')
     \right]
     A_{j}(\omega-\omega'),
  \label{eq:J-current1}
  \\
  \mathcal{J}^{[2]}_{i}(\omega)
  &= \frac{ie^{3}\hbar}{2}
     \sum_{\bm{k}}\sum_{jk}
     \int_{-\infty}^{\infty}\frac{d\omega'}{2\pi}
     \int_{-\infty}^{\infty}\frac{d\omega''}{2\pi}\,
     \mathrm{Tr}\!\left[
       V_{ijk}(\bm{k})\,G^{<}_{\bm{k}}(\omega')
     \right]
     A_{j}(\omega'')\,A_{k}(\omega-\omega'-\omega'').
  \label{eq:J-current2}
\end{align}
The full current expectation value is then obtained by
substituting the perturbative expansion of the lesser
Green's function $G^{<}_{\bm{k}}(\omega)$
[Eqs.~\eqref{eq:G0-omega}--\eqref{eq:G2-total}]
into Eqs.~\eqref{eq:J-current0}--\eqref{eq:J-current2}
and collecting terms order by order in $\bm{A}$:
\begin{align}
  \mathcal{J}_{i}(\omega)
  &= \mathcal{J}^{(0)}_{i}(\omega)
   + \mathcal{J}^{(1)}_{i}(\omega)
   + \mathcal{J}^{(2)}_{i}(\omega),
  \label{eq:J-total}
\end{align}
where the zeroth-order current,
\begin{align}
  \mathcal{J}^{(0)}_{i}(\omega)
  &= ie\hbar
     \sum_{\bm{k}}
     \mathrm{Tr}\!\left[
       V_{i}(\bm{k})\,G^{(0)<}_{\bm{k}}(\omega)
     \right],
  \label{eq:J0-def}
\end{align}
is the equilibrium current in the absence of the external
field, the first-order current,
\begin{align}
  \mathcal{J}^{(1)}_{i}(\omega)
  &= \mathcal{J}^{\mathrm{p}(1)}_{i}(\omega)
   + \mathcal{J}^{\mathrm{d}(1)}_{i}(\omega), \label{eq:J1p+J1d}
  \\
  \mathcal{J}^{\mathrm{p}(1)}_{i}(\omega)
  &= ie\hbar
     \sum_{\bm{k}}
     \mathrm{Tr}\!\left[
       V_{i}(\bm{k})\,G^{(1)<}_{\bm{k}}(\omega)
     \right],
  \label{eq:Jp1-def}
  \\
  \mathcal{J}^{\mathrm{d}(1)}_{i}(\omega)
  &= ie^{2}\hbar
     \sum_{\bm{k}}\sum_{j}
     \int_{-\infty}^{\infty}\frac{d\omega'}{2\pi}\,
     \mathrm{Tr}\!\left[
       V_{ij}(\bm{k})\,G^{(0)<}_{\bm{k}}(\omega')
     \right]
     A_{j}(\omega-\omega'),
  \label{eq:Jd1-def}
\end{align}
consists of the paramagnetic contribution
$\mathcal{J}^{\mathrm{p}(1)}_{i}$ from the one-photon
current vertex $V_{i}$, and the diamagnetic contribution
$\mathcal{J}^{\mathrm{d}(1)}_{i}$ from the two-photon
contact vertex $V_{ij}$, and the second-order current,
\begin{align}
  \mathcal{J}^{(2)}_{i}(\omega)
  &= \mathcal{J}^{\mathrm{p}(2a)}_{i}(\omega)
   + \mathcal{J}^{\mathrm{p}(2b)}_{i}(\omega)
   + \mathcal{J}^{\mathrm{d}(2)}_{i}(\omega)
   + \mathcal{J}^{\mathrm{t}(2)}_{i}(\omega),
  \\
  \mathcal{J}^{\mathrm{p}(2a)}_{i}(\omega)
  &= ie\hbar
     \sum_{\bm{k}}
     \mathrm{Tr}\!\left[
       V_{i}(\bm{k})\,G^{(2a)<}_{\bm{k}}(\omega)
     \right],
  \label{eq:Jp2a-def}
  \\
  \mathcal{J}^{\mathrm{p}(2b)}_{i}(\omega)
  &= ie\hbar
     \sum_{\bm{k}}
     \mathrm{Tr}\!\left[
       V_{i}(\bm{k})\,G^{(2b)<}_{\bm{k}}(\omega)
     \right],
  \label{eq:Jp2b-def}
  \\
  \mathcal{J}^{\mathrm{d}(2)}_{i}(\omega)
  &= ie^{2}\hbar
     \sum_{\bm{k}}\sum_{j}
     \int_{-\infty}^{\infty}\frac{d\omega'}{2\pi}\,
     \mathrm{Tr}\!\left[
       V_{ij}(\bm{k})\,G^{(1)<}_{\bm{k}}(\omega')
     \right]
     A_{j}(\omega-\omega'),
  \label{eq:Jd2-def}
  \\
  \mathcal{J}^{\mathrm{t}(2)}_{i}(\omega)
  &= \frac{ie^{3}\hbar}{2}
     \sum_{\bm{k}}\sum_{jk}
     \int_{-\infty}^{\infty}\frac{d\omega'}{2\pi}
     \int_{-\infty}^{\infty}\frac{d\omega''}{2\pi}\,
     \mathrm{Tr}\!\left[
       V_{ijk}(\bm{k})\,G^{(0)<}_{\bm{k}}(\omega')
     \right]
     A_{j}(\omega'')\,A_{k}(\omega-\omega'-\omega''),
  \label{eq:Jt2-def}
\end{align}
consists of the paramagnetic contributions
$\mathcal{J}^{\mathrm{p}(2a)}_{i}$ and
$\mathcal{J}^{\mathrm{p}(2b)}_{i}$ from the one-photon
current vertex $V_{i}$ combined with the second-order
Green's functions $G^{(2a)<}$ and $G^{(2b)<}$,
respectively, the diamagnetic contribution
$\mathcal{J}^{\mathrm{d}(2)}_{i}$ from the two-photon
contact vertex $V_{ij}$ combined with the first-order
Green's function $G^{(1)<}$, and the tadpole contribution
$\mathcal{J}^{\mathrm{t}(2)}_{i}$ from the third-rank
contact vertex $V_{ijk}$ combined with the equilibrium
Green's function $G^{(0)<}$.

\subsection{Zeroth order: equilibrium current}
\begin{figure}[!htbp]
  \centering
  \includegraphics[width=0.98\textwidth]{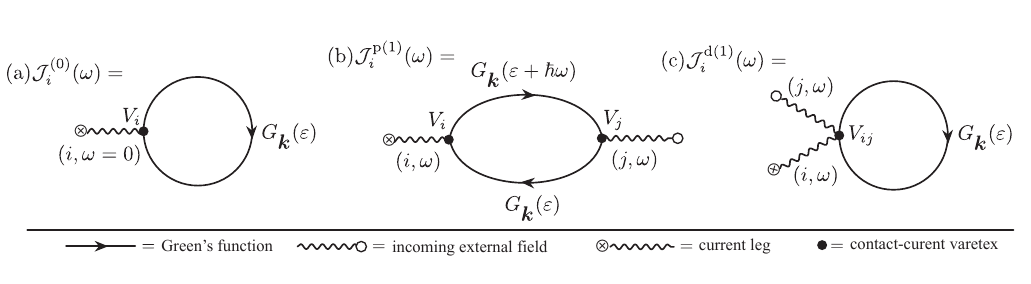}
  \caption{Diagrammatic representation of the zeroth- and first-order current contributions. Panel (a) shows the equilibrium current $\mathcal{J}_{i}^{(0)}$, panel (b) the paramagnetic linear current $\mathcal{J}_{i}^{\mathrm{p}(1)}$, and panel (c) the diamagnetic/contact linear current $\mathcal{J}_{i}^{\mathrm{d}(1)}$. The notation for propagators, external fields, and current legs is the same as in Fig.~\ref{fig:second_order_current_diagrams}.}
  \label{fig:zeroth_first_order_current_diagrams}
\end{figure}

The zeroth-order contribution arises from the unperturbed
lesser Green's function $G^{(0)<}_{\bm{k}}(\omega)$ and
represents the equilibrium current in the absence of the
external field.
Substituting Eq.~\eqref{eq:G0-omega} into Eq.~\eqref{eq:J0-def},
one obtains
\begin{align}
  \mathcal{J}^{(0)}_{i}(\omega)
  &= ie\hbar\sum_{\bm{k}}
     \mathrm{Tr}\!\left[
       V_{i}(\bm{k})\,G^{(0)<}_{\bm{k}}(\omega)
     \right]
  \nonumber\\
  &= -2\pi e\,\delta(\omega)
     \sum_{\bm{k}}
     \int_{-\infty}^{\infty}
     \frac{d\varepsilon}{2\pi i}\,
     \mathrm{Tr}\!\left[
       V_{i}(\bm{k})\,G^{<}_{\bm{k}}(\varepsilon)
     \right].
  \label{eq:J0}
\end{align}
The factor $\delta(\omega)$ reflects the fact that, in thermal
equilibrium, the current is time-independent and its Fourier
transform is concentrated at $\omega=0$.
For a system with time-reversal or inversion symmetry,
$J^{(0)}_{i}(\omega)$ vanishes identically, consistent with the
absence of a spontaneous equilibrium current. 
The diagrammatic representation of $J^{(0)}_{i}(\omega)$
is shown in Fig.~\ref{fig:zeroth_first_order_current_diagrams}.

\subsection{First order: linear nonequilibrium current}
The first-order current is the linear response to the
external field and decomposes into the paramagnetic and
diamagnetic contributions [Eq.~\eqref{eq:J1p+J1d}].
The paramagnetic contribution arises from the first-order
correction $G^{(1)<}_{\bm{k}}(\omega)$
[Eq.~\eqref{eq:G1-freq}] to the lesser Green's function
via the one-photon current vertex $V_{i}$
[Eq.~\eqref{eq:Jp1-def}; Fig.~\ref{fig:zeroth_first_order_current_diagrams}(b)],
\begin{align}
  \mathcal{J}^{\mathrm{p}(1)}_{i}(\omega)
  &= ie\hbar\sum_{\bm{k}}
     \mathrm{Tr}\!\left[
       V_{i}(\bm{k})\,G^{(1)<}_{\bm{k}}(\omega)
     \right]
  \nonumber\\
  &= \frac{e^{2}}{\omega}\sum_{j}\sum_{\bm{k}}
     \int_{-\infty}^{\infty}\frac{d\varepsilon}{2\pi}\,
     \mathrm{Tr}\!\Big[
       V_{i}(\bm{k})\,G_{\bm{k}}(\varepsilon+\hbar\omega)\,
       V_{j}(\bm{k})\,G_{\bm{k}}(\varepsilon)
     \Big]^{<}
     E_{j}(\omega),
  \label{eq:Jp1}
\end{align}
where the relation $A_{j}(\omega) = E_{j}(\omega)/(i\omega)$
has been used to convert from the vector potential to the
electric field. 
The diamagnetic contribution originates from the two-photon
contact vertex $V_{ij}$ acting on the equilibrium lesser
Green's function $G^{(0)<}_{\bm{k}}(\varepsilon)$
[Eq.~\eqref{eq:G0-omega};
Fig.~\ref{fig:zeroth_first_order_current_diagrams}(c)],
\begin{align}
  \mathcal{J}^{\mathrm{d}(1)}_{i}(\omega)
  &= -e^{2}\sum_{j}\sum_{\bm{k}}
     \int_{-\infty}^{\infty}\frac{d\varepsilon}{2\pi i}\,
     \mathrm{Tr}\!\Big[
       V_{ij}(\bm{k})\,G^{<}_{\bm{k}}(\varepsilon)
     \Big]A_{j}(\omega)
  \nonumber\\
  &= \frac{e^{2}}{\omega}\sum_{j}\sum_{\bm{k}}
     \int_{-\infty}^{\infty}\frac{d\varepsilon}{2\pi}\,
     \mathrm{Tr}\!\Big[
       V_{ij}(\bm{k})\,G^{<}_{\bm{k}}(\varepsilon)
     \Big]E_{j}(\omega).
  \label{eq:Jd1}
\end{align}
Combining Eqs.~\eqref{eq:Jp1} and \eqref{eq:Jd1},
the total first-order current defines the linear optical
conductivity $\sigma_{ij}(\omega)$ through
$J^{(1)}_{i}(\omega) = \sigma_{ij}(\omega)E_{j}(\omega)$,
giving
\begin{align}
  \sigma_{ij}(\omega)
  = e^2\sum_{\bm{k}}
    \int_{-\infty}^{\infty}\frac{d\varepsilon}{2\pi}\,
    \frac{1}{\omega}\mathrm{Tr}\!\Big[
      V_{i}(\bm{k})\,G_{\bm{k}}(\varepsilon+\hbar\omega)\,
      V_{j}(\bm{k})\,G_{\bm{k}}(\varepsilon)
      + V_{ij}(\bm{k})\,G_{\bm{k}}(\varepsilon)
    \Big]^{<}.
  \label{eq:sigma1}
\end{align}
Applying the Langreth rules to decompose the lesser
component of the product of two contour-ordered Green's
functions,
\begin{align}
  \bigl[
    G_{\bm{k}}(\varepsilon+\hbar\omega)\,
    G_{\bm{k}}(\varepsilon)
  \bigr]^{<}
  &= f(\varepsilon)\,
     G^{\mathrm{R}}_{\bm{k}}(\varepsilon+\hbar\omega)\,
     \Lambda_{\bm{k}}(\varepsilon)
   + f(\varepsilon+\hbar\omega)\,
     \Lambda_{\bm{k}}(\varepsilon+\hbar\omega)\,
     G^{\mathrm{A}}_{\bm{k}}(\varepsilon),
  \label{eq:Langreth-2G-linear}
\end{align}
where
\begin{align}
  \Lambda_{\bm{k}}(\varepsilon)
  \equiv G^{\mathrm{A}}_{\bm{k}}(\varepsilon)
       - G^{\mathrm{R}}_{\bm{k}}(\varepsilon)
  = -2i\,\mathrm{Im}\,G^{\mathrm{R}}_{\bm{k}}(\varepsilon)
  \label{eq:Lambda-linear}
\end{align}
is the spectral function,
and shifting the internal energy variable $\varepsilon$
so that both Fermi functions carry the same argument
$f(\varepsilon)$, the linear conductivity
Eq.~\eqref{eq:sigma1} becomes
\begin{align}
  \sigma_{ij}(\omega)
  &= \frac{e^{2}}{\omega}
     \sum_{\bm{k}}
     \int_{-\infty}^{\infty}\frac{d\varepsilon}{2\pi}\,
     f(\varepsilon)
     \mathrm{Tr}\!\Big[
       V_{i}G^{\mathrm{R}}(\varepsilon+\hbar\omega)\,
       V_{j}\Lambda(\varepsilon)
     + V_{i}\Lambda(\varepsilon)\,
       V_{j}G^{\mathrm{A}}(\varepsilon-\hbar\omega)
     + V_{ij}\Lambda(\varepsilon)
     \Big],
  \label{eq:sigma1-finite-frequency}
\end{align}
where the crystal-momentum argument $\bm{k}$ is suppressed
for brevity and all matrices are evaluated at the same
$\bm{k}$.

\subsection{Second-order nonequilibrium current}

\begin{figure}[!htbp]
  \centering
  \includegraphics[width=0.98\textwidth]{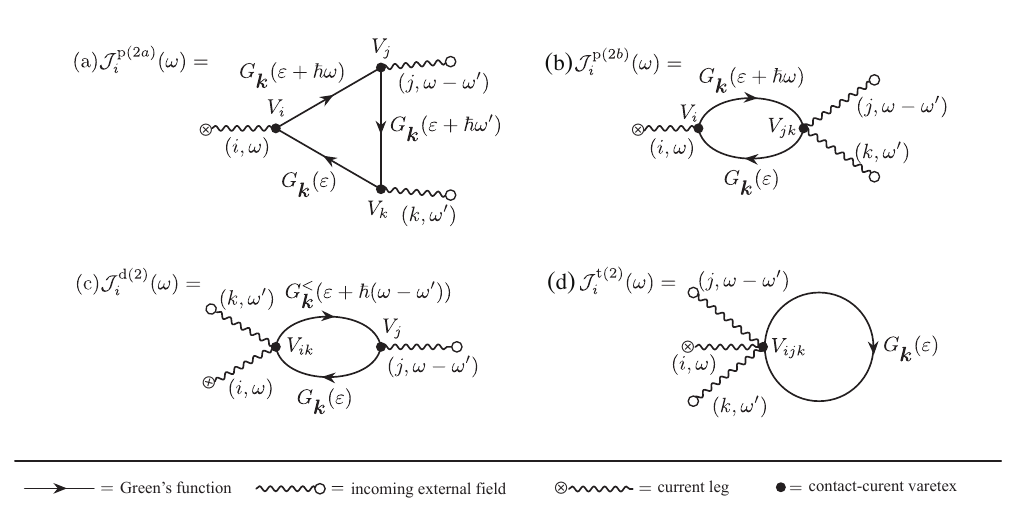}
  \caption{Diagrammatic representation of the four
    second-order current contributions:
    (a)~$\mathcal{J}_{i}^{\mathrm{p}(2a)}(\omega)$,
    (b)~$\mathcal{J}_{i}^{\mathrm{p}(2b)}(\omega)$,
    (c)~$\mathcal{J}_{i}^{\mathrm{d}(2)}(\omega)$, and
    (d)~$\mathcal{J}_{i}^{\mathrm{t}(2)}(\omega)$.
    The graphical conventions match
    Fig.~\ref{fig:zeroth_first_order_current_diagrams}.}
  \label{fig:second_order_current_diagrams}
\end{figure}

The Fourier component of the second-order current
$\mathcal{J}^{(2)}_{i}(\omega)$ decomposes as
\begin{align}
  \mathcal{J}^{(2)}_{i}(\omega)
  = \mathcal{J}^{\mathrm{p}(2a)}_{i}(\omega)
  + \mathcal{J}^{\mathrm{p}(2b)}_{i}(\omega)
  + \mathcal{J}^{\mathrm{d}(2)}_{i}(\omega)
  + \mathcal{J}^{\mathrm{t}(2)}_{i}(\omega),
  \label{eq:J2-decomp}
\end{align}
where the four contributions correspond, respectively, to
the paramagnetic current from $G^{(2a)<}(\omega)$
[Eq.~\eqref{eq:G2a-freq}], the paramagnetic current from
$G^{(2b)<}(\omega)$ [Eq.~\eqref{eq:G2b-freq}], the
diamagnetic current from $G^{(1)<}(\omega)$
[Eq.~\eqref{eq:G1-freq}], and the contact (tadpole)
current from $G^{(0)<}(\omega)$
[Eq.~\eqref{eq:G0-omega}].
Using $A_{j}(\omega) = E_{j}(\omega)/(i\omega)$, each
contribution reads
\begin{align}
  \mathcal{J}^{\mathrm{p}(2a)}_{i}(\omega)
  &= ie\hbar\sum_{\bm{k}}
     \mathrm{Tr}\!\left[
       V_{i}(\bm{k})\,G^{(2a)<}_{\bm{k}}(\omega)
     \right]
  \nonumber\\
  &= -ie^{3}\sum_{jk}\sum_{\bm{k}}
     \int_{-\infty}^{\infty}\frac{d\varepsilon}{2\pi}
     \int_{-\infty}^{\infty}\frac{d\omega'}{2\pi}\,
     \frac{1}{(\omega-\omega')\omega'}
  \nonumber\\
  &\quad\times
     \mathrm{Tr}\!\Big[
       V_{i}(\bm{k})\,
       G_{\bm{k}}(\varepsilon+\hbar\omega)\,
       V_{j}(\bm{k})\,
       G_{\bm{k}}(\varepsilon+\hbar\omega')\,
       V_{k}(\bm{k})\,
       G_{\bm{k}}(\varepsilon)
     \Big]^{<}
     E_{j}(\omega-\omega')\,E_{k}(\omega'),
  \label{eq:Jp2a}
  \\[4pt]
  \mathcal{J}^{\mathrm{p}(2b)}_{i}(\omega)
  &= ie\hbar\sum_{\bm{k}}
     \mathrm{Tr}\!\left[
       V_{i}(\bm{k})\,G^{(2b)<}_{\bm{k}}(\omega)
     \right]
  \nonumber\\
  &= -i\frac{e^{3}}{2}\sum_{jk}\sum_{\bm{k}}
     \int_{-\infty}^{\infty}\frac{d\varepsilon}{2\pi}
     \int_{-\infty}^{\infty}\frac{d\omega'}{2\pi}\,
     \frac{1}{(\omega-\omega')\omega'}
  \nonumber\\
  &\quad\times
     \mathrm{Tr}\!\Big[
       V_{i}(\bm{k})\,
       G_{\bm{k}}(\varepsilon+\hbar\omega)\,
       V_{jk}(\bm{k})\,
       G_{\bm{k}}(\varepsilon)
     \Big]^{<}
     E_{j}(\omega-\omega')\,E_{k}(\omega'),
  \label{eq:Jp2b}
  \\[4pt]
  \mathcal{J}^{\mathrm{d}(2)}_{i}(\omega)
  &= ie^{2}\hbar\sum_{j}\sum_{\bm{k}}
     \int_{-\infty}^{\infty}\frac{d\omega'}{2\pi}\,
     \mathrm{Tr}\!\left[
       V_{ij}(\bm{k})\,G^{(1)<}_{\bm{k}}(\omega')
     \right]A_{j}(\omega-\omega')
  \nonumber\\
  &= -ie^{3}\sum_{jk}\sum_{\bm{k}}
     \int_{-\infty}^{\infty}\frac{d\varepsilon}{2\pi}
     \int_{-\infty}^{\infty}\frac{d\omega'}{2\pi}\,
     \frac{E_{j}(\omega-\omega')\,E_{k}(\omega')}
          {(\omega-\omega')\omega'}
  \nonumber\\
  &\quad\times
     \mathrm{Tr}\!\left[
       V_{ij}(\bm{k})\,
       G_{\bm{k}}(\varepsilon+\hbar\omega')\,
       V_{k}(\bm{k})\,
       G_{\bm{k}}(\varepsilon)
     \right]^{<},
  \label{eq:Jd2}
  \\[6pt]
  \mathcal{J}^{\mathrm{t}(2)}_{i}(\omega)
  &= \frac{ie^{3}\hbar}{2}\sum_{jk}\sum_{\bm{k}}
     \int_{-\infty}^{\infty}\frac{d\omega'}{2\pi}
     \int_{-\infty}^{\infty}\frac{d\omega''}{2\pi}\,
     \mathrm{Tr}\!\left[
       V_{ijk}(\bm{k})\,G^{(0)<}_{\bm{k}}(\omega')
     \right]A_{j}(\omega'')A_{k}(\omega-\omega'-\omega'')
  \nonumber\\
  &= -i\frac{e^{3}}{2}\sum_{jk}\sum_{\bm{k}}
     \int_{-\infty}^{\infty}\frac{d\varepsilon}{2\pi}
     \int_{-\infty}^{\infty}\frac{d\omega'}{2\pi}\,
     \frac{E_{j}(\omega-\omega')\,E_{k}(\omega')}
          {(\omega-\omega')\omega'}
     \mathrm{Tr}\!\left[
       V_{ijk}(\bm{k})\,G^{<}_{\bm{k}}(\varepsilon)
     \right].
  \label{eq:Jt2}
\end{align}
The diagrammatic representation of these four contributions
is shown in Fig.~\ref{fig:second_order_current_diagrams}.
The one-photon-vertex diagram
[Fig.~\ref{fig:second_order_current_diagrams}(a)]
involves a triangular electron loop with three velocity
vertices, while panel~(b) shows a bubble diagram with
the two-photon contact vertex $V_{jk}$.
The contact diagrams
[Figs.~\ref{fig:second_order_current_diagrams}(c) and (d)]
contain the two-photon vertex $V_{ij}$ or the three-photon
vertex $V_{ijk}$ at a single node.

\FloatBarrier

Substituting $\omega_{1} = \omega - \omega'$ and
$\omega_{2} = \omega'$, the second-order current takes
the standard convolution form
\begin{align}
  \mathcal{J}_{i}^{(2)}(\omega)
  = \sum_{jk}
    \int_{-\infty}^{\infty}\frac{d\omega_{1}}{2\pi}
    \int_{-\infty}^{\infty}\frac{d\omega_{2}}{2\pi}\,
    2\pi\delta(\omega-\omega_{1}-\omega_{2})\,
    \sigma_{ijk}(\omega_{1},\omega_{2})\,
    E_{j}(\omega_{1})\,E_{k}(\omega_{2}),
  \label{eq:J2-sigma}
\end{align}
where $\sigma_{ijk}(\omega_{1},\omega_{2})$ is the
second-order conductivity tensor and
$\omega_{12} \equiv \omega_{1}+\omega_{2}$.
Collecting all four contributions
Eqs.~\eqref{eq:Jp2a-def}--\eqref{eq:Jt2-def},
$\sigma_{ijk}(\omega_{1},\omega_{2})$ is given by
\begin{align}
  \sigma_{ijk}(\omega_{1},\omega_{2})
  &= -\frac{ie^{3}}{2}
     \sum_{\bm{k}}
     \int_{-\infty}^{\infty}\frac{d\varepsilon}{2\pi}\,
     \frac{1}{\omega_{1}\omega_{2}}
  \bigg\{
  \nonumber\\
  &\quad
     \mathrm{Tr}\!\Big[
       V_{i}\,G_{\bm{k}}(\varepsilon+\hbar\omega_{12})\,
       V_{j}\,G_{\bm{k}}(\varepsilon+\hbar\omega_{2})\,
       V_{k}\,G_{\bm{k}}(\varepsilon)
     \Big]^{<}
  +  \mathrm{Tr}\!\Big[
       V_{i}\,G_{\bm{k}}(\varepsilon+\hbar\omega_{12})\,
       V_{k}\,G_{\bm{k}}(\varepsilon+\hbar\omega_{1})\,
       V_{j}\,G_{\bm{k}}(\varepsilon)
     \Big]^{<}
  \nonumber\\
  &\quad
  +  \mathrm{Tr}\!\Big[
       V_{i}\,G_{\bm{k}}(\varepsilon+\hbar\omega_{12})\,
       V_{jk}\,G_{\bm{k}}(\varepsilon)
     \Big]^{<}
  +  \mathrm{Tr}\!\Big[
       V_{ij}\,G_{\bm{k}}(\varepsilon+\hbar\omega_{2})\,
       V_{k}\,G_{\bm{k}}(\varepsilon)
     \Big]^{<}
  +  \mathrm{Tr}\!\Big[
       V_{ik}\,G_{\bm{k}}(\varepsilon+\hbar\omega_{1})\,
       V_{j}\,G_{\bm{k}}(\varepsilon)
     \Big]^{<}
  \nonumber\\
  &\quad
  +  \mathrm{Tr}\!\left[
       V_{ijk}\,G^{<}_{\bm{k}}(\varepsilon)
     \right]
  \bigg\}.
  \label{eq:sigma2-full}
\end{align}
The overall factor of $1/2$ ensures that
$\sigma_{ijk}(\omega_{1},\omega_{2})$ is symmetric under
the simultaneous interchange
$(j,\omega_{1})\leftrightarrow(k,\omega_{2})$, as required
by the permutation symmetry of the second-order response
tensor~\cite{Boyd2008}.
The first and second traces in Eq.~\eqref{eq:sigma2-full}
map onto each other under this interchange, as do the
fourth and fifth, while the third and sixth are
individually symmetric.

Applying the Langreth rules to decompose the lesser
component of a product of three contour-ordered Green's
functions,
\begin{align}
  &\Big[
    G_{\bm{k}}(\varepsilon+\hbar\omega_{12})\,
    G_{\bm{k}}(\varepsilon+\hbar\omega_{2})\,
    G_{\bm{k}}(\varepsilon)
  \Big]^{<}
  \nonumber\\
  &\quad
  = f(\varepsilon)\,
    G^{\mathrm{R}}_{\bm{k}}(\varepsilon+\hbar\omega_{12})\,
    G^{\mathrm{R}}_{\bm{k}}(\varepsilon+\hbar\omega_{2})\,
    \Lambda_{\bm{k}}(\varepsilon)
  \nonumber\\
  &\quad
  + f(\varepsilon+\hbar\omega_{2})\,
    G^{\mathrm{R}}_{\bm{k}}(\varepsilon+\hbar\omega_{12})\,
    \Lambda_{\bm{k}}(\varepsilon+\hbar\omega_{2})\,
    G^{\mathrm{A}}_{\bm{k}}(\varepsilon)
  \nonumber\\
  &\quad
  + f(\varepsilon+\hbar\omega_{12})\,
    \Lambda_{\bm{k}}(\varepsilon+\hbar\omega_{12})\,
    G^{\mathrm{A}}_{\bm{k}}(\varepsilon+\hbar\omega_{2})\,
    G^{\mathrm{A}}_{\bm{k}}(\varepsilon). 
  \label{eq:Langreth-3G}
\end{align}
Shifting the internal energy variable $\varepsilon$ in each
term so that all Fermi functions carry the same argument
$f(\varepsilon)$, one obtains the finite-frequency
second-order conductivity
\begin{align}
  \sigma_{ijk}(\omega_{1},\omega_{2})
  ={}&
  -\frac{ie^{3}}{2}
       \sum_{\bm{k}}
       \int_{-\infty}^{\infty}\frac{d\varepsilon}{2\pi}\,
       f(\varepsilon)\,\frac{1}{\omega_{1}\omega_{2}}
  \bigg\{
  \nonumber\\
  &\quad
     \mathrm{Tr}\!\big[
       V_{i}G^{\mathrm{R}}(\varepsilon+\hbar\omega_{12})\,
       V_{j}G^{\mathrm{R}}(\varepsilon+\hbar\omega_{2})\,
       V_{k}\Lambda(\varepsilon)
     \big]
  +  \mathrm{Tr}\!\big[
       V_{i}G^{\mathrm{R}}(\varepsilon+\hbar\omega_{12})\,
       V_{k}G^{\mathrm{R}}(\varepsilon+\hbar\omega_{1})\,
       V_{j}\Lambda(\varepsilon)
     \big]
  \nonumber\\
  &\quad
  +  \mathrm{Tr}\!\big[
       V_{i}G^{\mathrm{R}}(\varepsilon+\hbar\omega_{1})\,
       V_{j}\Lambda(\varepsilon)\,
       V_{k}G^{\mathrm{A}}(\varepsilon-\hbar\omega_{2})
     \big]
  +  \mathrm{Tr}\!\big[
       V_{i}G^{\mathrm{R}}(\varepsilon+\hbar\omega_{2})\,
       V_{k}\Lambda(\varepsilon)\,
       V_{j}G^{\mathrm{A}}(\varepsilon-\hbar\omega_{1})
     \big]
  \nonumber\\
  &\quad
  +  \mathrm{Tr}\!\big[
       V_{i}\Lambda(\varepsilon)\,
       V_{j}G^{\mathrm{A}}(\varepsilon-\hbar\omega_{1})\,
       V_{k}G^{\mathrm{A}}(\varepsilon-\hbar\omega_{12})
     \big]
  +  \mathrm{Tr}\!\big[
       V_{i}\Lambda(\varepsilon)\,
       V_{k}G^{\mathrm{A}}(\varepsilon-\hbar\omega_{2})\,
       V_{j}G^{\mathrm{A}}(\varepsilon-\hbar\omega_{12})
     \big]
  \nonumber\\
  &\quad
  +  \mathrm{Tr}\!\big[
       V_{i}G^{\mathrm{R}}(\varepsilon+\hbar\omega_{12})\,
       V_{jk}\Lambda(\varepsilon)
     \big]
  +  \mathrm{Tr}\!\big[
       V_{i}\Lambda(\varepsilon)\,
       V_{jk}G^{\mathrm{A}}(\varepsilon-\hbar\omega_{12})
     \big]
  \nonumber\\
  &\quad
  +  \mathrm{Tr}\!\big[
       V_{ij}G^{\mathrm{R}}(\varepsilon+\hbar\omega_{2})\,
       V_{k}\Lambda(\varepsilon)
     \big]
  +  \mathrm{Tr}\!\big[
       V_{ik}G^{\mathrm{R}}(\varepsilon+\hbar\omega_{1})\,
       V_{j}\Lambda(\varepsilon)
     \big]
  \nonumber\\
  &\quad
  +  \mathrm{Tr}\!\big[
       V_{ij}\Lambda(\varepsilon)\,
       V_{k}G^{\mathrm{A}}(\varepsilon-\hbar\omega_{2})
     \big]
  +  \mathrm{Tr}\!\big[
       V_{ik}\Lambda(\varepsilon)\,
       V_{j}G^{\mathrm{A}}(\varepsilon-\hbar\omega_{1})
     \big]
  \nonumber\\
  &\quad
  +  \mathrm{Tr}\!\big[
       V_{ijk}\,\Lambda(\varepsilon)
     \big]
  \bigg\},
  \label{eq:sigma2-finite-frequency}
\end{align}
Equation~\eqref{eq:sigma2-finite-frequency} is the
velocity-gauge second-order conductivity tensor at finite
frequency, valid prior to taking the dc limit.
The prefactor $1/(\omega_{1}\omega_{2})$ originates from
the conversion $A_{j}(\omega) = E_{j}(\omega)/(i\omega)$.
The apparent singularities at $\omega_{1}=0$ or
$\omega_{2}=0$ are unphysical: they cancel exactly
between the one-photon-vertex and contact-vertex
contributions once all terms in
Eq.~\eqref{eq:sigma2-finite-frequency} are combined,
as will be demonstrated explicitly in the dc limit below.

\section{Linear dc conductivity}
\label{sec:supp-dc-limit-linear}

In this section we derive the linear dc conductivity by extracting the zero-frequency limit
of $\sigma_{ij}(\omega)$ [Eq.~\eqref{eq:sigma1-finite-frequency}].

To extract the dc component of the linear conductivity
in the limit $\omega\to 0$, we expand the retarded and
advanced Green's functions to first order in $\omega$,
\begin{align}
  G^{\mathrm{R/A}}(\varepsilon+\hbar\omega)
  = G^{\mathrm{R/A}}(\varepsilon)
  + \hbar\omega\,\partial_{\varepsilon}G^{\mathrm{R/A}}(\varepsilon)
  + \mathcal{O}(\omega^{2}).
  \label{eq:GRA-expand-linear}
\end{align}
Substituting Eq.~\eqref{eq:GRA-expand-linear} into
Eq.~\eqref{eq:sigma1-finite-frequency} and collecting terms
in powers of $\omega$, one finds
\begin{align}
  \sigma_{ij}(\omega)
  = 
  e^2\frac{A_{0,ij}}{\omega}
  + \sigma^{\mathrm{DC}}_{ij}
  + \mathcal{O}(\omega),
  \label{eq:sigma1-expand}
\end{align}
where the zeroth-order coefficient is
\begin{align}
  A_{0,ij}
  &=\int_{-\infty}^{\infty}\frac{d\varepsilon}{2\pi}\,
     f(\varepsilon)\,\sum_{\bm{k}}
     \mathrm{Tr}\!\Big[
       V_{i}G^{\mathrm{R}}(\varepsilon)\,
       V_{j}\Lambda(\varepsilon)
     + V_{i}\Lambda(\varepsilon)\,
       V_{j}G^{\mathrm{A}}(\varepsilon)
     + V_{ij}\Lambda(\varepsilon)
     \Big]
  \nonumber\\
  &=       \int_{-\infty}^{\infty}\frac{d\varepsilon}{2\pi}\,
     f(\varepsilon)\,\sum_{\bm{k}}
     \mathrm{Tr}\!\Big[
       V_{i}G^{\mathrm{A}}(\varepsilon)\,V_{j}G^{\mathrm{A}}(\varepsilon)
     + V_{ij}G^{\mathrm{A}}(\varepsilon)
     - V_{i}G^{\mathrm{R}}(\varepsilon)\,V_{j}G^{\mathrm{R}}(\varepsilon)
     - V_{ij}G^{\mathrm{R}}(\varepsilon)
     \Big].
  \label{eq:A0ij}
\end{align}
This coefficient is a spurious $1/\omega$ pole that must
vanish identically by gauge invariance.
To show this, we use the Ward--Takahashi identity
\begin{align}
  \partial_{i}G^{\mathrm{R/A}}_{\bm{k}}(\varepsilon)
  = G^{\mathrm{R/A}}_{\bm{k}}(\varepsilon)\,
    \hbar V_{i}(\bm{k})\,
    G^{\mathrm{R/A}}_{\bm{k}}(\varepsilon),
  \label{eq:WT-linear}
\end{align}
which follows from differentiating
$G^{\mathrm{R/A}}(\varepsilon - H(\bm{k}))^{-1} = 1$
with respect to $k_{i}$.
Using Eq.~\eqref{eq:WT-linear} and
$\partial_{i}V_{j} = \hbar V_{ij}$, one recognizes 
$A_{0,ij}$ as a total crystal-momentum
derivative,
\begin{align}
\sum_{\bm k}  \mathrm{Tr}\!\Big[
    V_{i}G^{\mathrm{R/A}}\,V_{j}G^{\mathrm{R/A}}
  + V_{ij}G^{\mathrm{R/A}}
  \Big]
  = \sum_{\bm k} \mathrm{Tr}\!\Big[
    \partial_{i}\!\left(\hbar^{-1}V_{j}G^{\mathrm{R/A}}\right)
  \Big]
  =\hbar^{-1}\sum_{\bm k} \partial_{i}\,\mathrm{Tr}\!\Big[
    V_{j}G^{\mathrm{R/A}}
  \Big],
  \label{eq:A0-total-deriv}
\end{align}
whose integral over the Brillouin zone vanishes by the
periodicity of the Bloch functions.
Therefore $A_{0,ij} = 0$, and no $1/\omega$ divergence
appears in the dc limit. 

The finite dc linear conductivity is the coefficient of
$\omega^{0}$ in Eq.~\eqref{eq:sigma1-expand},
\begin{align}
  \sigma^{\mathrm{DC}}_{ij}
  &= e^{2}\hbar
     \sum_{\bm{k}}
     \int_{-\infty}^{\infty}\frac{d\varepsilon}{2\pi}\,
     f(\varepsilon)\,
     \mathrm{Tr}\!\Big[
       V_{i}\,\partial_{\varepsilon}G^{\mathrm{R}}\,V_{j}\,\Lambda
     - V_{i}\,\Lambda\,V_{j}\,\partial_{\varepsilon}G^{\mathrm{A}}
     \Big]\nonumber\\
   &=  e^2\hbar\sum_{\bm{k}}
     \int_{-\infty}^{\infty}\frac{d\varepsilon}{2\pi}\,
     f(\varepsilon)\,
     \mathrm{Tr}\!\Big[
       V_{i}\,\partial_{\varepsilon}G^{\mathrm{R}}\,V_{j}\,G^{\mathrm{A}}
     + V_{i}\,G^{\mathrm{R}}\,V_{j}\,\partial_{\varepsilon}G^{\mathrm{A}}
     - V_{i}\,\partial_{\varepsilon}G^{\mathrm{R}}\,V_{j}\,G^{\mathrm{R}}
     - V_{i}\,G^{\mathrm{A}}\,V_{j}\,\partial_{\varepsilon}G^{\mathrm{A}}
    \Big], 
  \label{eq:sigma-dc-linear-kubo}
\end{align}
which is the Kubo--Greenwood formula for the linear dc
conductivity~\cite{Kubo1957, Greenwood1958}.

\section{Second-order dc conductivity}
\label{sec:supp-dc-limit-second-order}

In this section we derive the second-order dc
conductivity by extracting the zero-frequency limits
of $\sigma_{ijk}(\omega_{1},\omega_{2})$ [Eq.~\eqref{eq:sigma2-finite-frequency}].

To extract the dc component of the second-order conductivity
in the limit $\omega_{1},\omega_{2}\to 0$, we expand the
Green's functions to second order in $\omega_{1}$ and
$\omega_{2}$, and write Eq.~\eqref{eq:sigma2-finite-frequency}
in the form
\begin{align}  
\sigma_{ijk}(\omega_{1},\omega_{2})
  =\frac{-ie^3}{2\omega_{1}\omega_{2}}\bigg[
  A_{0,ijk}
  + \hbar A^{(1)}_{1,ijk}\omega_{1}
        + \hbar A^{(2)}_{1,ijk}\omega_{2}
  + \hbar^{2} A^{(1)}_{2,ijk}\omega_{1}^{2}
        + \hbar^{2} A^{(2)}_{2,ijk}\omega_{2}^{2}\bigg]
  + \sigma^{\mathrm{DC}}_{ijk}.
  \label{eq:sigma2-dc-expansion}
\end{align}
Here $A_{0,ijk}$, $A^{(1)}_{n,ijk}$, and $A^{(2)}_{n,ijk}$
($n=1,2$) are spurious coefficients that must vanish
identically, as in the linear-response case, by virtue of
gauge invariance and the contact-vertex cancellation.
The dc second-order conductivity $\sigma^{\mathrm{DC}}_{ijk}$
is the finite remainder that survives in the limit
$\omega_{1},\omega_{2}\to 0$ after these cancellations
have taken place.

Specifically, $A_{0,ijk}$ is proportional to the zeroth-order
term in the frequency expansion, which vanishes by the same
total-momentum-derivative argument as in the linear case
[Eq.~\eqref{eq:A0-total-deriv}].
The terms linear in $\omega_{1}$ or $\omega_{2}$ in the
numerator, collected in $A^{(1)}_{1,ijk}$ and
$A^{(2)}_{1,ijk}$, cancel between the paramagnetic and
contact-vertex contributions by gauge invariance.
Similarly, $A^{(1)}_{2,ijk}$ and $A^{(2)}_{2,ijk}$ collect
the subleading terms that also cancel, leaving
$\sigma^{\mathrm{DC}}_{ijk}$ as the only surviving
contribution.
To verify this explicitly, we expand each trace in
Eq.~\eqref{eq:sigma2-finite-frequency} to second order in
$\omega_{1}$ and $\omega_{2}$.
The resulting coefficients are collected compactly below.

\subsection{Frequency expansion of the finite-frequency second-order conductivity}
Rather than displaying the frequency expansion of each of the thirteen
traces separately, we collect all contributions according to their powers
of $\omega_{1}$ and $\omega_{2}$.  We use
$\omega_{12}=\omega_{1}+\omega_{2}$ and
\begin{align}
  G^{\mathrm{R/A}}(\varepsilon+\hbar\omega)
  ={}& G^{\mathrm{R/A}}(\varepsilon)
  + \hbar\omega\,\partial_{\varepsilon}G^{\mathrm{R/A}}(\varepsilon)
  + \frac{\hbar^{2}\omega^{2}}{2}\,
    \partial_{\varepsilon}^{2}G^{\mathrm{R/A}}(\varepsilon)
  + \mathcal{O}(\omega^{3}).
  \label{eq:GRA-expand-second}
\end{align}

We write the finite-frequency conductivity as
\begin{align}
  \sigma_{ijk}(\omega_{1},\omega_{2})
  = -\frac{i e^{3}}{2\omega_{1}\omega_{2}}\,
    \mathcal{S}_{ijk}(\omega_{1},\omega_{2}),
  \label{eq:sigma-S-definition}
\end{align}
where
\begin{align}
\mathcal{S}_{ijk}(\omega_{1},\omega_{2})
={}& \int_{-\infty}^{\infty}\frac{d\varepsilon}{2\pi}\,
 f(\varepsilon)\sum_{\bm{k}}\operatorname{Tr}\bigg[
 \nonumber\\
& \quad V_{i}G^{\mathrm{R}}(\varepsilon+\hbar\omega_{12})
  V_{j}G^{\mathrm{R}}(\varepsilon+\hbar\omega_{2})
  V_{k}\Lambda(\varepsilon)
+V_{i}G^{\mathrm{R}}(\varepsilon+\hbar\omega_{12})
  V_{k}G^{\mathrm{R}}(\varepsilon+\hbar\omega_{1})
  V_{j}\Lambda(\varepsilon)
\nonumber\\
&+V_{i}G^{\mathrm{R}}(\varepsilon+\hbar\omega_{1})
  V_{j}\Lambda(\varepsilon)
  V_{k}G^{\mathrm{A}}(\varepsilon-\hbar\omega_{2})
  +V_{i}G^{\mathrm{R}}(\varepsilon+\hbar\omega_{2})
  V_{k}\Lambda(\varepsilon)
  V_{j}G^{\mathrm{A}}(\varepsilon-\hbar\omega_{1})
\nonumber\\
&+V_{i}\Lambda(\varepsilon)
  V_{j}G^{\mathrm{A}}(\varepsilon-\hbar\omega_{1})
  V_{k}G^{\mathrm{A}}(\varepsilon-\hbar\omega_{12})
+V_{i}\Lambda(\varepsilon)
  V_{k}G^{\mathrm{A}}(\varepsilon-\hbar\omega_{2})
  V_{j}G^{\mathrm{A}}(\varepsilon-\hbar\omega_{12})
\nonumber\\
&+V_{i}G^{\mathrm{R}}(\varepsilon+\hbar\omega_{12})
  V_{jk}\Lambda(\varepsilon)
+V_{i}\Lambda(\varepsilon)
  V_{jk}G^{\mathrm{A}}(\varepsilon-\hbar\omega_{12})
\nonumber\\
&+V_{ij}G^{\mathrm{R}}(\varepsilon+\hbar\omega_{2})
  V_{k}\Lambda(\varepsilon)
+V_{ik}G^{\mathrm{R}}(\varepsilon+\hbar\omega_{1})
  V_{j}\Lambda(\varepsilon)
\nonumber\\
&+V_{ij}\Lambda(\varepsilon)
  V_{k}G^{\mathrm{A}}(\varepsilon-\hbar\omega_{2})
+V_{ik}\Lambda(\varepsilon)
  V_{j}G^{\mathrm{A}}(\varepsilon-\hbar\omega_{1})
\nonumber\\
&+V_{ijk}\Lambda(\varepsilon)
\bigg].
\label{eq:Sijk-finite-frequency}
\end{align}
In the following, the common energy argument $\varepsilon$ of
$G^{\mathrm{R/A}}$, its energy derivatives, and $\Lambda$ is suppressed.
The expansion of $\mathcal{S}_{ijk}$ is written as
\begin{align}
\mathcal{S}_{ijk}(\omega_{1},\omega_{2})
={}& A_{0,ijk}
+\hbar\omega_{1}A^{(1)}_{1,ijk}
+\hbar\omega_{2}A^{(2)}_{1,ijk}
+\hbar^{2}\omega_{1}^{2}A^{(1)}_{2,ijk}
+\hbar^{2}\omega_{2}^{2}A^{(2)}_{2,ijk}
\nonumber\\
&+\hbar^{2}\omega_{1}\omega_{2}A_{12,ijk}
+\mathcal{O}(\omega^{3}).
\label{eq:Sijk-frequency-expansion}
\end{align}

The zeroth-order coefficient is
\begin{align}
A_{0,ijk}
={}&\mathcal{S}_{ijk}(0,0)
\nonumber\\
={}&\int_{-\infty}^{\infty}\frac{d\varepsilon}{2\pi}\,
 f(\varepsilon)\sum_{\bm{k}}\operatorname{Tr}\bigg[
\nonumber\\
&V_{i}G^{\mathrm{R}}V_{j}G^{\mathrm{R}}V_{k}\Lambda
+V_{i}G^{\mathrm{R}}V_{k}G^{\mathrm{R}}V_{j}\Lambda
+V_{i}G^{\mathrm{R}}V_{j}\Lambda V_{k}G^{\mathrm{A}}
+V_{i}G^{\mathrm{R}}V_{k}\Lambda V_{j}G^{\mathrm{A}}
+V_{i}\Lambda V_{j}G^{\mathrm{A}}V_{k}G^{\mathrm{A}}
+V_{i}\Lambda V_{k}G^{\mathrm{A}}V_{j}G^{\mathrm{A}}
\nonumber\\
&+V_{ij}G^{\mathrm{R}}V_{k}\Lambda
+V_{ik}G^{\mathrm{R}}V_{j}\Lambda
+V_{ij}\Lambda V_{k}G^{\mathrm{A}}
+V_{ik}\Lambda V_{j}G^{\mathrm{A}}
+V_{i}G^{\mathrm{R}}V_{jk}\Lambda
+V_{i}\Lambda V_{jk}G^{\mathrm{A}}
+V_{ijk}\Lambda
\bigg].
\label{eq:A0-compact}
\end{align}

The coefficients linear in frequency are
\begin{align}
A^{(1)}_{1,ijk}
\equiv{}&\frac{1}{\hbar}
\left.\partial_{\omega_{1}}\mathcal{S}_{ijk}(\omega_{1},\omega_{2})
\right|_{\omega_{1}=\omega_{2}=0}
\nonumber\\
={}&\int_{-\infty}^{\infty}\frac{d\varepsilon}{2\pi}\,
 f(\varepsilon)\sum_{\bm{k}}\operatorname{Tr}\bigg[
\nonumber\\
&V_{i}\partial_{\varepsilon}G^{\mathrm{R}}
 V_{j}G^{\mathrm{R}}V_{k}\Lambda
+V_{i}\partial_{\varepsilon}G^{\mathrm{R}}
 V_{j}\Lambda V_{k}G^{\mathrm{A}}
-V_{i}\Lambda V_{j}\partial_{\varepsilon}G^{\mathrm{A}}
 V_{k}G^{\mathrm{A}}
-V_{i}\Lambda V_{j}G^{\mathrm{A}}
 V_{k}\partial_{\varepsilon}G^{\mathrm{A}}
\nonumber\\
&+V_{i}\partial_{\varepsilon}G^{\mathrm{R}}
 V_{k}G^{\mathrm{R}}V_{j}\Lambda
+V_{i}G^{\mathrm{R}}V_{k}\partial_{\varepsilon}G^{\mathrm{R}}
 V_{j}\Lambda
-V_{i}G^{\mathrm{R}}V_{k}\Lambda
 V_{j}\partial_{\varepsilon}G^{\mathrm{A}}
-V_{i}\Lambda V_{k}G^{\mathrm{A}}
 V_{j}\partial_{\varepsilon}G^{\mathrm{A}}
\nonumber\\
&+V_{ik}\partial_{\varepsilon}G^{\mathrm{R}}V_{j}\Lambda
-V_{ik}\Lambda V_{j}\partial_{\varepsilon}G^{\mathrm{A}}
+V_{i}\partial_{\varepsilon}G^{\mathrm{R}}V_{jk}\Lambda
-V_{i}\Lambda V_{jk}\partial_{\varepsilon}G^{\mathrm{A}}
\bigg],
\label{eq:A11-compact}
\end{align}
\begin{align}
A^{(2)}_{1,ijk}
\equiv{}&\frac{1}{\hbar}
\left.\partial_{\omega_{2}}\mathcal{S}_{ijk}(\omega_{1},\omega_{2})
\right|_{\omega_{1}=\omega_{2}=0}
\nonumber\\
={}&\int_{-\infty}^{\infty}\frac{d\varepsilon}{2\pi}\,
 f(\varepsilon)\sum_{\bm{k}}\operatorname{Tr}\bigg[
\nonumber\\
&
V_{i}\partial_{\varepsilon}G^{\mathrm{R}}V_{j}G^{\mathrm{R}}V_{k}\Lambda
+V_{i}G^{\mathrm{R}}V_{j}\partial_{\varepsilon}G^{\mathrm{R}}V_{k}\Lambda
+V_{i}\partial_{\varepsilon}G^{\mathrm{R}}
 V_{k}G^{\mathrm{R}}V_{j}\Lambda
-V_{i}G^{\mathrm{R}}V_{j}\Lambda
 V_{k}\partial_{\varepsilon}G^{\mathrm{A}}
\nonumber\\&
+V_{i}\partial_{\varepsilon}G^{\mathrm{R}}
 V_{k}\Lambda V_{j}G^{\mathrm{A}}
-V_{i}\Lambda V_{j}G^{\mathrm{A}}
 V_{k}\partial_{\varepsilon}G^{\mathrm{A}}
-V_{i}\Lambda V_{k}\partial_{\varepsilon}G^{\mathrm{A}}
 V_{j}G^{\mathrm{A}}
-V_{i}\Lambda V_{k}G^{\mathrm{A}}
 V_{j}\partial_{\varepsilon}G^{\mathrm{A}}
\nonumber\\
&+V_{ij}\partial_{\varepsilon}G^{\mathrm{R}}V_{k}\Lambda
-V_{ij}\Lambda V_{k}\partial_{\varepsilon}G^{\mathrm{A}}
+V_{i}\partial_{\varepsilon}G^{\mathrm{R}}V_{jk}\Lambda
-V_{i}\Lambda V_{jk}\partial_{\varepsilon}G^{\mathrm{A}}
\bigg].
\label{eq:A12-compact}
\end{align}

For the quadratic coefficients, the common factor $1/2$ is pulled in front
of the integral.  Accordingly, the products of two first derivatives carry
a factor of $2$ inside the trace:
\begin{align}
A^{(1)}_{2,ijk}
\equiv{}&\frac{1}{2\hbar^{2}}
\left.\partial_{\omega_{1}}^{2}\mathcal{S}_{ijk}(\omega_{1},\omega_{2})
\right|_{\omega_{1}=\omega_{2}=0}
\nonumber\\
={}&\frac{1}{2}\int_{-\infty}^{\infty}\frac{d\varepsilon}{2\pi}\,
 f(\varepsilon)\sum_{\bm{k}}\operatorname{Tr}\bigg[
\nonumber\\
&V_{i}\partial_{\varepsilon}^{2}G^{\mathrm{R}}
 V_{j}G^{\mathrm{R}}V_{k}\Lambda
+V_{i}\partial_{\varepsilon}^{2}G^{\mathrm{R}}
 V_{k}G^{\mathrm{R}}V_{j}\Lambda
+2V_{i}\partial_{\varepsilon}G^{\mathrm{R}}
 V_{k}\partial_{\varepsilon}G^{\mathrm{R}}V_{j}\Lambda
+V_{i}G^{\mathrm{R}}V_{k}\partial_{\varepsilon}^{2}G^{\mathrm{R}}
 V_{j}\Lambda\nonumber\\
 &
+V_{i}\partial_{\varepsilon}^{2}G^{\mathrm{R}}
 V_{j}\Lambda V_{k}G^{\mathrm{A}}
+V_{i}G^{\mathrm{R}}V_{k}\Lambda
 V_{j}\partial_{\varepsilon}^{2}G^{\mathrm{A}}
+V_{i}\Lambda V_{j}\partial_{\varepsilon}^{2}G^{\mathrm{A}}
 V_{k}G^{\mathrm{A}}
+2V_{i}\Lambda V_{j}\partial_{\varepsilon}G^{\mathrm{A}}
 V_{k}\partial_{\varepsilon}G^{\mathrm{A}}
\nonumber\\
&+V_{i}\Lambda V_{j}G^{\mathrm{A}}
 V_{k}\partial_{\varepsilon}^{2}G^{\mathrm{A}}
+V_{i}\Lambda V_{k}G^{\mathrm{A}}
 V_{j}\partial_{\varepsilon}^{2}G^{\mathrm{A}}
\nonumber\\
&+V_{ik}\partial_{\varepsilon}^{2}G^{\mathrm{R}}V_{j}\Lambda
+V_{ik}\Lambda V_{j}\partial_{\varepsilon}^{2}G^{\mathrm{A}}
+V_{i}\partial_{\varepsilon}^{2}G^{\mathrm{R}}V_{jk}\Lambda
+V_{i}\Lambda V_{jk}\partial_{\varepsilon}^{2}G^{\mathrm{A}}
\bigg],
\label{eq:A21-compact}
\end{align}
\begin{align}
A^{(2)}_{2,ijk}
\equiv{}&\frac{1}{2\hbar^{2}}
\left.\partial_{\omega_{2}}^{2}\mathcal{S}_{ijk}(\omega_{1},\omega_{2})
\right|_{\omega_{1}=\omega_{2}=0}
\nonumber\\
={}&\frac{1}{2}\int_{-\infty}^{\infty}\frac{d\varepsilon}{2\pi}\,
 f(\varepsilon)\sum_{\bm{k}}\operatorname{Tr}\bigg[
\nonumber\\
&V_{i}\partial_{\varepsilon}^{2}G^{\mathrm{R}}
 V_{j}G^{\mathrm{R}}V_{k}\Lambda
+2V_{i}\partial_{\varepsilon}G^{\mathrm{R}}
 V_{j}\partial_{\varepsilon}G^{\mathrm{R}}V_{k}\Lambda
+V_{i}G^{\mathrm{R}}V_{j}\partial_{\varepsilon}^{2}G^{\mathrm{R}}
 V_{k}\Lambda
+V_{i}\partial_{\varepsilon}^{2}G^{\mathrm{R}}
 V_{k}G^{\mathrm{R}}V_{j}\Lambda
\nonumber\\
&+V_{i}G^{\mathrm{R}}V_{j}\Lambda
 V_{k}\partial_{\varepsilon}^{2}G^{\mathrm{A}}
+V_{i}\partial_{\varepsilon}^{2}G^{\mathrm{R}}
 V_{k}\Lambda V_{j}G^{\mathrm{A}}
+V_{i}\Lambda V_{j}G^{\mathrm{A}}
 V_{k}\partial_{\varepsilon}^{2}G^{\mathrm{A}}
+V_{i}\Lambda V_{k}\partial_{\varepsilon}^{2}G^{\mathrm{A}}
 V_{j}G^{\mathrm{A}}
\nonumber\\
&+2V_{i}\Lambda V_{k}\partial_{\varepsilon}G^{\mathrm{A}}
 V_{j}\partial_{\varepsilon}G^{\mathrm{A}}
+V_{i}\Lambda V_{k}G^{\mathrm{A}}
 V_{j}\partial_{\varepsilon}^{2}G^{\mathrm{A}}
\nonumber\\
&+V_{ij}\partial_{\varepsilon}^{2}G^{\mathrm{R}}V_{k}\Lambda
+V_{ij}\Lambda V_{k}\partial_{\varepsilon}^{2}G^{\mathrm{A}}
+V_{i}\partial_{\varepsilon}^{2}G^{\mathrm{R}}V_{jk}\Lambda
+V_{i}\Lambda V_{jk}\partial_{\varepsilon}^{2}G^{\mathrm{A}}
\bigg].
\label{eq:A22-compact}
\end{align}

The mixed coefficient is
\begin{align}
A_{12,ijk}
\equiv{}&\frac{1}{\hbar^{2}}
\left.\partial_{\omega_{1}}\partial_{\omega_{2}}
\mathcal{S}_{ijk}(\omega_{1},\omega_{2})
\right|_{\omega_{1}=\omega_{2}=0},
\end{align}
and hence
\begin{align}
\sigma^{\mathrm{DC}}_{ijk}
={}&-\frac{i e^{3}\hbar^{2}}{2}\,A_{12,ijk}
\nonumber\\
={}&-\frac{i e^{3}\hbar^{2}}{2}
\sum_{\bm{k}}\int_{-\infty}^{\infty}\frac{d\varepsilon}{2\pi}\,
 f(\varepsilon)\operatorname{Tr}\bigg[
\nonumber\\
&V_{i}\partial_{\varepsilon}^{2}G^{\mathrm{R}}
 V_{j}G^{\mathrm{R}}V_{k}\Lambda
+V_{i}\partial_{\varepsilon}G^{\mathrm{R}}
 V_{j}\partial_{\varepsilon}G^{\mathrm{R}}V_{k}\Lambda
+V_{i}\partial_{\varepsilon}^{2}G^{\mathrm{R}}
 V_{k}G^{\mathrm{R}}V_{j}\Lambda
+V_{i}\partial_{\varepsilon}G^{\mathrm{R}}
 V_{k}\partial_{\varepsilon}G^{\mathrm{R}}V_{j}\Lambda
\nonumber\\
&-V_{i}\partial_{\varepsilon}G^{\mathrm{R}}
 V_{j}\Lambda V_{k}\partial_{\varepsilon}G^{\mathrm{A}}
-V_{i}\partial_{\varepsilon}G^{\mathrm{R}}
 V_{k}\Lambda V_{j}\partial_{\varepsilon}G^{\mathrm{A}}
+V_{i}\Lambda V_{j}G^{\mathrm{A}}
 V_{k}\partial_{\varepsilon}^{2}G^{\mathrm{A}}
+V_{i}\Lambda V_{j}\partial_{\varepsilon}G^{\mathrm{A}}
 V_{k}\partial_{\varepsilon}G^{\mathrm{A}}
\nonumber\\
&+V_{i}\Lambda V_{k}G^{\mathrm{A}}
 V_{j}\partial_{\varepsilon}^{2}G^{\mathrm{A}}
+V_{i}\Lambda V_{k}\partial_{\varepsilon}G^{\mathrm{A}}
 V_{j}\partial_{\varepsilon}G^{\mathrm{A}}
\nonumber\\
&+V_{i}\partial_{\varepsilon}^{2}G^{\mathrm{R}}V_{jk}\Lambda
+V_{i}\Lambda V_{jk}\partial_{\varepsilon}^{2}G^{\mathrm{A}}
\bigg].
\label{eq:sigma-dc-before-rewrite0}
\end{align}

\subsection{\texorpdfstring{Vanishing of $A_{0,ijk}$}{Vanishing of A0ijk}}
We set $\hbar = 1$ throughout for brevity.
We repeatedly use the following elementary identities for $G=G^{\mathrm{R}}$ or $G=G^{\mathrm{A}}$:
\begin{align}
  &\partial_{k}G = G V_{k}G,\label{eq:basic-k-deriv-supp}\\
  &\partial_{\varepsilon}G = -G^{2},\label{eq:basic-e-deriv-supp}\\
  &\sum_{\bm{k}}\partial_{k}X(\bm{k}) =0.\label{eq:bz-total-deriv-supp}
\end{align}

We first consider the zeroth-order coefficient.  
\begin{align}
A_{0,ijk} = \int_{-\infty}^{\infty}\frac{d\varepsilon}{2\pi } f(\varepsilon)
\mathcal{I}_{0,ijk}, 
\end{align}
where the integrand  is given by 
\begin{align}
&\mathcal{I}_{0,ijk}
=\sum_{\bm k}{\rm Tr}\bigg[
V_iG^{\rm R}V_jG^{\rm R}V_k\Lambda
+V_iG^{\rm R}V_kG^{\rm R}V_j\Lambda
+V_iG^{\rm R}V_j\Lambda V_kG^{\rm A}
\nonumber\\
&\quad
+V_iG^{\rm R}V_k\Lambda V_jG^{\rm A}
+V_i\Lambda V_jG^{\rm A}V_kG^{\rm A}
+V_i\Lambda V_kG^{\rm A}V_jG^{\rm A}
\nonumber\\
&\quad
+V_{ij}G^{\rm R}V_k\Lambda+V_{ik}G^{\rm R}V_j\Lambda
+V_{ij}\Lambda V_kG^{\rm A}+V_{ik}\Lambda V_jG^{\rm A}
+V_iG^{\rm R}V_{jk}\Lambda+V_i\Lambda V_{jk}G^{\rm A}
+V_{ijk}\Lambda\bigg]
\nonumber\\
&=\sum_{\bm k}{\rm Tr}\bigg[
V_iG^{\rm A}V_jG^{\rm A}V_kG^{\rm A}
+V_iG^{\rm A}V_kG^{\rm A}V_jG^{\rm A}
+V_{ij}G^{\rm A}V_kG^{\rm A}+V_{ik}G^{\rm A}V_jG^{\rm A}
+V_iG^{\rm A}V_{jk}G^{\rm A}+V_{ijk}G^{\rm A}
\nonumber\\
&\quad
-V_iG^{\rm R}V_jG^{\rm R}V_kG^{\rm R}
-V_iG^{\rm R}V_kG^{\rm R}V_jG^{\rm R}
-V_{ij}G^{\rm R}V_kG^{\rm R}-V_{ik}G^{\rm R}V_jG^{\rm R}
-V_iG^{\rm R}V_{jk}G^{\rm R}-V_{ijk}G^{\rm R}\bigg]
\end{align}
Here we have used $\Lambda = G^{\rm A}-G^{\rm R}$. 
Let $G$ stand for either
$G^{\mathrm{R}}$ or $G^{\mathrm{A}}$.  From Eq.~\eqref{eq:basic-k-deriv-supp},
\begin{align}  \partial_{k}\operatorname{Tr}\bigl[V_{i}GV_{j}G\bigr]
  &=\operatorname{Tr}\bigl[
      V_{ik}GV_{j}G
      +V_{i}GV_{k}GV_{j}G
      +V_{i}GV_{jk}G
      +V_{i}GV_{j}GV_{k}G
    \bigr],
  \label{eq:A0-identity-1-supp}\\
  \partial_{k}\operatorname{Tr}\bigl[V_{ij}G\bigr]
  &=\operatorname{Tr}\bigl[
      V_{ijk}G+V_{ij}GV_{k}G
    \bigr].
  \label{eq:A0-identity-2-supp}
\end{align}
Adding these two identities gives
\begin{align}  &\operatorname{Tr}\bigl[
    V_{i}GV_{j}GV_{k}G
    +V_{i}GV_{k}GV_{j}G
    +V_{ij}GV_{k}G
    +V_{ik}GV_{j}G
    +V_{i}GV_{jk}G
    +V_{ijk}G
  \bigr]
  \nonumber\\
  &\hspace{30mm}
  =\partial_{k}\operatorname{Tr}\bigl[V_{i}GV_{j}G+V_{ij}G\bigr].
  \label{eq:A0-total-identity-supp}
\end{align}
Applying Eq.~\eqref{eq:A0-total-identity-supp} to $G=G^{\mathrm{A}}$ and $G=G^{\mathrm{R}}$,
the coefficient $A_{0,ijk}$ becomes
\begin{align}  A_{0,ijk}
  &=
    \int_{-\infty}^{\infty}\frac{\mathrm{d}\varepsilon}{2\pi}
    f(\varepsilon)
    \sum_{\bm{k}}\partial_{k}\operatorname{Tr}\bigl[
      V_{i}G^{\mathrm{A}} V_{j}G^{\mathrm{A}}+V_{ij}G^{\mathrm{A}}
      -V_{i}G^{\mathrm{R}} V_{j}G^{\mathrm{R}}-V_{ij}G^{\mathrm{R}}
    \bigr].
  \label{eq:A0-total-deriv-supp}
\end{align}
The right-hand side is a total derivative in the $k$ direction.  
Therefore,
by Eq.~\eqref{eq:bz-total-deriv-supp},
\begin{align}  A_{0,ijk}=0.
\label{eq:sm-auto-012}
\end{align}

\subsection{\texorpdfstring{Vanishing of $A^{(1)}_{1,ijk}$}{Vanishing of A1ijk (1)}}
We next examine the first-order term whose total derivative appears in the $k$ direction.  
The coefficient $A^{(1)}_{1,ijk}$ is 
written as 
\begin{align}
A^{(1)}_{1,ijk} = \int_{-\infty}^{\infty}\frac{d\varepsilon}{2\pi}f(\varepsilon)
\mathcal{I}^{(1)}_{1,ijk} , 
\end{align}
where the integrand  is 
\begin{align}
\mathcal{I}^{(1)}_{1,ijk}
={}&
\sum_{\bm k}\operatorname{Tr}\bigg[
V_{i}\partial_{\varepsilon}G^{\mathrm{R}} V_{j}G^{\mathrm{R}} V_{k}\Lambda
+V_{i}\partial_{\varepsilon}G^{\mathrm{R}} V_{j}\Lambda V_{k}G^{\mathrm{A}}
-V_{i}\Lambda V_{j}\partial_{\varepsilon}G^{\mathrm{A}} V_{k}G^{\mathrm{A}}
\nonumber\\
&-V_{i}\Lambda V_{j}G^{\mathrm{A}} V_{k}\partial_{\varepsilon}G^{\mathrm{A}}
+V_{i}\partial_{\varepsilon}G^{\mathrm{R}} V_{k}G^{\mathrm{R}} V_{j}\Lambda
+V_{i}G^{\mathrm{R}} V_{k}\partial_{\varepsilon}G^{\mathrm{R}} V_{j}\Lambda
\nonumber\\
&-V_{i}G^{\mathrm{R}} V_{k}\Lambda V_{j}\partial_{\varepsilon}G^{\mathrm{A}}
-V_{i}\Lambda V_{k}G^{\mathrm{A}} V_{j}\partial_{\varepsilon}G^{\mathrm{A}}
\nonumber\\
&+V_{ik}\partial_{\varepsilon}G^{\mathrm{R}} V_{j}\Lambda
-V_{ik}\Lambda V_{j}\partial_{\varepsilon}G^{\mathrm{A}}
\nonumber\\
&+V_{i}\partial_{\varepsilon}G^{\mathrm{R}} V_{jk}\Lambda
-V_{i}\Lambda V_{jk}\partial_{\varepsilon}G^{\mathrm{A}}\bigg].
\label{eq:I11-original-supp}
\end{align}
Collecting the purely
retarded, purely advanced, and mixed retarded--advanced terms, we use the
following identities.  For the purely retarded part,
\begin{align}  
&\partial_{k}\operatorname{Tr}\bigl[V_{i}G^{\mathrm{R}}G^{\mathrm{R}} V_{j}G^{\mathrm{R}}\bigr]\nonumber\\
  &={}  -\operatorname{Tr}\bigl[
V_{ik}\partial_{\varepsilon}G^{\mathrm{R}} V_{j}G^{\mathrm{R}}
+V_{i}\partial_{\varepsilon}G^{\mathrm{R}} V_{jk}G^{\mathrm{R}}
 +V_{i}G^{\mathrm{R}} V_{k}\partial_{\varepsilon}G^{\mathrm{R}} V_{j}G^{\mathrm{R}}
  +V_{i}\partial_{\varepsilon}G^{\mathrm{R}} V_{k}G^{\mathrm{R}} V_{j}G^{\mathrm{R}}
  +V_{i}\partial_{\varepsilon}G^{\mathrm{R}} V_{j}G^{\mathrm{R}} V_{k}G^{\mathrm{R}}\bigr].
  \label{eq:A11-RRR-identity-supp}
\end{align}
Indeed, before using Eq.~\eqref{eq:basic-e-deriv-supp}, the left-hand side is
\begin{align}
&\partial_{k}\operatorname{Tr}\bigl[V_{i}G^{\mathrm{R}}G^{\mathrm{R}} V_{j}G^{\mathrm{R}}\bigr]\nonumber\\
&=\operatorname{Tr}\bigl[
V_{ik}G^{\mathrm{R}}G^{\mathrm{R}} V_{j}G^{\mathrm{R}}
+V_{i}G^{\mathrm{R}} V_{k}G^{\mathrm{R}}G^{\mathrm{R}} V_{j}G^{\mathrm{R}}
+V_{i}G^{\mathrm{R}}G^{\mathrm{R}} V_{k}G^{\mathrm{R}} V_{j}G^{\mathrm{R}}
+V_{i}G^{\mathrm{R}}G^{\mathrm{R}} V_{jk}G^{\mathrm{R}}
+V_{i}G^{\mathrm{R}}G^{\mathrm{R}} V_{j}G^{\mathrm{R}} V_{k}G^{\mathrm{R}}
\bigr],
\label{eq:sm-auto-013}
\end{align}
and Eq.~\eqref{eq:A11-RRR-identity-supp} follows from
$G^{\mathrm{R}}G^{\mathrm{R}}=-\partial_{\varepsilon}G^{\mathrm{R}}$.  
Similarly, the purely advanced part obeys
\begin{align}  
&\partial_{k}\operatorname{Tr}\bigl[V_{i}G^{\mathrm{A}} V_{j}G^{\mathrm{A}}G^{\mathrm{A}}\bigr]
\nonumber\\  &=
  -\operatorname{Tr}\bigl[
  V_{ik}G^{\mathrm{A}} V_{j}\partial_{\varepsilon}G^{\mathrm{A}}
  +V_{i}G^{\mathrm{A}} V_{jk}\partial_{\varepsilon}G^{\mathrm{A}}
  V_{i}G^{\mathrm{A}} V_{k}G^{\mathrm{A}} V_{j}\partial_{\varepsilon}G^{\mathrm{A}}
  +V_{i}G^{\mathrm{A}} V_{j}G^{\mathrm{A}} V_{k}\partial_{\varepsilon}G^{\mathrm{A}}
  +V_{i}G^{\mathrm{A}} V_{j}\partial_{\varepsilon}G^{\mathrm{A}} V_{k}G^{\mathrm{A}}\bigr].
  \label{eq:A11-AAA-identity-supp}
\end{align}
The mixed terms are generated by differentiating the $RA$ trace:
\begin{align}  \partial_{k}\operatorname{Tr}\bigl[V_{i}G^{\mathrm{R}} V_{j}G^{\mathrm{A}}\bigr]
  ={}&\operatorname{Tr}\bigl[
     V_{ik}G^{\mathrm{R}} V_{j}G^{\mathrm{A}}
     +V_{i}G^{\mathrm{R}} V_{k}G^{\mathrm{R}} V_{j}G^{\mathrm{A}}
     +V_{i}G^{\mathrm{R}} V_{jk}G^{\mathrm{A}}
     +V_{i}G^{\mathrm{R}} V_{j}G^{\mathrm{A}} V_{k}G^{\mathrm{A}}
  \bigr].
  \label{eq:A11-RA-identity-supp}
\end{align}
Taking an energy derivative of Eq.~\eqref{eq:A11-RA-identity-supp} produces
exactly the mixed terms in Eq.~\eqref{eq:I11-original-supp}.  Hence
\begin{align}  \mathcal I^{(1)}_{1,ijk}
  ={}&
  \sum_{{\bm k}}\partial_{k}\operatorname{Tr}\bigl[V_{i}G^{\mathrm{R}}G^{\mathrm{R}} V_{j}G^{\mathrm{R}}+V_{i}G^{\mathrm{A}} V_{j}G^{\mathrm{A}}G^{\mathrm{A}}\bigr]
  +\sum_{{\bm k}}\partial_{\varepsilon}\partial_{k}\operatorname{Tr}\bigl[V_{i}G^{\mathrm{R}} V_{j}G^{\mathrm{A}}\bigr].
  \label{eq:I11-total-before-ibp-supp}
\end{align}
Therefore,
\begin{align}  
A^{(1)}_{1,ijk}
&=\int_{-\infty}^{\infty}\frac{\mathrm{d}\varepsilon}{2\pi }
 f(\varepsilon)\mathcal{I}^{(1)}_{1,ijk}
\nonumber\\
  &=
\int_{-\infty}^{\infty}\frac{\mathrm{d}\varepsilon}{2\pi }  \sum_{\bm{k}}
    \partial_{k}\Bigl\{
       f(\varepsilon)\operatorname{Tr}\bigl[V_{i}G^{\mathrm{R}}G^{\mathrm{R}} V_{j}G^{\mathrm{R}}+V_{i}G^{\mathrm{A}} V_{j}G^{\mathrm{A}}G^{\mathrm{A}}\bigr]
       -\partial_{\varepsilon} f(\varepsilon)\operatorname{Tr}\bigl[V_{i}G^{\mathrm{R}} V_{j}G^{\mathrm{A}}\bigr]
    \Bigr\}.
  \label{eq:A11-total-deriv-supp}
\end{align}
In the second line we integrated the $\varepsilon$ derivative by parts; the boundary
term vanishes because the Green functions decay at large $|\varepsilon|$.  The
remaining expression is a total derivative over the Brillouin zone, and
therefore
\begin{align}  A^{(1)}_{1,ijk}=0.
\label{eq:sm-auto-014}
\end{align}

\subsection{\texorpdfstring{Vanishing of $A^{(2)}_{1,ijk}$}{Vanishing of A1ijk (2)}}

The coefficient $A^{(2)}_{1,ijk}$ has the same structure as
$A^{(1)}_{1,ijk}$, but the final total derivative is in the $j$ direction.
This is given by 
\begin{align}
A^{(2)}_{1,ijk}= \int_{-\infty}^{\infty}\frac{d\varepsilon}{2\pi}f(\varepsilon)
\mathcal{I}^{(2)}_{1,ijk}, 
\end{align}
where the integrand  is 
\begin{align}
\mathcal{I}^{(2)}_{1,ijk}
={}&
\sum_{\bm k}\operatorname{Tr}\bigg[
V_{i}\partial_{\varepsilon}G^{\mathrm{R}} V_{j}G^{\mathrm{R}} V_{k}\Lambda
+V_{i}G^{\mathrm{R}} V_{j}\partial_{\varepsilon}G^{\mathrm{R}} V_{k}\Lambda
+V_{i}\partial_{\varepsilon}G^{\mathrm{R}} V_{k}G^{\mathrm{R}} V_{j}\Lambda
\nonumber\\
&-V_{i}G^{\mathrm{R}} V_{j}\Lambda V_{k}\partial_{\varepsilon}G^{\mathrm{A}}
+V_{i}\partial_{\varepsilon}G^{\mathrm{R}} V_{k}\Lambda V_{j}G^{\mathrm{A}}
-V_{i}\Lambda V_{j}G^{\mathrm{A}} V_{k}\partial_{\varepsilon}G^{\mathrm{A}}
\nonumber\\
&-V_{i}\Lambda V_{k}\partial_{\varepsilon}G^{\mathrm{A}} V_{j}G^{\mathrm{A}}
-V_{i}\Lambda V_{k}G^{\mathrm{A}} V_{j}\partial_{\varepsilon}G^{\mathrm{A}}
\nonumber\\
&+V_{ij}\partial_{\varepsilon}G^{\mathrm{R}} V_{k}\Lambda
-V_{ij}\Lambda V_{k}\partial_{\varepsilon}G^{\mathrm{A}}
\nonumber\\
&+V_{i}\partial_{\varepsilon}G^{\mathrm{R}} V_{jk}\Lambda
-V_{i}\Lambda V_{jk}\partial_{\varepsilon}G^{\mathrm{A}}
\bigg].
\label{eq:I12-original-supp}
\end{align}
After substituting $\Lambda=G^{\mathrm{A}}-G^{\mathrm{R}}$, the pure $R$ and pure $A$ pieces are
combined by
\begin{align}  
&\partial_{j}\operatorname{Tr}\bigl[V_{i}G^{\mathrm{R}}G^{\mathrm{R}} V_{k}G^{\mathrm{R}}\bigr]\nonumber\\
&=
  -\operatorname{Tr}\bigl[
  V_{ij}\partial_{\varepsilon}G^{\mathrm{R}} V_{k}G^{\mathrm{R}}
  +V_{i}\partial_{\varepsilon}G^{\mathrm{R}} V_{jk}G^{\mathrm{R}}
  +V_{i}G^{\mathrm{R}} V_{j}\partial_{\varepsilon}G^{\mathrm{R}} V_{k}G^{\mathrm{R}}
  +V_{i}\partial_{\varepsilon}G^{\mathrm{R}} V_{j}G^{\mathrm{R}} V_{k}G^{\mathrm{R}}
  +V_{i}\partial_{\varepsilon}G^{\mathrm{R}} V_{k}G^{\mathrm{R}} V_{j}G^{\mathrm{R}}\bigr],
  \label{eq:A12-RRR-identity-supp}\\[1mm]
&  \partial_{j}\operatorname{Tr}\bigl[V_{i}G^{\mathrm{A}} V_{k}G^{\mathrm{A}}G^{\mathrm{A}}\bigr]
\nonumber\\&=
  -\operatorname{Tr}\bigl[
  V_{ij}G^{\mathrm{A}} V_{k}\partial_{\varepsilon}G^{\mathrm{A}}
  +V_{i}G^{\mathrm{A}} V_{jk}\partial_{\varepsilon}G^{\mathrm{A}}
  +V_{i}G^{\mathrm{A}} V_{j}G^{\mathrm{A}} V_{k}\partial_{\varepsilon}G^{\mathrm{A}}
  +V_{i}G^{\mathrm{A}} V_{k}G^{\mathrm{A}} V_{j}\partial_{\varepsilon}G^{\mathrm{A}}
  +V_{i}G^{\mathrm{A}} V_{k}\partial_{\varepsilon}G^{\mathrm{A}} V_{j}G^{\mathrm{A}}\bigr].
  \label{eq:A12-AAA-identity-supp}
\end{align}
The mixed part is obtained from
\begin{align}  \partial_{j}\operatorname{Tr}\bigl[V_{i}G^{\mathrm{R}} V_{k}G^{\mathrm{A}}\bigr]
  ={}&\operatorname{Tr}\bigl[
     V_{ij}G^{\mathrm{R}} V_{k}G^{\mathrm{A}}
     +V_{i}G^{\mathrm{R}} V_{j}G^{\mathrm{R}} V_{k}G^{\mathrm{A}}
     +V_{i}G^{\mathrm{R}} V_{jk}G^{\mathrm{A}}
     +V_{i}G^{\mathrm{R}} V_{k}G^{\mathrm{A}} V_{j}G^{\mathrm{A}}
  \bigr].
  \label{eq:A12-RA-identity-supp}
\end{align}
Consequently,
\begin{align}  \mathcal I^{(2)}_{1,ijk}
  ={}&\sum_{{\bm k}}
  \partial_{j}\operatorname{Tr}\bigl[V_{i}G^{\mathrm{R}}G^{\mathrm{R}} V_{k}G^{\mathrm{R}}+V_{i}G^{\mathrm{A}} V_{k}G^{\mathrm{A}}G^{\mathrm{A}}\bigr]
  +\sum_{{\bm k}}\partial_{\varepsilon}\partial_{j}\operatorname{Tr}\bigl[V_{i}G^{\mathrm{R}} V_{k}G^{\mathrm{A}}\bigr].
  \label{eq:I12-total-before-ibp-supp}
\end{align}
Integrating the energy derivative by parts gives
\begin{align}  
A^{(2)}_{1,ijk}&= \int_{-\infty}^{\infty}\frac{d\varepsilon}{2\pi}f(\varepsilon)
\mathcal{I}^{(2)}_{1,ijk} \nonumber\\
  &=
   \int_{-\infty}^{\infty}\frac{\mathrm{d}\varepsilon}{2\pi} \sum_{\bm{k}}
    \partial_{j}\Bigl\{
       f(\varepsilon)\operatorname{Tr}\bigl[V_{i}G^{\mathrm{R}}G^{\mathrm{R}} V_{k}G^{\mathrm{R}}+V_{i}G^{\mathrm{A}} V_{k}G^{\mathrm{A}}G^{\mathrm{A}}\bigr]
       -\partial_{\varepsilon} f(\varepsilon)\operatorname{Tr}\bigl[V_{i}G^{\mathrm{R}} V_{k}G^{\mathrm{A}}\bigr]
    \Bigr\}
  \nonumber\\
  &=0.
\label{eq:sm-auto-015}
\end{align}
The last equality again follows from Eq.~\eqref{eq:bz-total-deriv-supp}.

\subsection{\texorpdfstring{Vanishing of $A^{(1)}_{2,ijk}$}{Vanishing of A2ijk (1)}}

We now consider the second-order term containing second energy derivatives.
This contribution becomes a total derivative in the $k$ direction.  Up to the
overall prefactor, the coefficient $A^{(1)}_{2,ijk}$ is given by 
\begin{align}
A^{(1)}_{2,ijk} = \frac{1}{2}\int_{-\infty}^{\infty}\frac{d\varepsilon}{2\pi}f(\varepsilon)\mathcal{I}^{(1)}_{2,ijk}, 
\end{align}
and its integrand is
\begin{align}\mathcal I^{(1)}_{2,ijk}
={}&\sum_{{\bm k}}
\operatorname{Tr}\bigg[
V_{i}\partial_{\varepsilon}^{2}G^{\mathrm{R}} V_{j}G^{\mathrm{R}} V_{k}\Lambda
+V_{i}\partial_{\varepsilon}^{2}G^{\mathrm{R}} V_{k}G^{\mathrm{R}} V_{j}\Lambda
 +2V_{i}\partial_{\varepsilon}G^{\mathrm{R}} V_{k}\partial_{\varepsilon}G^{\mathrm{R}} V_{j}\Lambda
\nonumber\\
&\quad
 +V_{i}G^{\mathrm{R}} V_{k}\partial_{\varepsilon}^{2}G^{\mathrm{R}} V_{j}\Lambda
 +V_{i}\partial_{\varepsilon}^{2}G^{\mathrm{R}} V_{j}\Lambda V_{k}G^{\mathrm{A}}
 +V_{i}G^{\mathrm{R}} V_{k}\Lambda V_{j}\partial_{\varepsilon}^{2}G^{\mathrm{A}}
\nonumber\\
&\quad
 +V_{i}\Lambda V_{j}\partial_{\varepsilon}^{2}G^{\mathrm{A}} V_{k}G^{\mathrm{A}}
 +2V_{i}\Lambda V_{j}\partial_{\varepsilon}G^{\mathrm{A}} V_{k}\partial_{\varepsilon}G^{\mathrm{A}}
 +V_{i}\Lambda V_{j}G^{\mathrm{A}} V_{k}\partial_{\varepsilon}^{2}G^{\mathrm{A}}
\nonumber\\
&\quad
 +V_{i}\Lambda V_{k}G^{\mathrm{A}} V_{j}\partial_{\varepsilon}^{2}G^{\mathrm{A}}
 +V_{ik}\partial_{\varepsilon}^{2}G^{\mathrm{R}} V_{j}\Lambda
 +V_{ik}\Lambda V_{j}\partial_{\varepsilon}^{2}G^{\mathrm{A}}
\nonumber\\
&\quad
 +V_{i}\partial_{\varepsilon}^{2}G^{\mathrm{R}} V_{jk}\Lambda
 +V_{i}\Lambda V_{jk}\partial_{\varepsilon}^{2}G^{\mathrm{A}}
\bigg].
\label{eq:I21-original-supp}
\end{align}
The following four identities show explicitly how the terms combine.  The
pure retarded part is
\begin{align}\partial_{k}\operatorname{Tr}\bigl[V_{i}\partial_{\varepsilon}^{2}G^{\mathrm{R}} V_{j}G^{\mathrm{R}}\bigr]
={}&\operatorname{Tr}\bigg[
 V_{ik}\partial_{\varepsilon}^{2}G^{\mathrm{R}} V_{j}G^{\mathrm{R}}
 +V_{i}\partial_{\varepsilon}^{2}G^{\mathrm{R}} V_{jk}G^{\mathrm{R}}
\nonumber\\
&\quad
 +V_{i}\partial_{\varepsilon}^{2}G^{\mathrm{R}} V_{j}G^{\mathrm{R}} V_{k}G^{\mathrm{R}}
 +V_{i}\partial_{\varepsilon}^{2}G^{\mathrm{R}} V_{k}G^{\mathrm{R}} V_{j}G^{\mathrm{R}}
\nonumber\\
&\quad
 +2V_{i}\partial_{\varepsilon}G^{\mathrm{R}} V_{k}\partial_{\varepsilon}G^{\mathrm{R}} V_{j}G^{\mathrm{R}}
 +V_{i}G^{\mathrm{R}} V_{k}\partial_{\varepsilon}^{2}G^{\mathrm{R}} V_{j}G^{\mathrm{R}}
\bigg],
\label{eq:A21-RR-identity-supp}
\end{align}
where we used $\partial_{\varepsilon}^{2}G^{\mathrm{R}}=2G^{\mathrm{R}}G^{\mathrm{R}}G^{\mathrm{R}}$.  The pure advanced part is
\begin{align}\partial_{k}\operatorname{Tr}\bigl[V_{i}G^{\mathrm{A}} V_{j}\partial_{\varepsilon}^{2}G^{\mathrm{A}}\bigr]
={}&\operatorname{Tr}\bigg[
 V_{ik}G^{\mathrm{A}} V_{j}\partial_{\varepsilon}^{2}G^{\mathrm{A}}
 +V_{i}G^{\mathrm{A}} V_{jk}\partial_{\varepsilon}^{2}G^{\mathrm{A}}
\nonumber\\
&\quad
 +V_{i}G^{\mathrm{A}} V_{k}G^{\mathrm{A}} V_{j}\partial_{\varepsilon}^{2}G^{\mathrm{A}}
 +V_{i}G^{\mathrm{A}} V_{j}G^{\mathrm{A}} V_{k}\partial_{\varepsilon}^{2}G^{\mathrm{A}}
\nonumber\\
&\quad
 +2V_{i}G^{\mathrm{A}} V_{j}\partial_{\varepsilon}G^{\mathrm{A}} V_{k}\partial_{\varepsilon}G^{\mathrm{A}}
 +V_{i}G^{\mathrm{A}} V_{j}\partial_{\varepsilon}^{2}G^{\mathrm{A}} V_{k}G^{\mathrm{A}}
\bigg].
\label{eq:A21-AA-identity-supp}
\end{align}
The mixed derivatives are
\begin{align}\partial_{k}\operatorname{Tr}\bigl[V_{i}\partial_{\varepsilon}^{2}G^{\mathrm{R}} V_{j}G^{\mathrm{A}}\bigr]
={}&\operatorname{Tr}\bigg[
 V_{ik}\partial_{\varepsilon}^{2}G^{\mathrm{R}} V_{j}G^{\mathrm{A}}
 +V_{i}\partial_{\varepsilon}^{2}G^{\mathrm{R}} V_{jk}G^{\mathrm{A}}
 +V_{i}\partial_{\varepsilon}^{2}G^{\mathrm{R}} V_{j}G^{\mathrm{A}} V_{k}G^{\mathrm{A}}
\nonumber\\
&\quad
 +V_{i}\partial_{\varepsilon}^{2}G^{\mathrm{R}} V_{k}G^{\mathrm{R}} V_{j}G^{\mathrm{A}}
 +2V_{i}\partial_{\varepsilon}G^{\mathrm{R}} V_{k}\partial_{\varepsilon}G^{\mathrm{R}} V_{j}G^{\mathrm{A}}
 + V_{i}G^{\mathrm{R}} V_{k}\partial_{\varepsilon}^{2}G^{\mathrm{R}} V_{j}G^{\mathrm{A}}
\bigg],
\label{eq:A21-RA1-identity-supp}\\[1mm]
\partial_{k}\operatorname{Tr}\bigl[V_{i}G^{\mathrm{R}} V_{j}\partial_{\varepsilon}^{2}G^{\mathrm{A}}\bigr]
={}&\operatorname{Tr}\bigg[
 V_{ik}G^{\mathrm{R}} V_{j}\partial_{\varepsilon}^{2}G^{\mathrm{A}}
 +V_{i}G^{\mathrm{R}} V_{jk}\partial_{\varepsilon}^{2}G^{\mathrm{A}}
 +V_{i}G^{\mathrm{R}} V_{k}G^{\mathrm{R}} V_{j}\partial_{\varepsilon}^{2}G^{\mathrm{A}}
\nonumber\\
&\quad
 +V_{i}G^{\mathrm{R}} V_{j}G^{\mathrm{A}} V_{k}\partial_{\varepsilon}^{2}G^{\mathrm{A}}
 +2V_{i}G^{\mathrm{R}} V_{j}\partial_{\varepsilon}G^{\mathrm{A}} V_{k}\partial_{\varepsilon}G^{\mathrm{A}}
 +V_{i}G^{\mathrm{R}} V_{j}\partial_{\varepsilon}^{2}G^{\mathrm{A}} V_{k}G^{\mathrm{A}}
\bigg].
\label{eq:A21-RA2-identity-supp}
\end{align}
Substituting $\Lambda=G^{\mathrm{A}}-G^{\mathrm{R}}$ into Eq.~\eqref{eq:I21-original-supp} and
matching the terms with Eqs.~\eqref{eq:A21-RR-identity-supp}--\eqref{eq:A21-RA2-identity-supp},
we obtain
\begin{align}  \mathcal I^{(1)}_{2,ijk}
  =\sum_{{\bm k}}\partial_{k}\operatorname{Tr}\bigl[
      V_{i}G^{\mathrm{A}} V_{j}\partial_{\varepsilon}^{2}G^{\mathrm{A}}
      -V_{i}\partial_{\varepsilon}^{2}G^{\mathrm{R}} V_{j}G^{\mathrm{R}}
      +V_{i}\partial_{\varepsilon}^{2}G^{\mathrm{R}} V_{j}G^{\mathrm{A}}
      -V_{i}G^{\mathrm{R}} V_{j}\partial_{\varepsilon}^{2}G^{\mathrm{A}}
    \bigr].
  \label{eq:I21-total-deriv-supp}
\end{align}
Thus, 
\begin{align}  A^{(1)}_{2,ijk}
  &=\frac{1}{2}\int_{-\infty}^{\infty}\frac{\mathrm{d}\varepsilon}{2\pi }
    f(\varepsilon)\mathcal{I}^{(1)}_{2,ijk}\nonumber\\
  &=\frac{1}{2}\int_{-\infty}^{\infty}\frac{\mathrm{d}\varepsilon}{2\pi }
    f(\varepsilon)\sum_{\bm{k}}
    \partial_{k}\operatorname{Tr}\bigl[
      V_{i}G^{\mathrm{A}} V_{j}\partial_{\varepsilon}^{2}G^{\mathrm{A}}
      -V_{i}\partial_{\varepsilon}^{2}G^{\mathrm{R}} V_{j}G^{\mathrm{R}}
      +V_{i}\partial_{\varepsilon}^{2}G^{\mathrm{R}} V_{j}G^{\mathrm{A}}
      -V_{i}G^{\mathrm{R}} V_{j}\partial_{\varepsilon}^{2}G^{\mathrm{A}}
    \bigr]\nonumber\\
  &=0.
\label{eq:sm-auto-016}
\end{align}
The vanishing follows solely from the Brillouin-zone total derivative.

\subsection{\texorpdfstring{Vanishing of $A^{(2)}_{2,ijk}$}{Vanishing of A2ijk (2)}}

Finally, $A^{(2)}_{2,ijk}$ is obtained from the previous calculation by
exchanging the role of the external momentum derivatives $j$ and $k$ in the
appropriate places.  
The coefficient $A^{(2)}_{2,ijk}$ is given by 
\begin{align}
A^{(2)}_{2,ijk} = \frac{1}{2}\int_{-\infty}^{\infty}\frac{d\varepsilon}{2\pi}f(\varepsilon)\mathcal{I}^{(2)}_{2,ijk}, 
\end{align} 
and its integrand is
\begin{align}\mathcal I^{(2)}_{2,ijk}
={}&\sum_{{\bm k}}
\operatorname{Tr}\bigg[
  V_{i}\partial_{\varepsilon}^{2}G^{\mathrm{R}} V_{j}G^{\mathrm{R}} V_{k}\Lambda
 +2V_{i}\partial_{\varepsilon}G^{\mathrm{R}} V_{j}\partial_{\varepsilon}G^{\mathrm{R}} V_{k}\Lambda
 + V_{i}G^{\mathrm{R}} V_{j}\partial_{\varepsilon}^{2}G^{\mathrm{R}} V_{k}\Lambda
\nonumber\\
&\quad
 +V_{i}\partial_{\varepsilon}^{2}G^{\mathrm{R}} V_{k}G^{\mathrm{R}} V_{j}\Lambda
 + V_{i}G^{\mathrm{R}} V_{j}\Lambda V_{k}\partial_{\varepsilon}^{2}G^{\mathrm{A}}
 + V_{i}\partial_{\varepsilon}^{2}G^{\mathrm{R}} V_{k}\Lambda V_{j}G^{\mathrm{A}}
\nonumber\\
&\quad
 + V_{i}\Lambda V_{j}G^{\mathrm{A}} V_{k}\partial_{\varepsilon}^{2}G^{\mathrm{A}}
 + V_{i}\Lambda V_{k}\partial_{\varepsilon}^{2}G^{\mathrm{A}} V_{j}G^{\mathrm{A}}
 +2V_{i}\Lambda V_{k}\partial_{\varepsilon}G^{\mathrm{A}} V_{j}\partial_{\varepsilon}G^{\mathrm{A}}
\nonumber\\
&\quad
 + V_{i}\Lambda V_{k}G^{\mathrm{A}} V_{j}\partial_{\varepsilon}^{2}G^{\mathrm{A}}
 + V_{ij}\partial_{\varepsilon}^{2}G^{\mathrm{R}} V_{k}\Lambda
 + V_{ij}\Lambda V_{k}\partial_{\varepsilon}^{2}G^{\mathrm{A}}
\nonumber\\
&\quad
 + V_{i}\partial_{\varepsilon}^{2}G^{\mathrm{R}} V_{jk}\Lambda
 + V_{i}\Lambda V_{jk}\partial_{\varepsilon}^{2}G^{\mathrm{A}}
\bigg].
\label{eq:I22-original-supp}
\end{align}
The terms are collected by the following $j$-derivative identities:
\begin{align}\partial_{j}\operatorname{Tr}\bigl[V_{i}\partial_{\varepsilon}^{2}G^{\mathrm{R}} V_{k}G^{\mathrm{R}}\bigr]
={}&\operatorname{Tr}\bigg[
 V_{ij}\partial_{\varepsilon}^{2}G^{\mathrm{R}} V_{k}G^{\mathrm{R}}
 +V_{i}\partial_{\varepsilon}^{2}G^{\mathrm{R}} V_{jk}G^{\mathrm{R}}
\nonumber\\
&\quad
 + V_{i}\partial_{\varepsilon}^{2}G^{\mathrm{R}} V_{k}G^{\mathrm{R}} V_{j}G^{\mathrm{R}}
 + V_{i}\partial_{\varepsilon}^{2}G^{\mathrm{R}} V_{j}G^{\mathrm{R}} V_{k}G^{\mathrm{R}}
\nonumber\\
&\quad
 +2V_{i}\partial_{\varepsilon}G^{\mathrm{R}} V_{j}\partial_{\varepsilon}G^{\mathrm{R}} V_{k}G^{\mathrm{R}}
 + V_{i}G^{\mathrm{R}} V_{j}\partial_{\varepsilon}^{2}G^{\mathrm{R}} V_{k}G^{\mathrm{R}}
\bigg],
\label{eq:A22-RR-identity-supp}\\[1mm]
\partial_{j}\operatorname{Tr}\bigl[V_{i}G^{\mathrm{A}} V_{k}\partial_{\varepsilon}^{2}G^{\mathrm{A}}\bigr]
={}&\operatorname{Tr}\bigg[
  V_{ij}G^{\mathrm{A}} V_{k}\partial_{\varepsilon}^{2}G^{\mathrm{A}}
 + V_{i}G^{\mathrm{A}} V_{jk}\partial_{\varepsilon}^{2}G^{\mathrm{A}}
\nonumber\\
&\quad
 + V_{i}G^{\mathrm{A}} V_{j}G^{\mathrm{A}} V_{k}\partial_{\varepsilon}^{2}G^{\mathrm{A}}
 + V_{i}G^{\mathrm{A}} V_{k}G^{\mathrm{A}} V_{j}\partial_{\varepsilon}^{2}G^{\mathrm{A}}
\nonumber\\
&\quad
 +2V_{i}G^{\mathrm{A}} V_{k}\partial_{\varepsilon}G^{\mathrm{A}} V_{j}\partial_{\varepsilon}G^{\mathrm{A}}
 + V_{i}G^{\mathrm{A}} V_{k}\partial_{\varepsilon}^{2}G^{\mathrm{A}} V_{j}G^{\mathrm{A}}
\bigg].
\label{eq:A22-AA-identity-supp}
\end{align}
For the mixed sector we use
\begin{align}\partial_{j}\operatorname{Tr}\bigl[V_{i}\partial_{\varepsilon}^{2}G^{\mathrm{R}} V_{k}G^{\mathrm{A}}\bigr]
={}&\operatorname{Tr}\bigg[
  V_{ij}\partial_{\varepsilon}^{2}G^{\mathrm{R}} V_{k}G^{\mathrm{A}}
 + V_{i}\partial_{\varepsilon}^{2}G^{\mathrm{R}} V_{jk}G^{\mathrm{A}}
 + V_{i}\partial_{\varepsilon}^{2}G^{\mathrm{R}} V_{k}G^{\mathrm{A}} V_{j}G^{\mathrm{A}}
\nonumber\\
&\quad
 +V_{i}\partial_{\varepsilon}^{2}G^{\mathrm{R}} V_{j}G^{\mathrm{R}} V_{k}G^{\mathrm{A}}
 +2V_{i}\partial_{\varepsilon}G^{\mathrm{R}} V_{j}\partial_{\varepsilon}G^{\mathrm{R}} V_{k}G^{\mathrm{A}}
 + V_{i}G^{\mathrm{R}} V_{j}\partial_{\varepsilon}^{2}G^{\mathrm{R}} V_{k}G^{\mathrm{A}}
\bigg],
\label{eq:A22-RA1-identity-supp}\\[1mm]
\partial_{j}\operatorname{Tr}\bigl[V_{i}G^{\mathrm{R}} V_{k}\partial_{\varepsilon}^{2}G^{\mathrm{A}}\bigr]
={}&\operatorname{Tr}\bigg[
  V_{ij}G^{\mathrm{R}} V_{k}\partial_{\varepsilon}^{2}G^{\mathrm{A}}
 +V_{i}G^{\mathrm{R}} V_{jk}\partial_{\varepsilon}^{2}G^{\mathrm{A}}
 + V_{i}G^{\mathrm{R}} V_{j}G^{\mathrm{R}} V_{k}\partial_{\varepsilon}^{2}G^{\mathrm{A}}
\nonumber\\
&\quad
 + V_{i}G^{\mathrm{R}} V_{k}G^{\mathrm{A}} V_{j}\partial_{\varepsilon}^{2}G^{\mathrm{A}}
 +2V_{i}G^{\mathrm{R}} V_{k}\partial_{\varepsilon}G^{\mathrm{A}} V_{j}\partial_{\varepsilon}G^{\mathrm{A}}
 + V_{i}G^{\mathrm{R}} V_{k}\partial_{\varepsilon}^{2}G^{\mathrm{A}} V_{j}G^{\mathrm{A}}
\bigg].
\label{eq:A22-RA2-identity-supp}
\end{align}
Combining these identities after inserting $\Lambda=G^{\mathrm{A}}-G^{\mathrm{R}}$ gives
\begin{align}  \mathcal I^{(2)}_{2,ijk}
  =\sum_{{\bm k}}\partial_{j}\operatorname{Tr}\bigl[
      V_{i}G^{\mathrm{A}} V_{k}\partial_{\varepsilon}^{2}G^{\mathrm{A}}
      -V_{i}\partial_{\varepsilon}^{2}G^{\mathrm{R}} V_{k}G^{\mathrm{R}}
      +V_{i}\partial_{\varepsilon}^{2}G^{\mathrm{R}} V_{k}G^{\mathrm{A}}
      -V_{i}G^{\mathrm{R}} V_{k}\partial_{\varepsilon}^{2}G^{\mathrm{A}}
    \bigr].
  \label{eq:I22-total-deriv-supp}
\end{align}
Therefore,
\begin{align}  A^{(2)}_{2,ijk}
  &=\frac{1}{2}\int_{-\infty}^{\infty}\frac{d\varepsilon}{2\pi}f(\varepsilon)\mathcal{I}^{(2)}_{2,ijk}\nonumber\\
  &=\frac{1}{2}\int_{-\infty}^{\infty}\frac{\mathrm{d}\varepsilon}{2\pi}
    f(\varepsilon)\sum_{\bm{k}}\partial_{j}\operatorname{Tr}\bigl[
      V_{i}G^{\mathrm{A}} V_{k}\partial_{\varepsilon}^{2}G^{\mathrm{A}}
      -V_{i}\partial_{\varepsilon}^{2}G^{\mathrm{R}} V_{k}G^{\mathrm{R}}
      +V_{i}\partial_{\varepsilon}^{2}G^{\mathrm{R}} V_{k}G^{\mathrm{A}}
      -V_{i}G^{\mathrm{R}} V_{k}\partial_{\varepsilon}^{2}G^{\mathrm{A}}
    \bigr]\nonumber\\
  &=0.
\label{eq:sm-auto-017}
\end{align}

\subsection{\texorpdfstring{Second-order dc conductivity $\sigma^{\mathrm{DC}}_{ijk}$}{Second-order dc conductivity sigma DC}}

Taking $\omega_{1},\omega_{2}\to 0$ in
Eq.~\eqref{eq:sigma2-dc-expansion}, all spurious
coefficients vanish and only $\sigma^{\mathrm{DC}}_{ijk}$
remains.
Collecting the $\hbar^{2}\omega_{1}\omega_{2}$ coefficient
from the frequency expansion of
Eq.~\eqref{eq:sigma2-finite-frequency}, one obtains
\begin{align}
  \sigma^{\mathrm{DC}}_{ijk}
  &= -\frac{ie^{3}\hbar^{2}}{2}
     \sum_{\bm{k}}
     \int_{-\infty}^{\infty}\frac{d\varepsilon}{2\pi}\,
     f(\varepsilon)\,
     \mathrm{Tr}\bigg[
  \nonumber\\
  &\quad
     V_{i}\partial_{\varepsilon}^{2}G^{\mathrm{R}}
     V_{j}G^{\mathrm{R}}V_{k}\Lambda
   + V_{i}\partial_{\varepsilon}G^{\mathrm{R}}
     V_{j}\partial_{\varepsilon}G^{\mathrm{R}}V_{k}\Lambda
   + V_{i}\partial_{\varepsilon}^{2}G^{\mathrm{R}}
     V_{k}G^{\mathrm{R}}V_{j}\Lambda
   + V_{i}\partial_{\varepsilon}G^{\mathrm{R}}
     V_{k}\partial_{\varepsilon}G^{\mathrm{R}}V_{j}\Lambda
  \nonumber\\
  &\quad
   - V_{i}\partial_{\varepsilon}G^{\mathrm{R}}
     V_{j}\Lambda\,V_{k}\partial_{\varepsilon}G^{\mathrm{A}}
   - V_{i}\partial_{\varepsilon}G^{\mathrm{R}}
     V_{k}\Lambda\,V_{j}\partial_{\varepsilon}G^{\mathrm{A}}
  \nonumber\\
  &\quad
   + V_{i}\Lambda\,V_{j}G^{\mathrm{A}}V_{k}\partial_{\varepsilon}^{2}G^{\mathrm{A}}
   + V_{i}\Lambda\,V_{j}\partial_{\varepsilon}G^{\mathrm{A}}V_{k}\partial_{\varepsilon}G^{\mathrm{A}}
   + V_{i}\Lambda\,V_{k}G^{\mathrm{A}}V_{j}\partial_{\varepsilon}^{2}G^{\mathrm{A}}
   + V_{i}\Lambda\,V_{k}\partial_{\varepsilon}G^{\mathrm{A}}V_{j}\partial_{\varepsilon}G^{\mathrm{A}}
  \nonumber\\
  &\quad
   + V_{i}\partial_{\varepsilon}^{2}G^{\mathrm{R}}V_{jk}\Lambda
   + V_{i}\Lambda\,V_{jk}\partial_{\varepsilon}^{2}G^{\mathrm{A}}
     \bigg].
  \label{eq:sigma-dc-before-rewrite}
\end{align}
Expanding $\Lambda = G^{\mathrm{A}} - G^{\mathrm{R}}$
and separating the purely retarded (RRR), purely advanced
(AAA), mixed retarded-advanced (RRA, RAA), and two-point
(RA) contributions, one obtains
\begin{align}
  \sigma^{\mathrm{DC}}_{ijk}
  &= -\frac{ie^{3}\hbar^{2}}{2}
     \sum_{\bm{k}}
     \int_{-\infty}^{\infty}\frac{d\varepsilon}{2\pi}\,
     f(\varepsilon)\,
     \mathrm{Tr}\bigg[
  \nonumber\\
  &\quad
   - V_{i}\partial_{\varepsilon}^{2}G^{\mathrm{R}}
     V_{j}G^{\mathrm{R}}V_{k}G^{\mathrm{R}}
   - V_{i}\partial_{\varepsilon}G^{\mathrm{R}}
     V_{j}\partial_{\varepsilon}G^{\mathrm{R}}V_{k}G^{\mathrm{R}}
   - V_{i}\partial_{\varepsilon}^{2}G^{\mathrm{R}}
     V_{k}G^{\mathrm{R}}V_{j}G^{\mathrm{R}}
   - V_{i}\partial_{\varepsilon}G^{\mathrm{R}}
     V_{k}\partial_{\varepsilon}G^{\mathrm{R}}V_{j}G^{\mathrm{R}}
   - V_{i}\partial_{\varepsilon}^{2}G^{\mathrm{R}}V_{jk}G^{\mathrm{R}}
  \nonumber\\
  &\quad
   + V_{i}G^{\mathrm{A}}V_{j}G^{\mathrm{A}}V_{k}\partial_{\varepsilon}^{2}G^{\mathrm{A}}
   + V_{i}G^{\mathrm{A}}V_{j}\partial_{\varepsilon}G^{\mathrm{A}}V_{k}\partial_{\varepsilon}G^{\mathrm{A}}
   + V_{i}G^{\mathrm{A}}V_{k}G^{\mathrm{A}}V_{j}\partial_{\varepsilon}^{2}G^{\mathrm{A}}
   + V_{i}G^{\mathrm{A}}V_{k}\partial_{\varepsilon}G^{\mathrm{A}}V_{j}\partial_{\varepsilon}G^{\mathrm{A}}
   + V_{i}G^{\mathrm{A}}V_{jk}\partial_{\varepsilon}^{2}G^{\mathrm{A}}
  \nonumber\\
  &\quad
   + V_{i}\partial_{\varepsilon}^{2}G^{\mathrm{R}}V_{j}G^{\mathrm{R}}V_{k}G^{\mathrm{A}}
   + V_{i}\partial_{\varepsilon}G^{\mathrm{R}}V_{j}\partial_{\varepsilon}G^{\mathrm{R}}V_{k}G^{\mathrm{A}}
   + V_{i}\partial_{\varepsilon}G^{\mathrm{R}}V_{j}G^{\mathrm{R}}V_{k}\partial_{\varepsilon}G^{\mathrm{A}}
  \nonumber\\
  &\quad
   + V_{i}\partial_{\varepsilon}^{2}G^{\mathrm{R}}V_{k}G^{\mathrm{R}}V_{j}G^{\mathrm{A}}
   + V_{i}\partial_{\varepsilon}G^{\mathrm{R}}V_{k}\partial_{\varepsilon}G^{\mathrm{R}}V_{j}G^{\mathrm{A}}
   + V_{i}\partial_{\varepsilon}G^{\mathrm{R}}V_{k}G^{\mathrm{R}}V_{j}\partial_{\varepsilon}G^{\mathrm{A}}
  \nonumber\\
  &\quad
   - V_{i}G^{\mathrm{R}}V_{j}G^{\mathrm{A}}V_{k}\partial_{\varepsilon}^{2}G^{\mathrm{A}}
   - V_{i}G^{\mathrm{R}}V_{j}\partial_{\varepsilon}G^{\mathrm{A}}V_{k}\partial_{\varepsilon}G^{\mathrm{A}}
   - V_{i}\partial_{\varepsilon}G^{\mathrm{R}}V_{j}G^{\mathrm{A}}V_{k}\partial_{\varepsilon}G^{\mathrm{A}}
  \nonumber\\
  &\quad
   - V_{i}G^{\mathrm{R}}V_{k}G^{\mathrm{A}}V_{j}\partial_{\varepsilon}^{2}G^{\mathrm{A}}
   - V_{i}G^{\mathrm{R}}V_{k}\partial_{\varepsilon}G^{\mathrm{A}}V_{j}\partial_{\varepsilon}G^{\mathrm{A}}
   - V_{i}\partial_{\varepsilon}G^{\mathrm{R}}V_{k}G^{\mathrm{A}}V_{j}\partial_{\varepsilon}G^{\mathrm{A}}
  \nonumber\\
  &\quad
   + V_{i}\partial_{\varepsilon}^{2}G^{\mathrm{R}}V_{jk}G^{\mathrm{A}}
   - V_{i}G^{\mathrm{R}}V_{jk}\partial_{\varepsilon}^{2}G^{\mathrm{A}}
     \bigg].
  \label{eq:sigma-dc-RA-separated}
\end{align}
To simplify the purely retarded and advanced terms, we use the
cyclic property of the trace and the Hermiticity relation
\begin{align}
  \mathrm{Tr}\!\left[ABC\right]
  &= \left(\mathrm{Tr}\!\left[C^{\dagger}B^{\dagger}A^{\dagger}\right]\right)^{*}
   = \left(\mathrm{Tr}\!\left[A^{\dagger}C^{\dagger}B^{\dagger}\right]\right)^{*},
  \label{eq:trace-cyclic-herm}
\end{align}
together with $(G^{\mathrm{A}})^{\dagger} = G^{\mathrm{R}}$
and $V_{i}^{\dagger} = V_{i}$, which give
\begin{align}
  \mathrm{Tr}\!\left[
    V_{i}G^{\mathrm{A}}V_{j}G^{\mathrm{A}}V_{k}
    \partial_{\varepsilon}^{2}G^{\mathrm{A}}
  \right]
  = \left(
    \mathrm{Tr}\!\left[
      V_{i}\partial_{\varepsilon}^{2}G^{\mathrm{R}}
      V_{k}G^{\mathrm{R}}V_{j}G^{\mathrm{R}}
    \right]
  \right)^{*}.
  \label{eq:AAA-to-RRR}
\end{align}
Using Eq.~\eqref{eq:AAA-to-RRR} and analogous relations
for all purely advanced terms, the RRR and AAA
contributions combine into imaginary parts, and the
mixed RRA and RAA terms can be reorganized as total
energy derivatives.
One finds
\begin{align}
  \sigma^{\mathrm{DC}}_{ijk}
  &= e^{3}\hbar^{2}
     \sum_{\bm{k}}
     \int_{-\infty}^{\infty}\frac{d\varepsilon}{2\pi}\,
     f(\varepsilon)
  \bigg[
  \nonumber\\
  &\quad
   - \mathrm{Im}\,\mathrm{Tr}\!\left[
       V_{i}\partial_{\varepsilon}^{2}G^{\mathrm{R}}
       V_{j}G^{\mathrm{R}}V_{k}G^{\mathrm{R}}
     \right]
   - \mathrm{Im}\,\mathrm{Tr}\!\left[
       V_{i}\partial_{\varepsilon}G^{\mathrm{R}}
       V_{j}\partial_{\varepsilon}G^{\mathrm{R}}V_{k}G^{\mathrm{R}}
     \right]
  \nonumber\\
  &\quad
   - \mathrm{Im}\,\mathrm{Tr}\!\left[
       V_{i}\partial_{\varepsilon}^{2}G^{\mathrm{R}}
       V_{k}G^{\mathrm{R}}V_{j}G^{\mathrm{R}}
     \right]
   - \mathrm{Im}\,\mathrm{Tr}\!\left[
       V_{i}\partial_{\varepsilon}G^{\mathrm{R}}
       V_{k}\partial_{\varepsilon}G^{\mathrm{R}}V_{j}G^{\mathrm{R}}
     \right]\nonumber\\
     &\quad
   - \mathrm{Im}\,\mathrm{Tr}\!\left[
       V_{i}\partial_{\varepsilon}^{2}G^{\mathrm{R}}V_{jk}G^{\mathrm{R}}
     \right]
  \nonumber\\
  &\quad
   + \frac{\partial}{\partial\varepsilon}
     \mathrm{Im}\,\mathrm{Tr}\!\left[
       V_{i}\partial_{\varepsilon}G^{\mathrm{R}}
       V_{j}G^{\mathrm{R}}V_{k}G^{\mathrm{A}}
     \right]
  \nonumber\\
  &\quad
   + \frac{\partial}{\partial\varepsilon}
     \mathrm{Im}\,\mathrm{Tr}\!\left[
       V_{i}\partial_{\varepsilon}G^{\mathrm{R}}
       V_{k}G^{\mathrm{R}}V_{j}G^{\mathrm{A}}
     \right]
  \nonumber\\
  &\quad
   + \frac{\partial}{\partial\varepsilon}
     \mathrm{Im}\,\mathrm{Tr}\!\left[
       V_{i}\partial_{\varepsilon}G^{\mathrm{R}}
       V_{jk}G^{\mathrm{A}}
     \right]
     \bigg].
  \label{eq:sigma-dc-with-deps}
\end{align}
Using 
\begin{align}
\partial_{\varepsilon}^2G^{\mathrm R}V_{k}G^{\mathrm R}+
\partial_{\varepsilon}G^{\mathrm R}V_{k}\partial_{\varepsilon}G^{\mathrm R}=
\partial_{\varepsilon}(\partial_{\varepsilon}G^{\mathrm R}V_{k}G^{\mathrm R})
,
\end{align}
the final form is
\begin{align}
  \sigma^{\mathrm{DC}}_{ijk}
  &= e^{3}\hbar^{2}
     \sum_{\bm{k}}
     \int_{-\infty}^{\infty}\frac{d\varepsilon}{2\pi}\,
     f(\varepsilon)\,
     \mathrm{Im}\,\mathrm{Tr}\bigg[
  \nonumber\\
  &\quad
     \partial_{\varepsilon}\!\left(
       V_{i}\partial_{\varepsilon}G^{\mathrm{R}}
       V_{j}G^{\mathrm{R}}V_{k}G^{\mathrm{A}}
     \right)
        - V_{i}\partial_{\varepsilon}\!\left(
       \partial_{\varepsilon}G^{\mathrm{R}}V_{j}G^{\mathrm{R}}
     \right)V_{k}G^{\mathrm{R}}
\nonumber\\
&\quad
   + \partial_{\varepsilon}\!\left(
       V_{i}\partial_{\varepsilon}G^{\mathrm{R}}
       V_{k}G^{\mathrm{R}}V_{j}G^{\mathrm{A}}
     \right)
 - V_{i}\partial_{\varepsilon}\!\left(
       \partial_{\varepsilon}G^{\mathrm{R}}V_{k}G^{\mathrm{R}}
     \right)V_{j}G^{\mathrm{R}}
   \nonumber\\
   &\quad
   + \partial_{\varepsilon}\!\left(
       V_{i}\partial_{\varepsilon}G^{\mathrm{R}}
       V_{jk}G^{\mathrm{A}}
     \right)
      - V_{i}\partial_{\varepsilon}^{2}G^{\mathrm{R}}
       V_{jk}G^{\mathrm{R}}
     \bigg],
  \label{eq:sigma-dc-trace}
\end{align}
which is the central result
Eq.~\eqref{eq:sigma-dc-trace} of the main text.

\section{Band-basis representation of the dc conductivities}
\label{sec:dc-conductivity-band}
This section collects the band-basis identities used to convert the
matrix Green's function expressions into a band-geometric form.  
We use standard conventions for the Berry connection, Berry curvature, and
quantum geometric tensor\cite{Berry1984,Provost1980,XiaoChangNiu2010,Resta2011}.

\subsection{Bloch band basis}
\label{subsec:Bloch band basis}

We evaluate the trace sums in Eq.~\eqref{eq:sigma-dc-trace}
by transforming to the band basis, in which the Green's
functions are diagonal.
Let $U(\bm{k})$ be the unitary matrix that diagonalizes
the Bloch Hamiltonian,
\begin{align}
  U^{\dagger}(\bm{k})\,H(\bm{k})\,U(\bm{k})
  = \mathcal{E}(\bm{k})
  \equiv \mathrm{diag}\bigl(
    \varepsilon_{1}(\bm{k}),\ldots,\varepsilon_{N}(\bm{k})
  \bigr),
  \label{eq:band-diagonalization-main}
\end{align}
whose columns $|u_{n}(\bm{k})\rangle$ are the Bloch
eigenstates satisfying
\begin{align}
  H(\bm{k})\,|u_{n}(\bm{k})\rangle
  = \varepsilon_{n}(\bm{k})\,|u_{n}(\bm{k})\rangle,
  \qquad n = 1,\ldots,N.
  \label{eq:bloch-eigen}
\end{align}
In the band basis, the retarded and advanced Green's
functions are diagonal,
\begin{align}
  g^{\mathrm{R}}_{n}(\varepsilon)
  &\equiv \langle u_{n}|G^{\mathrm{R}}(\varepsilon)|u_{n}\rangle
   = \frac{1}{\varepsilon - \varepsilon_{n} + i\gamma},
  \label{eq:gR-band}
  \\
  g^{\mathrm{A}}_{n}(\varepsilon)
  &\equiv \langle u_{n}|G^{\mathrm{A}}(\varepsilon)|u_{n}\rangle
   = \frac{1}{\varepsilon - \varepsilon_{n} - i\gamma}
   = \bigl(g^{\mathrm{R}}_{n}(\varepsilon)\bigr)^{*},
  \label{eq:green-functions-band}
\end{align}
where the chemical potential has been absorbed into the
energy variable $\varepsilon\to\varepsilon+\mu$ and
$\gamma = \hbar/(2\tau) > 0$ is the quasiparticle
broadening.
The corresponding spectral function is
\begin{align}
  \lambda_{n}(\varepsilon)
  \equiv g^{\mathrm{A}}_{n}(\varepsilon)
       - g^{\mathrm{R}}_{n}(\varepsilon)
  = \frac{2i\gamma}
         {(\varepsilon - \varepsilon_{n})^{2} + \gamma^{2}},
  \label{eq:lambda-band}
\end{align}
which reduces to
$\lambda_{n}(\varepsilon)\to 2\pi i\,\delta(\varepsilon
 - \varepsilon_{n})$ in the clean limit $\gamma\to 0^{+}$.

We work in the clean dc limit by retaining contributions
through $\mathcal{O}(\tau^{0})$ under the non-degenerate
band condition
\begin{align}
  \frac{|\varepsilon_{nm}|\,\tau}{\hbar} \gg 1,
  \qquad n \neq m,
  \label{eq:clean-limit-condition}
\end{align}
where $\varepsilon_{nm} \equiv \varepsilon_{n}-\varepsilon_{m}$
is the interband energy difference.
This condition ensures that interband coherences are
suppressed on the timescale $\tau$, allowing each band
to be treated independently.
Degenerate or nearly degenerate subspaces require a
non-Abelian generalization~\cite{Culcer2005} and are
left for future work.

\subsection{Linear and second-order dc conductivities in the band basis}
\label{subsec:velocity-geometric-tensors}
Inserting complete sets of eigenstates
$\sum_{n}|u_{n}\rangle\langle u_{n}| = I$
between each pair of matrices in
Eqs.~\eqref{eq:sigma-dc-linear-kubo} and \eqref{eq:sigma-dc-trace},
the linear and second-order dc conductivities take the form
\begin{align}
  \sigma^{\mathrm{DC}}_{ij} &= 
   2\frac{e^2}{\hbar}
     \int_{-\infty}^{\infty}\frac{d\varepsilon}{2\pi}\,
     f(\varepsilon)\,
     \sum_{\bm{k}}
     \mathrm{Re}\!\left[\mathcal{B}_{ij}\right],
  \label{eq:sigma-linear-dc-band}\\
  \sigma^{\mathrm{DC}}_{ijk}
  &= \frac{e^{3}}{\hbar}
     \int_{-\infty}^{\infty}\frac{d\varepsilon}{2\pi}\,
     f(\varepsilon)
     \sum_{\bm{k}}
     \mathrm{Im}\!\left[\mathcal{B}_{ijk}\right],
  \label{eq:sigma-dc-band}
\end{align}
where $\mathcal{B}_{ij}$ and $\mathcal{B}_{ijk}$ are a combination of
the band-basis velocity matrix elements
$v^{nm}_{i} = \langle u_{n}|\hbar V_{i}|u_{m}\rangle$ and
$v^{nm}_{ij} = \langle u_{n}|\hbar^{2}V_{ij}|u_{m}\rangle$
with the Green's function kernels
$\mathcal{K}_{nm}$, $\mathcal{K}^{(1)}_{nml}$ and $\mathcal{K}^{(2)}_{nm}$:
\begin{align}
  \mathcal{B}_{ij}&= \sum_{n,m}v^{nm}_{i}v^{mn}_{j}\mathcal{K}_{nm}
  \label{eq:Bij-band}\\
  \mathcal{B}_{ijk}
  &= \sum_{n,m,l}
     \bigl(
       v^{nm}_{i}v^{ml}_{j}v^{ln}_{k}
     + v^{nm}_{i}v^{ml}_{k}v^{ln}_{j}
     \bigr)
     \mathcal{K}^{(1)}_{nml}
  + \sum_{n,m}
    v^{nm}_{i}\,v^{mn}_{jk}\,
    \mathcal{K}^{(2)}_{nm}.
  \label{eq:Bijk-band}
\end{align}
The three-index kernel $\mathcal{K}^{(1)}_{nml}$ and
the two-index kernel $\mathcal{K}_{nm}, \mathcal{K}^{(2)}_{nm}$ are defined by
\begin{align}
  \mathcal{K}_{nm} &\equiv  ( \partial_{\varepsilon}g^{\mathrm{R}}_{m})\,g^{\mathrm{A}}_{n}
     - \,(\partial_{\varepsilon}g^{\mathrm{R}}_{m})\,g^{\mathrm{R}}_{n}
  \label{eq:K-def-band}\\
  \mathcal{K}^{(1)}_{nml}
  &\equiv
  \partial_{\varepsilon}
  \!\left[
    (\partial_{\varepsilon}g^{\mathrm{R}}_{m})\,
    g^{\mathrm{R}}_{l}\,g^{\mathrm{A}}_{n}
  \right]
  - \partial_{\varepsilon}
  \!\left[
    (\partial_{\varepsilon}g^{\mathrm{R}}_{m})\,
    g^{\mathrm{R}}_{l}
  \right]g^{\mathrm{R}}_{n},
  \label{eq:K1-def-band}
  \\
  \mathcal{K}^{(2)}_{nm}
  &\equiv
  \partial_{\varepsilon}
  \!\left[
    (\partial_{\varepsilon}g^{\mathrm{R}}_{m})\,
    g^{\mathrm{A}}_{n}
  \right]
  - (\partial_{\varepsilon}^{2}g^{\mathrm{R}}_{m})\,
    g^{\mathrm{R}}_{n},
  \label{eq:K2-def-band}
\end{align}
and encode the energy-resolved Green's function structure
after the band-basis projection.
Their explicit product forms and clean-limit expansion
are given in Sec.~\ref{sec:kernel-decomposition-clean-limit}.

\subsection{Band geometry in the velocity matrix}

The velocity vertex $V_{i} = \partial_{i}H/\hbar$ is represented in the band basis by the matrix elements
$v^{nm}_{i} = \langle u_{n}|\hbar V_{i}|u_{m}\rangle$,
which decompose as 
\begin{align}
  v^{nm}_{i}
  &\equiv
  \bigl[U^{\dagger}(\bm{k})\,\hbar V_i(\bm{k})\,U(\bm{k})\bigr]_{nm}\nonumber\\
  &
  = \langle u_n|\hbar V_i|u_m\rangle
  \nonumber\\
  &=\langle u_n|\partial_{i}H|u_m\rangle\nonumber\\
  &= \partial_i\varepsilon_n\,\delta_{nm}
   - \varepsilon_m\langle\partial_i u_n|u_m\rangle
   - \varepsilon_n\langle u_n|\partial_i u_m\rangle
  \nonumber\\
  &= \partial_i\varepsilon_n\,\delta_{nm}
   - \varepsilon_{nm}\langle u_n|\partial_i u_m\rangle\nonumber\\
   &=u^{n}_{i}\delta_{nm}+i\varepsilon_{nm}\mathcal{A}^{nm}_{i}
  \label{eq:velocity-decomp}
\end{align}
where $u^{n}_{i} \equiv \partial_{i}\varepsilon_{n}$
is the band-energy gradient (the group velocity multiplied by
$\hbar$), and
$\varepsilon_{nm}\equiv\varepsilon_{n}-\varepsilon_{m}$
is the interband energy difference.
The off-diagonal part is governed by the Berry connection
\begin{align}
  \mathcal{A}^{nm}_{i}(\bm{k})
  = i\langle u_{n}(\bm{k})|\partial_{i}u_{m}(\bm{k})\rangle,
  \label{eq:berry-connection-main}
\end{align}
where $\mathcal{A}^{nn}_{i}$ is the diagonal Berry
connection and the off-diagonal elements
$\mathcal{A}^{nm}_{i}$ ($n\neq m$) enter the interband
optical matrix elements.

The second-rank velocity vertex
$V_{ij} = \partial_{i}\partial_{j}H/\hbar^2$ is
represented in the band basis by the matrix elements
$v^{nm}_{ij} = \langle u_{n}|\hbar^{2}V_{ij}|u_{m}\rangle$,
whose general form is
\begin{align}
  v^{nm}_{ij} &\equiv
  \bigl[U^{\dagger}(\bm{k})\,\hbar^{2} V_{ij}(\bm{k})\,U(\bm{k})\bigr]_{nm}\nonumber\\
  &
  = \langle u_n|\hbar^2 V_{ij}|u_m\rangle
  \nonumber\\
  &=\langle u_n|\partial_{i}\partial_{j}H|u_m\rangle\nonumber\\
  &=\partial_{j}\langle u_n|\partial_{i}H|u_m\rangle
  -\langle \partial_{j}u_n|\partial_{i}H|u_m\rangle-\langle u_n|\partial_{i}H|\partial_{j}u_m\rangle\nonumber\\
  &= \partial_{j}v^{nm}_{i} +i\sum_{p}\left(
  i\langle \partial_{j}u_n|u_{p}\rangle\langle u_{p}|\partial_{i}H|u_m\rangle
 + i\langle u_n|\partial_{i}H|u_{p}\rangle\langle u_{p}|\partial_{j}u_m\rangle\right)\nonumber\\
 &= \partial_{j}v^{nm}_{i} + i\sum_{p}\left(v^{np}_{i}\mathcal{A}^{pm}_{j}-\mathcal{A}^{np}_{j}v^{pm}_{i}\right)\nonumber\\
 &= \partial_{j}v^{nm}_{i} +i\left[v_{i},\mathcal{A}_{j}\right]_{nm}.
  \label{eq:vnm-ij-general}
\end{align}
For the diagonal element, Eq.~\eqref{eq:vnm-ij-general} reduces to
\begin{align}
  v^{nn}_{ij}
  &= \partial_{j}v^{nn}_{i} +  i\sum_{p}\left(v^{np}_{i}\mathcal{A}^{pn}_{j}-\mathcal{A}^{np}_{j}v^{pn}_{i}\right)\nonumber\\
  &= \partial_i\partial_j\varepsilon_n
   - \sum_{p\neq n}\varepsilon_{np}
     \bigl(
       \mathcal A^{np}_{i}\mathcal A^{pn}_{j}
       +\mathcal A^{np}_{j}\mathcal A^{pn}_{i}
     \bigr)
  \nonumber\\
  &= 
  u^{n}_{ij} -2\sum_{p\neq n}\varepsilon_{np}\,\mathcal G^{np}_{ij}, 
  \label{eq:vnn-ij}
\end{align}
where
\begin{align}
  u^{n}_{ij} \equiv \partial_{i}\partial_{j}\varepsilon_{n}
  \label{eq:velocity-2-d-decomp}
\end{align}
is the second derivative of the band dispersion 
and $\mathcal{G}^{np}_{ij}$ is the
band-pair-resolved \textit{quantum metric} defined below. 
The off-diagonal part ($n\neq m$) is 
\begin{align}
  v^{nm}_{ij}
  &= \partial_{j}v^{nm}_{i} +  i\sum_{p}\left(v^{np}_{i}\mathcal{A}^{pm}_{j}-\mathcal{A}^{np}_{j}v^{pm}_{i}\right)
  \nonumber\\
  &=
  \partial_{j}\left(i\varepsilon_{nm}\mathcal{A}^{nm}_{i}\right)
   +i\left(v^{nn}_{i}\mathcal{A}^{nm}_{j}-\mathcal{A}^{nn}_{j}v^{nm}_{i}\right)
   +i\left(v^{nm}_{i}\mathcal{A}^{mm}_{j}-\mathcal{A}^{nm}_{j}v^{mm}_{i}\right)
  + i\sum_{p\neq n,m}\left(v^{np}_{i}\mathcal{A}^{pm}_{j}-\mathcal{A}^{np}_{j}v^{pm}_{i}\right)
  \nonumber\\
  &=i\left(\partial_{j}\varepsilon_{nm}\right)\mathcal{A}^{nm}_{i}+i\left(\partial_{i}\varepsilon_{nm}\right)\mathcal{A}^{nm}_{j}
  +i\varepsilon_{nm}\left\{\partial_{j}\mathcal{A}^{nm}_{i}+i\left(\mathcal{A}^{mm}_{j}-\mathcal{A}^{nn}_{j}
  \right)\mathcal{A}^{nm}_{i}\right\}
  + i\sum_{p\neq n,m}\left(v^{np}_{i}\mathcal{A}^{pm}_{j}-\mathcal{A}^{np}_{j}v^{pm}_{i}\right)\nonumber\\
  &= i u^{nm}_{i}\mathcal{A}^{nm}_{j}
   + i u^{nm}_{j}\mathcal{A}^{nm}_{i}
   + i\varepsilon_{nm}\,\mathcal{D}_{j}\mathcal{A}^{nm}_{i}
  - \sum_{p\neq n,m}
    \bigl(
      \varepsilon_{np}\,\mathcal{A}^{np}_{i}\mathcal{A}^{pm}_{j}
    + \varepsilon_{mp}\,\mathcal{A}^{np}_{j}\mathcal{A}^{pm}_{i}
    \bigr),
  \label{eq:vnm-ij}
\end{align}
where
\begin{align}
  u^{nm}_{i}
  \equiv \partial_{i}\varepsilon_{nm}
  = \partial_{i}\varepsilon_{n} - \partial_{i}\varepsilon_{m},
  \label{eq:velocity-2-od-decomp}
\end{align}
and $\mathcal{D}_{j}\mathcal{A}^{nm}_{i}$ is the 
\textit{covariant
derivative} defined below.
Although Eq.~\eqref{eq:vnm-ij} is not manifestly symmetric
under $i\leftrightarrow j$, both orderings are equivalent
since they originate from
$U^{\dagger}\partial_{i}\partial_{j}H\,U$, which is
symmetric by commutativity of partial derivatives.
Here $\mathcal{G}^{nm}_{ij}$ in Eq.~\eqref{eq:vnn-ij} 
denotes the band-pair-resolved
quantum metric, defined as the real part of the
band-pair-resolved \textit{quantum geometric tensor}
\begin{align}
  \mathcal{Q}_{ij}^{nm}
  \equiv \mathcal{A}_{i}^{nm}\mathcal{A}_{j}^{mn}
  = \mathcal{G}_{ij}^{nm} - \frac{i}{2}\Omega_{ij}^{nm},
  \qquad n\neq m.
  \label{eq:pair-qgt}
\end{align}
where $\Omega_{ij}^{nm}$ is the band-pair-resolved 
\textit{Berry curvature}. 
Explicitly,
\begin{align}
  \mathcal{G}_{ij}^{nm}
  &= \mathrm{Re}\,\mathcal{Q}_{ij}^{nm}
   = \frac{1}{2}
     \bigl(
       \mathcal{A}_{i}^{nm}\mathcal{A}_{j}^{mn}
     + \mathcal{A}_{j}^{nm}\mathcal{A}_{i}^{mn}
     \bigr),
  \label{eq:quantum-metric}
  \\
  \Omega_{ij}^{nm}
  &= -2\,\mathrm{Im}\,\mathcal{Q}_{ij}^{nm}
   = i\bigl(
       \mathcal{A}_{i}^{nm}\mathcal{A}_{j}^{mn}
     - \mathcal{A}_{j}^{nm}\mathcal{A}_{i}^{mn}
     \bigr).
  \label{eq:berry-curvature-pair}
\end{align}
The band-resolved quantum metric and Berry curvature are
obtained by summing over partner bands,
\begin{align}
  \mathcal{G}_{ij}^{n}
  = \sum_{m\neq n}\mathcal{G}_{ij}^{nm},
  \qquad
  \Omega_{ij}^{n}
  = \sum_{m\neq n}\Omega_{ij}^{nm}.
  \label{eq:metric-curvature-band}
\end{align}

The covariant derivative appearing in Eq.~\eqref{eq:vnm-ij}
is defined as 
\begin{align}
  \mathcal{D}_{j}\mathcal{A}^{nm}_{i}
  \equiv \partial_{j}\mathcal{A}^{nm}_{i}
  + i\bigl(
      \mathcal{A}^{mm}_{j} - \mathcal{A}^{nn}_{j}
    \bigr)\mathcal{A}^{nm}_{i}.
  \label{eq:covariant-derivative}
\end{align}
Under the band-dependent gauge transformation
$|u_{n}\rangle\to e^{i\chi_{n}}|u_{n}\rangle$,
the off-diagonal Berry connection transforms as
\begin{align}
  \mathcal{A}^{nm}_{i}
  \to e^{i(\chi_{m}-\chi_{n})}\mathcal{A}^{nm}_{i},
  \qquad n\neq m,
  \label{eq:gauge-transformation-offdiag-A}
\end{align}
while the diagonal connections shift as
$\mathcal{A}^{nn}_{j}\to\mathcal{A}^{nn}_{j}
 - \partial_{j}\chi_{n}$.
The ordinary derivative $\partial_{j}\mathcal{A}^{nm}_{i}$
acquires an extra term $\propto\partial_{j}(\chi_{m}-\chi_{n})$
under this transformation, which is cancelled by the
connection difference in Eq.~\eqref{eq:covariant-derivative},
so that $\mathcal{D}_{j}\mathcal{A}^{nm}_{i}$ transforms
covariantly,
\begin{align}
  \mathcal{D}_{j}\mathcal{A}^{nm}_{i}
  \to e^{i(\chi_{m}-\chi_{n})}
      \mathcal{D}_{j}\mathcal{A}^{nm}_{i}.
  \label{eq:gauge-transformation-covariant-derivative}
\end{align}
Consequently, the band-closed product
\begin{align}
  \mathcal{C}^{nm}_{ij|k}
  \equiv
  \mathcal{A}^{nm}_{i}\,\mathcal{D}_{k}\mathcal{A}^{mn}_{j}
  = \Gamma^{nm}_{ij|k} - i\widetilde{\Gamma}^{nm}_{ij|k}
  \label{eq:quantum-connection}
\end{align}
is gauge-invariant.
Although $\mathcal{D}_{k}\mathcal{A}^{mn}_{j}$ itself is
only gauge-covariant, carrying the phase
$e^{i(\chi_{n}-\chi_{m})}$, its product with
$\mathcal{A}^{nm}_{i}$, which carries the opposite phase
$e^{i(\chi_{m}-\chi_{n})}$, yields a band-closed object
whose phase factors cancel identically:
\begin{align}
  \mathcal{A}^{nm}_{i}\,\mathcal{D}_{k}\mathcal{A}^{mn}_{j}
  &\to
  e^{i(\chi_{m}-\chi_{n})}
  e^{i(\chi_{n}-\chi_{m})}
  \mathcal{A}^{nm}_{i}\,\mathcal{D}_{k}\mathcal{A}^{mn}_{j}
  \nonumber\\
  &= \mathcal{A}^{nm}_{i}\,\mathcal{D}_{k}\mathcal{A}^{mn}_{j}.
  \label{eq:quantum-connection-gauge-inv}
\end{align}
This gauge-invariant object provides a natural geometric
building block for second-order response.
Since $\sigma^{\mathrm{DC}}_{ijk}$ is a physical observable,
it must be gauge-invariant; the quantum connection
$\mathcal{C}^{nm}_{ij|k}$ is the appropriate
band-pair-resolved object that encodes the covariant
derivative of the Berry connection while remaining
gauge-invariant.
The real and imaginary parts are
\begin{align}
  \Gamma^{nm}_{ij|k}
  = \mathrm{Re}\,\mathcal{C}^{nm}_{ij|k},
  \qquad
  \widetilde{\Gamma}^{nm}_{ij|k}
  = -\mathrm{Im}\,\mathcal{C}^{nm}_{ij|k}.
  \label{eq:quantum-connection-real-imag}
\end{align}
These tensors are useful because they generate the momentum-space
gradients of the quantum metric and Berry curvature.  Differentiating
$\mathcal{Q}^{nm}_{ik}=\mathcal A^{nm}_{i}\mathcal A^{mn}_{k}$ and replacing
ordinary derivatives by covariant derivatives gives
\begin{align}
  \partial_j \mathcal{Q}^{nm}_{ik}
  &= \bigl(\mathcal D_j\mathcal A^{nm}_{i}\bigr)\mathcal A^{mn}_{k}
    +\mathcal A^{nm}_{i}\mathcal D_j\mathcal A^{mn}_{k}
  \nonumber\\
  &= \mathcal C^{mn}_{ki|j}+\mathcal C^{nm}_{ik|j}.
  \label{eq:dC-connection}
\end{align}
Using
$\mathcal C^{mn}_{ki|j}=(\mathcal C^{nm}_{ki|j})^{*}$,
Eq.~\eqref{eq:dC-connection} yields
\begin{align}
  \partial_j\mathcal G^{nm}_{ik}
  &= \Gamma^{nm}_{ik|j}+\Gamma^{nm}_{ki|j},
  \label{eq:dG-Gamma}
  \\
  \partial_j\Omega^{nm}_{ik}
  &= 2\bigl(
       \widetilde\Gamma^{nm}_{ik|j}
      -\widetilde\Gamma^{nm}_{ki|j}
     \bigr).
  \label{eq:dOmega-Gamma}
\end{align}
Thus $\Gamma^{nm}_{ik|j}$ and $\widetilde\Gamma^{nm}_{ik|j}$ are the
microscopic band-pair constituents of the quantum-metric dipole and the
Berry-curvature dipole, respectively.  The latter underlies the nonlinear
Hall response in time-reversal-invariant but inversion-broken systems
\cite{SodemannFu2015}; related band-geometric corrections in semiclassical
dynamics are discussed in Ref.~\cite{GaoYangNiu2014}.

\subsection{Useful covariant identities}
\label{subsec:useful-covariant-identities}
The following identities are used to reorganize derivatives of
$\Gamma^{nm}_{ik|j}$ into metric and curvature derivatives.  They follow
from the pure-gauge relation for the full non-Abelian Berry connection,
\begin{align}
  \partial_i\mathcal A^{nm}_{j}-\partial_j\mathcal A^{nm}_{i}
  = i[\mathcal A_i,\mathcal A_j]_{nm}.
  \label{eq:app-pure-gauge-identity}
\end{align}
After the external-band contributions are absorbed into covariant
derivatives, this becomes
\begin{align}
  \mathcal D_i\mathcal A^{nm}_{j}-\mathcal D_j\mathcal A^{nm}_{i}
  = i[\mathcal A_i,\mathcal A_j]'_{nm},
  \label{eq:app-covariant-curl}
\end{align}
where
\begin{align}
  [\mathcal A_i,\mathcal A_j]'_{nm}
  = \sum_{p\neq n,m}
    \bigl(
      \mathcal A^{np}_{i}\mathcal A^{pm}_{j}
      -\mathcal A^{np}_{j}\mathcal A^{pm}_{i}
    \bigr).
  \label{eq:app-primed-commutator}
\end{align}
The prime indicates that the two external bands $n$ and $m$ are excluded
from the intermediate-band sum.

Using Eq.~\eqref{eq:app-covariant-curl}, the real part of the quantum
connection can be written as
\begin{align}
  2\Gamma^{nm}_{ij|k}
  = \partial_k\mathcal G^{nm}_{ij}
   +\partial_j\mathcal G^{nm}_{ki}
   -\partial_i\mathcal G^{nm}_{jk}
   +X^{nm}_{ijk},
  \label{eq:Gamma-metric-X}
\end{align}
where the non-Abelian correction from intermediate bands is
\begin{align}
  X^{nm}_{ijk}
  &= \mathrm{Im}\!\bigl[
      \mathcal A^{nm}_{i}[\mathcal A_j,\mathcal A_k]'_{mn}
    \bigr]
   + \mathrm{Im}\!\bigl[
      \mathcal A^{nm}_{j}[\mathcal A_k,\mathcal A_i]'_{mn}
    \bigr]
   - \mathrm{Im}\!\bigl[
      \mathcal A^{nm}_{k}[\mathcal A_i,\mathcal A_j]'_{mn}
    \bigr].
  \label{eq:X-comm-app}
\end{align}
For completeness, we give the short derivation.  From
Eq.~\eqref{eq:dG-Gamma},
\begin{align}
  \partial_k\mathcal G^{nm}_{ij}
  &= \Gamma^{nm}_{ij|k}+\Gamma^{nm}_{ji|k},
  \label{eq:app-dG1}
  \\
  \partial_j\mathcal G^{nm}_{ki}
  &= \Gamma^{nm}_{ki|j}+\Gamma^{nm}_{ik|j},
  \label{eq:app-dG2}
  \\
  \partial_i\mathcal G^{nm}_{jk}
  &= \Gamma^{nm}_{jk|i}+\Gamma^{nm}_{kj|i}.
  \label{eq:app-dG3}
\end{align}
Therefore
\begin{align}
  &\partial_k\mathcal G^{nm}_{ij}
   +\partial_j\mathcal G^{nm}_{ki}
   -\partial_i\mathcal G^{nm}_{jk}
   -2\Gamma^{nm}_{ij|k}
  \nonumber\\
  &=
  \bigl(\Gamma^{nm}_{ji|k}-\Gamma^{nm}_{jk|i}\bigr)
 +\bigl(\Gamma^{nm}_{ki|j}-\Gamma^{nm}_{kj|i}\bigr)
 +\bigl(\Gamma^{nm}_{ik|j}-\Gamma^{nm}_{ij|k}\bigr).
  \label{eq:app-Gamma-difference}
\end{align}
Each difference is converted into a primed commutator by
Eq.~\eqref{eq:app-covariant-curl}:
\begin{align}
  \Gamma^{nm}_{ji|k}-\Gamma^{nm}_{jk|i}
  &=-\mathrm{Im}\!\bigl[
      \mathcal A^{nm}_{j}[\mathcal A_k,\mathcal A_i]'_{mn}
    \bigr],
  \label{eq:app-Gamma-diff-1}
  \\
  \Gamma^{nm}_{ki|j}-\Gamma^{nm}_{kj|i}
  &=\mathrm{Im}\!\bigl[
      \mathcal A^{nm}_{k}[\mathcal A_i,\mathcal A_j]'_{mn}
    \bigr],
  \label{eq:app-Gamma-diff-2}
  \\
  \Gamma^{nm}_{ik|j}-\Gamma^{nm}_{ij|k}
  &=-\mathrm{Im}\!\bigl[
      \mathcal A^{nm}_{i}[\mathcal A_j,\mathcal A_k]'_{mn}
    \bigr].
  \label{eq:app-Gamma-diff-3}
\end{align}
Substituting Eqs.~\eqref{eq:app-Gamma-diff-1}--\eqref{eq:app-Gamma-diff-3}
into Eq.~\eqref{eq:app-Gamma-difference} gives
Eq.~\eqref{eq:Gamma-metric-X}.

\subsection{Summary of band-geometric quantities and symmetries}
\label{subsec:band-geometry-summary-table}
The main band-geometric objects used in this section are summarized in
Table~\ref{tab:band-geometric-quantities}.  The table assumes isolated,
non-degenerate bands.  Under spatial inversion $\mathcal P$ or time
reversal $\mathcal T$, $\bar n$ denotes the band related to $n$ at
$-\bm k$; the listed parity refers to the comparison between the quantity
at $(n,m,\bm k)$ and the corresponding quantity at
$(\bar n,\bar m,-\bm k)$.

\begin{table*}[t]
\centering
\small
\renewcommand{\arraystretch}{1.15}
\begin{tabular}{@{}llll@{}}
\toprule
\parbox[t]{0.18\textwidth}{Quantity}
&
\parbox[t]{0.31\textwidth}{Definition}
&
\parbox[t]{0.22\textwidth}{Gauge property}
&
\parbox[t]{0.22\textwidth}{Useful symmetries}
\\
\midrule
\parbox[t]{0.18\textwidth}{Band dispersion derivatives}
&
\parbox[t]{0.31\textwidth}{
$u_i^n=\partial_i\varepsilon_n$,\quad
$u^n_{ij}=\partial_i\partial_j\varepsilon_n$
}
&
\parbox[t]{0.22\textwidth}{Gauge invariant}
&
\parbox[t]{0.22\textwidth}{
$u^n_{ij}=u^n_{ji}$.
Under $\mathcal P$ or $\mathcal T$,
$\varepsilon_n$ and $u^n_{ij}$ are even in $\bm k$,
whereas $u_i^n$ is odd.
}
\\
\addlinespace
\parbox[t]{0.18\textwidth}{Berry connection}
&
\parbox[t]{0.31\textwidth}{
$\mathcal A_i^{nm}=i\langle u_n|\partial_i u_m\rangle$
}
&
\parbox[t]{0.22\textwidth}{
$\mathcal A_i^{nm}\to
e^{i(\chi_m-\chi_n)}\mathcal A_i^{nm}$ for $n\neq m$;
$\mathcal A_i^{nn}\to
\mathcal A_i^{nn}-\partial_i\chi_n$
}
&
\parbox[t]{0.22\textwidth}{
$\mathcal A_i^{mn}=(\mathcal A_i^{nm})^{*}$.
Gauge dependent, hence no standalone
$\mathcal P/\mathcal T$ parity is assigned.
}
\\
\addlinespace
\parbox[t]{0.18\textwidth}{Rank-one velocity matrix}
&
\parbox[t]{0.31\textwidth}{
$v_i^{nm}=u_i^n\delta_{nm}
+i\varepsilon_{nm}\mathcal A_i^{nm}(1-\delta_{nm})$
}
&
\parbox[t]{0.22\textwidth}{
Covariant for $n\neq m$; diagonal part gauge invariant
}
&
\parbox[t]{0.22\textwidth}{
$v_i^{mn}=(v_i^{nm})^{*}$
}
\\
\addlinespace
\parbox[t]{0.18\textwidth}{Pair quantum geometric tensor}
&
\parbox[t]{0.31\textwidth}{
$\mathcal{Q}_{ij}^{nm}=\mathcal A_i^{nm}\mathcal A_j^{mn}$,
$n\neq m$
}
&
\parbox[t]{0.22\textwidth}{Gauge invariant}
&
\parbox[t]{0.22\textwidth}{
$\mathcal{Q}_{ji}^{nm}=(\mathcal{Q}_{ij}^{nm})^{*}$,
$\mathcal{Q}_{ij}^{mn}=\mathcal{Q}_{ji}^{nm}$
}
\\
\addlinespace
\parbox[t]{0.18\textwidth}{Quantum metric}
&
\parbox[t]{0.31\textwidth}{
$\mathcal G_{ij}^{nm}=\mathrm{Re}\,\mathcal{Q}_{ij}^{nm}$,\quad
$\mathcal G_{ij}^{n}=\sum_{m\neq n}\mathcal G_{ij}^{nm}$
}
&
\parbox[t]{0.22\textwidth}{Gauge invariant}
&
\parbox[t]{0.22\textwidth}{
$\mathcal G_{ij}^{nm}
=\mathcal G_{ji}^{nm}
=\mathcal G_{ij}^{mn}$.
Even under both $\mathcal P$ and $\mathcal T$.
}
\\
\addlinespace
\parbox[t]{0.18\textwidth}{Berry curvature}
&
\parbox[t]{0.31\textwidth}{
$\Omega_{ij}^{nm}=-2\,\mathrm{Im}\,\mathcal{Q}_{ij}^{nm}$,\quad
$\Omega_{ij}^{n}=\sum_{m\neq n}\Omega_{ij}^{nm}$
}
&
\parbox[t]{0.22\textwidth}{Gauge invariant}
&
\parbox[t]{0.22\textwidth}{
$\Omega_{ij}^{nm}
=-\Omega_{ji}^{nm}
=-\Omega_{ij}^{mn}$.
Even under $\mathcal P$, odd under $\mathcal T$.
}
\\
\addlinespace
\parbox[t]{0.18\textwidth}{Covariant derivative}
&
\parbox[t]{0.31\textwidth}{
$\mathcal D_j\mathcal A_i^{nm}
=\partial_j\mathcal A_i^{nm}
+i(\mathcal A_j^{mm}-\mathcal A_j^{nn})
\mathcal A_i^{nm}$
}
&
\parbox[t]{0.22\textwidth}{
Same covariance as $\mathcal A_i^{nm}$
}
&
\parbox[t]{0.22\textwidth}{
$(\mathcal D_j\mathcal A_i^{mn})
=(\mathcal D_j\mathcal A_i^{nm})^{*}$.
No simple $i\leftrightarrow j$ symmetry.
}
\\
\addlinespace
\parbox[t]{0.18\textwidth}{Rank-two velocity matrix}
&
\parbox[t]{0.31\textwidth}{
$v_{ij}^{nm}
=\langle u_n|\partial_i\partial_jH|u_m\rangle$
}
&
\parbox[t]{0.22\textwidth}{
Covariant for $n\neq m$; diagonal part gauge invariant
}
&
\parbox[t]{0.22\textwidth}{
$v_{ij}^{nm}=v_{ji}^{nm}$,
$v_{ij}^{mn}=(v_{ij}^{nm})^{*}$
}
\\
\addlinespace
\parbox[t]{0.18\textwidth}{Quantum connection}
&
\parbox[t]{0.31\textwidth}{
$\mathcal C_{ij|k}^{nm}
=\mathcal A_i^{nm}\mathcal D_k\mathcal A_j^{mn}
=\Gamma_{ij|k}^{nm}
-i\widetilde\Gamma_{ij|k}^{nm}$
}
&
\parbox[t]{0.22\textwidth}{Gauge invariant}
&
\parbox[t]{0.22\textwidth}{
$\mathcal C_{ij|k}^{mn}
=(\mathcal C_{ij|k}^{nm})^{*}$;
$\Gamma_{ij|k}^{mn}=\Gamma_{ij|k}^{nm}$,
$\widetilde\Gamma_{ij|k}^{mn}
=-\widetilde\Gamma_{ij|k}^{nm}$.
}
\\
\addlinespace
\parbox[t]{0.18\textwidth}{Metric and curvature gradients}
&
\parbox[t]{0.31\textwidth}{
$\partial_k\mathcal G_{ij}^{nm}
=\Gamma_{ij|k}^{nm}+\Gamma_{ji|k}^{nm}$,\quad
$\partial_k\Omega_{ij}^{nm}
=2(\widetilde\Gamma_{ij|k}^{nm}
-\widetilde\Gamma_{ji|k}^{nm})$
}
&
\parbox[t]{0.22\textwidth}{Gauge invariant}
&
\parbox[t]{0.22\textwidth}{
$\partial_k\mathcal G_{ij}$ is symmetric in $i,j$;
$\partial_k\Omega_{ij}$ is antisymmetric in $i,j$.
The metric gradient is odd under both $\mathcal P$ and
$\mathcal T$, whereas the curvature gradient is odd under
$\mathcal P$ and even under $\mathcal T$.
}
\\
\bottomrule
\end{tabular}
\caption{Band-geometric quantities appearing in the velocity-matrix expansion and their basic gauge and symmetry properties.}
\label{tab:band-geometric-quantities}
\end{table*}

\subsection{Band-geometric decomposition of the dc conductivities}
\label{subsec:band-geom-decomp}

Substituting the band-basis velocity decompositions
Eqs.~\eqref{eq:velocity-decomp},
\eqref{eq:vnn-ij}, and
\eqref{eq:vnm-ij} into
$\mathcal{B}_{ij}$ [Eq.~\eqref{eq:Bij-band}] and
$\mathcal{B}_{ijk}$ [Eq.~\eqref{eq:Bijk-band}],
every contribution can be classified according to the
number of distinct band indices involved.
For $\mathcal{B}_{ij}$, only a closed two-band product
appears; for $\mathcal{B}_{ijk}$, both closed two-band
and three-band products arise.
We treat each in turn.

The closed two-band velocity product $v^{nm}_{i}v^{mn}_{j}$
represents a two-band interband process in which two
one-photon current vertices $V_{i}$ and $V_{j}$ connect
the same band pair $(n,m)$; it corresponds to the closed
electron loop of
Fig.~\ref{fig:zeroth_first_order_current_diagrams}(b).
The decomposition reads
\begin{align}
  v_{i}^{nm}v_{j}^{mn}
  &= \bigl(
       u^{n}_{i}\delta_{nm}
     + i\varepsilon_{nm}\mathcal{A}^{nm}_{i}
     \bigr)
     \bigl(
       u^{m}_{j}\delta_{mn}
     + i\varepsilon_{mn}\mathcal{A}^{mn}_{j}
     \bigr)
\nonumber\\
  &= \underbrace{u^{n}_{i}u^{n}_{j}\,\delta_{nm}}_{\sigma^{\mathrm D}_{ij}}
   \underbrace{+ \varepsilon_{nm}^{2}\,\mathcal{Q}^{nm}_{ij}}_{\sigma^{\mathrm{QG}}_{ij}}.
  \label{eq:vv}
\end{align}
The first term ($n=m$, purely intraband) is the product
$u^{n}_{i}u^{n}_{j}$ of two group velocities of band $n$ 
and generates the Drude sector $\sigma^{\mathrm{D}}_{ij}$.
The second term ($n\neq m$) involves the product of two
interband Berry connections and generates the quantum
geometric tensor $\mathcal{Q}^{nm}_{ij}$, contributing to the
quantum-geometric sector $\sigma^{\mathrm{QG}}_{ij}$.

The closed three-band product
$v^{nm}_{i}v^{ml}_{j}v^{ln}_{k}$ represents a
three-band interband process in which three one-photon
current vertices $V_{i}$, $V_{j}$, and $V_{k}$ connect
three distinct bands $n$, $m$, and $l$ in a closed cycle;
it corresponds to the triangular electron loop of
Fig.~\ref{fig:second_order_current_diagrams}(a).
The decomposition reads
\begin{align}
v_{i}^{nm}v_{j}^{ml}v_{k}^{ln}
  + v_{i}^{nm}v_{k}^{ml}v_{j}^{ln}
  &=\bigl(
       u^{n}_{i}\delta_{nm}
     + i\varepsilon_{nm}\mathcal{A}^{nm}_{i}
     \bigr)
     \bigl(
       u^{m}_{j}\delta_{ml}
     + i\varepsilon_{ml}\mathcal{A}^{ml}_{j}
     \bigr)
     \bigl(
       u^{l}_{k}\delta_{ln}
     + i\varepsilon_{ln}\mathcal{A}^{ln}_{k}
     \bigr)+ (j \leftrightarrow k)\nonumber\\
  &= \underbrace{2u_{i}^{n}u_{j}^{n}u_{k}^{n}
     \,\delta_{nm}\delta_{ml}\delta_{ln}}_{\sigma^{\mathrm ND}_{ijk}}
  \nonumber\\
  &\quad
  \underbrace{+
  \varepsilon_{nl}^{2}u_{i}^{n}
    \bigl(\mathcal{Q}^{nl}_{jk}+\mathcal{Q}^{nl}_{kj}\bigr)\delta_{nm}
  + \varepsilon_{nm}^{2}
    \bigl(u_{j}^{m}\mathcal{Q}^{nm}_{ik}+u_{k}^{m}\mathcal{Q}^{nm}_{ij}\bigr)
    \delta_{ml}
  + \varepsilon_{nm}^{2}
    \bigl(u_{j}^{n}\mathcal{Q}^{nm}_{ik}+u_{k}^{n}\mathcal{Q}^{nm}_{ij}\bigr)
    \delta_{ln}}_{\sigma^{\mathrm QG}_{ijk}}
  \nonumber\\
  &\quad
   \underbrace{- i\varepsilon_{nm}\varepsilon_{ml}\varepsilon_{ln}
    \bigl(\mathcal{T}^{nml}_{ijk}+\mathcal{T}^{nml}_{ikj}\bigr)}_{\sigma^{\mathcal T}_{ijk}},
  \label{eq:vvv}
\end{align}
where
\begin{align}
  \mathcal{T}^{nml}_{ijk}
  \equiv
  \mathcal{A}_{i}^{nm}\mathcal{A}_{j}^{ml}\mathcal{A}_{k}^{ln}
  \label{eq:three-Berry-connection}
\end{align}
is the \textit{three-Berry-connection} factor, in which the band
indices form a closed cycle $n\to m\to l\to n$.
It is gauge-invariant because the three relative phases
$e^{i(\chi_{m}-\chi_{n})}$,
$e^{i(\chi_{l}-\chi_{m})}$, and
$e^{i(\chi_{n}-\chi_{l})}$ cancel in the product
[cf.~Eq.~\eqref{eq:gauge-transformation-offdiag-A}].
Physically, $\mathcal{T}^{nml}_{ijk}$ represents a
three-band interband process corresponding to the
triangular electron loop of
Fig.~\ref{fig:second_order_current_diagrams}(a).

Each line of Eq.~\eqref{eq:vvv} contributes to a distinct
sector of the second-order dc conductivity.
The first line ($n=m=l$, purely intraband) contains the
product $2u^{n}_{i}u^{n}_{j}u^{n}_{k}$ of three group
velocities of band $n$ and generates the nonlinear Drude
sector $\sigma^{\mathrm{ND}}_{ijk}$.
The second through fourth lines each involve one intraband
and one interband pair, coupling a group velocity to the
quantum geometric tensor $\mathcal{Q}^{nm}_{ij}$; together
they contribute to the
quantum-geometric sector $\sigma^{\mathrm{QG}}_{ijk}$.
The fifth line ($n,m,l$ all distinct) is proportional to
the three-Berry-connection factor $\mathcal{T}^{nml}_{ijk}$
and defines the three-connection sector
$\sigma^{\mathcal{T}}_{ijk}$.

The closed two-band product $v^{nm}_{i}v^{mn}_{jk}$
represents a two-band process in which a one-photon
current vertex $V_{i}$ and a two-photon contact vertex
$V_{jk}$ connect the same band pair $(n,m)$; it
corresponds to the diagrams of
Figs.~\ref{fig:second_order_current_diagrams}(b) and (c).
The decomposition is
\begin{align}
  v_{i}^{nm}v_{jk}^{mn}
  &=\bigl(
       u^{n}_{i}\delta_{nm}
     + i\varepsilon_{nm}\mathcal{A}^{nm}_{i}
     \bigr)
     \bigg\{ \bigg(u^{n}_{jk} -2\sum_{p\neq n}\varepsilon_{np}\,\mathcal G^{np}_{jk}\bigg)\delta_{mn}\nonumber\\
     &\quad
     +i u^{mn}_{j}\mathcal{A}^{mn}_{k}
   + i u^{mn}_{k}\mathcal{A}^{mn}_{j}
   + i\varepsilon_{mn}\,\mathcal{D}_{k}\mathcal{A}^{mn}_{j}
  - \sum_{p\neq n,m}
    \bigl(
      \varepsilon_{mp}\,\mathcal{A}^{mp}_{j}\mathcal{A}^{pn}_{k}
    + \varepsilon_{np}\,\mathcal{A}^{mp}_{k}\mathcal{A}^{pn}_{j}
    \bigr)
     \bigg\}\nonumber\\
  &= 
 \underbrace{u_{i}^{n}u_{jk}^{n}\delta_{nm}}_{\sigma^{\rm ND}_{ijk}}\nonumber\\
 &
 \underbrace{
 - 2u_{i}^{n}\sum_{p\neq n}\varepsilon_{np}\mathcal{G}^{np}_{jk}\delta_{nm}
  + \varepsilon_{nm}(\partial_{j}\varepsilon_{nm})\,\mathcal{Q}^{nm}_{ik}
  + \varepsilon_{nm}(\partial_{k}\varepsilon_{nm})\,\mathcal{Q}^{nm}_{ij}}_{\sigma^{\rm QG}_{ijk}}
  \nonumber\\
  &\quad
\underbrace{  + \varepsilon_{nm}^{2}\,\mathcal{C}^{nm}_{ij|k}}_{\sigma^{\mathcal C}_{ijk}}
  \nonumber\\
  &\quad
\underbrace{   - i\sum_{p\neq n,m}
    \Bigl(
      \varepsilon_{nm}\varepsilon_{mp}\,\mathcal{T}^{nmp}_{ijk}
    + \varepsilon_{nm}\varepsilon_{np}\,\mathcal{T}^{nmp}_{ikj}
    \Bigr)}_{\sigma^{\mathcal T}_{ijk}}.
  \label{eq:v-v2}
\end{align}
The first term ($\delta_{nm}$, intraband) collects the
single-band Drude term $u^{n}_{i}u^{n}_{jk}$, which
contributes to $\sigma^{\mathrm{ND}}_{ijk}$ together with
the first line of Eq.~\eqref{eq:vvv}, and the
quantum-metric correction
$-2u^{n}_{i}\sum_{p\neq n}\varepsilon_{np}\mathcal{G}^{np}_{jk}$
from the off-diagonal part of the diagonal second-rank
velocity [Eq.~\eqref{eq:vnn-ij}], which contributes to
$\sigma^{\mathrm{QG}}_{ijk}$.
The next two terms couple gradients of the interband
energy difference to the quantum geometric tensor and also
contribute to $\sigma^{\mathrm{QG}}_{ijk}$.
The following term,
$\varepsilon_{nm}^{2}\mathcal{C}^{nm}_{ij|k}
 = \varepsilon_{nm}^{2}
   \mathcal{A}^{nm}_{i}\mathcal{D}_{k}\mathcal{A}^{mn}_{j}$,
is the quantum-connection term that defines the
covariant-connection sector $\sigma^{\mathcal{C}}_{ijk}$.
The final term generates an additional three-connection
contribution $\sigma^{\mathcal{T}}_{ijk}$ involving an
intermediate band $p$ ($p\neq n,m$).

Collecting all contributions, the linear and the second-order dc conductivities 
decompose as
\begin{align}
  \sigma^{\mathrm{DC}}_{ij} &= \sigma^{\mathrm{D}}_{ij}+\sigma^{\mathrm{QG}}_{ij}, 
  \label{eq:sigma-dc-linear}\\
  \sigma^{\mathrm{DC}}_{ijk}
  &= \sigma^{\mathrm{ND}}_{ijk}
  + \sigma^{\mathrm{QG}}_{ijk}
  + \sigma^{\mathcal{C}}_{ijk}
  + \sigma^{\mathcal{T}}_{ijk}.
  \label{eq:sigma-dc-intermediate}
\end{align}

The Drude sector $\sigma^{\mathrm{D}}_{ij}$ originates from the single-band contributions and is given by 
\begin{align}
  \sigma^{\mathrm{D}}_{ij}
  &= 2\frac{e^{2}}{\hbar}
     \int_{-\infty}^{\infty}\frac{d\varepsilon}{2\pi}\,
     f(\varepsilon)
     \sum_{\bm{k}}
     \mathrm{Re}\!\left[\mathcal{B}^{\mathrm{D}}_{ij}\right],
  \label{eq:sigma-D}
  \\
  \mathcal{B}^{\mathrm{D}}_{ij}
  &= \sum_{n}
       (\partial_{i}\varepsilon_{n})
       (\partial_{j}\varepsilon_{n})
       \mathcal{K}_{nn}. 
  \label{eq:B-D}
\end{align}
The quantum-geometric sector $\sigma^{\mathrm{QG}}_{ij}$
originates from interband contributions and is given by
\begin{align}
  \sigma^{\mathrm{QG}}_{ij}
  &= 2\frac{e^{2}}{\hbar}
     \int_{-\infty}^{\infty}\frac{d\varepsilon}{2\pi}\,
     f(\varepsilon)
     \sum_{\bm{k}}
     \mathrm{Re}\!\left[\mathcal{B}^{\mathrm{QG}}_{ij}\right],
  \label{eq:sigma-linear-QG}
  \\
  \mathcal{B}^{\mathrm{QG}}_{ij}
  &= {\sum_{n,m}}^{\prime}\varepsilon_{nm}^{2}\mathcal{Q}^{nm}_{ij}\mathcal{K}_{nm}. 
  \label{eq:B-linear-QG}
\end{align}

The nonlinear Drude sector $\sigma^{\mathrm{ND}}_{ijk}$
originates from the single-band contributions and is
given by
\begin{align}
  \sigma^{\mathrm{ND}}_{ijk}
  &= \frac{e^{3}}{\hbar}
     \int_{-\infty}^{\infty}\frac{d\varepsilon}{2\pi}\,
     f(\varepsilon)
     \sum_{\bm{k}}
     \mathrm{Im}\!\left[\mathcal{B}^{\mathrm{ND}}_{ijk}\right],
  \label{eq:sigma-ND}
  \\
  \mathcal{B}^{\mathrm{ND}}_{ijk}
  &= \sum_{n}
     \biggl[
       2(\partial_{i}\varepsilon_{n})
       (\partial_{j}\varepsilon_{n})
       (\partial_{k}\varepsilon_{n})\,
       \mathcal{K}^{(1)}_{nnn}
     + (\partial_{i}\varepsilon_{n})
       (\partial_{j}\partial_{k}\varepsilon_{n})\,
       \mathcal{K}^{(2)}_{nn}
     \biggr].
  \label{eq:B-ND}
\end{align}
The quantum-geometric sector $\sigma^{\mathrm{QG}}_{ijk}$
originates from products involving the quantum geometric
tensor $\mathcal{Q}^{nm}_{ij}$ and is given by
\begin{align}
  \sigma^{\mathrm{QG}}_{ijk}
  &= \frac{e^{3}}{\hbar}
     \int_{-\infty}^{\infty}\frac{d\varepsilon}{2\pi}\,
     f(\varepsilon)
     \sum_{\bm{k}}
     \mathrm{Im}\!\left[\mathcal{B}^{\mathrm{QG}}_{ijk}\right],
  \label{eq:sigma-QG}
\end{align}
with
\begin{align}
  \mathcal{B}^{\mathrm{QG}}_{ijk}
  &= {\sum_{n,m}}^{\prime}
     \varepsilon_{nm}(\partial_{i}\varepsilon_{n})
     \bigl(\mathcal{Q}^{nm}_{jk} + \mathcal{Q}^{nm}_{kj}\bigr)
     \bigl[\varepsilon_{nm}\mathcal{K}^{(1)}_{nnm}
           - \mathcal{K}^{(2)}_{nn}\bigr]
  \nonumber\\
  &\quad
  + {\sum_{n,m}}^{\prime}
    \varepsilon_{nm}\mathcal{Q}^{nm}_{ik}
    \Bigl[
      (\partial_{j}\varepsilon_{m})
      \bigl(\varepsilon_{nm}\mathcal{K}^{(1)}_{nmm}
            - \mathcal{K}^{(2)}_{nm}\bigr)
    + (\partial_{j}\varepsilon_{n})
      \bigl(\varepsilon_{nm}\mathcal{K}^{(1)}_{nmn}
            + \mathcal{K}^{(2)}_{nm}\bigr)
    \Bigr]
  \nonumber\\
  &\quad
  + {\sum_{n,m}}^{\prime}
    \varepsilon_{nm}\mathcal{Q}^{nm}_{ij}
    \Bigl[
      (\partial_{k}\varepsilon_{m})
      \bigl(\varepsilon_{nm}\mathcal{K}^{(1)}_{nmm}
            - \mathcal{K}^{(2)}_{nm}\bigr)
    + (\partial_{k}\varepsilon_{n})
      \bigl(\varepsilon_{nm}\mathcal{K}^{(1)}_{nmn}
            + \mathcal{K}^{(2)}_{nm}\bigr)
    \Bigr].
  \label{eq:B-QG}
\end{align}
The imaginary part of $\mathcal{B}^{\mathrm{QG}}_{ijk}$
decomposes according to the real and imaginary parts of
the quantum geometric tensor
$\mathcal{Q}^{nm}_{ij}
 = \mathcal{G}^{nm}_{ij} - (i/2)\Omega^{nm}_{ij}$,
giving rise to the BCD and quantum-metric contributions
after the clean-limit kernel expansion.
The covariant-connection sector
$\sigma^{\mathcal{C}}_{ijk}$ arises from the
$\mathcal{D}_{k}\mathcal{A}^{mn}_{j}$ term in
Eq.~\eqref{eq:vnm-ij} and is given by
\begin{align}
  \sigma^{\mathcal{C}}_{ijk}
  &= \frac{e^{3}}{\hbar}
     \int_{-\infty}^{\infty}\frac{d\varepsilon}{2\pi}\,
     f(\varepsilon)
     \sum_{\bm{k}}
     \mathrm{Im}\!\left[\mathcal{B}^{\mathcal{C}}_{ijk}\right],
  \label{eq:sigma-C}
  \\
  \mathcal{B}^{\mathcal{C}}_{ijk}
  &= {\sum_{n,m}}^{\prime}
     \varepsilon_{nm}^{2}\,
     \mathcal{C}^{nm}_{ij|k}\,
     \mathcal{K}^{(2)}_{nm}.
  \label{eq:B-C}
\end{align}
Finally, the three-connection sector
$\sigma^{\mathcal{T}}_{ijk}$ originates from products of
three off-diagonal Berry connections and is given by
\begin{align}
  \sigma^{\mathcal{T}}_{ijk}
  &= \frac{e^{3}}{\hbar}
     \int_{-\infty}^{\infty}\frac{d\varepsilon}{2\pi}\,
     f(\varepsilon)
     \sum_{\bm{k}}
     \mathrm{Im}\!\left[\mathcal{B}^{\mathcal{T}}_{ijk}\right],
  \label{eq:sigma-T}
  \\
  \mathcal{B}^{\mathcal{T}}_{ijk}
  &= -i{\sum_{n,m,l}}^{\prime}
     \biggl\{
       \varepsilon_{nm}\varepsilon_{ml}\varepsilon_{ln}
       \mathcal{K}^{(1)}_{nml}
       \bigl(
         \mathcal{T}^{nml}_{ijk}
       + \mathcal{T}^{nml}_{ikj}
       \bigr)
     + \varepsilon_{nm}\mathcal{K}^{(2)}_{nm}
       \bigl(
         \varepsilon_{ml}\mathcal{T}^{nml}_{ijk}
       - \varepsilon_{nl}\mathcal{T}^{nml}_{ikj}
       \bigr)
     \biggr\}.
  \label{eq:B-T}
\end{align}

\section{Kernel decomposition and clean-limit evaluation}
\label{sec:kernel-decomposition-clean-limit}
\label{sec:dc-calculations}

To evaluate the linear conductivity $\sigma^{\mathrm{DC}}_{ij}$ from
Eq.~\eqref{eq:sigma-dc-linear} and the second-order conductivity
$\sigma^{\mathrm{DC}}_{ijk}$ from Eqs.~\eqref{eq:sigma-dc-intermediate},
we first expand the Green's function kernels
$\mathcal{K}_{nm}$, $\mathcal{K}^{(1)}_{nml}$ and
$\mathcal{K}^{(2)}_{nm}$ in the clean limit.  In the following,
all terms through $\mathcal{O}(\tau^{0})$ are retained under the
non-degenerate band condition
$|\varepsilon_{nm}|\tau/\hbar\gg 1$ for $n\neq m$.
Throughout we set $\mu = 0$ for notational brevity, suppress the
crystal-momentum argument $\bm{k}$, and introduce the shorthand 
\begin{align}
  R_{n} \equiv g^{\mathrm{R}}_{n}(\varepsilon)
  &= \frac{1}{\varepsilon - \varepsilon_{n} + i\gamma},
  \label{eq:Rn-def}
  \\
  A_{n} \equiv g^{\mathrm{A}}_{n}(\varepsilon)
  &= \frac{1}{\varepsilon - \varepsilon_{n} - i\gamma},
  \qquad \gamma = \frac{\hbar}{2\tau},
  \label{eq:An-def}
\end{align}
together with
\begin{align}
  r   &\equiv \frac{\tau}{i\hbar} = \frac{1}{2i\gamma},
  \label{eq:r-def}
  \\
  \delta_{n} &\equiv \delta(\varepsilon - \varepsilon_{n}).
  \label{eq:delta-n-def}
\end{align}

\subsection{Algebraic identities for products and clean-limit behavior of powers of Green's functions}

Three identities are used repeatedly to decompose products
of scalar retarded and advanced propagators.

For the same-band retarded--advanced product,
\begin{align}
  R_{n}A_{n} = r(A_{n} - R_{n}),
  \label{eq:RA-same}
\end{align}
which follows from $A_{n} - R_{n} = 2i\gamma R_{n}A_{n}$
and $r = 1/(2i\gamma)$.
For two retarded propagators at different band energies,
partial fractions give
\begin{align}
  R_{m}R_{n} = \frac{R_{n} - R_{m}}{\varepsilon_{nm}}.
  \label{eq:RR-diff}
\end{align}
For a mixed retarded--advanced product at different bands,
\begin{align}
  R_{m}A_{n}
  &= \frac{A_{n} - R_{m}}{\varepsilon_{nm} + 2i\gamma}
  \nonumber\\
  &= \frac{A_{n} - R_{m}}{\varepsilon_{nm}}
   - \frac{2i\gamma}{\varepsilon_{nm}^{2}}(A_{n} - R_{m})
   + \mathcal{O}(\tau^{-2}).
  \label{eq:RA-diff}
\end{align}
The second line holds in the non-degenerate clean limit
$|\varepsilon_{nm}|\tau/\hbar \gg 1$, under which the
$2i\gamma$ correction is $\mathcal{O}(\tau^{-1})$.

In the clean limit $\gamma \to 0^{+}$, the Sokhotski--Plemelj
formula gives 
\begin{align}
  \mathrm{Im}\,R_{n}^{p}
  &= -\frac{\pi(-1)^{p-1}}{(p-1)!}
     \partial_{\varepsilon}^{p-1}\delta_{n},
  \label{eq:Im-Rp-clean}
  \\
  \mathrm{Im}\,A_{n}^{p}
  &= +\frac{\pi(-1)^{p-1}}{(p-1)!}
     \partial_{\varepsilon}^{p-1}\delta_{n}.
  \label{eq:Im-Ap-clean}
\end{align}
In particular,
\begin{align}
  A_{n} - R_{n}
  &\xrightarrow{\gamma\to 0^{+}} 2\pi i\,\delta_{n},
  \label{eq:AminusR-clean}
  \\
  A_{n}^{2} - R_{n}^{2}
  &\xrightarrow{\gamma\to 0^{+}} -2\pi i\,\partial_{\varepsilon}\delta_{n}.
  \label{eq:A2minusR2-clean}
\end{align}
Note that $r^{2} = -\tau^{2}/\hbar^{2}$ since
$r = \tau/(i\hbar)$; this sign is important for the
$\tau$-power counting below.

\subsection{Product forms of the kernels}

Using $\partial_{\varepsilon}R_{m} = -R_{m}^{2}$ and
$\partial_{\varepsilon}^{2}R_{m} = 2R_{m}^{3}$, the
kernels defined in Eqs.~\eqref{eq:K1-def-band} and
\eqref{eq:K2-def-band} take the product forms
\begin{align}
&\mathcal{K}_{nm}
=  -R^{2}_{m}(A_{n}-R_{n}),
\label{eq:K-product}
\\
&\mathcal{K}^{(1)}_{nml}
= \bigl(2R_{m}^{3}R_{l} + R_{m}^{2}R_{l}^{2}\bigr)
(A_{n} - R_{n})
+ R_{m}^{2}R_{l}\,A_{n}^{2},
\label{eq:K1-product}
\\
&\mathcal{K}^{(2)}_{nm}
= 2R_{m}^{3}(A_{n} - R_{n}) + R_{m}^{2}\,A_{n}^{2}.
\label{eq:K2-product}
\end{align}
The following subsections expand each kernel in powers of
$\tau$ using the identities
Eqs.~\eqref{eq:RA-same}--\eqref{eq:RA-diff}.
For compactness in the partial-fraction manipulations below, we also write
\begin{align}
  s \equiv 2i\gamma,
  \qquad r=\frac{1}{s}.
  \label{eq:s-def}
\end{align}

\subsection{\texorpdfstring{Fully diagonal kernels:
$\mathcal{K}_{nn}$, $\mathcal{K}^{(1)}_{nnn}$ and $\mathcal{K}^{(2)}_{nn}$}
{Fully diagonal kernels}}

Throughout this subsection, the retarded and advanced Green
functions and the broadening convention are
\begin{align}
R_n&=\frac{1}{\varepsilon-\varepsilon_n+i\gamma},
\qquad
A_n=\frac{1}{\varepsilon-\varepsilon_n-i\gamma},
\qquad
\gamma>0,
\\
r&=\frac{1}{2i\gamma}=\frac{\tau}{i\hbar},
\qquad
\tau=\frac{\hbar}{2\gamma}.
\end{align}
The energy variable is $\varepsilon$, whereas $\varepsilon_n$
denotes the band energy at fixed crystal momentum.
We use $\delta_n=\delta(\varepsilon-\varepsilon_n)$ for the
Dirac delta distribution.  Energy derivatives act at fixed
$\varepsilon_n$ and fixed $\gamma$, so that
\begin{align}
\partial_\varepsilon^p R_n&=(-1)^p p!R_n^{p+1},
\quad
\partial_\varepsilon^p A_n=(-1)^p p!A_n^{p+1}.
\end{align}
These identities explain the conversion from products of
Green functions to the derivative expressions below.
For the energy integrals, we take fixed $\mu$ and fixed $T>0$
and define
\begin{align}
f(\varepsilon)&=
\frac{1}{\exp[(\varepsilon-\mu)/(k_{\mathrm B}T)]+1},
\qquad
f_n^{(p)}=
\left.\partial_\varepsilon^p f(\varepsilon)
\right|_{\varepsilon=\varepsilon_n}.
\end{align}
Here $f_n=f_n^{(0)}$, and primes denote these energy derivatives.

Setting $m=n$ in Eqs.~\eqref{eq:K-product} and
\eqref{eq:K2-product}, and $m=l=n$ in
Eq.~\eqref{eq:K1-product}, we obtain
\begin{align}
&\mathcal{K}_{nn}=-R_n^2(A_n-R_n),
\label{eq:K-nn-first}
\\
&\mathcal{K}^{(1)}_{nnn}
=3R_n^4(A_n-R_n)+R_n^3A_n^2,
\label{eq:K1-nnn-first}
\\
&\mathcal{K}^{(2)}_{nn}
=2R_n^3(A_n-R_n)+R_n^2A_n^2.
\label{eq:K2-nn-first}
\end{align}

For $\mathcal{K}_{nn}$, applying Eq.~\eqref{eq:RA-same}
repeatedly gives
\begin{align}
\mathcal{K}_{nn}
&=-R_n^2(A_n-R_n)
\nonumber\\
&=-R_n(R_nA_n-R_n^2)
\nonumber\\
&=-rR_n(A_n-R_n)+R_n^3
\nonumber\\
&=-r\{R_nA_n-R_n^2\}+R_n^3
\nonumber\\
&=-r^2(A_n-R_n)+rR_n^2+R_n^3
\nonumber\\
&=-r^2(A_n-R_n)
-r\partial_\varepsilon R_n
+\frac12\partial_\varepsilon^2R_n
\nonumber\\
&=\left(\frac{\tau}{\hbar}\right)^2(A_n-R_n)
+\left(\frac{\tau}{\hbar}\right)
\partial_\varepsilon(iR_n)
+\frac12\partial_\varepsilon^2R_n.
\label{eq:K-nn-expanded}
\end{align}
The leading real and imaginary parts of $\mathcal{K}_{nn}$ are
\begin{align}
\left.\mathrm{Re}\,\mathcal{K}_{nn}\right|_{\mathcal O(\tau)}
&=\frac{\tau}{\hbar}
\partial_\varepsilon\mathrm{Re}(iR_n)
=-\frac{\tau}{\hbar}
\partial_\varepsilon\mathrm{Im}\,R_n
\nonumber\\
&=\pi\frac{\tau}{\hbar}\partial_\varepsilon\delta_n,
\label{eq:K-nn-leading-expanded-real}
\\
\left.\mathrm{Im}\,\mathcal{K}_{nn}\right|_{\mathcal O(\tau^2)}
&=\left(\frac{\tau}{\hbar}\right)^2
\mathrm{Im}(A_n-R_n)
=2\pi\frac{\tau^2}{\hbar^2}\delta_n.
\label{eq:K-nn-leading-expanded-img}
\end{align}

For $\mathcal{K}^{(1)}_{nnn}$, we similarly obtain
\begin{align}
\mathcal{K}^{(1)}_{nnn}
&=3R_n^4(A_n-R_n)+R_n^3A_n^2
\nonumber\\
&=3rR_n^3(A_n-R_n)-3R_n^5
+r^2R_n(A_n^2+R_n^2-2R_nA_n)
\nonumber\\
&=3r^2R_n^2(A_n-R_n)
-3rR_n^4-3R_n^5
+r^3(A_n-R_n)A_n
+r^2R_n^3
-2r^3R_n(A_n-R_n)
\nonumber\\
&=3r^3R_n(A_n-R_n)
-3r^2R_n^3-3rR_n^4-3R_n^5
+r^3A_n^2-r^4(A_n-R_n)
\nonumber\\
&\quad
+r^2R_n^3-2r^4(A_n-R_n)+2r^3R_n^2
\nonumber\\
&=3r^4(A_n-R_n)-3r^3R_n^2
-3r^2R_n^3-3rR_n^4-3R_n^5+r^3A_n^2
\nonumber\\
&\quad
-3r^4(A_n-R_n)+r^2R_n^3+2r^3R_n^2
\nonumber\\
&=r^3(A_n^2-R_n^2)-2r^2R_n^3-3rR_n^4-3R_n^5
\nonumber\\
&=-r^3\partial_\varepsilon(A_n-R_n)
-r^2\partial_\varepsilon^2R_n
+\frac r2\partial_\varepsilon^3R_n
-\frac18\partial_\varepsilon^4R_n
\nonumber\\
&=-\left(\frac{\tau}{\hbar}\right)^3
\partial_\varepsilon\{i(A_n-R_n)\}
+\left(\frac{\tau}{\hbar}\right)^2
\partial_\varepsilon^2R_n
-\frac12\left(\frac{\tau}{\hbar}\right)
\partial_\varepsilon^3(iR_n)
-\frac18\partial_\varepsilon^4R_n.
\label{eq:K1-nnn-expanded}
\end{align}
The leading real and imaginary parts of
$\mathcal{K}^{(1)}_{nnn}$ are
\begin{align}
\left.\mathrm{Re}\,\mathcal{K}^{(1)}_{nnn}
\right|_{\mathcal O(\tau^3)}
&=-\left(\frac{\tau}{\hbar}\right)^3
\partial_\varepsilon\mathrm{Re}\{i(A_n-R_n)\}
\nonumber\\
&=\left(\frac{\tau}{\hbar}\right)^3
\partial_\varepsilon\mathrm{Im}(A_n-R_n)
=2\pi\left(\frac{\tau}{\hbar}\right)^3
\partial_\varepsilon\delta_n,
\label{eq:K1-nnn-leading-expanded-real}
\\
\left.\mathrm{Im}\,\mathcal{K}^{(1)}_{nnn}
\right|_{\mathcal O(\tau^2)}
&=\left(\frac{\tau}{\hbar}\right)^2
\partial_\varepsilon^2\mathrm{Im}\,R_n
=-\pi\left(\frac{\tau}{\hbar}\right)^2
\partial_\varepsilon^2\delta_n.
\label{eq:K1-nnn-leading-expanded-img}
\end{align}

For $\mathcal{K}^{(2)}_{nn}$, applying Eq.~\eqref{eq:RA-same}
repeatedly gives
\begin{align}
\mathcal{K}^{(2)}_{nn}
&=2R_n^3(A_n-R_n)+R_n^2A_n^2
\nonumber\\
&=2rR_n^2(A_n-R_n)-2R_n^4+r^2(A_n-R_n)^2
\nonumber\\
&=2r^2R_n(A_n-R_n)-2rR_n^3-2R_n^4
+r^2(A_n^2+R_n^2-2R_nA_n)
\nonumber\\
&=r^2(A_n^2-R_n^2)-2rR_n^3-2R_n^4
\nonumber\\
&=-r^2\partial_\varepsilon(A_n-R_n)
-r\partial_\varepsilon^2R_n
+\frac13\partial_\varepsilon^3R_n
\nonumber\\
&=\left(\frac{\tau}{\hbar}\right)^2
\partial_\varepsilon(A_n-R_n)
+\left(\frac{\tau}{\hbar}\right)
\partial_\varepsilon^2(iR_n)
+\frac13\partial_\varepsilon^3R_n.
\label{eq:K2-nn-expanded}
\end{align}
The leading real and imaginary parts of
$\mathcal{K}^{(2)}_{nn}$ are
\begin{align}
\left.\mathrm{Re}\,\mathcal{K}^{(2)}_{nn}
\right|_{\mathcal O(\tau)}
&=\left(\frac{\tau}{\hbar}\right)
\partial_\varepsilon^2\mathrm{Re}(iR_n)
=-\left(\frac{\tau}{\hbar}\right)
\partial_\varepsilon^2\mathrm{Im}\,R_n
\nonumber\\
&=\pi\left(\frac{\tau}{\hbar}\right)
\partial_\varepsilon^2\delta_n,
\label{eq:K2-nn-leading-expanded-real}
\\
\left.\mathrm{Im}\,\mathcal{K}^{(2)}_{nn}
\right|_{\mathcal O(\tau^2)}
&=\left(\frac{\tau}{\hbar}\right)^2
\partial_\varepsilon\mathrm{Im}(A_n-R_n)
\nonumber\\
&=2\pi\left(\frac{\tau}{\hbar}\right)^2
\partial_\varepsilon\delta_n.
\label{eq:K2-nn-leading-expanded-img}
\end{align}

The leading expressions above are distributional clean-limit
results.  They determine the displayed leading powers of
$\tau$, but do not by themselves determine the coefficients
at order $\tau^0$.  A finite-width correction multiplied by
an explicit power of $\tau$ can contribute at that order.
We therefore examine the finite-broadening kernels before
performing the energy integrals.

\subsubsection{\texorpdfstring{Exact finite-broadening form of $\mathrm{Re}\,\mathcal{K}_{nn}$ and the absence of an $\mathcal{O}(\tau^{0})$ Drude term}
{Exact finite-broadening form of Re K nn and the absence of a tau0 Drude term}}
\label{ssec:Knn-parity}

Equation~\eqref{eq:K-nn-expanded} contains two terms with a
nonzero real part.  Its last term,
$\tfrac12\partial_\varepsilon^2R_n$, has no explicit factor
of $\tau$ and has the clean-limit real part
$\tfrac12\partial_\varepsilon^2
\mathcal P(\varepsilon-\varepsilon_n)^{-1}$.
Here $\mathcal P(\varepsilon-\varepsilon_n)^{-1}$ denotes the
Cauchy principal-value distribution.  Its derivatives are
defined distributionally by integration by parts; the
corresponding higher inverse powers are interpreted as
Hadamard finite-part distributions.  They are not ordinary
integrable functions at $\varepsilon=\varepsilon_n$.
This apparent order-$\tau^0$ term must therefore be retained
until it is combined with the subleading part of the other
real term in Eq.~\eqref{eq:K-nn-expanded}.

Introduce the dimensionless variable
\begin{align}
s&\equiv\frac{\varepsilon-\varepsilon_n}{\gamma},
\qquad
R_n=\frac{1}{\gamma(s+i)},
\qquad
A_n=\frac{1}{\gamma(s-i)}.
\label{eq:sm-s-variable}
\end{align}
The real part of the full kernel is, for every $\gamma>0$,
\begin{align}
\mathrm{Re}\,\mathcal K_{nn}
&=\frac{1}{\gamma^3}\frac{-4s}{(1+s^2)^3}.
\label{eq:ReKnn-exact}
\end{align}
For comparison, the two contributions in
Eq.~\eqref{eq:K-nn-expanded} are separately
\begin{align}
\mathrm{Re}\biggl[
\frac{\tau}{\hbar}\partial_\varepsilon(iR_n)\biggr]
&=\frac{1}{\gamma^3}a(s),
\qquad
a(s)=-\frac{s}{(1+s^2)^2},
\label{eq:ReKnn-a}
\\
\mathrm{Re}\biggl[
\frac12\partial_\varepsilon^2R_n\biggr]
&=\frac{1}{\gamma^3}b(s),
\qquad
b(s)=\frac{s^3-3s}{(1+s^2)^3}.
\label{eq:ReKnn-b}
\end{align}
Both are odd functions of $s$, and their sum satisfies
\begin{align}
a(s)+b(s)&=-\frac{4s}{(1+s^2)^3}.
\end{align}
The separate kernels decay as $s^{-3}$, whereas their sum
decays as $s^{-5}$.  This cancellation improves the
convergence of the moments of the combined kernel.

For clarity, the special functions used to evaluate the
moments are Euler's beta and gamma functions, defined for
positive real arguments by
\begin{align}
B(x,y)&\equiv\int_0^1dt\,t^{x-1}(1-t)^{y-1}
=\frac{\Gamma(x)\Gamma(y)}{\Gamma(x+y)},
\\
\Gamma(z)&\equiv\int_0^\infty dt\,t^{z-1}e^{-t}.
\end{align}
The gamma function obeys $\Gamma(z+1)=z\Gamma(z)$ and
$\Gamma(1/2)=\sqrt{\pi}$.
Using the evenness of the integrand and the substitution
$t=s^2/(1+s^2)$ on the positive half-axis gives
\begin{align}
\int_{-\infty}^{\infty}ds\,
\frac{s^{2a}}{(1+s^2)^b}
&=B\biggl(a+\frac12,b-a-\frac12\biggr)
=\frac{\Gamma(a+\frac12)\Gamma(b-a-\frac12)}{\Gamma(b)}.
\label{eq:sm-beta}
\end{align}
Here $a$ is a nonnegative integer and $b>a+1/2$.
The latter condition ensures convergence at infinity,
where the integrand scales as $|s|^{2a-2b}$.

The zeroth, first, second, and third moments of the scaled
kernel $\gamma^3\mathrm{Re}\,\mathcal K_{nn}$ are
\begin{subequations}
\begin{align}
\int_{-\infty}^{\infty}ds\,
\frac{-4s}{(1+s^2)^3}&=0,
\\
\int_{-\infty}^{\infty}ds\,
\frac{-4s^2}{(1+s^2)^3}
&=-4B\biggl(\frac32,\frac32\biggr)
=-\frac{\pi}{2},
\\
\int_{-\infty}^{\infty}ds\,
\frac{-4s^3}{(1+s^2)^3}&=0,
\\
\int_{-\infty}^{\infty}ds\,
\frac{-4s^4}{(1+s^2)^3}
&=-4B\biggl(\frac52,\frac12\biggr)
=-\frac{3\pi}{2}.
\end{align}
\label{eq:Knn-moments}
\end{subequations}
All four integrals are absolutely convergent.  The zeroth
and second moments vanish because their integrands are odd.

With $d\varepsilon=\gamma ds$, the exact energy integral is
\begin{align}
\int_{-\infty}^{\infty}\frac{d\varepsilon}{2\pi}\,
f(\varepsilon)\mathrm{Re}\,\mathcal K_{nn}
&=\frac{1}{2\pi\gamma^2}
\int_{-\infty}^{\infty}ds\,
f(\varepsilon_n+\gamma s)\frac{-4s}{(1+s^2)^3}.
\end{align}
For fixed $s$, the local Taylor expansion is
\begin{align}
f(\varepsilon_n+\gamma s)
&=\sum_{p=0}^{3}\frac{(\gamma s)^p}{p!}f_n^{(p)}
+\mathcal O\bigl((\gamma s)^4\bigr).
\end{align}
The contribution associated with a convergent $p$-th moment
scales as $\gamma^{p-2}$.  Thus $p=0,1,2,3$ correspond to
$\tau^2,\tau^1,\tau^0,\tau^{-1}$, respectively.
Only the $p=1$ and $p=3$ moments contribute, giving
\begin{align}
\int_{-\infty}^{\infty}\frac{d\varepsilon}{2\pi}\,
f(\varepsilon)\mathrm{Re}\,\mathcal K_{nn}
&=\frac{1}{2\pi\gamma^2}
\biggl[-\frac{\pi\gamma}{2}f_n'
-\frac{3\pi\gamma^3}{2\cdot3!}f_n^{(3)}\biggr]
+\mathcal O(\gamma^2)
\nonumber\\
&=-\frac{f_n'}{4\gamma}
-\frac{\gamma}{8}f_n^{(3)}+\mathcal O(\gamma^2).
\label{eq:Knn-integrated-through-gamma}
\end{align}
The first term is $-\tau f_n'/(2\hbar)$, in agreement with
Eq.~\eqref{eq:K-nn-leading-expanded-real} after the energy
integration, and generates the linear Drude conductivity
in Eq.~\eqref{eq:sigma-D-linear}.
The second moment is zero, so there is no order-$\tau^0$
linear Drude contribution.
The remainder is evaluated at fixed $T>0$.  It is not
obtained by integrating the local Taylor remainder over
the entire real axis: the higher moments cease to be
absolutely convergent.  In particular, the combined scaled
kernel decays as $s^{-5}$, and its large-$|s|$ contribution
can produce an order-$\gamma^2$ nonlocal remainder.

The absence of a constant term must not be inferred from
separately assigning zero second moments to $a(s)$ and
$b(s)$.  Indeed,
\begin{align}
s^2a(s)&=-\frac1s+\mathcal O(s^{-3}),
\quad
s^2b(s)=\frac1s+\mathcal O(s^{-3}).
\end{align}
These individual second moments are not absolutely
convergent.  Their symmetric principal values do not
suffice to determine the asymptotics of the energy
integrals against a general Fermi function.

The cancellation can instead be exhibited using the
finite-width distributional expansions
\begin{align}
\mathrm{Re}\,R_n
&=\mathcal P(\varepsilon-\varepsilon_n)^{-1}
+\pi\gamma\partial_\varepsilon\delta_n
-\frac{\gamma^2}{2}\partial_\varepsilon^2
\mathcal P(\varepsilon-\varepsilon_n)^{-1}
+\mathcal O(\gamma^3),
\\
\mathrm{Im}\,R_n
&=-\pi\delta_n
+\gamma\partial_\varepsilon
\mathcal P(\varepsilon-\varepsilon_n)^{-1}
+\frac{\pi\gamma^2}{2}\partial_\varepsilon^2\delta_n
+\mathcal O(\gamma^3).
\end{align}
These are distributional asymptotics, not pointwise
expansions at $\varepsilon=\varepsilon_n$.
They are used in the differentiated kernel combinations
before the energy integration.
Since $\mathrm{Re}(iR_n)=-\mathrm{Im}\,R_n$ and
$\tau/\hbar=1/(2\gamma)$, the two order-$\tau^0$ terms are
\begin{align}
\left.\mathrm{Re}\biggl[
\frac{\tau}{\hbar}\partial_\varepsilon(iR_n)
\biggr]\right|_{\tau^0}
&=-\frac12\partial_\varepsilon^2
\mathcal P(\varepsilon-\varepsilon_n)^{-1},
\\
\left.\mathrm{Re}\biggl[
\frac12\partial_\varepsilon^2R_n
\biggr]\right|_{\tau^0}
&=+\frac12\partial_\varepsilon^2
\mathcal P(\varepsilon-\varepsilon_n)^{-1}.
\label{eq:Knn-finite-part-cancellation}
\end{align}
The first line originates from the order-$\gamma$ correction
to the spectral factor multiplied by $\tau/\hbar$.
It cancels the finite-part distribution in the second line.
Neither contribution vanishes separately as a distribution.
Thus Eq.~\eqref{eq:sigma-D-linear} is complete through
order $\tau^0$, but is not an exact finite-broadening
formula at all subsequent orders.
This conclusion concerns the diagonal Drude channel and
does not remove an intrinsic interband Berry-curvature
response.

Parity controls the local terms only while the required
moments converge.  It does not imply that the complete
asymptotic expansion contains only odd powers of $\tau$.
At second order in the field, the constant contribution
instead involves the convergent third moment of
$\mathrm{Im}\,\mathcal K^{(2)}_{nn}$, evaluated in
Eq.~\eqref{eq:sm-M2-3} below.  This different moment counting
allows a kinetic order-$\tau^0$ term in the second-order
response even though the linear Drude response has none.

\subsubsection{\texorpdfstring{Exact finite-broadening evaluation and the
$\mathcal{O}(\tau^{0})$ remainder}{Exact finite-broadening evaluation and the tau0 remainder}}
\label{ssec:diag-tau0}

Because the main text retains contributions through
$\mathcal O(\tau^0)$, the subleading parts of
$\mathrm{Im}\,\mathcal K^{(2)}_{nn}$ and
$\mathrm{Im}\,\mathcal K^{(1)}_{nnn}$ must be retained.
As in the linear channel, they cannot be obtained by keeping
only the last term of the exact derivative expressions
in Eqs.~\eqref{eq:K2-nn-expanded} and
\eqref{eq:K1-nnn-expanded} while replacing all other
spectral factors by their leading distributional limits.
Throughout this calculation, $\delta_n$ is the Dirac delta
distribution.  The finite-width spectral function has a
Lorentzian profile; $\delta_n$ itself does not represent
that finite-width profile.

Using the variable defined in Eq.~\eqref{eq:sm-s-variable},
direct substitution into Eqs.~\eqref{eq:K2-nn-first} and
\eqref{eq:K1-nnn-first} yields
\begin{align}
\mathrm{Im}\,\mathcal K^{(2)}_{nn}
&=\frac{1}{\gamma^4}
\frac{4s(s^2-3)}{(1+s^2)^4},
\label{eq:sm-ImK2-exact}
\\
\mathrm{Im}\,\mathcal K^{(1)}_{nnn}
&=\frac{1}{\gamma^5}
\frac{5s^4-38s^2+5}{(1+s^2)^5}.
\label{eq:sm-ImK1-exact}
\end{align}
Both formulas hold for every $\gamma>0$.
The first scaled kernel is odd and the second is even.
Among the convergent moments, only odd moments of the first
and even moments of the second can therefore be nonzero.

Using the beta-function identity in Eq.~\eqref{eq:sm-beta},
the first and third moments of
$\gamma^4\mathrm{Im}\,\mathcal K^{(2)}_{nn}$ are
\begin{align}
\int_{-\infty}^{\infty}ds\,s\,
\bigl(\gamma^4\mathrm{Im}\,\mathcal K^{(2)}_{nn}\bigr)
&=\int_{-\infty}^{\infty}ds\,
\frac{4s^2(s^2-3)}{(1+s^2)^4}
\nonumber\\
&=4\biggl[B\biggl(\frac52,\frac32\biggr)
-3B\biggl(\frac32,\frac52\biggr)\biggr]
\nonumber\\
&=4\biggl(\frac{\pi}{16}-\frac{3\pi}{16}\biggr)
=-\frac{\pi}{2},
\label{eq:sm-M2-1}
\\
\int_{-\infty}^{\infty}ds\,s^3\,
\bigl(\gamma^4\mathrm{Im}\,\mathcal K^{(2)}_{nn}\bigr)
&=\int_{-\infty}^{\infty}ds\,
\frac{4s^4(s^2-3)}{(1+s^2)^4}
\nonumber\\
&=4\biggl[B\biggl(\frac72,\frac12\biggr)
-3B\biggl(\frac52,\frac32\biggr)\biggr]
\nonumber\\
&=4\biggl(\frac{5\pi}{16}-\frac{3\pi}{16}\biggr)
=\frac{\pi}{2}.
\label{eq:sm-M2-3}
\end{align}
The zeroth, second, and fourth moments of
$\gamma^5\mathrm{Im}\,\mathcal K^{(1)}_{nnn}$ are
\begin{align}
\int_{-\infty}^{\infty}ds\,
\bigl(\gamma^5\mathrm{Im}\,\mathcal K^{(1)}_{nnn}\bigr)
&=\int_{-\infty}^{\infty}ds\,
\frac{5s^4-38s^2+5}{(1+s^2)^5}
\nonumber\\
&=5B\biggl(\frac52,\frac52\biggr)
-38B\biggl(\frac32,\frac72\biggr)
+5B\biggl(\frac12,\frac92\biggr)
\nonumber\\
&=\frac{\pi}{128}(15-190+175)=0,
\label{eq:sm-M1-0}
\\
\int_{-\infty}^{\infty}ds\,s^2\,
\bigl(\gamma^5\mathrm{Im}\,\mathcal K^{(1)}_{nnn}\bigr)
&=\int_{-\infty}^{\infty}ds\,
\frac{s^2(5s^4-38s^2+5)}{(1+s^2)^5}
\nonumber\\
&=5B\biggl(\frac72,\frac32\biggr)
-38B\biggl(\frac52,\frac52\biggr)
+5B\biggl(\frac32,\frac72\biggr)
\nonumber\\
&=\frac{\pi}{128}(25-114+25)=-\frac{\pi}{2},
\label{eq:sm-M1-2}
\\
\int_{-\infty}^{\infty}ds\,s^4\,
\bigl(\gamma^5\mathrm{Im}\,\mathcal K^{(1)}_{nnn}\bigr)
&=\int_{-\infty}^{\infty}ds\,
\frac{s^4(5s^4-38s^2+5)}{(1+s^2)^5}
\nonumber\\
&=5B\biggl(\frac92,\frac12\biggr)
-38B\biggl(\frac72,\frac32\biggr)
+5B\biggl(\frac52,\frac52\biggr)
\nonumber\\
&=\frac{\pi}{128}(175-190+15)=0.
\label{eq:sm-M1-4}
\end{align}
The zeros in Eqs.~\eqref{eq:sm-M1-0} and
\eqref{eq:sm-M1-4} are cancellations between the three
polynomial terms in the numerator.  They do not follow
from parity, since both integrands are even.

After setting $\varepsilon=\varepsilon_n+\gamma s$ and
$d\varepsilon=\gamma ds$, the exact energy integrals are
\begin{align}
\int_{-\infty}^{\infty}\frac{d\varepsilon}{2\pi}\,
f(\varepsilon)\mathrm{Im}\,\mathcal K^{(2)}_{nn}
&=\frac{1}{2\pi\gamma^3}
\int_{-\infty}^{\infty}ds\,
f(\varepsilon_n+\gamma s)\frac{4s(s^2-3)}{(1+s^2)^4},
\\
\int_{-\infty}^{\infty}\frac{d\varepsilon}{2\pi}\,
f(\varepsilon)\mathrm{Im}\,\mathcal K^{(1)}_{nnn}
&=\frac{1}{2\pi\gamma^4}
\int_{-\infty}^{\infty}ds\,
f(\varepsilon_n+\gamma s)
\frac{5s^4-38s^2+5}{(1+s^2)^5}.
\end{align}
For fixed $s$, we use the finite local expansion
\begin{align}
f(\varepsilon_n+\gamma s)
&=\sum_{p=0}^{4}\frac{(\gamma s)^p}{p!}f_n^{(p)}
+\mathcal O\bigl((\gamma s)^5\bigr).
\end{align}
This expansion extracts the local terms associated with
the convergent moments listed above.  It is not integrated
term by term to arbitrary order over the full real axis.
For $\mathcal K^{(2)}_{nn}$, a convergent $p$-th moment
produces a factor $\gamma^{p-3}$, so $p=1,3$ generate the
$\tau^2,\tau^0$ terms.  For $\mathcal K^{(1)}_{nnn}$, the
factor is $\gamma^{p-4}$, so $p=0,2,4$ would generate
$\tau^4,\tau^2,\tau^0$, respectively.  Its zeroth and fourth
moments vanish, leaving only the $\tau^2$ local term at
these orders.

The numerical coefficients follow explicitly as
\begin{align}
\int_{-\infty}^{\infty}\frac{d\varepsilon}{2\pi}\,
f(\varepsilon)\mathrm{Im}\,\mathcal K^{(2)}_{nn}
&=\frac{1}{2\pi\gamma^3}
\biggl[-\frac{\pi\gamma}{2}f_n'
+\frac{\pi\gamma^3}{2\cdot3!}f_n'''\biggr]
+\mathcal O(\gamma)
\nonumber\\
&=-\frac{f_n'}{4\gamma^2}
+\frac{1}{24}f_n'''+\mathcal O(\gamma)
\nonumber\\
&=-\frac{\tau^2}{\hbar^2}f_n'
+\frac{1}{24}f_n'''+\mathcal O(\tau^{-1}),
\label{eq:sm-K2-eint}
\\
\int_{-\infty}^{\infty}\frac{d\varepsilon}{2\pi}\,
f(\varepsilon)\mathrm{Im}\,\mathcal K^{(1)}_{nnn}
&=\frac{1}{2\pi\gamma^4}
\biggl[f_n\times0-\frac{\pi\gamma^2}{2\cdot2!}f_n''
+\frac{\gamma^4}{4!}f_n^{(4)}\times0\biggr]
+\mathcal O(\gamma)
\nonumber\\
&=-\frac{f_n''}{8\gamma^2}+\mathcal O(\gamma)
\nonumber\\
&=-\frac{\tau^2}{2\hbar^2}f_n''
+\mathcal O(\tau^{-1}).
\label{eq:sm-K1-eint}
\end{align}
The two explicit zeros in the intermediate line of
Eq.~\eqref{eq:sm-K1-eint} are the zeroth and fourth moments
in Eqs.~\eqref{eq:sm-M1-0} and \eqref{eq:sm-M1-4}.
The $\tau^2$ terms reproduce
Eqs.~\eqref{eq:K1-nnn-leading-expanded-img} and
\eqref{eq:K2-nn-leading-expanded-img} after integration,
using
\begin{align}
\int_{-\infty}^{\infty}d\varepsilon\,
f(\varepsilon)\partial_\varepsilon^p\delta_n
&=(-1)^p f_n^{(p)}.
\end{align}
At order $\tau^0$, only the $\mathcal K^{(2)}_{nn}$ channel
survives.  Its coefficient $1/24$ is the product of the
energy-integral normalization $1/(2\pi)$, the Taylor
coefficient $1/3!$, and the third moment $\pi/2$.

It is also useful to track the constant terms directly in
the exact derivative representations.  Applying the
finite-width distributional expansions given above to
Eq.~\eqref{eq:K2-nn-expanded}, the three terms contribute,
in their displayed order,
\begin{align}
\left.\mathrm{Im}\,\mathcal K^{(2)}_{nn}\right|_{\tau^0}
&=\pi\biggl(-\frac14+\frac12-\frac13\biggr)
\partial_\varepsilon^3\delta_n
=-\frac{\pi}{12}\partial_\varepsilon^3\delta_n.
\end{align}
After energy integration, the three corresponding
coefficients of $f_n'''$ are $1/8$, $-1/4$, and $1/6$,
whose sum is $1/24$.  Keeping only the last derivative term
would therefore give the incorrect coefficient $1/6$.
For Eq.~\eqref{eq:K1-nnn-expanded}, the term proportional
to $\tau^3$ is purely real at every $\gamma>0$.
The remaining three derivative terms give
\begin{align}
\left.\mathrm{Im}\,\mathcal K^{(1)}_{nnn}\right|_{\tau^0}
&=\pi\biggl(\frac18-\frac14+\frac18\biggr)
\partial_\varepsilon^4\delta_n=0.
\end{align}
These cancellations independently reproduce the convergent
moment calculation.  They exhibit explicitly how
subleading spectral factors multiplied by powers of
$\tau$ alter the constant term.

Two remarks specify the remainder and the applicability
of the expansion.  First, the scaled kernels have the tails
\begin{align}
\gamma^4\mathrm{Im}\,\mathcal K^{(2)}_{nn}
&=\frac4{s^5}+\mathcal O(s^{-7}),
\qquad
\gamma^5\mathrm{Im}\,\mathcal K^{(1)}_{nnn}
=\frac5{s^6}+\mathcal O(s^{-8}).
\end{align}
For the first scaled kernel, the next moment allowed by
parity is the fifth moment; for the second, it is the sixth.
Both diverge.  These divergent moments do not imply
a divergence of the original finite-$\gamma$ energy
integrals.  Rather, they prevent an unrestricted Taylor
expansion under those integrals.  Each kernel integral
generally has a nonlocal correction of order $\gamma$,
which depends on the Fermi function away from
$\varepsilon_n$ and cannot be fixed by finitely many
energy derivatives at that point.
In the fully diagonal current sector, these order-$\gamma$
terms cancel after the current terms are combined and
integrated over the periodic Brillouin zone.
For the one-band response, this is established by the exact
combined-kernel identity in the main-text subsection
``Exact one-band check of the $\tau^0$ hierarchy.''
The next correction to that response is
$\mathcal O(\gamma^2)=\mathcal O(\tau^{-2})$.
This diagonal-sector result does not establish the
subleading orders of every multiband geometric channel.

Second, the finite-temperature local expansion is
controlled when $\gamma\ll k_{\mathrm B}T$, since the
characteristic scale of $f_n^{(p)}$ is
$(k_{\mathrm B}T)^{-p}$.
In a multiband application, this thermal condition is
additional to the non-degeneracy condition.
All remainder estimates here refer to $\gamma\to0^+$ at
fixed $T>0$.  They are not asserted to be uniform in a
simultaneous $T\to0$ limit; the zero-temperature limit
requires a separate treatment.

\subsection{\texorpdfstring{Kernels $\mathcal{K}_{nm}$, $\mathcal{K}^{(1)}_{nmn}$ and 
$\mathcal{K}^{(2)}_{nm}$ with $n\neq m$}{Kernels Knm K1nmn K2nm with n not equal m}}

For $n \neq m$, the kernel $\mathcal{K}_{nm}$ in
Eq.~\eqref{eq:K-product} is expanded as follows.  First,
\begin{align}
R_m^2A_n
&=R_m(R_mA_n)
\nonumber\\
&=\frac{R_m(A_n-R_m)}{\varepsilon_{nm}+s}
\nonumber\\
&=
\frac{A_n}{(\varepsilon_{nm}+s)^2}
-\frac{R_m}{(\varepsilon_{nm}+s)^2}
-\frac{R_m^2}{\varepsilon_{nm}+s},
\label{eq:Rm2-An-Knm}
\\
R_m^2R_n
&=R_m(R_mR_n)
\nonumber\\
&=
\frac{R_n}{\varepsilon_{nm}^{2}}
-\frac{R_m}{\varepsilon_{nm}^{2}}
-\frac{R_m^2}{\varepsilon_{nm}}.
\label{eq:Rm2-Rn-Knm}
\end{align}
Therefore,
\begin{align}
\mathcal{K}_{nm} 
&= -R_{m}^{2}(A_{n}-R_{n})
\nonumber\\
&= -R_m^2A_n+R_m^2R_n
\nonumber\\
&=
-\frac{A_n}{(\varepsilon_{nm}+s)^2}
+\frac{R_n}{\varepsilon_{nm}^{2}}
-\left[
\frac{1}{\varepsilon_{nm}^{2}}
-\frac{1}{(\varepsilon_{nm}+s)^2}
\right]R_m
-\left[
\frac{1}{\varepsilon_{nm}}
-\frac{1}{\varepsilon_{nm}+s}
\right]R_m^2
\nonumber\\
&= -\frac{1}{\varepsilon^2_{nm}}(A_{n}-R_{n})
+\mathcal{O}(\tau^{-1}).
\label{eq:K-nm-expanded}
\end{align}
The leading real and imaginary parts of $\mathcal{K}_{nm}$ are 
\begin{align}
\left.\mathrm{Re}\mathcal{K}_{nm}\right|_{\mathcal{O}(\tau^0)}
&=0, 
\label{eq:K-nm-leading-expanded-real}
\\
\left.\mathrm{Im}\mathcal{K}_{nm}\right|_{\mathcal{O}(\tau^0)}
&=
-\frac{1}{\varepsilon_{nm}^{2}}\mathrm{Im}(A_{n}-R_{n})
=
-\frac{2\pi}{\varepsilon_{nm}^{2}}\delta_{n}.
\label{eq:K-nm-leading-expanded-img}
\end{align}

Setting $l = n$ in Eq.~\eqref{eq:K1-product}, with $n\neq m$,
we obtain 
\begin{align}
\mathcal{K}^{(1)}_{nmn}
&= (2R^{3}_{m}R_{n}+R^{2}_{m}R^{2}_{n})(A_{n}-R_{n})
+R^{2}_{m}R_{n}A^{2}_{n}. 
\label{eq:K1-nmn-first}
\end{align}
We first separate the same-band retarded--advanced products:
\begin{align}
R_{n}(A_{n}-R_{n})
&=R_{n}A_{n}-R_{n}^{2}
=r(A_{n}-R_{n})-R_{n}^{2},
\label{eq:Rn-AminusR-nmn}
\\
R_{n}^{2}(A_{n}-R_{n})
&=R_{n}\{r(A_{n}-R_{n})-R_{n}^{2}\}
\nonumber\\
&=r^{2}(A_{n}-R_{n})-rR_{n}^{2}-R_{n}^{3},
\label{eq:Rn2-AminusR-nmn}
\\
R_{n}A_{n}^{2}
&=A_{n}(R_{n}A_{n})
=rA_{n}(A_{n}-R_{n})
\nonumber\\
&=rA_{n}^{2}-r^{2}(A_{n}-R_{n}).
\label{eq:Rn-An2-nmn}
\end{align}
Substituting Eqs.~\eqref{eq:Rn-AminusR-nmn}--\eqref{eq:Rn-An2-nmn}
into Eq.~\eqref{eq:K1-nmn-first}, we find
\begin{align}
\mathcal{K}^{(1)}_{nmn}
&=2R_{m}^{3}\{r(A_{n}-R_{n})-R_{n}^{2}\}
+R_{m}^{2}\{r^{2}(A_{n}-R_{n})-rR_{n}^{2}-R_{n}^{3}\}
+R_{m}^{2}\{rA_{n}^{2}-r^{2}(A_{n}-R_{n})\}
\nonumber\\
&=2rR_{m}^{3}(A_{n}-R_{n})
-2R_{m}^{3}R_{n}^{2}
+rR_{m}^{2}(A_{n}^{2}-R_{n}^{2})
-R_{m}^{2}R_{n}^{3}.
\label{eq:K1-nmn-sameband-separated}
\end{align}
The remaining products are reduced by partial fractions:
\begin{align}
R_{m}^{2}A_{n}^{2}
&=
\frac{A_{n}^{2}+R_{m}^{2}}{(\varepsilon_{nm}+s)^{2}}
-\frac{2A_{n}}{(\varepsilon_{nm}+s)^{3}}
+\frac{2R_{m}}{(\varepsilon_{nm}+s)^{3}},
\label{eq:Rm2-An2-pf}
\\
R_{m}^{2}R_{n}^{2}
&=
\frac{R_{m}^{2}+R_{n}^{2}}{\varepsilon_{nm}^{2}}
+\frac{2R_{m}}{\varepsilon_{nm}^{3}}
-\frac{2R_{n}}{\varepsilon_{nm}^{3}},
\label{eq:Rm2-Rn2-pf}
\\
R_{m}^{3}(A_{n}-R_{n})
&=
\frac{A_{n}}{(\varepsilon_{nm}+s)^{3}}
-\frac{R_{n}}{\varepsilon_{nm}^{3}}
+\frac{s}{\varepsilon_{nm}(\varepsilon_{nm}+s)}R_{m}^{3}
\nonumber\\
&\quad
+\frac{s(2\varepsilon_{nm}+s)}
{\varepsilon_{nm}^{2}(\varepsilon_{nm}+s)^{2}}R_{m}^{2}
+\frac{s(3\varepsilon_{nm}^{2}+3\varepsilon_{nm}s+s^{2})}
{\varepsilon_{nm}^{3}(\varepsilon_{nm}+s)^{3}}R_{m},
\label{eq:Rm3-AminusR-pf}
\\
R_{m}^{3}R_{n}^{2}
&=
\frac{R_{m}^{3}}{\varepsilon_{nm}^{2}}
+\frac{2R_{m}^{2}}{\varepsilon_{nm}^{3}}
+\frac{R_{n}^{2}}{\varepsilon_{nm}^{3}}
+\frac{3R_{m}}{\varepsilon_{nm}^{4}}
-\frac{3R_{n}}{\varepsilon_{nm}^{4}},
\label{eq:Rm3-Rn2-pf}
\\
R_{m}^{2}R_{n}^{3}
&=
\frac{R_{n}^{3}}{\varepsilon_{nm}^{2}}
-\frac{R_{m}^{2}}{\varepsilon_{nm}^{3}}
-\frac{2R_{n}^{2}}{\varepsilon_{nm}^{3}}
-\frac{3R_{m}}{\varepsilon_{nm}^{4}}
+\frac{3R_{n}}{\varepsilon_{nm}^{4}}.
\label{eq:Rm2-Rn3-pf}
\end{align}
Using Eqs.~\eqref{eq:Rm2-An2-pf}--\eqref{eq:Rm2-Rn3-pf}
in Eq.~\eqref{eq:K1-nmn-sameband-separated}, and using $rs=1$,
all terms proportional to $A_n$ cancel.  Thus the exact
decomposition is
\begin{align}
\mathcal{K}^{(1)}_{nmn}
&=
\frac{r}{(\varepsilon_{nm}+s)^{2}}A_{n}^{2}
-\frac{r}{\varepsilon_{nm}^{2}}R_{n}^{2}
-\frac{1}{\varepsilon_{nm}^{2}}R_{n}^{3}
\nonumber\\
&\quad
-\frac{2s}{\varepsilon_{nm}^{2}(\varepsilon_{nm}+s)}R_{m}^{3}
-\frac{\varepsilon_{nm}^{2}+5\varepsilon_{nm}s+3s^{2}}
{\varepsilon_{nm}^{3}(\varepsilon_{nm}+s)^{2}}R_{m}^{2}
-\frac{3}{\varepsilon_{nm}^{4}}R_{m}
+\frac{3}{\varepsilon_{nm}^{4}}R_{n}.
\label{eq:K1-nmn-exact-s}
\end{align}
Expanding in the non-degenerate clean limit
$|\varepsilon_{nm}|\tau/\hbar\gg1$, we obtain
\begin{align}
\mathcal{K}^{(1)}_{nmn}
&=
\frac{r}{\varepsilon_{nm}^{2}}(A_{n}^{2}-R_{n}^{2})
-\frac{2}{\varepsilon_{nm}^{3}}A_{n}^{2}
-\frac{1}{\varepsilon_{nm}^{2}}R_{n}^{3}
-\frac{1}{\varepsilon_{nm}^{3}}R_{m}^{2}
-\frac{3}{\varepsilon_{nm}^{4}}(R_{m}-R_{n})
+\mathcal{O}(\tau^{-1})
\nonumber\\
&=
\frac{\tau}{\hbar}\frac{1}{\varepsilon_{nm}^{2}}
\partial_{\varepsilon}\{i(A_{n}-R_{n})\}
-\frac{2}{\varepsilon_{nm}^{3}}A_{n}^{2}
-\frac{1}{\varepsilon_{nm}^{2}}R_{n}^{3}
-\frac{1}{\varepsilon_{nm}^{3}}R_{m}^{2}
-\frac{3}{\varepsilon_{nm}^{4}}(R_{m}-R_{n})
+\mathcal{O}(\tau^{-1}).
\label{eq:K1-nmn-expanded}
\end{align}
In the last line we used
$A_{n}^{2}-R_{n}^{2}=-\partial_{\varepsilon}(A_{n}-R_{n})$
and $r=\tau/(i\hbar)$.
The leading real part is
\begin{align}
\left.\mathrm{Re}\mathcal{K}^{(1)}_{nmn}\right|_{\mathcal{O}(\tau)}
&=
\frac{\tau}{\hbar}\frac{1}{\varepsilon_{nm}^{2}}
\partial_{\varepsilon}
\mathrm{Re}\{i(A_{n}-R_{n})\}
\nonumber\\
&=
-2\pi\frac{\tau}{\hbar}
\frac{1}{\varepsilon_{nm}^{2}}
\partial_{\varepsilon}\delta_{n}.
\label{eq:K1-nmn-leading-expanded-real}
\end{align}
The $\mathcal{O}(\tau)$ term is real in the clean limit, so the
leading imaginary part is of order $\tau^{0}$:
\begin{align}
\left.\mathrm{Im}\mathcal{K}^{(1)}_{nmn}\right|_{\mathcal{O}(\tau^{0})}
&=
-\frac{2}{\varepsilon_{nm}^{3}}\mathrm{Im}A_{n}^{2}
-\frac{1}{\varepsilon_{nm}^{2}}\mathrm{Im}R_{n}^{3}
-\frac{1}{\varepsilon_{nm}^{3}}\mathrm{Im}R_{m}^{2}
-\frac{3}{\varepsilon_{nm}^{4}}
\{\mathrm{Im}R_{m}-\mathrm{Im}R_{n}\}
\nonumber\\
&=
\frac{\pi}{2\varepsilon_{nm}^{2}}
\partial_{\varepsilon}^{2}\delta_{n}
+\frac{2\pi}{\varepsilon_{nm}^{3}}
\partial_{\varepsilon}\delta_{n}
-\frac{\pi}{\varepsilon_{nm}^{3}}
\partial_{\varepsilon}\delta_{m}
+\frac{3\pi}{\varepsilon_{nm}^{4}}
(\delta_{m}-\delta_{n}).
\label{eq:K1-nmn-leading-expanded-img}
\end{align}

For $n\neq m$, the kernel $\mathcal{K}^{(2)}_{nm}$ is
\begin{align}
\mathcal{K}^{(2)}_{nm}
&= 2R_{m}^{3}(A_{n} - R_{n}) + R_{m}^{2}A_{n}^{2}.
\label{eq:K2-nm-first}
\end{align}
Using Eqs.~\eqref{eq:Rm3-AminusR-pf} and
\eqref{eq:Rm2-An2-pf}, we obtain
\begin{align}
\mathcal{K}^{(2)}_{nm}
&=
2\left[
\frac{A_n}{(\varepsilon_{nm}+s)^3}
-\frac{R_n}{\varepsilon_{nm}^{3}}
+\frac{s}{\varepsilon_{nm}(\varepsilon_{nm}+s)}R_m^3
+\frac{s(2\varepsilon_{nm}+s)}
{\varepsilon_{nm}^{2}(\varepsilon_{nm}+s)^2}R_m^2
+\frac{s(3\varepsilon_{nm}^{2}
+3\varepsilon_{nm}s+s^2)}
{\varepsilon_{nm}^{3}(\varepsilon_{nm}+s)^3}R_m
\right]
\nonumber\\
&\quad
+\frac{A_n^2+R_m^2}{(\varepsilon_{nm}+s)^2}
-\frac{2A_n}{(\varepsilon_{nm}+s)^3}
+\frac{2R_m}{(\varepsilon_{nm}+s)^3}
\nonumber\\
&=
\frac{A_n^2}{(\varepsilon_{nm}+s)^2}
+\frac{2s}{\varepsilon_{nm}(\varepsilon_{nm}+s)}R_m^3
+\frac{\varepsilon_{nm}^{2}
+4\varepsilon_{nm}s+2s^2}
{\varepsilon_{nm}^{2}(\varepsilon_{nm}+s)^2}R_m^2
+\frac{2}{\varepsilon_{nm}^{3}}(R_m-R_n).
\label{eq:K2-nm-exact-s}
\end{align}
Therefore,
\begin{align}
\mathcal{K}^{(2)}_{nm}
&=
\frac{1}{\varepsilon_{nm}^{2}}
\left(A_n^2+R_m^2\right)
+\frac{2}{\varepsilon_{nm}^{3}}(R_m-R_n)
+\mathcal{O}(\tau^{-1}).
\label{eq:K2-nm-expanded}
\end{align}
Thus there is no positive power of $\tau$ in
$\mathcal{K}^{(2)}_{nm}$ for $n\neq m$.

The leading real part is of order $\tau^0$.  In the clean limit,
\begin{align}
\left.\mathrm{Re}\mathcal{K}^{(2)}_{nm}\right|_{\mathcal{O}(\tau^0)}
&=
\frac{1}{\varepsilon_{nm}^{2}}
\left[
\mathcal{P}\frac{1}{(\varepsilon-\varepsilon_n)^2}
+
\mathcal{P}\frac{1}{(\varepsilon-\varepsilon_m)^2}
\right]
\nonumber\\
&\quad
+\frac{2}{\varepsilon_{nm}^{3}}
\left[
\mathcal{P}\frac{1}{\varepsilon-\varepsilon_m}
-
\mathcal{P}\frac{1}{\varepsilon-\varepsilon_n}
\right],
\label{eq:K2-nm-leading-expanded-real}
\end{align}
where $\mathcal{P}$ denotes the Cauchy principal value.
The leading imaginary part is
\begin{align}
\left.\mathrm{Im}\mathcal{K}^{(2)}_{nm}\right|_{\mathcal{O}(\tau^0)}
&=
\frac{1}{\varepsilon_{nm}^{2}}
\left(
\mathrm{Im}A_n^2+\mathrm{Im}R_m^2
\right)
+\frac{2}{\varepsilon_{nm}^{3}}
\left(
\mathrm{Im}R_m-\mathrm{Im}R_n
\right)
\nonumber\\
&=
\frac{\pi}{\varepsilon_{nm}^{2}}
\left(
\partial_{\varepsilon}\delta_m
-
\partial_{\varepsilon}\delta_n
\right)
+\frac{2\pi}{\varepsilon_{nm}^{3}}
\left(
\delta_n-\delta_m
\right).
\label{eq:K2-nm-leading-expanded-img}
\end{align}

\subsection{\texorpdfstring{Kernel $\mathcal{K}^{(1)}_{nnm}$ with $n\neq m$}{Kernel K1 nnm with n not equal m}}

Setting the second index equal to $n$ and the third index equal
to $m$ in Eq.~\eqref{eq:K1-product}, with $n\neq m$, gives
\begin{align}
\mathcal{K}^{(1)}_{nnm}
&=
\left(2R_n^3R_m+R_n^2R_m^2\right)(A_n-R_n)
+R_n^2R_mA_n^2 .
\label{eq:K1-nnm-first}
\end{align}
We now decompose this expression by applying
Eqs.~\eqref{eq:RA-same}, \eqref{eq:RR-diff} and
\eqref{eq:RA-diff}.  For compactness, we write
\begin{align}
s \equiv 2i\gamma,
\qquad
r=\frac{1}{s}.
\label{eq:s-def-nnm}
\end{align}
Then the mixed retarded--advanced identity is
\begin{align}
R_mA_n=\frac{A_n-R_m}{\varepsilon_{nm}+s}.
\label{eq:RA-diff-s-nnm}
\end{align}

First, we separate the same-band retarded--advanced products
in Eq.~\eqref{eq:K1-nnm-first}.  The useful identities are
\begin{align}
R_n(A_n-R_n)
&=R_nA_n-R_n^2
\nonumber\\
&=r(A_n-R_n)-R_n^2,
\label{eq:Rn-AminusR-nnm}
\\
R_n^2(A_n-R_n)
&=R_n\{R_n(A_n-R_n)\}
\nonumber\\
&=R_n\{r(A_n-R_n)-R_n^2\}
\nonumber\\
&=rR_n(A_n-R_n)-R_n^3
\nonumber\\
&=r^2(A_n-R_n)-rR_n^2-R_n^3,
\label{eq:Rn2-AminusR-nnm}
\\
R_n^3(A_n-R_n)
&=R_n\{R_n^2(A_n-R_n)\}
\nonumber\\
&=r^2R_n(A_n-R_n)-rR_n^3-R_n^4
\nonumber\\
&=r^3(A_n-R_n)-r^2R_n^2-rR_n^3-R_n^4.
\label{eq:Rn3-AminusR-nnm}
\end{align}
Similarly,
\begin{align}
R_n^2A_n^2
&=(R_nA_n)^2
\nonumber\\
&=r^2(A_n-R_n)^2
\nonumber\\
&=r^2(A_n^2+R_n^2-2R_nA_n)
\nonumber\\
&=r^2(A_n^2+R_n^2)-2r^3(A_n-R_n),
\label{eq:Rn2-An2-nnm}
\end{align}
where Eq.~\eqref{eq:RA-same} has been used in the last line.

Using Eqs.~\eqref{eq:Rn2-AminusR-nnm},
\eqref{eq:Rn3-AminusR-nnm} and \eqref{eq:Rn2-An2-nnm},
Eq.~\eqref{eq:K1-nnm-first} becomes
\begin{align}
\mathcal{K}^{(1)}_{nnm}
&=
2R_m\{R_n^3(A_n-R_n)\}
+R_m^2\{R_n^2(A_n-R_n)\}
+R_m\{R_n^2A_n^2\}
\nonumber\\
&=
2R_m\{r^3(A_n-R_n)-r^2R_n^2-rR_n^3-R_n^4\}
\nonumber\\
&\quad +R_m^2\{r^2(A_n-R_n)-rR_n^2-R_n^3\}
\nonumber\\
&\quad +R_m\{r^2(A_n^2+R_n^2)-2r^3(A_n-R_n)\}
\nonumber\\
&=
r^2R_m(A_n^2-R_n^2)
-2rR_mR_n^3
-2R_mR_n^4
\nonumber\\
&\quad +r^2R_m^2(A_n-R_n)
-rR_m^2R_n^2
-R_m^2R_n^3 .
\label{eq:K1-nnm-sameband-separated}
\end{align}

We next reduce the remaining products containing both $m$ and $n$.
From Eq.~\eqref{eq:RA-diff-s-nnm},
\begin{align}
R_mA_n^2
&=A_n(R_mA_n)
\nonumber\\
&=
\frac{A_n(A_n-R_m)}{\varepsilon_{nm}+s}
\nonumber\\
&=
\frac{A_n^2}{\varepsilon_{nm}+s}
-\frac{R_mA_n}{\varepsilon_{nm}+s}
\nonumber\\
&=
\frac{A_n^2}{\varepsilon_{nm}+s}
-\frac{A_n-R_m}{(\varepsilon_{nm}+s)^2},
\label{eq:Rm-An2-nnm}
\\
R_mR_n^2
&=R_n(R_mR_n)
\nonumber\\
&=
\frac{R_n(R_n-R_m)}{\varepsilon_{nm}}
\nonumber\\
&=
\frac{R_n^2}{\varepsilon_{nm}}
-\frac{R_mR_n}{\varepsilon_{nm}}
\nonumber\\
&=
\frac{R_n^2}{\varepsilon_{nm}}
-\frac{R_n-R_m}{\varepsilon_{nm}^2},
\label{eq:Rm-Rn2-nnm}
\\
R_mR_n^3
&=R_n(R_mR_n^2)
\nonumber\\
&=
\frac{R_n^3}{\varepsilon_{nm}}
-\frac{R_n^2}{\varepsilon_{nm}^2}
+\frac{R_n-R_m}{\varepsilon_{nm}^3},
\label{eq:Rm-Rn3-nnm}
\\
R_mR_n^4
&=R_n(R_mR_n^3)
\nonumber\\
&=
\frac{R_n^4}{\varepsilon_{nm}}
-\frac{R_n^3}{\varepsilon_{nm}^2}
+\frac{R_n^2}{\varepsilon_{nm}^3}
-\frac{R_n-R_m}{\varepsilon_{nm}^4}.
\label{eq:Rm-Rn4-nnm}
\end{align}
For the products with $R_m^2$, we similarly obtain
\begin{align}
R_m^2A_n
&=R_m(R_mA_n)
\nonumber\\
&=
\frac{R_m(A_n-R_m)}{\varepsilon_{nm}+s}
\nonumber\\
&=
\frac{A_n}{(\varepsilon_{nm}+s)^2}
-\frac{R_m}{(\varepsilon_{nm}+s)^2}
-\frac{R_m^2}{\varepsilon_{nm}+s},
\label{eq:Rm2-An-nnm}
\\
R_m^2R_n
&=R_m(R_mR_n)
\nonumber\\
&=
\frac{R_m(R_n-R_m)}{\varepsilon_{nm}}
\nonumber\\
&=
\frac{R_n}{\varepsilon_{nm}^2}
-\frac{R_m}{\varepsilon_{nm}^2}
-\frac{R_m^2}{\varepsilon_{nm}},
\label{eq:Rm2-Rn-nnm}
\\
R_m^2R_n^2
&=R_n(R_m^2R_n)
\nonumber\\
&=
\frac{R_n^2}{\varepsilon_{nm}^2}
-\frac{R_mR_n}{\varepsilon_{nm}^2}
-\frac{R_m^2R_n}{\varepsilon_{nm}}
\nonumber\\
&=
\frac{R_n^2}{\varepsilon_{nm}^2}
-\frac{R_n-R_m}{\varepsilon_{nm}^3}
-\frac{1}{\varepsilon_{nm}}
\left(
\frac{R_n}{\varepsilon_{nm}^2}
-\frac{R_m}{\varepsilon_{nm}^2}
-\frac{R_m^2}{\varepsilon_{nm}}
\right)
\nonumber\\
&=
\frac{R_n^2}{\varepsilon_{nm}^2}
-\frac{2R_n}{\varepsilon_{nm}^3}
+\frac{2R_m}{\varepsilon_{nm}^3}
+\frac{R_m^2}{\varepsilon_{nm}^2},
\label{eq:Rm2-Rn2-nnm}
\\
R_m^2R_n^3
&=R_n(R_m^2R_n^2)
\nonumber\\
&=
\frac{R_n^3}{\varepsilon_{nm}^2}
-\frac{2R_n^2}{\varepsilon_{nm}^3}
+\frac{2R_mR_n}{\varepsilon_{nm}^3}
+\frac{R_m^2R_n}{\varepsilon_{nm}^2}
\nonumber\\
&=
\frac{R_n^3}{\varepsilon_{nm}^2}
-\frac{2R_n^2}{\varepsilon_{nm}^3}
+\frac{2(R_n-R_m)}{\varepsilon_{nm}^4}
\nonumber\\
&\quad
+\frac{1}{\varepsilon_{nm}^2}
\left(
\frac{R_n}{\varepsilon_{nm}^2}
-\frac{R_m}{\varepsilon_{nm}^2}
-\frac{R_m^2}{\varepsilon_{nm}}
\right)
\nonumber\\
&=
\frac{R_n^3}{\varepsilon_{nm}^2}
-\frac{2R_n^2}{\varepsilon_{nm}^3}
+\frac{3R_n}{\varepsilon_{nm}^4}
-\frac{3R_m}{\varepsilon_{nm}^4}
-\frac{R_m^2}{\varepsilon_{nm}^3}.
\label{eq:Rm2-Rn3-nnm}
\end{align}
Furthermore,
\begin{align}
R_m^2(A_n-R_n)
&=R_m^2A_n-R_m^2R_n
\nonumber\\
&=
\frac{A_n}{(\varepsilon_{nm}+s)^2}
-\frac{R_n}{\varepsilon_{nm}^2}
+\left(
\frac{1}{\varepsilon_{nm}^2}
-\frac{1}{(\varepsilon_{nm}+s)^2}
\right)R_m
\nonumber\\
&\quad
+\left(
\frac{1}{\varepsilon_{nm}}
-\frac{1}{\varepsilon_{nm}+s}
\right)R_m^2 .
\label{eq:Rm2-AminusR-nnm}
\end{align}

Substituting Eqs.~\eqref{eq:Rm-An2-nnm}--\eqref{eq:Rm2-AminusR-nnm}
into Eq.~\eqref{eq:K1-nnm-sameband-separated}, and using
$rs=1$, all terms proportional to $A_n$ cancel.  We obtain
\begin{align}
\mathcal{K}^{(1)}_{nnm}
&=
\frac{r^2}{\varepsilon_{nm}+s}A_n^2
+\left(
-\frac{r^2}{\varepsilon_{nm}}
+\frac{r}{\varepsilon_{nm}^{2}}
\right)R_n^2
+\left(
-\frac{2r}{\varepsilon_{nm}}
+\frac{1}{\varepsilon_{nm}^{2}}
\right)R_n^3
\nonumber\\
&\quad
-\frac{2}{\varepsilon_{nm}}R_n^4
+\frac{s}{\varepsilon_{nm}^{3}(\varepsilon_{nm}+s)}R_m^2
+\frac{R_m-R_n}{\varepsilon_{nm}^{4}}.
\label{eq:K1-nnm-exact-s}
\end{align}
Restoring $s=2i\gamma$ and expanding in the non-degenerate
clean limit $|\varepsilon_{nm}|\tau/\hbar\gg1$, we use
\begin{align}
\frac{r^2}{\varepsilon_{nm}+s}
&=
\frac{r^2}{\varepsilon_{nm}}
-\frac{r}{\varepsilon_{nm}^{2}}
+\frac{1}{\varepsilon_{nm}^{3}}
+\mathcal{O}(\tau^{-1}),
\label{eq:r2-over-denom-nnm}
\\
\frac{s}{\varepsilon_{nm}^{3}(\varepsilon_{nm}+s)}
&=\mathcal{O}(\tau^{-1}).
\label{eq:s-over-denom-nnm}
\end{align}
Therefore,
\begin{align}
\mathcal{K}^{(1)}_{nnm}
&=
\frac{r^2}{\varepsilon_{nm}}(A_n^2-R_n^2)
-\frac{r}{\varepsilon_{nm}^{2}}(A_n^2-R_n^2)
-\frac{2r}{\varepsilon_{nm}}R_n^3
\nonumber\\
&\quad
+\frac{1}{\varepsilon_{nm}^{3}}A_n^2
+\frac{1}{\varepsilon_{nm}^{2}}R_n^3
-\frac{2}{\varepsilon_{nm}}R_n^4
+\frac{R_m-R_n}{\varepsilon_{nm}^{4}}
+\mathcal{O}(\tau^{-1}).
\label{eq:K1-nnm-expanded-r}
\end{align}
Finally, using
\begin{align}
A_n^2-R_n^2
&=-\partial_{\varepsilon}(A_n-R_n),
\label{eq:A2minusR2-derivative-nnm}
\\
R_n^3
&=\frac{1}{2}\partial_{\varepsilon}^{2}R_n,
\label{eq:Rn3-derivative-nnm}
\\
R_n^4
&=-\frac{1}{6}\partial_{\varepsilon}^{3}R_n,
\label{eq:Rn4-derivative-nnm}
\end{align}
and $r=\tau/(i\hbar)$, Eq.~\eqref{eq:K1-nnm-expanded-r}
can be written as
\begin{align}
\mathcal{K}^{(1)}_{nnm}
&=
\left(\frac{\tau}{\hbar}\right)^2
\frac{1}{\varepsilon_{nm}}
\partial_{\varepsilon}(A_n-R_n)
-\left(\frac{\tau}{\hbar}\right)
\frac{1}{\varepsilon_{nm}^{2}}
\partial_{\varepsilon}\{i(A_n-R_n)\}
\nonumber\\
&\quad
+\left(\frac{\tau}{\hbar}\right)
\frac{1}{\varepsilon_{nm}}
\partial_{\varepsilon}^{2}(iR_n)
+\frac{1}{\varepsilon_{nm}^{3}}A_n^2
+\frac{1}{2\varepsilon_{nm}^{2}}\partial_{\varepsilon}^{2}R_n
\nonumber\\
&\quad
+\frac{1}{3\varepsilon_{nm}}\partial_{\varepsilon}^{3}R_n
+\frac{R_m-R_n}{\varepsilon_{nm}^{4}}
+\mathcal{O}(\tau^{-1}).
\label{eq:K1-nnm-expanded}
\end{align}

The leading real part is of order $\tau$:
\begin{align}
\left.\mathrm{Re}\mathcal{K}^{(1)}_{nnm}\right|_{\mathcal{O}(\tau)}
&=
-\left(\frac{\tau}{\hbar}\right)
\frac{1}{\varepsilon_{nm}^{2}}
\partial_{\varepsilon}\mathrm{Re}\{i(A_n-R_n)\}
+
\left(\frac{\tau}{\hbar}\right)
\frac{1}{\varepsilon_{nm}}
\partial_{\varepsilon}^{2}\mathrm{Re}(iR_n)
\nonumber\\
&=
2\pi\frac{\tau}{\hbar}
\frac{1}{\varepsilon_{nm}^{2}}
\partial_{\varepsilon}\delta_n
+\pi\frac{\tau}{\hbar}
\frac{1}{\varepsilon_{nm}}
\partial_{\varepsilon}^{2}\delta_n.
\label{eq:K1-nnm-leading-expanded-real}
\end{align}
The leading imaginary part is of order $\tau^2$:
\begin{align}
\left.\mathrm{Im}\mathcal{K}^{(1)}_{nnm}\right|_{\mathcal{O}(\tau^2)}
&=
\left(\frac{\tau}{\hbar}\right)^2
\frac{1}{\varepsilon_{nm}}
\partial_{\varepsilon}\mathrm{Im}(A_n-R_n)
\nonumber\\
&=
2\pi\left(\frac{\tau}{\hbar}\right)^2
\frac{1}{\varepsilon_{nm}}
\partial_{\varepsilon}\delta_n.
\label{eq:K1-nnm-leading-expanded-img}
\end{align}
The $\mathcal{O}(\tau^0)$ part of the imaginary contribution is
\begin{align}
\left.\mathrm{Im}\mathcal{K}^{(1)}_{nnm}\right|_{\mathcal{O}(\tau^0)}
&=
\frac{1}{\varepsilon_{nm}^{3}}\mathrm{Im}A_n^2
+\frac{1}{\varepsilon_{nm}^{2}}\mathrm{Im}R_n^3
-\frac{2}{\varepsilon_{nm}}\mathrm{Im}R_n^4 
+\frac{1}{\varepsilon_{nm}^{4}}
\{\mathrm{Im}R_m-\mathrm{Im}R_n\}
\nonumber\\
&=
\pi\left[
-\frac{\partial_{\varepsilon}\delta_n}{\varepsilon_{nm}^{3}}
-\frac{\partial_{\varepsilon}^{2}\delta_n}{2\varepsilon_{nm}^{2}}
-\frac{\partial_{\varepsilon}^{3}\delta_n}{3\varepsilon_{nm}}
+\frac{\delta_n-\delta_m}{\varepsilon_{nm}^{4}}
\right].
\label{eq:K1-nnm-subleading-expanded-img}
\end{align}

It is often useful to combine this kernel with
$\mathcal{K}^{(2)}_{nn}$.  Multiplying Eq.~\eqref{eq:K1-nnm-expanded}
by $\varepsilon_{nm}$ and subtracting Eq.~\eqref{eq:K2-nn-expanded},
the $\mathcal{O}(\tau^2)$ terms cancel and one obtains
\begin{align}
\varepsilon_{nm}\mathcal{K}^{(1)}_{nnm}
-\mathcal{K}^{(2)}_{nn}
&=
\frac{r}{\varepsilon_{nm}}
\partial_{\varepsilon}(A_n-R_n)
+\frac{A_n^2}{\varepsilon_{nm}^{2}}
+\frac{R_n^3}{\varepsilon_{nm}}
+\frac{R_m-R_n}{\varepsilon_{nm}^{3}}
+\mathcal{O}(\tau^{-1}).
\label{eq:K1-nnm-minus-K2nn}
\end{align}
Its imaginary part is
\begin{align}
\mathrm{Im}\!\left[
\varepsilon_{nm}\mathcal{K}^{(1)}_{nnm}
-\mathcal{K}^{(2)}_{nn}
\right]
&=
\pi\left[
-\frac{\partial_{\varepsilon}\delta_n}{\varepsilon_{nm}^{2}}
-\frac{\partial_{\varepsilon}^{2}\delta_n}{2\varepsilon_{nm}}
+\frac{\delta_n-\delta_m}{\varepsilon_{nm}^{3}}
\right]
+\mathcal{O}(\tau^{-1}).
\label{eq:Im-K1-nnm-minus-K2nn}
\end{align}

\subsection{\texorpdfstring{Kernel $\mathcal{K}^{(1)}_{nmm}$ with $n\neq m$}
{Kernel K1 nmm with n not equal m}}

Setting $l=m$ in Eq.~\eqref{eq:K1-product}, with $n\neq m$, gives
\begin{align}
\mathcal{K}^{(1)}_{nmm}
&=
\left(2R_m^4+R_m^4\right)(A_n-R_n)
+R_m^3A_n^2
\nonumber\\
&=
3R_m^4(A_n-R_n)+R_m^3A_n^2.
\label{eq:K1-nmm-first}
\end{align}
Using Eqs.~\eqref{eq:RR-diff} and \eqref{eq:RA-diff}, the exact
partial-fraction form is
\begin{align}
\mathcal{K}^{(1)}_{nmm}
&=
\frac{A_n^2}{(\varepsilon_{nm}+s)^3}
+\frac{3s}{\varepsilon_{nm}(\varepsilon_{nm}+s)}R_m^4
+\frac{\varepsilon_{nm}^{2}+6\varepsilon_{nm}s+3s^2}
{\varepsilon_{nm}^{2}(\varepsilon_{nm}+s)^2}R_m^3
\nonumber\\
&\quad
+\frac{
2\varepsilon_{nm}^{3}
+9\varepsilon_{nm}^{2}s
+9\varepsilon_{nm}s^2
+3s^3}
{\varepsilon_{nm}^{3}(\varepsilon_{nm}+s)^3}R_m^2
+\frac{3(R_m-R_n)}{\varepsilon_{nm}^{4}}.
\label{eq:K1-nmm-exact-s}
\end{align}
Thus, in the non-degenerate clean limit,
\begin{align}
\mathcal{K}^{(1)}_{nmm}
&=
\frac{A_n^2}{\varepsilon_{nm}^{3}}
+\frac{R_m^3}{\varepsilon_{nm}^{2}}
+\frac{2R_m^2}{\varepsilon_{nm}^{3}}
+\frac{3(R_m-R_n)}{\varepsilon_{nm}^{4}}
+\mathcal{O}(\tau^{-1}).
\label{eq:K1-nmm-expanded}
\end{align}
The leading real part is of order $\tau^0$:
\begin{align}
\left.\mathrm{Re}\mathcal{K}^{(1)}_{nmm}\right|_{\mathcal{O}(\tau^0)}
&=
\frac{1}{\varepsilon_{nm}^{3}}
\mathcal{P}\frac{1}{(\varepsilon-\varepsilon_n)^2}
+\frac{1}{\varepsilon_{nm}^{2}}
\mathcal{P}\frac{1}{(\varepsilon-\varepsilon_m)^3}
\nonumber\\
&\quad
+\frac{2}{\varepsilon_{nm}^{3}}
\mathcal{P}\frac{1}{(\varepsilon-\varepsilon_m)^2}
+\frac{3}{\varepsilon_{nm}^{4}}
\left[
\mathcal{P}\frac{1}{\varepsilon-\varepsilon_m}
-\mathcal{P}\frac{1}{\varepsilon-\varepsilon_n}
\right].
\label{eq:K1-nmm-leading-expanded-real}
\end{align}
The leading imaginary part is
\begin{align}
\left.\mathrm{Im}\mathcal{K}^{(1)}_{nmm}\right|_{\mathcal{O}(\tau^0)}
&=
\frac{1}{\varepsilon_{nm}^{3}}\mathrm{Im}A_n^2
+\frac{1}{\varepsilon_{nm}^{2}}\mathrm{Im}R_m^3
+\frac{2}{\varepsilon_{nm}^{3}}\mathrm{Im}R_m^2
+\frac{3}{\varepsilon_{nm}^{4}}
\{\mathrm{Im}R_m-\mathrm{Im}R_n\}
\nonumber\\
&=
\pi\left[
-\frac{\partial_{\varepsilon}\delta_n}{\varepsilon_{nm}^{3}}
-\frac{\partial_{\varepsilon}^{2}\delta_m}{2\varepsilon_{nm}^{2}}
+\frac{2\partial_{\varepsilon}\delta_m}{\varepsilon_{nm}^{3}}
+\frac{3(\delta_n-\delta_m)}{\varepsilon_{nm}^{4}}
\right].
\label{eq:K1-nmm-leading-expanded-img}
\end{align}

\subsection{\texorpdfstring{Three-distinct-band kernel $\mathcal{K}^{(1)}_{nml}$ with $n,m,l$ all distinct}{Three-distinct-band kernel}}

Finally, consider $\mathcal{K}^{(1)}_{nml}$ for $n$, $m$ and
$l$ all distinct.  From Eq.~\eqref{eq:K1-product},
\begin{align}
\mathcal{K}^{(1)}_{nml}
&=
\left(2R_m^3R_l+R_m^2R_l^2\right)(A_n-R_n)
+R_m^2R_lA_n^2.
\label{eq:K1-nml-first}
\end{align}
The useful partial-fraction identities are
\begin{align}
R_m^2R_l
&=
\frac{R_m^2}{\varepsilon_{ml}}
-\frac{R_m}{\varepsilon_{ml}^{2}}
+\frac{R_l}{\varepsilon_{ml}^{2}},
\label{eq:Rm2-Rl-pf}
\\
R_m^3R_l
&=
\frac{R_m^3}{\varepsilon_{ml}}
-\frac{R_m^2}{\varepsilon_{ml}^{2}}
+\frac{R_m}{\varepsilon_{ml}^{3}}
-\frac{R_l}{\varepsilon_{ml}^{3}},
\label{eq:Rm3-Rl-pf}
\\
R_m^2R_l^2
&=
\frac{R_m^2+R_l^2}{\varepsilon_{ml}^{2}}
-\frac{2R_m}{\varepsilon_{ml}^{3}}
+\frac{2R_l}{\varepsilon_{ml}^{3}}.
\label{eq:Rm2-Rl2-pf}
\end{align}
Applying these identities together with Eqs.~\eqref{eq:RR-diff}
and \eqref{eq:RA-diff}, one obtains
\begin{align}
\left(2R_m^3R_l+R_m^2R_l^2\right)(A_n-R_n)
&=
\left(
\frac{2}{\varepsilon_{nm}^{3}\varepsilon_{nl}}
+\frac{1}{\varepsilon_{nm}^{2}\varepsilon_{nl}^{2}}
\right)(A_n-R_n)
+\mathcal{O}(\tau^{-1}),
\label{eq:F-nml-AminusR}
\\
R_m^2R_lA_n^2
&=
\frac{A_n^2}{\varepsilon_{nm}^{2}\varepsilon_{nl}}
-\frac{\varepsilon_{nm}+2\varepsilon_{nl}}
{\varepsilon_{nm}^{3}\varepsilon_{nl}^{2}}A_n
+\frac{R_m^2}{\varepsilon_{nm}^{2}\varepsilon_{ml}}
\nonumber\\
&\quad
+\frac{2\varepsilon_{nl}-3\varepsilon_{nm}}
{\varepsilon_{nm}^{3}\varepsilon_{ml}^{2}}R_m
+\frac{R_l}{\varepsilon_{nl}^{2}\varepsilon_{ml}^{2}}
+\mathcal{O}(\tau^{-1}).
\label{eq:Rm2-Rl-An2-pf}
\end{align}
Substituting Eqs.~\eqref{eq:F-nml-AminusR} and
\eqref{eq:Rm2-Rl-An2-pf} into Eq.~\eqref{eq:K1-nml-first}, we find
\begin{align}
\mathcal{K}^{(1)}_{nml}
&=
\left(
\frac{2}{\varepsilon_{nm}^{3}\varepsilon_{nl}}
+\frac{1}{\varepsilon_{nm}^{2}\varepsilon_{nl}^{2}}
\right)(A_n-R_n)
+\frac{A_n^2}{\varepsilon_{nm}^{2}\varepsilon_{nl}}
-\frac{\varepsilon_{nm}+2\varepsilon_{nl}}
{\varepsilon_{nm}^{3}\varepsilon_{nl}^{2}}A_n
+\frac{R_m^2}{\varepsilon_{nm}^{2}\varepsilon_{ml}}
\nonumber\\
&\quad
+\frac{2\varepsilon_{nl}-3\varepsilon_{nm}}
{\varepsilon_{nm}^{3}\varepsilon_{ml}^{2}}R_m
+\frac{R_l}{\varepsilon_{nl}^{2}\varepsilon_{ml}^{2}}
+\mathcal{O}(\tau^{-1}).
\label{eq:K1-nml-distinct-expanded}
\end{align}
Thus $\mathcal{K}^{(1)}_{nml}$ has no positive power of $\tau$
when all three band indices are distinct.

The leading real part is of order $\tau^0$:
\begin{align}
\left.\mathrm{Re}\mathcal{K}^{(1)}_{nml}\right|_{\mathcal{O}(\tau^0)}
&=
\frac{1}{\varepsilon_{nm}^{2}\varepsilon_{nl}}
\mathcal{P}\frac{1}{(\varepsilon-\varepsilon_n)^2}
-\frac{\varepsilon_{nm}+2\varepsilon_{nl}}
{\varepsilon_{nm}^{3}\varepsilon_{nl}^{2}}
\mathcal{P}\frac{1}{\varepsilon-\varepsilon_n}
\nonumber\\
&\quad
+\frac{1}{\varepsilon_{nm}^{2}\varepsilon_{ml}}
\mathcal{P}\frac{1}{(\varepsilon-\varepsilon_m)^2}
+\frac{2\varepsilon_{nl}-3\varepsilon_{nm}}
{\varepsilon_{nm}^{3}\varepsilon_{ml}^{2}}
\mathcal{P}\frac{1}{\varepsilon-\varepsilon_m}
\nonumber\\
&\quad
+\frac{1}{\varepsilon_{nl}^{2}\varepsilon_{ml}^{2}}
\mathcal{P}\frac{1}{\varepsilon-\varepsilon_l}.
\label{eq:K1-nml-distinct-leading-real}
\end{align}
The leading imaginary part is
\begin{align}
\left.\mathrm{Im}\mathcal{K}^{(1)}_{nml}\right|_{\mathcal{O}(\tau^0)}
&=
\pi\bigg[
-\frac{\partial_{\varepsilon}\delta_n}
{\varepsilon_{nm}^{2}\varepsilon_{nl}}
+\left(
\frac{2}{\varepsilon_{nm}^{3}\varepsilon_{nl}}
+\frac{1}{\varepsilon_{nm}^{2}\varepsilon_{nl}^{2}}
\right)\delta_n
+\frac{\partial_{\varepsilon}\delta_m}
{\varepsilon_{nm}^{2}\varepsilon_{ml}}
+\frac{3\varepsilon_{nm}-2\varepsilon_{nl}}
{\varepsilon_{nm}^{3}\varepsilon_{ml}^{2}}\delta_m
-\frac{\delta_l}{\varepsilon_{nl}^{2}\varepsilon_{ml}^{2}}
\bigg].
\label{eq:K1-nml-distinct-leading-img}
\end{align}

\section{Band-basis evaluation of the linear and second-order dc conductivities}
\label{sec:conductivity-contributions}
We now evaluate the linear dc conductivity and the second-order dc conductivity using the
velocity decompositions and clean-limit kernel expansions derived above.
Throughout this section we use
\begin{align}
&\delta_{n} = \delta(\varepsilon+\mu-\varepsilon_{n}), \\
&f_n \equiv f(\varepsilon_n-\mu),\\
&f'_n
\equiv \frac{\partial f(\varepsilon_n-\mu)}{\partial \varepsilon_n},\\
&f''_n\equiv
\frac{\partial^2 f(\varepsilon_n-\mu)}{\partial \varepsilon_n^2},\\
&f^{(p)}_n\equiv
\frac{\partial^p f(\varepsilon_n-\mu)}{\partial \varepsilon_n^p},
\qquad
f'''_n\equiv f^{(3)}_n .
\label{eq:f-derivatives-def}
\end{align}

The quantum-geometric tensor $\mathcal{Q}^{nm}_{ij}$,
the quantum metric $\mathcal{G}^{nm}_{ij}$, and the Berry curvature $\Omega^{nm}_{ij}$
are
\begin{align}
\mathcal{Q}^{nm}_{ij}
&=
\mathcal{A}^{nm}_{i}\mathcal{A}^{mn}_{j},
\qquad n\neq m,
\label{eq:Cij-def}
\\
\mathcal{G}^{nm}_{ij}
&=
\mathrm{Re}\,\mathcal{Q}^{nm}_{ij},
\label{eq:Gij-def}
\\
\Omega^{nm}_{ij}
&=
-2\,\mathrm{Im}\,\mathcal{Q}^{nm}_{ij}.
\label{eq:Omega-nm-def}
\end{align}
Thus
\begin{align}
\mathcal{Q}^{nm}_{ij}
=
\mathcal{G}^{nm}_{ij}
-\frac{i}{2}\Omega^{nm}_{ij}, \quad
\mathcal{G}^{n}_{ij}= \sum_{m\neq n}\mathcal{G}^{nm}_{ij},
\quad
\Omega^n_{ij}
=
\sum_{m\neq n}\Omega^{nm}_{ij}.
\label{eq:C-G-Omega-relation}
\end{align}
The covariant derivative of an off-diagonal matrix element $\mathcal{A}^{nm}_{i}$
with respect to the crystal momentum $\bm{k}$ is defined as
\begin{align}
\mathcal{D}_{k}\mathcal{A}^{nm}_{i}
\equiv
\partial_{k}\mathcal{A}^{nm}_{i}
-i(\mathcal{A}^{nn}_{k}-\mathcal{A}^{mm}_{k})\mathcal{A}^{nm}_{i},
\qquad n\neq m,
\label{eq:covariant-deriv-def}
\end{align}
and the three-index quantum connection $\mathcal{C}^{nm}_{ij|k}$
and the closed three-band Berry connection $\mathcal{T}^{nm}_{ijk}$
are defined as
\begin{align}
&\mathcal{C}^{nm}_{ij|k}
\equiv
\mathcal{A}^{nm}_{i}\,\mathcal{D}_{k}\mathcal{A}^{mn}_{j},
\qquad n\neq m,
\label{eq:quantum-connection-def}\\
&\mathcal{T}^{nm}_{ijk}
\equiv
\sum_{l}\mathcal{A}^{nm}_{i}\mathcal{A}^{ml}_{j}\mathcal{A}^{ln}_{k},
\qquad n\neq m.
\label{eq:three-berry-connection-def}
\end{align}

\subsection{\texorpdfstring{Linear dc conductivity: Drude and quantum-geometric terms
$\sigma^{\mathrm{D}}_{ij}$ and $\sigma^{\mathrm{QG}}_{ij}$}{Linear dc conductivity: Drude and quantum-geometric terms}}
The linear response naturally separates into the intraband Drude term and
the interband quantum-geometric term,
\begin{align}
\sigma^{\mathrm{DC}}_{ij}=\sigma^{\mathrm{D}}_{ij}+\sigma^{\mathrm{QG}}_{ij}.
\label{eq:sigma-linear-decomposition}
\end{align}
The Drude sector $\sigma^{\mathrm{D}}_{ij}$ originates from the single-band contributions and is given by
\begin{align}
  \sigma^{\mathrm{D}}_{ij}
  &= 2\frac{e^{2}}{\hbar}
     \int_{-\infty}^{\infty}\frac{d\varepsilon}{2\pi}\,
     f(\varepsilon)
     \sum_{\bm{k}}
     \mathrm{Re}\!\left[\mathcal{B}^{\mathrm{D}}_{ij}\right],
  \label{eq:sigma-D2}
  \\
  \mathcal{B}^{\mathrm{D}}_{ij}
  &= \sum_{n}
       (\partial_{i}\varepsilon_{n})
       (\partial_{j}\varepsilon_{n})
       \mathcal{K}_{nn}.
  \label{eq:B-D2}
\end{align}
The leading real part of $\mathcal{K}_{nn}$ is evaluated in Eq.~\eqref{eq:K-nn-leading-expanded-real},
\begin{align}
\left.\mathrm{Re}\,\mathcal{K}_{nn}\right|_{\mathcal{O}(\tau)}
&= \pi\frac{\tau}{\hbar} \partial_{\varepsilon}\delta_{n}.
\label{eq:K-nn-leading-expanded-real2}
\end{align}
Substituting Eq.~\eqref{eq:K-nn-leading-expanded-real2} into Eq.~\eqref{eq:B-D2},
one obtains
\begin{align}
\sigma^{\mathrm{D}}_{ij}
&=\frac{e^2}{\hbar^2}\tau\int_{-\infty}^{\infty}d\varepsilon\, f(\varepsilon)
\sum_{\bm{k}}\sum_{n}(\partial_{i}\varepsilon_{n})(\partial_{j}\varepsilon_{n})\partial_{\varepsilon}\delta_{n}
\nonumber\\
&=
-\frac{e^2}{\hbar^2}\tau
\sum_{\bm{k}}\sum_n
f'_n\,
\partial_i\varepsilon_n\,
\partial_j\varepsilon_n.
\label{eq:sigma-D-linear}
\end{align}
Since $f'_n<0$ near the Fermi surface, this expression has the
usual positive longitudinal Drude weight.

The quantum-geometric sector $\sigma^{\mathrm{QG}}_{ij}$
originates from the interband transitions and is given by
\begin{align}
  \sigma^{\mathrm{QG}}_{ij}
  &= 2\frac{e^{2}}{\hbar}
     \int_{-\infty}^{\infty}\frac{d\varepsilon}{2\pi}\,
     f(\varepsilon)
     \sum_{\bm{k}}
     \mathrm{Re}\!\left[\mathcal{B}^{\mathrm{QG}}_{ij}\right],
  \label{eq:sigma-linear-QG2}
  \\
  \mathcal{B}^{\mathrm{QG}}_{ij}
  &= {\sum_{n,m}}^{\prime}\varepsilon_{nm}^{2}\mathcal{Q}^{nm}_{ij}\mathcal{K}_{nm}.
  \label{eq:B-linear-QG2}
\end{align}
The leading real and imaginary parts of $\mathcal{K}_{nm}$ are evaluated in
Eqs.~\eqref{eq:K-nm-leading-expanded-real} and \eqref{eq:K-nm-leading-expanded-img},
\begin{align}
\left.\mathrm{Re}\,\mathcal{K}_{nm}\right|_{\mathcal{O}(\tau^0)}
&=0,
\label{eq:K-nm-leading-expanded-real2}
\\
\left.\mathrm{Im}\,\mathcal{K}_{nm}\right|_{\mathcal{O}(\tau^0)}
&=
-\frac{2\pi}{\varepsilon_{nm}^{2}}\delta_{n}.
\label{eq:K-nm-leading-expanded-img2}
\end{align}
Substituting Eq.~\eqref{eq:K-nm-leading-expanded-img2} into Eq.~\eqref{eq:B-linear-QG2},
one obtains
\begin{align}
\sigma^{\mathrm{QG}}_{ij}
&= 2\frac{e^2}{\hbar}\int_{-\infty}^{\infty}\frac{d\varepsilon}{2\pi}\,f(\varepsilon)
\sum_{\bm{k}}{\sum_{n,m}}^{\prime}\varepsilon_{nm}^{2}\,\mathrm{Re}(\mathcal{Q}^{nm}_{ij}
\mathcal{K}_{nm})\nonumber\\
&= -2\frac{e^2}{\hbar}\int_{-\infty}^{\infty}\frac{d\varepsilon}{2\pi}\,f(\varepsilon)
\sum_{\bm{k}}{\sum_{n,m}}^{\prime}\varepsilon_{nm}^{2}(\mathrm{Im}\,\mathcal{Q}^{nm}_{ij})(\mathrm{Im}\,\mathcal{K}_{nm})\nonumber\\
&= -\frac{e^2}{\hbar}\int_{-\infty}^{\infty}d\varepsilon\, f(\varepsilon)
\sum_{\bm{k}}{\sum_{n,m}}^{\prime}\Omega^{nm}_{ij}
\delta_{n}\nonumber\\
&=-\frac{e^2}{\hbar}
\sum_{\bm{k}}{\sum_{n,m}}^{\prime}\Omega^{nm}_{ij}
f_{n}\nonumber\\
&=-\frac{e^2}{\hbar}\sum_{\bm{k}}\sum_{n}f_{n}\Omega^{n}_{ij}\equiv \sigma^{\mathrm{BC}}_{ij}.
\label{eq:sigma-QG-linear}
\end{align}
Thus the interband part of the linear response is controlled by the Berry curvature.
Combining Eqs.~\eqref{eq:sigma-D-linear} and \eqref{eq:sigma-QG-linear},
the full linear dc conductivity is
\begin{align}
\sigma^{\mathrm{DC}}_{ij}&=\sigma^{\mathrm{D}}_{ij}+\sigma^{\mathrm{BC}}_{ij}\nonumber\\
&=
-\frac{e^2\tau}{\hbar^2}
\sum_{\bm{k}}\sum_n
f'_n\,
\partial_i\varepsilon_n
\partial_j\varepsilon_n
-
\frac{e^2}{\hbar}
\sum_{\bm{k}}\sum_n
f_n\,\Omega^n_{ij}.
\label{eq:sigma-linear-final}
\end{align}

\subsection{\texorpdfstring{Detailed calculation of the nonlinear-Drude term
$\sigma^{\mathrm{ND}}_{ijk}$}{Detailed calculation of the nonlinear-Drude term}}

This term is the second-order analogue of the Drude response:
all band indices in the Green's function kernels are diagonal,
and therefore the leading powers of $\tau$ come from the
fully diagonal kernels
$\mathcal{K}^{(1)}_{nnn}$ and $\mathcal{K}^{(2)}_{nn}$.
The starting point is
\begin{align}
\sigma^{\mathrm{ND}}_{ijk}
&=
\frac{e^{3}}{\hbar}
\int_{-\infty}^{\infty}
\frac{d\varepsilon}{2\pi}
f(\varepsilon)
\sum_{\bm{k}}
\operatorname{Im}
\mathcal{B}^{\mathrm{ND}}_{ijk},
\label{eq:sigma-ND-def}
\\
\mathcal{B}^{\mathrm{ND}}_{ijk}
&=
\sum_{n}
\left[
2\partial_{i}\varepsilon_{n}
 \partial_{j}\varepsilon_{n}
 \partial_{k}\varepsilon_{n}
\mathcal{K}^{(1)}_{nnn}
+
\partial_{i}\varepsilon_{n}
\partial_{j}\partial_{k}\varepsilon_{n}
\mathcal{K}^{(2)}_{nn}
\right].
\label{eq:B-ND-def}
\end{align}
The coefficients multiplying the kernels are real.  Hence the
imaginary part acts only on the Green's function kernels:
\begin{align}
\operatorname{Im}\mathcal{B}^{\mathrm{ND}}_{ijk}
&=
\sum_{n}
\left[
2\partial_{i}\varepsilon_{n}
 \partial_{j}\varepsilon_{n}
 \partial_{k}\varepsilon_{n}
\operatorname{Im}\mathcal{K}^{(1)}_{nnn}
+
\partial_{i}\varepsilon_{n}
\partial_{j}\partial_{k}\varepsilon_{n}
\operatorname{Im}\mathcal{K}^{(2)}_{nn}
\right].
\label{eq:Im-B-ND}
\end{align}
Using the diagonal-kernel energy integrals
Eqs.~\eqref{eq:sm-K2-eint} and \eqref{eq:sm-K1-eint} derived
in Sec.~\ref{ssec:diag-tau0},
\begin{align}
\int_{-\infty}^{\infty}\!\frac{d\varepsilon}{2\pi}\,
f(\varepsilon)\,
\operatorname{Im}\mathcal{K}^{(1)}_{nnn}
&=
-\frac{\tau^2}{2\hbar^2}f''_n+\mathcal{O}(\tau^{-1}),
\label{eq:Im-K1-nnn-for-ND}
\\
\int_{-\infty}^{\infty}\!\frac{d\varepsilon}{2\pi}\,
f(\varepsilon)\,
\operatorname{Im}\mathcal{K}^{(2)}_{nn}
&=
-\frac{\tau^2}{\hbar^2}f'_n
+\frac{1}{24}f'''_n+\mathcal{O}(\tau^{-1}),
\label{eq:Im-K2-nn-for-ND}
\end{align}
the nonlinear Drude sector splits into an
$\mathcal{O}(\tau^{2})$ and an $\mathcal{O}(\tau^{0})$ part,
\begin{align}
\sigma^{\mathrm{ND}}_{ijk}
=
\sigma^{\mathrm{ND},\tau^{2}}_{ijk}
+
\sigma^{\mathrm{ND},\tau^{0}}_{ijk}.
\label{eq:sm-ND-split}
\end{align}
Substituting Eqs.~\eqref{eq:Im-K1-nnn-for-ND} and
\eqref{eq:Im-K2-nn-for-ND} into
Eq.~\eqref{eq:sigma-ND-def}, the $\mathcal{O}(\tau^{2})$ part
gives
\begin{align}
\sigma^{\mathrm{ND},\tau^{2}}_{ijk}
&=
-\frac{e^3\tau^2}{\hbar^3}
\sum_{\bm{k}}\sum_n
\left[
f''_n
\partial_i\varepsilon_n
\partial_j\varepsilon_n
\partial_k\varepsilon_n
+
f'_n
\partial_i\varepsilon_n
\partial_j\partial_k\varepsilon_n
\right].
\label{eq:sigma-ND-before-total-derivative}
\end{align}
This expression can be written more compactly by observing that
\begin{align}
\partial_j\partial_k f_n
&=
\partial_j
\left[
(\partial_k\varepsilon_n)f'_n
\right]
\nonumber\\
&=
(\partial_j\partial_k\varepsilon_n)f'_n
+
(\partial_j\varepsilon_n)
(\partial_k\varepsilon_n)f''_n.
\label{eq:dj-dk-fn}
\end{align}
Therefore the bracket in
Eq.~\eqref{eq:sigma-ND-before-total-derivative} is
$\partial_i\varepsilon_n\,\partial_j\partial_k f_n$, and
\begin{align}
\sigma^{\mathrm{ND},\tau^{2}}_{ijk}
&=
-\frac{e^3\tau^2}{\hbar^3}
\sum_{\bm{k}}\sum_n
\partial_i\varepsilon_n\,
\partial_j\partial_k f_n .
\label{eq:sigma-ND-compact}
\end{align}
Finally, integrating by parts over the Brillouin zone and
dropping the boundary term, we obtain
\begin{align}
\sigma^{\mathrm{ND},\tau^{2}}_{ijk}
&=
-\frac{e^3\tau^2}{\hbar^3}
\sum_{\bm{k}}\sum_n
f_n\,
\partial_i\partial_j\partial_k\varepsilon_n .
\label{eq:sigma-ND-final}
\end{align}

The $\mathcal{O}(\tau^{0})$ part of
Eq.~\eqref{eq:Im-K2-nn-for-ND} contributes only through the
$\partial_i\varepsilon_n\,\partial_j\partial_k\varepsilon_n$
vertex of Eq.~\eqref{eq:Im-B-ND}, because the corresponding
$\mathcal{O}(\tau^{0})$ moment of
$\operatorname{Im}\mathcal{K}^{(1)}_{nnn}$ vanishes
[Eq.~\eqref{eq:sm-M1-4}].
Hence
\begin{align}
\sigma^{\mathrm{ND},\tau^{0}}_{ijk}
&=
\frac{e^3}{24\hbar}
\sum_{\bm{k}}\sum_n
\partial_i\varepsilon_n\,
\partial_j\partial_k\varepsilon_n\,
f'''_n .
\label{eq:sm-ND-tau0-raw}
\end{align}
Since
$\partial_i f''_n=(\partial_i\varepsilon_n)f'''_n$, the
integrand is
$(\partial_j\partial_k\varepsilon_n)\,\partial_i f''_n$, and
one integration by parts over the Brillouin zone gives
\begin{align}
\sigma^{\mathrm{ND},\tau^{0}}_{ijk}
&=
-\frac{e^3}{24\hbar}
\sum_{\bm{k}}\sum_n
f''_n\,
\partial_i\partial_j\partial_k\varepsilon_n .
\label{eq:sm-ND-tau0}
\end{align}
Two further integrations by parts, using
$\partial_i f'''_n=(\partial_i\varepsilon_n)f^{(4)}_n$,
convert Eq.~\eqref{eq:sm-ND-tau0} into the equivalent
Fermi-surface form
\begin{align}
\sigma^{\mathrm{ND},\tau^{0}}_{ijk}
&=
-\frac{e^3}{48\hbar}
\sum_{\bm{k}}\sum_n
(\partial_i\varepsilon_n)
(\partial_j\varepsilon_n)
(\partial_k\varepsilon_n)\,
f^{(4)}_n .
\label{eq:sm-ND-tau0-alt}
\end{align}
Equation~\eqref{eq:sm-ND-tau0} is the purely kinetic
$\mathcal{O}(\tau^{0})$ contribution quoted in the main text.
It carries no band-geometric content and survives in the
single-band limit, where every Berry-curvature and
quantum-metric sector vanishes identically.

\subsection{\texorpdfstring{Detailed calculation of the quantum-geometric tensor term
$\sigma^{\mathrm{QG}}_{ijk}$}{Detailed calculation of the quantum-geometric tensor term}}

We now turn to the second-order contribution containing the
two-band quantum-geometric tensor
$\mathcal{Q}^{nm}_{ij}=\mathcal{A}^{nm}_i\mathcal{A}^{mn}_j$.
This term contains both the Berry-curvature dipole contribution,
coming from $\mathrm{Im}\,\mathcal{Q}^{nm}_{ij}$, and the quantum-metric
contribution, coming from $\mathrm{Re}\,\mathcal{Q}^{nm}_{ij}$.
We write
\begin{align}
\sigma^{\mathrm{QG}}_{ijk}
&=
\frac{e^{3}}{\hbar}
\int_{-\infty}^{\infty}
\frac{d\varepsilon}{2\pi}
f(\varepsilon)
\sum_{\bm{k}}
\operatorname{Im}
\mathcal{B}^{\mathrm{QG}}_{ijk},
\label{eq:sigma-QG2-def}
\end{align}
with
\begin{align}
\mathcal{B}^{\mathrm{QG}}_{ijk}
&=
{\sum_{n,m}}^{\prime}
\varepsilon_{nm}
(\partial_i\varepsilon_n)
\left(
\mathcal{Q}^{nm}_{jk}+\mathcal{Q}^{nm}_{kj}
\right)
\left[
\varepsilon_{nm}
\mathcal{K}^{(1)}_{nnm}
-
\mathcal{K}^{(2)}_{nn}
\right]
\nonumber\\
&\quad
+
{\sum_{n,m}}^{\prime}
\varepsilon_{nm}\mathcal{Q}^{nm}_{ik}
\left[
(\partial_j\varepsilon_m)
\left(
\varepsilon_{nm}
\mathcal{K}^{(1)}_{nmm}
-
\mathcal{K}^{(2)}_{nm}
\right)
\right.
\nonumber\\
&\qquad\qquad\qquad\left.
+
(\partial_j\varepsilon_n)
\left(
\varepsilon_{nm}
\mathcal{K}^{(1)}_{nmn}
+
\mathcal{K}^{(2)}_{nm}
\right)
\right]
\nonumber\\
&\quad
+
{\sum_{n,m}}^{\prime}
\varepsilon_{nm}\mathcal{Q}^{nm}_{ij}
\left[
(\partial_k\varepsilon_m)
\left(
\varepsilon_{nm}
\mathcal{K}^{(1)}_{nmm}
-
\mathcal{K}^{(2)}_{nm}
\right)
\right.
\nonumber\\
&\qquad\qquad\qquad\left.
+
(\partial_k\varepsilon_n)
\left(
\varepsilon_{nm}
\mathcal{K}^{(1)}_{nmn}
+
\mathcal{K}^{(2)}_{nm}
\right)
\right].
\label{eq:B-QG2-def}
\end{align}
The first line contains the symmetrized tensor
$\mathcal{Q}^{nm}_{jk}+\mathcal{Q}^{nm}_{kj}=2\mathcal{G}^{nm}_{jk}$, and hence
only the quantum metric contributes.  The remaining two lines
contain both real and imaginary parts of $\mathcal{Q}^{nm}_{ij}$ and
generate both the quantum-metric and Berry-curvature-dipole
pieces.

We begin with the first part,
\begin{align}
\operatorname{Im}\mathcal{B}^{\mathrm{QG(1)}}_{ijk}
&=
2{\sum_{n,m}}^{\prime}
\varepsilon_{nm}
(\partial_i\varepsilon_n)
\mathcal{G}^{nm}_{jk}
\operatorname{Im}
\left[
\varepsilon_{nm}
\mathcal{K}^{(1)}_{nnm}
-
\mathcal{K}^{(2)}_{nn}
\right].
\label{eq:Im-B-QG1}
\end{align}
The required kernel combination is
\begin{align}
\varepsilon_{nm}
\mathcal{K}^{(1)}_{nnm}
-
\mathcal{K}^{(2)}_{nn}
&=
\frac{r}{\varepsilon_{nm}}
\partial_{\varepsilon}(A_n-R_n)
+
\frac{A_n^2}{\varepsilon_{nm}^{2}}
+
\frac{R_n^3}{\varepsilon_{nm}}
+
\frac{R_m-R_n}{\varepsilon_{nm}^{3}}
+
\mathcal{O}(\tau^{-1}).
\label{eq:K-nnm-minus-K2nn-for-QG}
\end{align}
Taking the imaginary part and using the clean-limit
Sokhotski--Plemelj formulas gives
\begin{align}
\operatorname{Im}
\left[
\varepsilon_{nm}
\mathcal{K}^{(1)}_{nnm}
-
\mathcal{K}^{(2)}_{nn}
\right]
&=
\pi
\left[
-\frac{\partial_{\varepsilon}\delta_n}{\varepsilon_{nm}^{2}}
-
\frac{\partial_{\varepsilon}^{2}\delta_n}{2\varepsilon_{nm}}
+
\frac{\delta_n-\delta_m}{\varepsilon_{nm}^{3}}
\right]
+
\mathcal{O}(\tau^{-1}).
\label{eq:Im-K-nnm-minus-K2nn-for-QG}
\end{align}
After the $\varepsilon$ integration, this gives
\begin{align}
\sigma^{\mathrm{QG(1)}}_{ijk}
&=
\frac{e^3}{2\hbar}
\sum_{\bm{k}}{\sum_{n,m}}^{\prime}
2\varepsilon_{nm}
(\partial_i\varepsilon_n)
\mathcal{G}^{nm}_{jk}
\left[
\frac{f'_n}{\varepsilon_{nm}^{2}}
-
\frac{f''_n}{2\varepsilon_{nm}}
+
\frac{f_n-f_m}{\varepsilon_{nm}^{3}}
\right]
\nonumber\\
&=
\frac{e^3}{2\hbar}
\sum_{\bm{k}}{\sum_{n,m}}^{\prime}
2
(\partial_i\varepsilon_n)
\mathcal{G}^{nm}_{jk}
\left[
\frac{f'_n}{\varepsilon_{nm}}
-
\frac{f''_n}{2}
+
\frac{f_n-f_m}{\varepsilon_{nm}^{2}}
\right].
\label{eq:sigma-QG1-before-rearrange}
\end{align}
The first two terms can be written as derivatives of $f_n$:
\begin{align}
\partial_i f_n
=
(\partial_i\varepsilon_n)f'_n,\quad
\partial_i f'_n
=
(\partial_i\varepsilon_n)f''_n.
\label{eq:dfn-identities-QG}
\end{align}
For the last term, the double sum may be symmetrized under
$n\leftrightarrow m$, yielding
\begin{align}
{\sum_{n,m}}^{\prime}
\mathcal{G}^{nm}_{jk}
(\partial_i\varepsilon_n)
\frac{f_n-f_m}{\varepsilon_{nm}^{2}}
&=
{\sum_{n,m}}^{\prime}
f_n\mathcal{G}^{nm}_{jk}
\frac{\partial_i\varepsilon_{nm}}
{\varepsilon_{nm}^{2}}.
\label{eq:QG1-sym-last-term}
\end{align}
Thus
\begin{align}
\sigma^{\mathrm{QG(1)}}_{ijk}
&=
\frac{e^3}{2\hbar}
\sum_{\bm{k}}{\sum_{n,m}}^{\prime}
2\mathcal{G}^{nm}_{jk}
\left[
\frac{\partial_i f_n}{\varepsilon_{nm}}
-
\frac{\partial_i f'_n}{2}
+
f_n\frac{\partial_i\varepsilon_{nm}}{\varepsilon_{nm}^{2}}
\right].
\label{eq:sigma-QG1-derivative-form}
\end{align}
We now integrate by parts over the Brillouin zone.  The term
containing $\partial_i f'_n$ gives
\begin{align}
-\sum_{\bm{k}}{\sum_{n,m}}^{\prime}
\mathcal{G}^{nm}_{jk}\partial_i f'_n
=
\sum_{\bm{k}}{\sum_{n,m}}^{\prime}
f'_n\partial_i\mathcal{G}^{nm}_{jk},
\label{eq:QG1-ibp-fprime}
\end{align}
while
\begin{align}
\sum_{\bm{k}}{\sum_{n,m}}^{\prime}
\mathcal{G}^{nm}_{jk}
\frac{\partial_i f_n}{\varepsilon_{nm}}
&=
-\sum_{\bm{k}}{\sum_{n,m}}^{\prime}
f_n
\partial_i
\left(
\frac{\mathcal{G}^{nm}_{jk}}{\varepsilon_{nm}}
\right)
\nonumber\\
&=
-\sum_{\bm{k}}{\sum_{n,m}}^{\prime}
f_n
\frac{\partial_i\mathcal{G}^{nm}_{jk}}{\varepsilon_{nm}}
+
\sum_{\bm{k}}{\sum_{n,m}}^{\prime}
f_n
\mathcal{G}^{nm}_{jk}
\frac{\partial_i\varepsilon_{nm}}{\varepsilon_{nm}^{2}}.
\label{eq:QG1-ibp-f}
\end{align}
Combining these terms gives
\begin{align}
&\sigma^{\mathrm{QG(1)}}_{ijk}
=\sigma^{\mathrm{intra\text{-}QMD}}_{ijk} + \sigma^{\mathrm{QG}(1)\mathcal{G}}_{ijk},
\label{eq:sigma-QG1-final}\\
&\sigma^{\mathrm{intra\text{-}QMD}}_{ijk} =
\frac{e^3}{2\hbar}
\sum_{\bm{k}}{\sum_{n,m}}^{\prime}
f'_n
\partial_i\mathcal{G}^{nm}_{jk}, 
\label{eq:sigma-intra-QMD-final}\\
&\sigma^{\mathrm{QG}(1)\mathcal{G}}_{ijk}
=
\frac{e^3}{\hbar}
\sum_{\bm{k}}{\sum_{n,m}}^{\prime}
f_n
\left[
-\frac{\partial_i\mathcal{G}^{nm}_{jk}}{\varepsilon_{nm}}
+
2\frac{\partial_i\varepsilon_{nm}}{\varepsilon_{nm}^{2}}
\mathcal{G}^{nm}_{jk}
\right].
\label{eq:sigma-QG1-G-final}
\end{align}
The intraband quantum-metric-dipole term $\sigma^{\mathrm{intra\text{-}QMD}}_{ijk}$
thus arises directly from Eq.~\eqref{eq:Im-B-QG1}, evaluated at
the Fermi surface.

We next evaluate the part proportional to $\mathcal{Q}^{nm}_{ik}$.
The corresponding contribution is
\begin{align}
\operatorname{Im}\mathcal{B}^{\mathrm{QG(2)}}_{ijk}
&=
\operatorname{Im}
\left\{
{\sum_{n,m}}^{\prime}
\varepsilon_{nm}\mathcal{Q}^{nm}_{ik}
\left[
(\partial_j\varepsilon_m)
\left(
\varepsilon_{nm}\mathcal{K}^{(1)}_{nmm}
-\mathcal{K}^{(2)}_{nm}
\right)
\right.\right.
\nonumber\\
&\qquad\qquad\qquad\left.\left.
+
(\partial_j\varepsilon_n)
\left(
\varepsilon_{nm}\mathcal{K}^{(1)}_{nmn}
+\mathcal{K}^{(2)}_{nm}
\right)
\right]
\right\}.
\label{eq:Im-B-QG2-start}
\end{align}
For a product of a complex geometric factor and a complex kernel,
we separate real and imaginary parts:
\begin{align}
\operatorname{Im}\left(CK\right)
=
\operatorname{Re}C\,\operatorname{Im}K
+
\operatorname{Im}C\,\operatorname{Re}K.
\label{eq:Im-product-CK}
\end{align}
The two kernel combinations needed here are
\begin{align}
\varepsilon_{nm}\mathcal{K}^{(1)}_{nmm}
-\mathcal{K}^{(2)}_{nm}
&=
\frac{R_m^3}{\varepsilon_{nm}}
+
\frac{R_m^2}{\varepsilon_{nm}^{2}}
+
\frac{R_m-R_n}{\varepsilon_{nm}^{3}}
+
\mathcal{O}(\tau^{-1}),
\label{eq:K-nmm-minus-K2nm}
\\
\operatorname{Im}
\left[
\varepsilon_{nm}\mathcal{K}^{(1)}_{nmm}
-\mathcal{K}^{(2)}_{nm}
\right]
&=
\pi
\left[
\frac{\partial_{\varepsilon}\delta_m}{\varepsilon_{nm}^{2}}
-
\frac{\partial_{\varepsilon}^{2}\delta_m}{2\varepsilon_{nm}}
+
\frac{\delta_n-\delta_m}{\varepsilon_{nm}^{3}}
\right]
+
\mathcal{O}(\tau^{-1}),
\label{eq:Im-K-nmm-minus-K2nm}
\\
\varepsilon_{nm}\mathcal{K}^{(1)}_{nmn}
+\mathcal{K}^{(2)}_{nm}
&=
-\frac{r}{\varepsilon_{nm}}
\partial_{\varepsilon}(A_n-R_n)
-
\frac{A_n^2}{\varepsilon_{nm}^{2}}
-
\frac{R_n^3}{\varepsilon_{nm}}
+
\frac{R_n-R_m}{\varepsilon_{nm}^{3}}
+
\mathcal{O}(\tau^{-1}),
\label{eq:K-nmn-plus-K2nm}
\\
\operatorname{Re}
\left[
\varepsilon_{nm}\mathcal{K}^{(1)}_{nmn}
+\mathcal{K}^{(2)}_{nm}
\right]
&=
-2\pi\frac{\tau}{\hbar}
\frac{\partial_{\varepsilon}\delta_n}{\varepsilon_{nm}}
+
\mathcal{O}(\tau^{0}),
\label{eq:Re-K-nmn-plus-K2nm}
\\
\operatorname{Im}
\left[
\varepsilon_{nm}\mathcal{K}^{(1)}_{nmn}
+\mathcal{K}^{(2)}_{nm}
\right]
&=
\pi
\left[
\frac{\partial_{\varepsilon}^{2}\delta_n}{2\varepsilon_{nm}}
+
\frac{\partial_{\varepsilon}\delta_n}{\varepsilon_{nm}^{2}}
+
\frac{\delta_m-\delta_n}{\varepsilon_{nm}^{3}}
\right]
+
\mathcal{O}(\tau^{-1}).
\label{eq:Im-K-nmn-plus-K2nm}
\end{align}
Using these relations and carrying out the energy integral, the
part involving $\operatorname{Re}\mathcal{Q}^{nm}_{ik}=\mathcal{G}^{nm}_{ik}$
is
\begin{align}
\sigma^{\mathrm{QG}(2)\mathcal{G}}_{ijk}
&=
\frac{e^3}{2\hbar}
\sum_{\bm{k}}{\sum_{n,m}}^{\prime}
\mathcal{G}^{nm}_{ik}
\bigg[
(\partial_j\varepsilon_m)
\left(
-\frac{f'_m}{\varepsilon_{nm}}
-f''_m
+\frac{f_n-f_m}{\varepsilon_{nm}^{2}}
\right)
\nonumber\\
&\qquad\qquad
+
(\partial_j\varepsilon_n)
\left(
f''_n
-\frac{f'_n}{\varepsilon_{nm}}
+\frac{f_m-f_n}{\varepsilon_{nm}^{2}}
\right)
\bigg].
\label{eq:sigma-QG2-G-before-sym}
\end{align}
In the first bracket, we exchange $n$ and $m$ in the double sum.
Using
\begin{align}
\mathcal{G}^{mn}_{ik}
=
\mathcal{G}^{nm}_{ik},
\qquad
\varepsilon_{mn}
=
-\varepsilon_{nm},
\label{eq:G-sym-QG2}
\end{align}
the terms containing $f'_n$ and $f''_n$ cancel between the two
brackets.  The remaining contribution is
\begin{align}
\sigma^{\mathrm{QG}(2)\mathcal{G}}_{ijk}
&=
-\frac{e^3}{\hbar}
\sum_{\bm{k}}{\sum_{n,m}}^{\prime}
f_n
\mathcal{G}^{nm}_{ik}
\frac{\partial_j\varepsilon_{nm}}
{\varepsilon_{nm}^{2}}.
\label{eq:sigma-QG2-G-final}
\end{align}
The part involving $\operatorname{Im}\mathcal{Q}^{nm}_{ik}$ is controlled
by Eq.~\eqref{eq:Re-K-nmn-plus-K2nm}.  Since
$\operatorname{Im}\mathcal{Q}^{nm}_{ik}=-\Omega^{nm}_{ik}/2$, we obtain
\begin{align}
\sigma^{\mathrm{QG}(2)\Omega}_{ijk}
&=
-\frac{e^3}{2\hbar^2}\tau
\sum_{\bm{k}}{\sum_{n,m}}^{\prime}
f'_n
(\partial_j\varepsilon_n)
\Omega^{nm}_{ik}
\nonumber\\
&=
-\frac{e^3}{2\hbar^2}\tau
\sum_{\bm{k}}{\sum_{n,m}}^{\prime}
(\partial_j f_n)
\Omega^{nm}_{ik}.
\label{eq:sigma-QG2-Omega-before-ibp}
\end{align}
After summing over the intermediate band $m$,
\begin{align}
\Omega^n_{ik}
=
\sum_{m\neq n}\Omega^{nm}_{ik}
=
\sum_l\varepsilon_{ikl}\Omega^n_l,
\label{eq:Omega-band-sum}
\end{align}
and integrating by parts over the Brillouin zone, we find
\begin{align}
\sigma^{\mathrm{QG}(2)\Omega}_{ijk}
&=
\frac{e^3}{2\hbar^2}\tau
\sum_{\bm{k}}\sum_n
f_n
\partial_j\Omega^n_{ik}
\nonumber\\
&=
\frac{e^3}{2\hbar^2}\tau
\sum_{\bm{k}}\sum_{n,l}
f_n
\varepsilon_{ikl}
\mathcal{D}^n_{jl},
\label{eq:sigma-QG2-Omega-final-j}
\end{align}
where
\begin{align}
\mathcal{D}^{n}_{jl}
=
\partial_j\Omega^n_l.
\label{eq:Berry-curvature-dipole-def}
\end{align}
The same calculation applied to the term proportional to
$\mathcal{Q}^{nm}_{ij}$ gives the partner contribution with
$j\leftrightarrow k$:
\begin{align}
\sigma^{\mathrm{QG}(3)\Omega}_{ijk}
&=
\frac{e^3}{2\hbar^2}\tau
\sum_{\bm{k}}\sum_{n,l}
f_n
\varepsilon_{ijl}
\mathcal{D}^n_{kl},
\label{eq:sigma-QG3-Omega-final-k}
\end{align}
and
\begin{align}
\sigma^{\mathrm{QG}(3)\mathcal{G}}_{ijk}
&=
-\frac{e^3}{\hbar}
\sum_{\bm{k}}{\sum_{n,m}}^{\prime}
f_n
\mathcal{G}^{nm}_{ij}
\frac{\partial_k\varepsilon_{nm}}
{\varepsilon_{nm}^{2}}.
\label{eq:sigma-QG3-G-final}
\end{align}

Collecting Eqs.~
\eqref{eq:sigma-intra-QMD-final}, 
\eqref{eq:sigma-QG1-G-final}, 
\eqref{eq:sigma-QG2-G-final},
\eqref{eq:sigma-QG2-Omega-final-j},
\eqref{eq:sigma-QG3-Omega-final-k} and
\eqref{eq:sigma-QG3-G-final}, the second-order
Berry and quantum-geometric contribution can be written as
\begin{align}
\sigma^{\mathrm{QG}}_{ijk}
&=
\sigma^{\mathrm{BCD}}_{ijk}
+
\sigma^{\mathcal{G}}_{ijk},
\label{eq:sigma-QG2-decomposition}
\\
\sigma^{\mathrm{BCD}}_{ijk}
&=
\frac{e^3}{2\hbar^2}\tau
\sum_{\bm{k}}\sum_{n,l}
f_n
\left[
\varepsilon_{ijl}\mathcal{D}^{n}_{kl}
+
\varepsilon_{ikl}\mathcal{D}^{n}_{jl}
\right],
\label{eq:sigma-Omega-final}
\\
\sigma^{\mathcal{G}}_{ijk}
&=\sigma^{\textrm{intra-}\mathrm{QMD}}_{ijk}+\sigma^{\mathrm{QG},\mathcal{G}}_{ijk}, \\
\sigma^{\mathrm{QG},\mathcal{G}}_{ijk}&=
\sigma^{\mathrm{QG}(1)\mathcal{G}}_{ijk}+\sigma^{\mathrm{QG}(2)\mathcal{G}}_{ijk}+\sigma^{\mathrm{QG}(3)\mathcal{G}}_{ijk},\\
\sigma^{\textrm{intra-}\mathrm{QMD}}_{ijk}&=
\frac{e^3}{2\hbar}
\sum_{\bm{k}}{\sum_{n,m}}^{\prime}
f'_n
\partial_i\mathcal{G}^{nm}_{jk}, \\
\sigma^{\mathrm{QG},\mathcal{G}}_{ijk}
&=
\frac{e^3}{\hbar}
\sum_{\bm{k}}{\sum_{n,m}}^{\prime}
f_n
\left[
-\frac{\partial_i\mathcal{G}^{nm}_{jk}}{\varepsilon_{nm}}
+
2\frac{\partial_i\varepsilon_{nm}}
{\varepsilon_{nm}^{2}}
\mathcal{G}^{nm}_{jk}
\right]
\nonumber\\
&\quad
-
\frac{e^3}{\hbar}
\sum_{\bm{k}}{\sum_{n,m}}^{\prime}
f_n
\left[
\mathcal{G}^{nm}_{ij}
\frac{\partial_k\varepsilon_{nm}}{\varepsilon_{nm}^{2}}
+
\mathcal{G}^{nm}_{ik}
\frac{\partial_j\varepsilon_{nm}}{\varepsilon_{nm}^{2}}
\right].
\label{eq:sigma-G-final}
\end{align}
Here $\sigma^{\mathrm{BCD}}_{ijk}$ is the Berry-curvature-dipole
contribution, which is proportional to $\tau$, whereas
$\sigma^{\mathcal{G}}_{ijk}$ is the quantum-metric contribution
and remains of order $\tau^0$ in the clean limit.

\subsection{\texorpdfstring{Detailed calculation of the quantum-connection term
$\sigma^{\mathcal{C}}_{ijk}$}{Detailed calculation of the quantum-connection term}}
\label{subsec:sigma-C-evaluation}
We next evaluate the quantum-connection contribution.
In the intermediate band-basis expression, this term contains the
covariant derivative of the interband Berry connection.
To make the differentiated component explicit, we use the notation
\begin{align}
\mathcal{C}^{nm}_{ij|k}
&\equiv
\mathcal{A}^{nm}_{i}
\mathcal{D}_{k}\mathcal{A}_{j}^{mn}
=\Gamma^{nm}_{ij|k}-i\widetilde{\Gamma}^{nm}_{ij|k}
,\qquad \mathcal{C}^{mn}_{ij|k}=(\mathcal{C}^{nm}_{ij|k}
)^{*},
\qquad n\neq m,
\label{eq:Cbar-def-eval}
\end{align}
where the index after the vertical bar denotes the momentum
derivative acting on the Berry-connection component.
With this notation, the quantum-connection contribution is written as
\begin{align}
\sigma^{\mathcal{C}}_{ijk}
&=
\frac{e^{3}}{\hbar}
\int_{-\infty}^{\infty}
\frac{d\varepsilon}{2\pi}
f(\varepsilon)
\sum_{\bm{k}}
\operatorname{Im}
\mathcal{B}^{\mathcal{C}}_{ijk},
\label{eq:sigma-C-def-eval}
\\
\mathcal{B}^{\mathcal{C}}_{ijk}
&=
{\sum_{n,m}}^{\prime}
\varepsilon_{nm}^{2}
\mathcal{C}^{nm}_{ij|k}
\mathcal{K}^{(2)}_{nm}.
\label{eq:B-C-def-eval}
\end{align}
Since both $\mathcal{C}^{nm}_{ij|k}$ and
$\mathcal{K}^{(2)}_{nm}$ are in general complex, we separate
their real and imaginary parts as
\begin{align}
\operatorname{Im}
\mathcal{B}^{\mathcal{C}}_{ijk}
&=
{\sum_{n,m}}^{\prime}
\varepsilon_{nm}^{2}
\operatorname{Im}
\left[
\mathcal{C}^{nm}_{ij|k}
\mathcal{K}^{(2)}_{nm}
\right]
\nonumber\\
&=
{\sum_{n,m}}^{\prime}
\varepsilon_{nm}^{2}
\left[
\operatorname{Re}
\mathcal{C}^{nm}_{ij|k}\,
\operatorname{Im}
\mathcal{K}^{(2)}_{nm}
+
\operatorname{Im}
\mathcal{C}^{nm}_{ij|k}\,
\operatorname{Re}
\mathcal{K}^{(2)}_{nm}
\right].
\label{eq:Im-B-C-separated-eval}
\end{align}
The kernel $\mathcal{K}^{(2)}_{nm}$ [Eq.~\eqref{eq:K2-nm-expanded}] is given by
\begin{align}
\mathcal{K}^{(2)}_{nm}
&=
\frac{1}{\varepsilon_{nm}^{2}}
\left(A_n^2+R_m^2\right)
+\frac{2}{\varepsilon_{nm}^{3}}(R_m-R_n)
+\mathcal{O}(\tau^{-1}),
\label{eq:K2-nm-expanded2}
\end{align}
and the real part of $\mathcal{K}^{(2)}_{nm}$ is symmetric under $n\leftrightarrow m$.
On the other hand, the imaginary part of $\mathcal{C}^{nm}_{ij|k}$ is antisymmetric under
$n\leftrightarrow m$ because $\mathcal{C}^{mn}_{ij|k}=(\mathcal{C}^{nm}_{ij|k})^{*}$.
Thus the second term in Eq.~\eqref{eq:Im-B-C-separated-eval} vanishes identically, 
\begin{align}
\operatorname{Im}
\mathcal{B}^{\mathcal{C}}_{ijk}
&=
{\sum_{n,m}}^{\prime}
\varepsilon_{nm}^{2}
\Gamma^{nm}_{ij|k}
\operatorname{Im}
\mathcal{K}^{(2)}_{nm}.
\label{eq:Im-B-C-Gamma-eval}
\end{align}
Using the clean-limit result
\begin{align}
\operatorname{Im}\mathcal{K}^{(2)}_{nm}
&=
\pi
\left[
\frac{
\partial_{\varepsilon}\delta_m
-
\partial_{\varepsilon}\delta_n
}
{\varepsilon_{nm}^{2}}
+
\frac{2(\delta_n-\delta_m)}
{\varepsilon_{nm}^{3}}
\right]
+
\mathcal{O}(\tau^{-1}),
\label{eq:Im-K2-nm-for-C-eval}
\end{align}
we obtain
\begin{align}
\sigma^{\mathcal{C}}_{ijk}
&=
\frac{e^{3}}{2\hbar}
\sum_{\bm{k}}{\sum_{n,m}}^{\prime}
\varepsilon_{nm}^{2}
\Gamma^{nm}_{ij|k}
\left[
-\frac{f'_m-f'_n}{\varepsilon_{nm}^{2}}
+
2\frac{f_n-f_m}{\varepsilon_{nm}^{3}}
\right].
\label{eq:sigma-C-before-sym-eval}
\end{align}
The term proportional to $f'_m-f'_n$ vanishes after the
$n\leftrightarrow m$ symmetrization, while the second term gives
\begin{align}
\sigma^{\mathcal{C}}_{ijk}
&=
2\frac{e^{3}}{\hbar}
\sum_{\bm{k}}{\sum_{n,m}}^{\prime}
\frac{f_n}{\varepsilon_{nm}}\Gamma^{nm}_{ij|k}.
\label{eq:sigma-C-Gamma-eval}
\end{align}
The real coefficient $\Gamma^{nm}_{ij|k}$ can be expressed in
terms of the quantum metric and the commutator of Berry
connections [Eq.~\eqref{eq:Gamma-metric-X}] as
\begin{align}
2\Gamma^{nm}_{ij|k}
&=
\partial_k\mathcal{G}^{nm}_{ij}
+
\partial_j\mathcal{G}^{nm}_{ki}
-
\partial_i\mathcal{G}^{nm}_{jk}
+
X^{nm}_{ijk},
\label{eq:Gamma-metric-Xi-eval}
\end{align}
where
\begin{align}
X^{nm}_{ijk}
&\equiv
\operatorname{Im}
\left[
\mathcal{A}_{i}^{nm}
[\mathcal{A}_{j},\mathcal{A}_{k}]^{\prime}_{mn}
\right]
+
\operatorname{Im}
\left[
\mathcal{A}_{j}^{nm}
[\mathcal{A}_{k},\mathcal{A}_{i}]^{\prime}_{mn}
\right]
-
\operatorname{Im}
\left[
\mathcal{A}_{k}^{nm}
[\mathcal{A}_{i},\mathcal{A}_{j}]^{\prime}_{mn}
\right],
\label{eq:Xi-def-eval}
\\
[\mathcal{A}_{j},\mathcal{A}_{k}]^{\prime}_{mn}
&\equiv
{\sum_l}^{\prime}
\left(
\mathcal{A}^{ml}_{j}\mathcal{A}^{ln}_{k}
-
\mathcal{A}^{ml}_{k}\mathcal{A}^{ln}_{j}
\right).
\label{eq:commutator-prime-def-eval}
\end{align}
Here the prime on the $l$ sum means that $l$ is distinct from
both $m$ and $n$.  Substituting Eq.~\eqref{eq:Gamma-metric-Xi-eval}
into Eq.~\eqref{eq:sigma-C-Gamma-eval}, we decompose
$\sigma^{\mathcal{C}}_{ijk}$ into a metric-derivative part and
a commutator part:
\begin{align}
\sigma^{\mathcal{C}}_{ijk}
&=
\sigma^{\mathcal{C},\mathcal{G}}_{ijk}
+
\sigma^{\mathcal{C},\mathcal{T}}_{ijk},
\label{eq:sigma-C-decomp-eval}
\\
\sigma^{\mathcal{C},\mathcal{G}}_{ijk}
&=
\frac{e^{3}}{\hbar}
\sum_{\bm{k}}{\sum_{n,m}}^{\prime}
\frac{f_n}{\varepsilon_{nm}}\left(
\partial_k\mathcal{G}^{nm}_{ij}
+
\partial_j\mathcal{G}^{nm}_{ki}
-
\partial_i\mathcal{G}^{nm}_{jk}
\right),
\label{eq:sigma-C-G-final-eval}
\\
\sigma^{\mathcal{C},\mathcal{T}}_{ijk}
&=
\frac{e^{3}}{\hbar}
\sum_{\bm{k}}{\sum_{n,m}}^{\prime}
f_n
\frac{X^{nm}_{ijk}}
{\varepsilon_{nm}}.
\label{eq:sigma-C-Tpart-final-eval}
\end{align}

\subsection{\texorpdfstring{Detailed calculation of the three-Berry-connection term
$\sigma^{\mathcal{T}}_{ijk}$}{Detailed calculation of the three-Berry-connection term}}
\label{subsec:sigma-T-evaluation}

We now evaluate the three-connection contribution.  This term
was denoted by an $\mathcal{A}^{3}$ structure in the intermediate
algebra, but here we denote it by $\mathcal{T}$ because it is
built from the three-connection tensor
\begin{align}
\mathcal{T}^{nml}_{ijk}
&\equiv
\mathcal{A}^{nm}_{i}
\mathcal{A}^{ml}_{j}
\mathcal{A}^{ln}_{k}.
\label{eq:Tijk-def-eval}
\end{align}
The corresponding conductivity is
\begin{align}
\sigma^{\mathcal{T}}_{ijk}
&=
\frac{e^{3}}{\hbar}
\int_{-\infty}^{\infty}
\frac{d\varepsilon}{2\pi}
f(\varepsilon)
\sum_{\bm{k}}
\operatorname{Im}
\mathcal{B}^{\mathcal{T}}_{ijk},
\label{eq:sigma-T-def-eval}
\\
\mathcal{B}^{\mathcal{T}}_{ijk}
&=
-i{\sum_{n,m,l}}^{\prime}
\bigg\{
\varepsilon_{nm}\varepsilon_{ml}\varepsilon_{ln}
\mathcal{K}^{(1)}_{nml}
\left(
\mathcal{T}^{nml}_{ijk}
+
\mathcal{T}^{nml}_{ikj}
\right)
+
\varepsilon_{nm}\mathcal{K}^{(2)}_{nm}
\left(
\varepsilon_{ml}\mathcal{T}^{nml}_{ijk}
-
\varepsilon_{ln}\mathcal{T}^{nml}_{ikj}
\right)
\bigg\}.
\label{eq:B-T-def-eval}
\end{align}
The prime on the triple sum means that $n$, $m$ and $l$ are all
distinct.  To reduce this expression, we use the three-distinct
kernel identity
\begin{align}
\mathcal{K}^{(1)}_{nml}
&=
-\varepsilon_{ln}^{-1}
\mathcal{K}^{(2)}_{nm}
-\varepsilon_{nm}^{-2}
 \varepsilon_{ln}^{-2}(R_n-R_m)
-\varepsilon_{ml}^{-2}
 \varepsilon_{ln}^{-2}(R_m-R_l)
-\varepsilon_{nm}^{-1}
 \varepsilon_{ml}^{-1}
 \varepsilon_{ln}^{-1}R_m^2,
\label{eq:K1-nml-for-T-eval}
\\
\mathcal{K}^{(2)}_{nm}
&=
-2\varepsilon_{nm}^{-3}(R_n-R_m)
+\varepsilon_{nm}^{-2}(A_n^2+R_m^2)
+\mathcal{O}(\tau^{-1}).
\label{eq:K2-nm-for-T-eval}
\end{align}
Substitution into Eq.~\eqref{eq:B-T-def-eval} gives
\begin{align}
\mathcal{B}^{\mathcal{T}}_{ijk}
&=
-iS_{ijk},
\label{eq:B-T-minus-iS-eval}
\\
S_{ijk}
&=
{\sum_{n,m,l}}^{\prime}
\left[
\Lambda^{nml}_{ijk}\mathcal{T}^{nml}_{ijk}
+
\Lambda^{nml}_{ikj}\mathcal{T}^{nml}_{ikj}
\right],
\label{eq:Sijk-def-eval}
\end{align}
where the scalar coefficients are
\begin{align}
\Lambda^{nml}_{ijk}
&=
-\varepsilon_{nm}^{-1}
 \varepsilon_{ml}
 \varepsilon_{ln}^{-1}(R_n-R_m)
-\varepsilon_{nm}
 \varepsilon_{ml}^{-1}
 \varepsilon_{ln}^{-1}(R_m-R_l)
-R_m^2,
\label{eq:Lambda-ijk-eval}
\\
\Lambda^{nml}_{ikj}
&=
(\varepsilon_{ln}^{-1}-\varepsilon_{nm}^{-1})(R_n-R_m)
-\varepsilon_{nm}
 \varepsilon_{ml}^{-1}
 \varepsilon_{ln}^{-1}(R_m-R_l)
+A_n^2.
\label{eq:Lambda-ikj-eval}
\end{align}
It is useful to add and subtract the auxiliary quantity
\begin{align}
G_{ijk}
&=
{\sum_{n,m,l}}^{\prime}
\bigg[
\left(
-R_nR_m-R_mR_l+R_nR_l
\right)\mathcal{T}^{nml}_{ijk}
+
\left(
R_nR_m-R_mR_l+R_nR_l
\right)\mathcal{T}^{nml}_{ikj}
\bigg].
\label{eq:Gijk-def-eval}
\end{align}
Using 
\begin{align}
R_aR_b
=
\frac{R_a-R_b}{\varepsilon_{ab}},
\label{eq:RR-for-T-eval}
\end{align}
and writing 
\begin{align}
a=\varepsilon_{nm},
\qquad
b=\varepsilon_{ml},
\qquad
c=\varepsilon_{ln},
\qquad
a+b+c=0,
\label{eq:abc-eval}
\end{align}
one obtains 
\begin{align}
\Lambda^{nml}_{ijk}
&= -\frac{b}{ac}(R_n-R_m)-\frac{a}{bc}(R_m-R_l)-R^2_m\nonumber\\
&= \frac{a+c}{ac}(R_n-R_m)+\frac{b+c}{bc}(R_m-R_l)-R^2_m\nonumber\\
&= \left(\frac{1}{a}+\frac{1}{c}\right)(R_n-R_m)+\left(\frac{1}{b}+\frac{1}{c}\right)(R_m-R_l)-R^2_m. \\
\Lambda^{nml}_{ikj}
&=
\left(\frac{1}{c}-\frac{1}{a}\right)(R_n-R_m)-\frac{a}{bc}(R_m-R_l)+A_n^2\nonumber\\
&=\left(\frac{1}{c}-\frac{1}{a}\right)(R_n-R_m)+\frac{b+c}{bc}(R_m-R_l)+A_n^2\nonumber\\
&=\left(\frac{1}{c}-\frac{1}{a}\right)(R_n-R_m)+\left(\frac{1}{b}+\frac{1}{c}\right)(R_m-R_l)+A_n^2, \\
\mp R_nR_m-R_mR_l+R_nR_l 
&=\mp \frac{1}{a}(R_n-R_m)-\frac{1}{b}(R_m-R_l)+\frac{1}{c}(R_l-R_n), 
\end{align}
Thus one obtains 
\begin{align}
&\Lambda^{nml}_{ijk}
+
\left(
-R_nR_m-R_mR_l+R_nR_l
\right)\nonumber\\
&\quad=
\left(\frac{1}{a}+\frac{1}{c}\right)(R_n-R_m)+\left(\frac{1}{b}+\frac{1}{c}\right)(R_m-R_l)
-R^2_m-\frac{1}{a}(R_n-R_m)-\frac{1}{b}(R_m-R_l)+\frac{1}{c}(R_l-R_n)
\nonumber\\
&\quad=
-R_m^2,
\label{eq:Lambda-plus-Dijk-eval}\\
&\Lambda^{nml}_{ikj}
+
\left(
R_nR_m-R_mR_l+R_nR_l
\right)\nonumber\\
&\quad=
\left(\frac{1}{c}-\frac{1}{a}\right)(R_n-R_m)+\left(\frac{1}{b}+\frac{1}{c}\right)(R_m-R_l)+A_n^2
+\frac{1}{a}(R_n-R_m)-\frac{1}{b}(R_m-R_l)+\frac{1}{c}(R_l-R_n)\nonumber\\
&\quad=
A_n^2.
\label{eq:Lambda-plus-Dikj-eval}
\end{align}
Therefore
\begin{align}
S_{ijk}+G_{ijk}
&=
{\sum_{n,m,l}}^{\prime}
\left[
-R_m^2\mathcal{T}^{nml}_{ijk}
+
A_n^2\mathcal{T}^{nml}_{ikj}
\right]
\equiv
D_{ijk},
\label{eq:SplusG-D-eval}
\end{align}
or equivalently
\begin{align}
-iS_{ijk}
=
iG_{ijk}
-iD_{ijk}.
\label{eq:SminusGplusD-eval}
\end{align}
The second term does not contribute to the imaginary part.  To
see this, relabel $n\leftrightarrow m$ in the first term of
$D_{ijk}$ and use
\begin{align}
\mathcal{T}^{mnl}_{ijk}
&=
\left(
\mathcal{T}^{nml}_{ikj}
\right)^{*},
\qquad
A_n^2=(R_n^2)^{*}.
\label{eq:T-conj-D-eval}
\end{align}
Then
\begin{align}
\operatorname{Im}(-iD_{ijk})
&=
-{\sum_{n,m,l}}^{\prime}
\operatorname{Re}
\left[
-R_m^2\mathcal{T}^{nml}_{ijk}
+
A_n^2\mathcal{T}^{nml}_{ikj}
\right]
\nonumber\\
&=
{\sum_{n,m,l}}^{\prime}
\operatorname{Re}
\left[
R_n^2
\left(\mathcal{T}^{nml}_{ikj}\right)^{*}
-
(R_n^2)^{*}
\mathcal{T}^{nml}_{ikj}
\right]
\nonumber\\
&=0.
\label{eq:Im-minus-iD-zero-eval}
\end{align}
Thus only $iG_{ijk}$ remains.

We next fold the triple sum in $G_{ijk}$ into a double sum.
The products in Eq.~\eqref{eq:Gijk-def-eval} can be written as
\begin{align}
{\sum_{n,m,l}}^{\prime}
\mathcal{T}^{nml}_{ijk}
\left(
-R_nR_m-R_mR_l+R_nR_l
\right)
&=
{\sum_{n,m,l}}^{\prime}
\varepsilon_{nm}^{-1}
(R_n-R_m)\left(
-\mathcal{T}^{nml}_{ijk}
-\mathcal{T}^{nml}_{jki}
+\mathcal{T}^{nml}_{kij}
\right), 
\label{eq:G1fold-eval}
\\
{\sum_{n,m,l}}^{\prime}
\mathcal{T}^{nml}_{ikj}
\left(
R_nR_m-R_mR_l+R_nR_l
\right)
&=
{\sum_{n,m,l}}^{\prime}
\varepsilon_{nm}^{-1}
(R_n-R_m)\left(
\mathcal{T}^{nml}_{ikj}
-\mathcal{T}^{nml}_{kji}
+\mathcal{T}^{nml}_{jik}
\right). 
\label{eq:G2fold-eval}
\end{align}
Combining these two relations gives
\begin{align}
G_{ijk}
&=
-{\sum_{n,m}}^{\prime}
\varepsilon_{nm}^{-1}
(R_n-R_m)Y^{nm}_{ijk},
\label{eq:G-double-Y-eval}
\end{align}
where 
\begin{align}
Y^{nm}_{ijk}
&={\sum_{l}}^{\prime}\left(
\mathcal{T}^{nml}_{ijk}
+\mathcal{T}^{nml}_{jki}
-\mathcal{T}^{nml}_{kij}
-\mathcal{T}^{nml}_{ikj}
+\mathcal{T}^{nml}_{kji}
-\mathcal{T}^{nml}_{jik}\right)\nonumber\\
&={\sum_{l}}^{\prime}\left(
\mathcal{A}^{nm}_{i}\mathcal{A}^{ml}_{j}\mathcal{A}^{ln}_{k}
+\mathcal{A}^{nm}_{j}\mathcal{A}^{ml}_{k}\mathcal{A}^{ln}_{i}
-\mathcal{A}^{nm}_{k}\mathcal{A}^{ml}_{i}\mathcal{A}^{ln}_{j}
-\mathcal{A}^{nm}_{i}\mathcal{A}^{ml}_{k}\mathcal{A}^{ln}_{j}
+\mathcal{A}^{nm}_{k}\mathcal{A}^{ml}_{j}\mathcal{A}^{ln}_{i}
-\mathcal{A}^{nm}_{j}\mathcal{A}^{ml}_{i}\mathcal{A}^{ln}_{k}
\right)
\nonumber\\
&=
\mathcal{A}^{nm}_{i}{\sum_{l}}^{\prime}\bigl(\mathcal{A}^{ml}_{j}\mathcal{A}^{ln}_{k}-\mathcal{A}^{ml}_{k}\mathcal{A}^{ln}_{j}\bigr)
+\mathcal{A}^{nm}_{j}{\sum_{l}}^{\prime}\bigl(
\mathcal{A}^{ml}_{k}\mathcal{A}^{ln}_{i}-\mathcal{A}^{ml}_{i}\mathcal{A}^{ln}_{k}\bigr)
-\mathcal{A}^{nm}_{k}{\sum_{l}}^{\prime}
\bigl(\mathcal{A}^{ml}_{i}\mathcal{A}^{ln}_{j}-\mathcal{A}^{ml}_{j}\mathcal{A}^{ln}_{i}\bigr)
\nonumber\\
&=
\mathcal{A}_{i}^{nm}
[\mathcal{A}_{j},\mathcal{A}_{k}]^{\prime}_{mn}
+\mathcal{A}_{j}^{nm}
[\mathcal{A}_{k},\mathcal{A}_{i}]^{\prime}_{mn}
-\mathcal{A}_{k}^{nm}
[\mathcal{A}_{i},\mathcal{A}_{j}]^{\prime}_{mn}.
\label{eq:Y-def-eval}
\end{align}
Its imaginary part is precisely the real commutator combination introduced
in Eq.~\eqref{eq:Xi-def-eval},
\begin{align}
  \operatorname{Im}Y^{nm}_{ijk}=X^{nm}_{ijk}.
  \label{eq:ImY-equals-X-eval}
\end{align}
The Hermiticity of the Berry connection implies
\begin{align}
Y^{mn}_{ijk}
&=
-\left(Y^{nm}_{ijk}\right)^{*}.
\label{eq:Y-conj-eval}
\end{align}
Hence, after symmetrizing the double sum under $n\leftrightarrow m$,
the real-principal-value part cancels and
\begin{align}
\operatorname{Im}(iG_{ijk})
&=
2{\sum_{n,m}}^{\prime}
\varepsilon_{nm}^{-1}
\operatorname{Im}R_n\,
\operatorname{Im}Y^{nm}_{ijk}.
\label{eq:Im-iG-Y-eval}
\end{align}
In the clean limit, using $\mathrm{Im}R_{n} = -\pi \delta_{n}$, 
one obtains  
\begin{align}
\operatorname{Im}(iG_{ijk})
&=
-2\pi{\sum_{n,m}}^{\prime}
\varepsilon_{nm}^{-1}\delta_{n}\,
\operatorname{Im}Y^{nm}_{ijk}
\nonumber\\
&=
-2\pi{\sum_{n,m}}^{\prime}
\varepsilon_{nm}^{-1}\delta_{n}\,
X^{nm}_{ijk},
\label{eq:Im-iG-Xi-eval}
\end{align}
Therefore, we obtain 
\begin{align}
\sigma^{\mathcal{T}}_{ijk}
&=
\frac{e^{3}}{\hbar}
\int_{-\infty}^{\infty}
\frac{d\varepsilon}{2\pi}
f(\varepsilon)
\sum_{\bm{k}}
\operatorname{Im}(iG_{ijk})
\nonumber\\
&=-
\frac{e^{3}}{\hbar}
\sum_{\bm{k}}{\sum_{n,m}}^{\prime}\frac{f_{n}}{\varepsilon_{nm}}X^{nm}_{ijk}.
\label{eq:sigma-T-final-eval}
\end{align}

\subsection{Cancellation of the three-connection sector}
\label{subsec:sigma-T-cancellation}

Equations~\eqref{eq:sigma-C-Tpart-final-eval} and
\eqref{eq:sigma-T-final-eval} show that the commutator part of
the connection-geometric contribution and the three-connection
contribution are exactly opposite:
\begin{align}
\sigma^{\mathcal{C},\mathcal{T}}_{ijk}
&=
\frac{e^{3}}{\hbar}
\sum_{\bm{k}}{\sum_{n,m}}^{\prime}
\frac{f_n}
{\varepsilon_{nm}}X^{nm}_{ijk},
\label{eq:sigma-C-Tpart-repeat-eval}
\\
\sigma^{\mathcal{T}}_{ijk}
&=
-\frac{e^{3}}{\hbar}
\sum_{\bm{k}}{\sum_{n,m}}^{\prime}
\frac{f_n}
{\varepsilon_{nm}}X^{nm}_{ijk}.
\label{eq:sigma-T-repeat-eval}
\end{align}
Therefore all explicit three-connection terms cancel:
\begin{align}
\sigma^{\mathcal{C},\mathcal{T}}_{ijk}
+
\sigma^{\mathcal{T}}_{ijk}
&=0.
\label{eq:sigma-T-cancel-eval}
\end{align}
After this cancellation, the only surviving part of
$\sigma^{\mathcal{C}}_{ijk}+\sigma^{\mathcal{T}}_{ijk}$ is the
metric-derivative contribution
\begin{align}
\sigma^{\mathcal{C}}_{ijk}
+
\sigma^{\mathcal{T}}_{ijk}
&=
\sigma^{\mathcal{C},\mathcal{G}}_{ijk}
\nonumber\\
&=
\frac{e^{3}}{\hbar}
\sum_{\bm{k}}{\sum_{n,m}}^{\prime}
\frac{f_n}{\varepsilon_{nm}}
\left(
\partial_k\mathcal{G}^{nm}_{ij}
+
\partial_j\mathcal{G}^{nm}_{ki}
-
\partial_i\mathcal{G}^{nm}_{jk}
\right).
\label{eq:sigma-C-plus-T-surviving-eval}
\end{align}

\subsection{Final expression for the second-order dc conductivity}
\label{subsec:sigma-DC-final-summary}

We now collect the contributions to the second-order dc
conductivity.  The nonlinear Drude term is 
\begin{align}
&\sigma^{\mathrm{ND}}_{ijk}
=\sigma^{\mathrm{ND},\tau^{2}}_{ijk} + \sigma^{\mathrm{ND},\tau^{0}}_{ijk},
\label{eq:sigma-ND-final-summary-eval}\\
&\sigma^{\mathrm{ND},\tau^{2}}_{ijk}=
-\frac{e^{3}\tau^{2}}{\hbar^{3}}
\sum_{\bm{k}}\sum_{n}
f_n\,
\partial_i\partial_j\partial_k\varepsilon_n,
\label{eq:sigma-ND-tau2-final-summary-eval}\\
&\sigma^{\mathrm{ND},\tau^{0}}_{ijk} =  
-\frac{e^3}{24\hbar}
\sum_{\bm{k}}\sum_n
f''_n\,
\partial_i\partial_j\partial_k\varepsilon_n 
= -\frac{e^3}{48\hbar}
\sum_{\bm{k}}\sum_n
(\partial_i\varepsilon_n)
(\partial_j\varepsilon_n)
(\partial_k\varepsilon_n)\,
f^{(4)}_n.
\label{eq:sigma-ND-tau0-final-summary-eval}
\end{align}
The Berry-curvature-dipole contribution is
\begin{align}
\sigma^{\mathrm{BCD}}_{ijk}
&=
\frac{e^{3}\tau}{2\hbar^{2}}
\sum_{\bm{k}}\sum_{n,l}
f_n
\left[
\varepsilon_{ijl}\mathcal{D}^{n}_{kl}
+
\varepsilon_{ikl}\mathcal{D}^{n}_{jl}
\right],
\label{eq:sigma-Omega-final-summary-eval}
\\
\mathcal{D}^{n}_{jl}
&=
\partial_j\Omega^n_l.
\label{eq:BCD-def-summary-eval}
\end{align}
The quantum-metric part before adding
$\sigma^{\mathcal{C}}+\sigma^{\mathcal{T}}$ is
\begin{align}
&\sigma^{\mathcal{G}}_{ijk}=\sigma^{\textrm{intra-}\mathrm{QMD}}_{ijk}
+\sigma^{\mathrm{QG},\mathcal{G}}_{ijk}, \\
&\sigma^{\textrm{intra-}\mathrm{QMD}}_{ijk}=
\frac{e^{3}}{2\hbar}
\sum_{\bm{k}}{\sum_{n,m}}^{\prime}
f'_n
\partial_i\mathcal{G}^{nm}_{jk}
\label{eq:sigma-intra-QMD-final-summary-eval}\\
&\sigma^{\mathrm{QG},\mathcal{G}}_{ijk}=
\frac{e^{3}}{\hbar}
\sum_{\bm{k}}{\sum_{n,m}}^{\prime}
f_n
\left[
-\frac{\partial_i\mathcal{G}^{nm}_{jk}}{\varepsilon_{nm}}
+
2\frac{\partial_i\varepsilon_{nm}}
{\varepsilon_{nm}^{2}}
\mathcal{G}^{nm}_{jk}
\right]
\nonumber\\
&\qquad\qquad
-
\frac{e^{3}}{\hbar}
\sum_{\bm{k}}{\sum_{n,m}}^{\prime}
f_n
\left[
\mathcal{G}^{nm}_{ij}
\frac{\partial_k\varepsilon_{nm}}{\varepsilon_{nm}^{2}}
+
\mathcal{G}^{nm}_{ik}
\frac{\partial_j\varepsilon_{nm}}{\varepsilon_{nm}^{2}}
\right].
\label{eq:sigma-G-before-CT-summary-eval}
\end{align}
Adding the surviving part
Eq.~\eqref{eq:sigma-C-plus-T-surviving-eval}, this metric sector
can be written in a compact total-derivative form:
\begin{align}
&\sigma^{\mathrm{QG},\mathcal{G}}_{ijk}
+
\sigma^{\mathcal{C}}_{ijk}
+
\sigma^{\mathcal{T}}_{ijk}
=\sigma^{\mathrm{QG},\mathcal{G}}_{ijk}
+
\sigma^{\mathcal{C},\mathcal{G}}_{ijk}
=\sigma^{\textrm{inter-}\mathrm{QMD}}_{ijk}
\\
&\sigma^{\textrm{inter-}\mathrm{QMD}}_{ijk}=
\frac{e^{3}}{\hbar}
\sum_{\bm{k}}{\sum_{n,m}}^{\prime}
f_n
\left[
-2\partial_i
\left(
\frac{\mathcal{G}^{nm}_{jk}}{\varepsilon_{nm}}
\right)
+
\partial_j
\left(
\frac{\mathcal{G}^{nm}_{ki}}{\varepsilon_{nm}}
\right)
+
\partial_k
\left(
\frac{\mathcal{G}^{nm}_{ij}}{\varepsilon_{nm}}
\right)
\right]. 
\label{eq:sigma-GCT-final-summary-eval}
\end{align}
Therefore the full second-order dc conductivity is
\begin{align}
\sigma^{\rm DC}_{ijk}
&=
\sigma^{\mathrm{ND}}_{ijk}
+
\sigma^{\mathrm{BCD}}_{ijk}
+
\sigma^{\mathcal{G}}_{ijk}
+
\sigma^{\mathcal{C}}_{ijk}
+
\sigma^{\mathcal{T}}_{ijk}
\nonumber\\
&=
-\frac{e^{3}\tau^{2}}{\hbar^{3}}
\sum_{\bm{k}}\sum_{n}
f_n\,
\partial_i\partial_j\partial_k\varepsilon_n
-\frac{e^3}{24\hbar}
\sum_{\bm{k}}\sum_n
f''_n\,
\partial_i\partial_j\partial_k\varepsilon_n 
\nonumber\\
&\quad
+
\frac{e^{3}\tau}{2\hbar^{2}}
\sum_{\bm{k}}\sum_{n,l}
f_n
\left[
\varepsilon_{ijl}\mathcal{D}^{n}_{kl}
+
\varepsilon_{ikl}\mathcal{D}^{n}_{jl}
\right]
\nonumber\\
&\quad
+
\frac{e^{3}}{2\hbar}
\sum_{\bm{k}}{\sum_{n,m}}^{\prime}
f'_n
\partial_i\mathcal{G}^{nm}_{jk}
\nonumber\\
&\quad
+
\frac{e^{3}}{\hbar}
\sum_{\bm{k}}{\sum_{n,m}}^{\prime}
f_n
\left[
-2\partial_i
\left(
\frac{\mathcal{G}^{nm}_{jk}}{\varepsilon_{nm}}
\right)
+
\partial_j
\left(
\frac{\mathcal{G}^{nm}_{ki}}{\varepsilon_{nm}}
\right)
+
\partial_k
\left(
\frac{\mathcal{G}^{nm}_{ij}}{\varepsilon_{nm}}
\right)
\right].
\label{eq:sigma-DC-final-eval}
\end{align}
In this final expression the explicit three-connection sector
does not appear separately, because it cancels exactly against
the commutator part of the connection-geometric contribution.

\section{Comments and consistency checks}
\label{sec:comments-checks}

Several points are worth emphasizing when using the final formula.
\begin{enumerate}
  \item The velocity-gauge derivation must retain the contact vertices
  $V_{ij}$ and $V_{ijk}$.  They cancel the apparent
  $1/(\omega_{1}\omega_{2})$ singularities generated when the vector
  potentials are converted to electric fields.

  \item The leading $\tau^{2}$ contribution in
  Eq.~\eqref{eq:sigma-ND-tau2-final-summary-eval} is the nonlinear
  Drude term.  It is a single-band contribution and agrees, after
  integration by parts over the Brillouin zone, with the
  constant-relaxation-time Boltzmann result.  The same diagonal
  sector also contains the kinetic $\tau^{0}$ contribution in
  Eq.~\eqref{eq:sigma-ND-tau0-final-summary-eval}, which depends
  only on the band dispersion.  The full $\tau^{0}$ response is
  the sum of this kinetic term and the intraband- and interband-QMD
  terms.

  \item The $\tau^{1}$ contribution in
  Eq.~\eqref{eq:sigma-Omega-final-summary-eval} is controlled by the
  Berry-curvature dipole.  In the convention of
  Eq.~\eqref{eq:sigma-DC-final-eval}, it is symmetric under
  $j\leftrightarrow k$.

  \item The closed combinations $\mathcal{C}^{nm}_{ij|k}$ and
  $\mathcal{T}^{nml}_{ijk}$ are individually invariant under
  bandwise $U(1)$ phase transformations.  Combining their response
  sectors cancels the explicit three-connection structures and reduces
  the geometric $\tau^{0}$ response to metric derivatives.
  The cancellation in Eq.~\eqref{eq:sigma-T-cancel-eval} is an
  essential internal check of this reduction, distinct from the
  electromagnetic gauge consistency enforced by the contact vertices.

  \item The intra-QMD contribution in
  Eq.~\eqref{eq:sigma-intra-QMD-final-summary-eval} is a Fermi-surface term
  involving the dipole of the ordinary band quantum metric
  $\mathcal{G}_{ij}^{n}$.  The real two-band model presented in the main
  text provides an explicit example in which this contribution remains
  finite even though the Berry curvature, and hence the BCD response,
  vanishes identically.  This isolates the intra-QMD channel as
  a numerical diagnostic; attributing a measured total $\tau^{0}$
  response to that channel also requires separating the kinetic and
  interband-QMD terms.

  \item The formulas are written for Peierls velocity-gauge coupling
  with a constant phenomenological relaxation time.  Additional dipole
  matrix elements, vertex corrections, and microscopic disorder
  corrections must be included separately in models for which they are
  important.
\end{enumerate}